\documentclass[12pt]{article}

\usepackage{graphicx}
\usepackage{url} 
\usepackage{mathtools}
\usepackage{amsmath,amssymb}
\usepackage[separate-uncertainty=true]{siunitx}
\usepackage{multirow}
\usepackage[percent]{overpic}
\usepackage{microtype}
\usepackage{hyperref}
\usepackage[dvipsnames,table,xcdraw]{xcolor}
\usepackage{makecell}
\definecolor{linkcolor}{rgb}{0.5,0.1,0.1}
\definecolor{urlcolor} {rgb}{0.1,0.1,0.5}
\definecolor{citecolor}{rgb}{0.1,0.5,0.1}
\hypersetup{colorlinks=true,linkcolor=linkcolor,urlcolor=urlcolor,citecolor=citecolor,pdfborder={0 0 0}}
\usepackage{cite}
\usepackage{geometry}
\usepackage[makeroom]{cancel}

\usepackage[usestackEOL]{stackengine}
\newcommand\scaledinset[6]{%
  \setbox0=\hbox{#6}%
  \stackinset{#1}{#2\wd0}{#3}{#4\ht0}{#5}{#6}%
}

\graphicspath{{figs/}}
\newcommand{\inl}[1]{\,{#1}\,}
\newcommand{\eq}{\inl{=}}

\newcommand{\delete}[1]{}
\newcommand{\add}[1]{#1}
\newcommand{\replace}[2]{#2}
\renewcommand{\cancel}[1]{}

\title{Algorithmic Design of Disordered Networks with Arbitrary Coordination: Application to Biophotonics}

\author{
Florin Hemmann\thanks{\href{mailto:florin.hemmann@unifr.ch}{florin.hemmann@unifr.ch}} $^{1,2}$,
Vincent Glauser$^{1,2}$,
Ullrich Steiner$^{1,2}$,
Matthias Saba\thanks{\href{mailto:matthias.saba@unifr.ch}{matthias.saba@unifr.ch}} $^{1,2}$
}

\date{
\small
$^{1}$Adolphe Merkle Institute, University of Fribourg, Chemin des Verdiers 4, 1700 Fribourg, Switzerland\\
$^{2}$NCCR Bio-inspired Materials, University of Fribourg, Chemin des Verdiers 4, 1700 Fribourg, Switzerland
}

\begin{document}

\maketitle

\begin{abstract}
Disordered spatial networks describe structures and interactions across multiple length scales. The scattering and interference of waves within these networks result in structural phase transitions, localization, diffusion, and band gaps. Studying these phenomena requires efficient numerical methods for generating disordered networks with specific structural properties. The Wooten-Weaire-Winer algorithm is an established method that introduces disorder into an initial network through a series of bond switch moves. However, the strain energies that govern this evolution are conventionally limited to three-dimensional networks with coordination numbers of no more than four.
We here introduce a maximum bond repulsion to produce networks with an arbitrary coordination number. We control the degree and type of disorder by adjusting the bond-bending force constant in the strain energy and the temperature profile. The effects of these variables are quantified through a list of order metrics that capture both direct and reciprocal space. A feedforward neural network predicts the structural characteristics from the algorithm inputs, enabling efficient targeted network generation. As a case study, we statistically reproduce four disordered biophotonic networks that exhibit structural color. This work presents a versatile method for generating disordered networks with tailored structural properties. It will provide new insights into structure-property relations.
\end{abstract}

\section{Introduction}

The analysis of disordered spatial networks is an active field of research in the natural and social sciences. 
Studied systems span a wide range of lengths and timescales and include atomic \cite{medvedeva_2017}, polymeric \cite{gu2020}, photonic \cite{vynck2023}, brain \cite{lynn2019}, geological \cite{wang2023}, geographical \cite{anderson2020}, social \cite{akbari2021}, and economic systems \cite{gan2021}. Investigating their topology, spatial correlations, and temporal correlations is crucial for understanding fundamental phenomena such as phase transitions, localization, and transport. Examples of these phenomena at the microscopic level include the glass transition in molecular liquids and colloidal systems \cite{jackle1986, berthier2023}, ionic transport in amorphous solid-state batteries \cite{zhang2023, zhu2024}, and the localization of light and photonic band gap materials \cite{edagawa2008, scheffold2022}.

Due to their statistical nature and complexity, numerical methods are essential for generating and analyzing disordered networks. Among these methods, Monte Carlo algorithms play a crucial role due to their inherently random nature, versatility, and computational efficiency \cite{newman1999}. 
The Wooten-Weaire-Winer (WWW) algorithm is a well-established Monte Carlo method for computer-generating continuous random networks -- that is, disordered networks with a fixed coordination number \cite{wooten1985, zachariasen1932}. The WWW algorithm introduces disorder into an initial network by switching bond chains, relaxing introduced local strain by translating vertices, and accepting the state with the Metropolis acceptance probability, thereby lifting the system out of metastable states.

The WWW algorithm was introduced to model the structural correlations of amorphous semiconductors \cite{wooten1985}. Following the original study, efforts were made to accelerate network generation by parallelizing the process, optimizing the annealing procedure, and rejecting unpromising bond switches early on \cite{omard1993,barkema2000}. These advances have enabled the analysis of the melting transition of 2D continuous random networks \cite{tu1998}, the crystallization of an initially amorphous silicon model \cite{nakhmanson2002}, and the correct sampling of the Boltzmann distribution at finite temperatures \cite{vink2014}.

The temperature profile affects network evolution via the Metropolis acceptance probability and is tailored to the networks under study. Previous work has focused on generating networks with minimal local strain to reproduce the experimental diffraction data of amorphous semiconductors \cite{wooten1985, omard1993, djordjevic1995, barkema2000, vink2001, vink2003}. The temperature profile typically includes alternating periods of heating to a high, constant temperature ($T$) and quenching to $ T\eq 0$, followed by annealing with gradually decreasing $T$. The resulting low-strain networks exhibit full electronic and photonic band gaps \cite{barkema2000, edagawa2008}. A second class of studies has investigated the phase transition between the crystalline and amorphous states by observing a phase interface at a constant temperature or by gradually increasing the temperature \cite{tu1998, tu2000, nakhmanson2002, cabriolu2009, vink2014}.

The WWW algorithm includes additional degrees of freedom, such as the initial network and the functional form of the strain energy. These are linked through the coordination number statistic of the network. Early studies on this method involved randomizing four-valent 3D crystalline cubic diamond networks through a series of bond switches at high temperatures \cite{wooten1985, omard1993, djordjevic1995}. Later studies bypassed this randomization by starting with a random, homogeneously four-valent 3D configuration that remained within the framework of continuous random networks \cite{barkema2000, vink2001, nakhmanson2002}. Conventional choices for the strain energy include the Keating energy \cite{keating1966} and the Stillinger-Weber potential \cite{stillinger1985}. Both contain two-body bond-stretching terms with equilibrium lengths and three-body bond-bending terms with equilibrium angles. 

In three dimensions, networks with valencies of up to four are exceptional cases, in which all bonds in the ground state possess the same equilibrium angle. The same applies to 2D networks with coordination numbers up to three. However, for higher or mixed coordination numbers, strain potentials with fixed equilibrium angles generally do not favor bonds that uniformly cover the sphere. Tu et al.\ \cite{tu2000} modeled 3D silica with mixed valencies of 2 and 4 by setting the equilibrium angles to $\SI{180}{\degree}$ for two-valent vertices and $\SI{109.5}{\degree}$ for four-valent vertices. Bayley et al.\ \cite{bailey2020} extended this approach to arbitrary coordination numbers in 2D biological networks by setting the vertex-dependent equilibrium angle to $\SI{360}{\degree}/Z_i$, where $Z_i$ is the valency of vertex $i$. In three dimensions, however, this method generally does not yield isotropic bond arrangements and is therefore not favorable. Thus, this limitation restricts the WWW algorithm to networks with coordination numbers of four or less.

Here, we \replace{overcome the limitations of the strain potentials used previously}{extend the WWW algorithm to 3D networks with arbitrary coordination statistics, including valencies above four} by setting the equilibrium angle to $\SI{180}{\degree}$ for all vertices in the Keating energy. This yields angle-dependent repulsion. Since this angle can only be achieved by two-valent vertices, the bond angle energy generally increases compared to the equilibrium angles used previously. To compensate, we adjust the ratio of bond stretching to bond-bending energies in the Keating energy by modifying the bond-bending force constant $\beta$. 

We analyzed the effect of $\beta$ under the new equilibrium angle by evolving six different crystalline networks with valencies up to eight, \add{which are only possible through our modification of the Keating energy}. We selected the diamond \cite{wooten1985} and gyroid \cite{sellers2017} networks because the WWW algorithm has previously been applied to them, albeit with different equilibrium angles. The other four networks are less well-known and have never been investigated using WWW evolution.
We adjust the level of disorder in the melted networks by altering their temperature profiles during the transition from the crystalline state. We use a triangular heating and cooling sequence with adjustable temperature gradient and maximum temperature, followed by a quench. 
\add{In previous work, different temperature profiles were used to generate low-strain networks
\cite{wooten1985, omard1993, djordjevic1995, barkema2000, vink2001, vink2003} or to study the phase transition between the crystalline and disordered state \cite{tu1998, tu2000, nakhmanson2002, cabriolu2009, vink2014}. For the first time, we use the temperature profile of WWW evolution as a tuning knob for the degree of disorder.}

We compiled a list of 42 order metrics and measured the effects of the initial network, the bond-bending force constant $\beta$, and the temperature profile parameters on the resulting networks. Specifically, we measure network primitive similarity by comparing bonds and angles, quantify homogeneity and isotropy in direct and reciprocal spaces, and characterize the network topology by determining coordination numbers and ring statistics. \add{This combination of metrics provides a novel and thorough statistical classification of disordered networks. While different initial networks, temperature profiles, and values of $\beta$ have been explored in previous studies \cite{tu1998, tu2000, nakhmanson2002, vink2003, alfthan2003, hudson2007, cabriolu2009, sellers2017}, we systematically investigate their combined effects on the resulting networks for the first time. Furthermore}, we trained a feedforward neural network to predict the order metric values as a function of the algorithm's input parameters for one of the initial crystalline networks, the three- and four-valent \textbf{ctn} network \cite{rcsr_ctn2025}, \add{enabling targeted network generation}.

We demonstrate the capabilities of our extended WWW algorithm by statistically reproducing disordered biophotonic networks. 
These photonic materials exhibit structural color in nature at low refractive index contrasts and without long-range order \cite{yin2012, sellers2017, rothammer2021, djeghdi2022, vogler-neuling2023, bauernfeind2023, bauernfeind2024}. Computational studies of these networks are essential for elucidating the relationship between their structural characteristics and color.
Using a dataset of thousands of generated networks, we identified networks with the highest similarity to three biological structures, as determined by our list of order metrics. There is good agreement between generated and biological networks, particularly with regard to small-length-scale order metrics, which are directly controlled by the Keating strain energy. \add{Furthermore, we find that all analyzed biophotonic structures are hyperuniform.}

\add{In summary, this paper introduces the following key aspects to the computer-generation of disordered networks, made available through an open-source Julia code \cite{hemmann_github_2026}:
    \begin{description}
        \item[a) Networks with Arbitrary Coordination Number:] Previous studies have all been limited to coordination numbers of $Z\eq3$ or 4 using the locally self-uniform equilibrium bond angle of the associated ordered networks \cite{sellers2017}. We propose using a maximum equilibrium bond angle of \SI{180}{\degree}, which yields networks with an equally uniform bond-angle distribution up to at least $Z\eq12$.
        \item[b) Comprehensive Statistical Enumeration of Disordered Nets:] We present an efficient toolset for characterising and comparing networks with a large number of degrees of freedom, using only 12 quantifiers -- out of 42 investigated order metrics -- with well-defined physical and mathematical meanings. These quantifiers have been grouped into four main categories (Appendix \ref{sec:list_of_symbols}): Network primitives, homogeneity, isotropy, and topology.
        \item[c) Neural Network Prediction:] We have trained a neural network that can determine the type and strength of network disorder -- characterised by the 12 parameters discussed above -- based on WWW input parameters, such as the temperature profile and bond-angle weight $\beta$. This significantly speeds up the process of identifying experimental target networks.
        \item[d) Hyperuniformity in Biophotonic Networks:] Hyperuniformity is well known to be associated with the formation of photonic band gaps. Although the index contrast in biophotonics is insufficient to form complete band gaps, hyperuniformity is believed to play a pivotal role in creating structural colour in biological photonic crystals \cite{rothammer2021}. Our analysis shows that hyperuniformity is present in all four disordered photonic beetle and weevil structures we investigated. This adds weight to the concept of hyperuniformity in biophotonics and encourages further research into its role in generating structural color.
    \end{description}}

The article is organized as follows: Section~\ref{sec:WWW} introduces the well-established WWW algorithm and its generalization to arbitrary coordination numbers and disorder control via the temperature profile. Then, Section~\ref{sec:order_metrics} discusses the various order metrics to characterize the generated networks. Section~\ref{sec:targeted} presents a case study outlining the computer-generated networks resembling biophotonic structures. After introducing the biological and initial networks for the extended WWW algorithm in Section~\ref{sec:biophotonic}, we examine how algorithm inputs affect generated networks in Section~\ref{sec:effect_www_parameters}. Section~\ref{sec:neural_network} uses a neural network to predict the structural characteristics from algorithm inputs and evaluates its accuracy to investigate the statistical variance of the WWW algorithm. Section~\ref{sec:reproducing_biophotonic} identifies networks from the database that resemble biological structures and compares their order metrics. 
\section{Disordered network generation}
\label{sec:WWW}
The Wooten-Weaire-Winer (WWW) algorithm is a well-established Monte Carlo method for generating \textit{continuous random networks}. These networks are disordered with \textit{fixed valency}, meaning all vertices have the same coordination number $Z$ \cite{wooten1985}.
First, we will discuss the original method used to produce disordered networks resembling amorphous semiconductors. Next, we will introduce our extensions that handle networks with \textit{varying valency}, where multiple coordination numbers are present. Lastly, we will explain how adjusting WWW algorithm parameters can alter the degree of disorder.

\subsection{Established Wooten-Weaire-Winer algorithm}
\label{sec:established_www}
This algorithm introduces disorder into an initial configuration by performing a series of bond switches that leave the coordination numbers invariant but alter the network's overall topology.

The initial network configuration may be crystalline \cite{wooten1985} or amorphous \cite{barkema2000}. The WWW algorithm is typically applied to networks with $10^3$ to $10^4$ vertices and periodic boundary conditions, which are used throughout this article. Most of the literature focuses on networks with valency $Z=4$. The initial network is therefore either the cubic diamond network \textbf{dia} \cite{wooten1985}\footnote{For all network structures, we use the nomenclature of the Reticular Chemistry Structure Resource \cite{rcsr2025}} or a disordered network with fixed $Z=4$ \add{\cite{barkema2000}}. We define a strain energy for this initial network. A common choice is the Keating energy for networks with $Z=4$ \cite{keating1966},
\begin{align}
    \mathcal{E} &=
    \frac{3}{16} 
    \frac{A}{d^2}
    \Sigma_{\langle ij \rangle} 
    \left( 
    \mathbf{R}_{ij} \cdot \mathbf{R}_{ij}-d^2 
    \right)^2 
    +
    \frac{3}{8} 
    \frac{B}{d^2}
    \Sigma_{\langle jik \rangle} 
    \left(
    \mathbf{R}_{ij} \cdot \mathbf{R}_{ik}+\frac{1}{3}d^2
    \right)^2\quad.
    \label{eqn:keating}
\end{align}
Here, $d$ is the strain-free equilibrium bond length, $i$, $j$, and $k$ are labels for the vertices. The bond connecting vertices $i$ and $j$ is $\mathbf{R}_{ij}=\mathbf{R}_j-\mathbf{R}_i$. Summation over all neighboring vertices $i$ and $j$ is denoted by $\sum_{\langle ij \rangle}$, and summation over all connected vertex chains $j-i-k$ is denoted by $\sum_{\langle jik \rangle}$. Furthermore, $A$ and $B$ are the force constants for stretching and bending bonds, respectively. 

We define the corresponding dimensionless Keating energy, length, and force weight, 
\begin{equation*}
    E=\frac{\mathcal{E}}{A d^2} 
    \quad , \quad \mathbf{r}_{ij}=\frac{\mathbf{R}_{ij}}{d} 
    \quad , \quad \beta= \frac{B}{A} \quad .
\end{equation*}
Thus, the non-dimensionalized form of \eqref{eqn:keating} becomes
\begin{align}
    E &=
    \frac{3}{16} 
    \Sigma_{\langle ij \rangle} 
    \left( 
    r_{ij}^2-1 
    \right)^2 
    +
    \frac{3}{8} 
    \beta
    \Sigma_{\langle jik \rangle}  
    \left( \mathbf{r}_{ij} \cdot \mathbf{r}_{ik}+\frac{1}{3}
    \right)^2
    =
    E_r
    + E_\theta^\mathrm{orig}
    \quad .
    \label{eqn:keating_dimensionless}
\end{align}
\add{The bond-stretching energy $E_r$ describes two-body interactions between coordinated vertices, while the bond-bending energy $E_\theta^\mathrm{orig}$ represents three-body interactions. For a network with fixed valency $Z\eq4$, the given Keating energy (Equations \eqref{eqn:keating}~and~\eqref{eqn:keating_dimensionless}) is the simplest strain energy that includes only two- and three-body interactions and is invariant under rotations and translations of the rigid network \cite{keating1966}.}

\begin{figure*}
\begin{minipage}{0.32\textwidth}
\centering
\scaledinset{l}{-0.08}{b}{0.9}{\textbf{a}}{\includegraphics[width=0.8\textwidth]{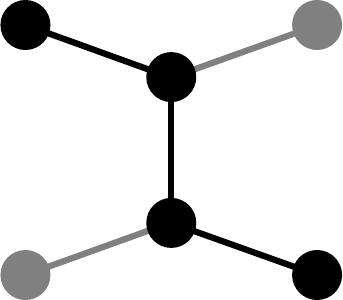}}
\end{minipage}
%
\begin{minipage}{0.32\textwidth}
\centering
\scaledinset{l}{-0.08}{b}{0.9}{\textbf{b}}{\includegraphics[width=0.8\textwidth]{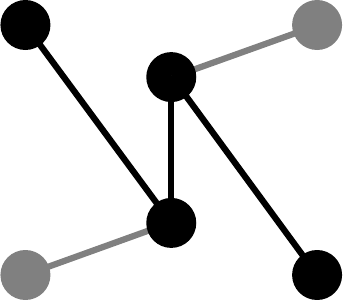}}
\end{minipage}
%
\begin{minipage}{0.32\textwidth}
\centering
\scaledinset{l}{-0.08}{b}{0.9}{\textbf{c}}{\includegraphics[width=0.8\textwidth]{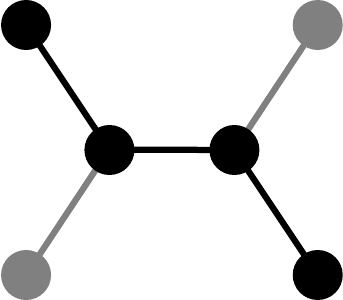}}
\end{minipage}
\caption{\label{fig:bond_switch}
A Monte Carlo move is shown for a 2D, three-valent network.
\textbf{a} A chain of four vertices (black) is randomly selected.
\textbf{b} A bond switch reconnects the vertices while maintaining all coordination numbers.
\textbf{c} The stress introduced by the bond switch is relaxed by translating the vertices.
}
\end{figure*}

The original network configuration has a total strain energy $E_\mathrm{i}$. This configuration is modified by randomly reconnecting a chain of bonds, as illustrated in Figure~\ref{fig:bond_switch}.
This \textit{bond switch} preserves the coordination numbers. Thus, the initial configuration defines the coordination number statistics and correlations. However, the bond switch introduces local strain into the network. This strain is relaxed iteratively by successively moving the vertices, starting from the switched chain and moving outward to higher neighbor shells \cite{omard1993}. Each vertex is translated to a position of minimal stress, which is approximated using the strain energy gradient as described in ref.\ \cite{weaire1987}. The local relaxation of all vertices marks the completion of one \textit{relaxation cycle}. This cycle is repeated approximately 25 times to reach \textit{relaxation} or convergence toward the true equilibrium positions of all vertices. The new network configuration, which has a total strain energy of $E_\mathrm{f}$, is accepted using the Metropolis acceptance probability \cite{metropolis1953}
\begin{equation}
    P_\mathrm{accept}=\min \left[1, \exp \left( \frac{\mathcal{E}_\mathrm{f}-\mathcal{E}_\mathrm{i}}{k_\mathrm{B}\mathcal{T}} \right) \right]
    = \min \left[1, \exp \left( \frac{E_\mathrm{f}-E_\mathrm{i}}{T} \right) \right]
    \label{eqn:metropolis}
    \quad.
\end{equation}
Here, $\mathcal{T}$ is the physical temperature, and $T\eq\frac{k_\mathrm{B}\mathcal{T}}{A d^2}$ is the dimensionless temperature. A \textit{Monte Carlo move} involves random bond switching, relaxation, and acceptance or rejection. \add{Hudson et al. \cite{hudson2007} demonstrated that the WWW Monte Carlo move exceeds other topological modifications in relaxation efficiency and exploration of the configuration space.}

We use one \textit{Monte Carlo step} as the unit of simulation time. A step is finished when an attempt to switch a bond is made on average on every available bond chain. The number of available bond chains is estimated based on the number of vertices $N_\mathrm{vertices}$ and the mean coordination number $\overline{Z}$. There are $N_\mathrm{vertices}$ choices for the first vertex of the chain, $\overline{Z}$ for the second vertex connected to the first, and $\overline{Z}-1$ choices for the third and fourth vertices, respectively. Since the order of the chain is irrelevant, we divide by two to obtain the number of chains \cite{barkema2000}
\begin{align*}
    N_\mathrm{chains}=N_\mathrm{vertices} \cdot \overline{Z} \cdot (\overline{Z}-1)^2/2 \quad .
\end{align*}

In~ Ref.\ \cite{barkema2000}, the WWW algorithm is accelerated by using a local/nonlocal relaxation procedure. Before performing a relaxation, a threshold energy
\begin{equation*}
    E_\mathrm{t}= E_\mathrm{i} - T \ln(s) 
\end{equation*}
with random $s\in (0,1]$ is set. After the bond switch, only the three innermost neighbor shells are relaxed for the first ten relaxation cycles. If it becomes clear during local relaxation that the final energy $E_\mathrm{f}$ will not fall below the threshold energy $E_\mathrm{t}$, the bond switch is rejected early. To estimate $E_\mathrm{f}$ during local relaxation, we use the fact that the energy is harmonic around the minimum and can be expressed as a function of force $F$ and a proportionality constant $c_\mathrm{f}$:
\begin{equation*}
    E_\mathrm{f} \approx E - c_\mathrm{f} |F|^2 \quad . 
\end{equation*}
Fitting this equation to the total energy and strain force of the last two relaxation cycles yields $E_\mathrm{f}$ and $c_\mathrm{f}$. If $E_\mathrm{f} > E_\mathrm{t}$ during local relaxation, the bond switch is rejected. To account for anharmonicities, no moves are rejected during the first five relaxation cycles. After ten cycles, relaxation continues globally without early rejections, as in the original WWW algorithm.

A nearly fully local relaxation scheme is introduced in Ref.\ \cite{vink2001}, wherein \add{the network is just relaxed up to the fourth neighbor shell \footnote{Here, the vertices of the switched bond chain are counted as the first shell.}.} Global relaxation only occurs if $E_\mathrm{f}$ is slightly above $E_\mathrm{t}$. This allows for the formation of large networks with up to 20,000 vertices. While global relaxation can reduce stress in a locally relaxed network, it introduces dependence on system size in the change of total strain energy. Therefore, network evolution depends on both temperature and the number of vertices. Since our goal is to achieve evolution independent of system size, we use an entirely local relaxation scheme \add{up to the fourth neighbor shell}, which results in slightly higher strain.

\subsection{Extension to networks with arbitrary valency}
We modify the Keating energy~\cite{keating1966} to generate networks with any desired coordination number statistic. This is necessary when modeling arbitrary 3D disordered networks.
\add{Due to a limitation in previously used strain energies,} to our knowledge, the WWW algorithm has only been used for 3D systems with coordination numbers of four or fewer \cite{wooten1985, djordjevic1995, tu2000, barkema2000, vink2001, vink2003, sellers2017}. Reported networks with higher coordination numbers are two-dimensional only \cite{bailey2020}. Furthermore, when evolving networks with varying valency, the bond-bending energy in Equation \eqref{eqn:keating_dimensionless} depends on coordination \cite{tu2000, vink2003, alfthan2003, bailey2020}.

\replace{To relax}{We enable the relaxation of} a network with arbitrary coordination numbers, \add{by modifying} the bond-bending term $E_\theta^\mathrm{orig}$ \delete{must be modified} in the Keating energy in Equation \eqref{eqn:keating_dimensionless}.  Since the equilibrium angle is typically set to a value corresponding to the coordination number, this term requires adjustment. In the above example, the equilibrium angle for diamond-like networks with $Z\,{=}\,4$ is $\theta_\mathrm{eqm}\eq\arccos(-1/3)\inl\approx \SI{109.5}{\degree}$, resulting in
\begin{align*}
    &E_\theta^\mathrm{orig} \sim \left(\mathbf{r}_{ij} \cdot \mathbf{r}_{ik}+\frac{1}{3}\right)^2  
    = \left(r_{ij}r_{ik} \cos(\theta_{jik}) +\frac{1}{3}\right)^2 \overset{!}{=} 0 \\
    \Rightarrow \quad &r_{ij}=r_{ik}=d \quad 
    \mathrm{and} 
    \quad \theta_{jik}=\theta_\mathrm{eqm}=\arccos \left(-\frac{1}{3} \right) \quad .
\end{align*}
The equilibrium angle is the angle between all bonds in the 4-valent cubic diamond network \textbf{dia}, with $\theta_\textbf{dia}=\theta_\mathrm{eqm}=\arccos \left(-\frac{1}{3} \right) $ \cite{GrossMarx2012}. Another crystallographic network in which all bonds span the same angle is the 3-valent gyroid network \textbf{srs}, with $\theta_\mathrm{eqm}=\theta_\textbf{srs}=\SI{120}{\degree}$. Therefore, analogously to \textbf{dia}, the \textbf{srs} is the ground state of the Keating energy with appropriately chosen equilibrium angle.
However, this strategy fails for networks with mixed coordination numbers, where each vertex requires an individual equilibrium angle \cite{tu2000, alfthan2003, bailey2020}.
There is no configuration with a single bond angle for vertices with valency $Z>4$, which are present in many other networks \cite{medvedeva_2017, gu2020, lynn2019, wang2023, anderson2020, akbari2021, gan2021, bailey2020}.
Consequently, there is no value of $\theta_\mathrm{eqm}$ that yields a vanishing ground state bond-bending energy $E_\theta^\mathrm{orig}=0$.

To favor bonds that uniformly cover the sphere for any coordination number, we propose bonds that energetically repel each other. Here, \textit{sphere} refers to the idealized surface surrounding the vertex, on which the bond directions are arranged as evenly as possible.
This is achieved by setting the equilibrium angle to $\theta_\mathrm{eqm}\eq\SI{180}{\degree}$, regardless of the coordination number.
Thus, the lowest bending energy at a given vertex is positive.
For a small coordination number ($Z\eq3$ and $4$), the energy is minimized by the well-known \textbf{srs} (equilateral triangle) and \textbf{dia} (regular tetrahedron) bonds.
For larger $Z\inl<10$, the energy is minimized by the solutions to Feje's problem \cite{whyte1952}, which maximizes the distance between points on a sphere.
The energy is also minimized by connecting the center to the vertices of Platonic solids.
\footnote{However, for $Z\inl>5$, a higher-dimensional manifold of minimum energy can be found in the configuration space. For example, when $Z\eq6$, the regular octahedron can transform into a triangular antiprism with an arbitrary torsion angle in the $[111]$ direction while remaining within the global energy minimum. When $Z\eq8$, the cube (a Platonic solid corresponding to a \textbf{bcu} network \cite{rcsr2025}, or an IWP minimal surface structure \cite{bauernfeind2023}) and the quadratic antiprism (Feje's solution) lie on a similar torsion-degenerate line of minimum energy in configuration space.}
Thus, we propose the following generalized Keating energy, which favors uniformly covering the unit sphere with bond angles for arbitrary valency,
\begin{equation}
    E=
    \frac{3}{16} 
    \Sigma_{\langle ij \rangle} 
    \left( 
    r_{ij}^2-1 
    \right)^2 
    +
    \frac{3}{8} 
    \beta
    \Sigma_{\langle jik \rangle}  
    \left( r_{ij} r_{ik} \cos(\theta_{jik}) + 1 
    \right)^2
    = E_r + E_\theta
    \quad .
    \label{eqn:keating_generalized}
\end{equation}

\add{While the new proposed equilibrium angle might appear like a minor modification, it requires careful adjustments in the energy and network evolution parameters.} For vertices with a valency $Z\inl>2$, there is no bond arrangement in which all angles are $\theta_\mathrm{eqm}\eq180^{\circ}$. Therefore, $E_\theta\inl>E_\theta^\mathrm{orig}$ increases monotonically with $Z$. 
One method to counteract the increase in angle energy is to vary the bond-bending force constant $\beta$, which sets the ratio of \replace{bond bending to bond-stretching energy}{$E_\theta$ to $E_r$}. In this work, we use a global $\beta$ to control the relative degrees of disorder in bond lengths and angles. For polyvalent networks, one could introduce a valency-dependent $\beta$ to accommodate the $Z$-dependence of energy and expand the range of accessible disordered network morphologies.

\subsection{Heating profile and melting temperature}
\label{sec:temperature_profile}
\begin{figure}
\centering
\includegraphics[width=0.4 \linewidth]{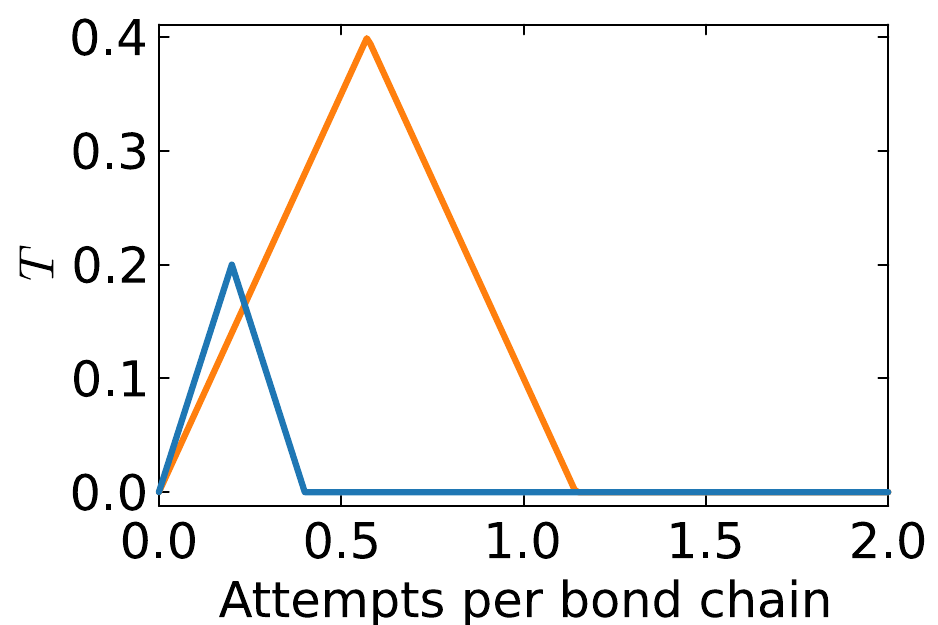}
\caption{\label{fig:temperature_profile} In our extended WWW algorithm, we introduce disorder to networks by successively heating, cooling, and quenching them. The quench continues until it is no longer possible to reduce the strain energy through additional Monte Carlo moves.}
\end{figure}

In addition to the bond-bending constant $\beta$, \add{we propose to tune} the degree of disorder in the generated networks \delete{can be adjusted} by altering the temperature profile during network evolution. 
We found that the triangular profile shown in Figure~\ref{fig:temperature_profile} provides reasonable control over the structural properties of the resulting networks.
This profile consists of constant heating with gradient $\Delta T$ up to maximum temperature $T_\mathrm{max}$, followed by constant cooling with gradient $-\Delta T$ down to $T\eq0$, concluding with a quench. Evolution occurs at $T\eq0$ until no single bond switch can decrease the total strain energy. This temperature profile has two degrees of freedom: $\Delta T$ and $T_\mathrm{max}$\replace{. These can be used}{which we use} to adjust the degree of disorder introduced into the initial network.

When a low-strain initial network is heated, bond switches are only likely to be accepted once the \textit{melting temperature} $T_\text{melt}$ is reached. We define $T_\text{melt}$ as the temperature at which the energetically lowest bond switch and relaxation are accepted with probability $P_\mathrm{accept}\inl>P_\mathrm{melt}\inl{:=} \SI{0.1}{\percent}$. This causes the initial periodic configuration to begin melting. Isolating $T$ in the Metropolis acceptance probability in Equation \eqref{eqn:metropolis} yields the melting temperature
\begin{equation}
    T_\text{melt}= \frac{E_\mathrm{f}-E_\mathrm{i}}{\ln(P_\mathrm{melt})} 
    \quad.
    \label{eqn:t_melt}
\end{equation}
Due to the Boltzmann distribution in $P_\mathrm{accept}$ \add{(Equation \eqref{eqn:metropolis})}, Monte Carlo moves can be accepted at any finite temperature $T\inl>0$, including temperatures below $T_\text{melt}$. However, $T_\text{melt}$ serves as an estimate when melting is likely to occur and as a reference for rescaling temperatures when changing the bond-bending force constant $\beta$ or the equilibrium angle $\theta_\mathrm{eqm}$.

Section~\ref{sec:targeted} discusses how variations of $\beta$ and the temperature profiles affect the degree of disorder in the generated networks. The next section introduces the metrics used to quantify structural disorder.

\section{Order metrics}
\label{sec:order_metrics}
\replace{Network structure disorder can be quantified in various ways.}{In this section, we compile a list of order metrics that enable a thorough characterization of structural disorder in networks. We complement existing measures with novel ones to quantify various facets of disorder.} These aspects fall into four main categories, which are discussed in detail below: Similarity of primitives, homogeneity, isotropy, and topological classification.

\subsection{Similarity of network primitives}
\label{sec:similarity_primitives}
A variety of short-range order metrics are suitable for measuring the similarity of the network primitives consisting of multiple bonds that meet at a vertex. We calculate the standard deviations of the bond length and bond angle, $\sigma_r$ and $\sigma_\theta$, respectively. Additionally, we determine the mean value $\overline{q}_{l}$ and standard deviation $\sigma_{q_l}$ of the Steinhardt local bond order parameters $q_{l}$ for all vertices \cite{steinhardt1983}. We compute the parameters to order $l \inl\leq 12$ because higher orders do not provide meaningful information for coordination numbers $Z \inl\lesssim 10$.

Additionally, we determine the order of neighboring primitive orientations by computing the distribution of dihedral angles. For all chains of four connected vertices $j-i-k-l$, we consider the dihedral angles between the planes spanned by $\mathbf{r}_{ij}$ and $\mathbf{r}_{jk}$, and by $\mathbf{r}_{jk}$ and $\mathbf{r}_{kl}$ \cite{blondel1996},
\begin{align*}
    \phi_{m} = \operatorname{arctan2}\big( \mathbf{r}_{jk} \cdot \big( (\mathbf{r}_{ij} \times \mathbf{r}_{jk}) \times (\mathbf{r}_{jk} \times \mathbf{r}_{kl}) \big), \; |\mathbf{r}_{jk}| \big( (\mathbf{r}_{ij} \times \mathbf{r}_{jk}) \cdot (\mathbf{r}_{jk} \times \mathbf{r}_{kl}) \big) \big) \quad.
\end{align*}
The index $m$ labels all four-vertex chains, and $\operatorname{arctan2}$ is the two-argument arctangent function. Since the dihedral angle distribution usually has multiple peaks, we create an 18-bin ($N_\mathrm{bins}\eq 18$) histogram $p(\phi_m)$, with $10^{\circ}$ bin widths and centers $\phi_m$.  We then determine the normalized information entropy of the resulting distribution, called \textit{dihedral angle entropy},\footnote{$h_\phi$ is a simple measure of uniformity in the dihedral angle distribution. The contribution of dihedral angles to the thermodynamic entropy of networks and chains is studied in \cite{cukier2015}.}
\begin{align*}
    h_\phi = -\frac{1}{\log(N_\mathrm{bins})} \sum_{i=1}^{N_\mathrm{bins}} p(\phi_m) \log(p(\phi_m)) \quad.
\end{align*}

\subsection{Homogeneity}
\label{sec:homogeneity}
To extend the order quantification beyond individual primitives, we analyze the statistical homogeneity of the network by examining the interparticle distances. For statistically isotropic and homogeneous materials, the dimensionless pair correlation function $g_2(r)$ represents the mean particle density at distance $r$ from a reference particle divided by the total mean particle density, $g_2(r) \eq n(r)/n_0$ \cite{GrossMarx2012}. Another commonly used function to describe particle densities is the three-dimensional radial distribution function, defined as the mean number of particles per unit length, $\rho(r)\eq 4\pi r^2 n_0 g_2(r)$. The cumulative coordination number $Z(r) \eq \int_0^r dr' \, \rho(r')$
is the expected number of particles within a sphere of radius $r$ centered at a reference particle \cite{torquato2018}. 
\replace{We define the \textit{coordinated neighbor distance} $r_\mathrm{c}$ as the radius at which $Z(r_\mathrm{c})\eq 1$. $r_\mathrm{c}\eq1$ indicates a homogeneous network whereas $r_\mathrm{c} \inl\lesssim 0.7$ may indicate vertex clustering.}{We define the \textit{nearest-neighbor distance} $r_\mathrm{nn}$ as the smallest radius at which $Z(r)$ exceeds 1,
\begin{align*}
    r_\mathrm{nn} = \inf \{ r \, : \, Z(r) \geq 1 \} \quad.
\end{align*}
Many crystalline networks like the diamond exhibit a uniform bond length throughout the network, giving rise to a discontinuity in the cumulative coordination number from $Z(r\inl<1)\eq 0$ to the mean coordination number $Z(r\eq1)\eq \overline{Z}$. Such networks exhibit $r_\mathrm{nn}\eq 1$ and homogeneous disordered networks show $r_\mathrm{nn} \inl\sim 1$  whereas $r_\mathrm{nn} \inl\lesssim 0.7$ may indicate vertex clustering. As in the dimensionless form of the Keating energy~\eqref{eqn:keating_dimensionless}, here and in the following, lengths are given in units of the strain-free equilibrium bond length $d$.}

Since \replace{$r_\mathrm{c}$}{$r_\mathrm{nn}$} only captures distances between connected vertices, we use $r_\mathrm{u}$, \replace{\textit{uncoordinated neighbor distance}}{\textit{nearest-uncoordinated-neighbor distance}}, as an additional measure of network homogeneity. $r_\mathrm{u}$ is the average distance from vertex $i$ to the closest unconnected vertex $j$. Unlike connected vertices, it is not included in the bond-stretching term of the Keating energy (Equation \eqref{eqn:keating_generalized}). Therefore, we expect less control over $r_\mathrm{u}$ via the tuning knobs of the extended WWW algorithm.
$r_\mathrm{u}$ is obtained by calculating the pair correlation function $g_2^\mathrm{u}(r)$, the radial distribution function $\rho_\mathrm{u}(r)$, and the cumulative coordination number $Z_\mathrm{u}(r)$. This calculation takes into account only vertex pairs that are not connected by a bond. $r_\mathrm{u}$ is defined as the distance at which \replace{$Z_\mathrm{u}(r_\mathrm{u})\eq 1$. }{$Z_\mathrm{u}(r)$ exceeds 1,
\begin{align*}
    r_\mathrm{u} = \inf \{ r \, : \, Z_\mathrm{u}(r) \geq 1 \} \quad.
\end{align*}
}Vertex clustering may occur when $r_\mathrm{u} \inl\lesssim 1$.

Furthermore, the statistical homogeneity of a network can be quantified by calculating the geometrical pore size distribution $P(r)$ \cite{gelb1999}. $P(r)$ represents the typical size scale(s) of the void space, that is, the volume of the network excluding its vertices and infinitely thin bonds. The pore size distribution is linked to the fraction of the void space accessible to a sphere of radius $r$ without bond overlap via the integral $V(r)\eq\int_0^r dr' \, P(r')$.
We use the algorithm developed by Song et al.\ \cite{song2019} to determine $P(r)$ numerically and sample points on a 3D grid with a grid size of $0.2$. For each grid, we determine the radius of the largest sphere that contains the grid point and does not intersect the mathematical network. $P(r)$ is the histogram of these sphere radii. The \textit{critical pore radius} $\delta_\text{c}$ is the maximum radius of a sphere that can percolate through the void space \cite{klatt2021}. It can be accurately estimated using the second moment of the pore size distribution \cite{klatt2021},\footnote{Note that Klatt et al.\ \cite{klatt2021} use the definition of the pore size distribution of Torquato et al.\ \cite{torquato2013}, which differs from Gelb et al.'s \cite{gelb1999} definition of $P(r)$ used in this article. The two definitions are compared in Agrawal et al.\ \cite{agrawal2023}.}
\begin{align*}
    \delta_\text{c} \approx \left( \int_0^\infty dr \, r^2 \, P(r) \right)^{\frac{1}{2}} \quad .
\end{align*}
A critical pore radius $\delta_\text{c} \inl\gtrsim 0.5$ indicates the presence of large pores, which can result from vertex clustering.

Additionally, information from reciprocal space can be used to determine the homogeneity of a structure at large length scales. This concept is related to hyperuniformity \cite{torquato2003, torquato2018}. As before, we use dimensionless units where positions are given in units of $d$, and wave vectors and wave numbers are scaled by $1/d$. A hyperuniform system is characterized by a vanishing structure factor in the limit of infinitely small wave vectors \cite{torquato2003},
\begin{align}
\lim_{\mathbf{k} \rightarrow 0} S (\mathbf{k} ) = 0 \quad .
\label{eqn:hyperuniform_crit}
\end{align}
In direct space, this property translates to a suppression of large-length-scale density fluctuations \cite{torquato2018}. When testing the hyperuniformity criterion of Equation \eqref{eqn:hyperuniform_crit} for finite-size systems, the structure factor must be extrapolated to $\mathbf{k} \rightarrow 0$ since the minimal wave number $k_\mathrm{min}\equiv |\mathbf{k}|_\mathrm{min}$ depends on the system size. For a system with edge length $L$ and periodic boundary conditions, the largest length scale is $L/2$, and the smallest feasible wave number is $k_\mathrm{min}\eq 4\pi / L$.

We calculate two types of structure factors. First, we consider only the point cloud of vertices at positions $\mathbf{x}_i$ and use the scattering intensity estimator \add{to obtain the \textit{vertex structure factor}} \cite{hawat2023}
\begin{align}
    S_\mathrm{v}(\mathbf{k}) = \frac{1}{N_\mathrm{v}} \left|  \sum_{i=1}^{N_\mathrm{v}} \exp \left( -i \mathbf{k} \cdot \mathbf{x}_i \right) \right|^2 \quad .
    \label{eqn:struct_fact_vertices}
\end{align} 

\delete{Second, we consider infinitely thin bonds and obtain} 
\begin{align*} 
\cancel{S_\mathrm{b}(\mathbf{k}) 
= \frac{1}{N_\mathrm{b}} \left| \sum_{i=1}^{N_\mathrm{b}} 2 \operatorname{sinc} \left( \mathbf{k} \cdot \mathbf{r}_i \right) \exp \left( -i \mathbf{k} \cdot \mathbf{x}_i \right) \right|^2 \quad . }
\end{align*}
\delete{Here, $\mathbf{x}_i$ is the position of the center of bond $i$, and $\mathbf{r}_i$ is the vector connecting the endpoints (vertices).
 }
\add{Second, we determine the \textit{bond structure factor} $S_\mathrm{b}$ that represents the mathematical network in reciprocal space. $S_\mathrm{b}$ is obtained by replacing the vertex positions in Equation~\ref{eqn:struct_fact_vertices}, by a parametrization of the bonds
\begin{align} 
\mathbf{x}_i + t\, \frac{\mathbf{r}_i}{r_i} \quad \text{with } t \in \left[-\frac{r_i}{2},\,\frac{r_i}{2}\right] \quad , \label{eqn:struct_fact_bonds_derivation}
\end{align}
where $\mathbf{x}_i$ is the position of the center of bond $i$, and $\mathbf{r}_i$ is the vector connecting the endpoints (vertices). $r_i$ is the length of bond $i$ in units of $d$. By summing over all bonds, we can write the bond structure factor as
\begin{align}
S_\mathrm{b}(\mathbf{k})&= \frac{1}{N_\mathrm{b}}
   \left| \sum_{i=1}^{N_\mathrm{b}}
          \int_{-r_i/2}^{r_i/2} dt\,
          e^{-i\,\mathbf{k}\cdot\left(\mathbf{x}_i + t\, \frac{\mathbf{r}_i}{r_i}\right)} 
   \right|^{2} \nonumber
   \\
&= \frac{1}{N_\mathrm{b}}
   \left| \sum_{i=1}^{N_\mathrm{b}}
          \frac{r_i}{-i\,\mathbf{k}\cdot\mathbf{r}_i}
          \left(
              e^{-i\,\frac{1}{2}\mathbf{k}\cdot\mathbf{r}_i}
            - e^{+i\,\frac{1}{2}\mathbf{k}\cdot\mathbf{r}_i}
          \right)
          \, e^{-i\,\mathbf{k}\cdot\mathbf{x}_i}
   \right|^{2} \nonumber
   \\
&= \frac{1}{N_\mathrm{b}}
   \left| \sum_{i=1}^{N_\mathrm{b}}
          \frac{2 \, r_i}{\mathbf{k}\cdot\mathbf{r}_i}
          \sin\!\left( \frac{\mathbf{k}\cdot\mathbf{r}_i}{2} \right)
          \, e^{-i\,\mathbf{k}\cdot\mathbf{x}_i}
   \right|^{2} \nonumber
   \\
&= \frac{1}{N_\mathrm{b}}
   \left| \sum_{i=1}^{N_\mathrm{b}}
          r_i \,
          \mathrm{sinc}\!\left( \frac{\mathbf{k}\cdot\mathbf{r}_i}{2} \right)
          \, e^{-i\,\mathbf{k}\cdot\mathbf{x}_i}
   \right|^{2}.
   \label{eqn:struct_fact_bonds}
\end{align}}
$S_\mathrm{b}(\mathbf{k})$ provides more information about the network than $S_\mathrm{v}(\mathbf{k})$ but comes at the cost of increased computational complexity.
\add{
$S_\mathrm{b}(\mathbf{k})$ is the natural generalization of the structure factor to networks with one‑dimensional edges.
Note that the normalization $1/N_\mathrm{b}$ in Equation~\eqref{eqn:struct_fact_bonds} is, however, somewhat arbitrary, and one could equally well normalize by the sum over the bond lengths.
However, none of the metrics that are calculated from the bond structure factor depend on this coefficient.
$S_\mathrm{b}(\mathbf{k})$ may alternatively be interpreted as the spectral density of a two‑phase medium \cite{torquato2018} in which the edges of the network are replaced by cylindrical rods of radius $a$, after factoring out the single‑rod form factor and taking the limit $a\inl\rightarrow 0$.}

To investigate the hyperuniformity of a network, we calculate the angle-average of the \add{bond} structure factor \delete{including bonds} $S(k)\inl\equiv \overline{S}_\mathrm{b}(k)$ and write it in the limit of small wave numbers as \cite{torquato2018}
\begin{align}
\lim_{k \rightarrow 0} S (k) = k^\alpha \quad .
\label{eqn:hyperuniform_alpha}
\end{align}
The exponent $\alpha$ serves as a \textit{hyperuniformity metric}. By comparing Equations \eqref{eqn:hyperuniform_crit} and \eqref{eqn:hyperuniform_alpha}, we see that non-hyperuniform systems have an exponent of $\alpha \inl\leq 0$. Hyperuniform systems have $\alpha \inl> 0$ and are divided into three classes \cite{torquato2018}. Class I systems possess $\alpha \inl> 1$. This class contains all crystals, most quasicrystals, and some disordered systems. The two weaker hyperuniform classes are class II, with $\alpha \eq 1$, and class III with $0 \inl<\alpha \inl< 1$. A power law describes the decay in density fluctuations in classes I and III at large length scales. Class II is a special case in which fluctuations fall off logarithmically \cite{torquato2018}. 

Excess spreadability is a robust method for extrapolating the scaling of the \add{bond} structure factor as $k$ approaches zero and for determining the hyperuniformity metric $\alpha$ \cite{torquato2021}. In a diffusion process involving a solute initially present in one phase but not in the other, the diffusion spreadability $\mathcal{S}(t)$ is the amount of solute in the initially empty phase over time. This assumes that both phases have the same diffusion constant $D$ \cite{prager1963}. Using dimensionless units, time $t$ is written in units of $d^2/D$. The excess spreadability $\mathcal{S}(t)\inl-\mathcal{S}(\infty)$ at time $t$ can be calculated using the  \add{bond} structure factor \delete{including bonds} (Equation~\eqref{eqn:struct_fact_bonds}). In 3D, it takes the form \cite{torquato2021}
\begin{align}
\mathcal{S}(t)-\mathcal{S}(\infty) = \frac{1}{(2\pi)^3}\int_{\mathbb{R}^3} d\mathbf{k} \, S_\mathrm{b}(\mathbf{k}) \, \exp(-k^2 t) \quad .
\label{eqn:excess_spread}
\end{align}
For long times $t$, the excess spreadability scales as \cite{torquato2021}
\begin{align}
\lim_{t \rightarrow \infty} \mathcal{S}(t)-\mathcal{S}(\infty) \sim 1/t^{(3-\alpha)/2} \quad .
\label{eqn:alpha_spread}
\end{align}
Using Equation~\eqref{eqn:excess_spread} to calculate the excess spreadability from the  \add{bond} structure factor \delete{including bonds} for various times $t$, the hyperuniformity metric $\alpha$ can thus be estimated by fitting Equation~\eqref{eqn:alpha_spread}. To determine a feasible range of times for the fit, consider Equation~\eqref{eqn:excess_spread} for the angle-averaged  \add{bond} structure factor,
\begin{align}
\mathcal{S}(t)-\mathcal{S}(\infty) = \frac{1}{2\pi^2}\int_0^\infty dk \, S(k) \, k^2 \, \exp(-k^2 t) \quad .
\label{eqn:excess_spread_angle_average}
\end{align}
For our system sizes, the structure factor can be calculated for wave numbers in the range $k \inl\in [\pi/4, 4\pi]$. In Equation~\eqref{eqn:excess_spread_angle_average}, this $k$-range is sampled by $t \inl\in [0.1, 1]$.
For this range of $t$, we estimate $\alpha$ by calculating the average slope of $\mathcal{S}(t) \inl-\mathcal{S}(\infty)$ in a double-logarithmic plot. This robust method is effective for relatively small system sizes and does not require us to assume a functional form of the structure factor as $k$ approaches zero.

\subsection{Isotropy}
\label{sec:isotropy}
We quantify the statistical isotropy of a network using information from direct and reciprocal spaces. First, we define a metric for direct space isotropy based on variation in bond orientations. When categorizing the network's bonds by their direction vectors $\mathbf{b}$, variance in bin counts indicates isotropy. We obtain forty orientation bins ($N_\mathbf{b}\eq40$ ) by subdividing each triangular face of a regular icosahedron into four equilateral triangles and consider only one hemisphere of the subdivided icosahedron. Each resulting bin covers a solid angle of $4\pi / 80 \eq \pi / 20$ and the centers of the bins are approximately $20.9 ^{\circ}$ apart in direction $\mathbf{b}_i$. All network bonds are then assigned to their closest bin direction $\mathbf{b}_i$. We call the normalized information entropy of the bond orientation distribution $p(\mathbf{b}_i)$ \textit{bond orientation entropy},
\begin{align}
    h_\mathbf{b} = -\frac{1}{\log(N_\mathbf{b})} \sum_{i=1}^{N_\mathbf{b}} p(\mathbf{b}_i) \log(p(\mathbf{b}_i)) \quad.
    \label{eqn:bond_orientation_entropy}
\end{align}
We find empirically that the values of $h_\mathbf{b} \inl\gtrsim 0.975$ correspond to statistically isotropic networks, while networks with smaller values of $h_\mathbf{b}$ exhibit crystalline domains.

A second class of isotropy metrics is obtained from information in reciprocal space using the structure factors given by Equations \eqref{eqn:struct_fact_vertices} and \eqref{eqn:struct_fact_bonds}. 
Unlike anisotropic systems, such as crystals, which exhibit Bragg peaks, statistically isotropic structures have angle-independent structure factors. We quantify the angle dependence of the two structure factors $S_j(\mathbf{k})$, $j \inl\in \{\mathrm{v}, \mathrm{b}\}$, by first averaging them over angles to obtain the means $\overline{S}_j(k)$ and standard deviations $\sigma_{S_j}(k)$. We then calculate the normalized \textit{anistropy metrics}
\begin{align}
    A_j = \frac{1}{N_l} \sum_{k_l} \frac{1}{\overline{S}_j(k_l)/\sigma_{S_j}(k_l) + 1}
    \label{eqn:anisotropy_metric}
\end{align}
for $N_l\eq 5$ wave numbers $k_l \inl\in \{\pi, \frac{5\pi}{4}, \frac{3\pi}{2}, \frac{7\pi}{4}, 2\pi \}$. We add $1$ to the denominator to normalize the metric to the range $A_j \in [0,1]$. Anisotropic networks exhibit values of $A_j \inl\gtrsim 0.6$. Generally, $A_\mathrm{v}$ and $A_\mathrm{b}$ are similar. For the remainder of the article, we will focus on the bond metric $A_\mathrm{b}$ because it contains information about the mathematical network that the vertex metric  $A_\mathrm{v}$ does not.

\subsection{Topology}
\label{sec:topology}
We classify disordered networks based on their topological properties, such as coordination number and ring statistics. First, we calculate the mean $\overline{Z}$ and standard deviation $\sigma_Z$ of the coordination number distribution. These are inherited from the crystallographic starting networks because the WWW algorithm conserves coordination numbers. Next, we determine the ring size distribution by considering all very strong rings, that is, rings containing at least one bond that does not belong to a shorter ring \cite{goetzke1991}. The ring size $s_i$ is the number of bonds in the ring $i$. Then, we calculate the mean ring size $\overline{s}$ and its standard deviation $\sigma_s$ from the ring size distribution.

To estimate the geometric ring sizes, we first project the 3D rings onto two-dimensional planes. We treat each ring as a polygon and use Newell's method to estimate the normal vector $\mathbf{n}$ that maximizes the polygon's area when projected along $\mathbf{n}$ \cite{tampieri1992}. Then, we project the polygon onto the plane defined by $\mathbf{n}$ and determine its pole of inaccessibility. This point is maximally distant from all edges of the projected polygon. We call the minimum distance from the pole of inaccessibility to any edge of the projected ring \textit{ring radius} $r_s$, as it estimates the radius of the largest circle that can fit inside the ring. We determine the ring radii of all very strong rings in the network and calculate their mean $\overline{r}_s$ and standard deviation $\sigma_{r_s}$.

We have now introduced a comprehensive list of metrics that quantify disorder in network structures across different length scales. \replace{The list is}{This list,} summarized in Table~\ref{tab:list_of_metrics}, Appendix~\ref{sec:list_of_symbols}\add{, combines various structural information in direct and reciprocal space to provide a new classification of disorder in spatial networks}.
In the next section, we determine the parameter space of metrics that can be reached using the extended WWW algorithm introduced in Section~\ref{sec:WWW}. We also demonstrate how our network generation and characterization method can be used to statistically reproduce disordered biophotonic networks.


\section{Case study: reproducing biophotonic networks}
\label{sec:targeted}
In this section, we investigate how the parameters of the extended WWW algorithm (Section~\ref {sec:WWW}) impact generated networks\add{, demonstrating vast tunability in structural characteristics. We link the WWW algorithm inputs to the average generated order metrics through a feedforward neural network}. As a test scenario, we consider biophotonic networks that produce structural color.  We statistically reproduce these networks using the order metric list from Section~\ref{sec:order_metrics}.

\begin{figure*}
\begin{minipage}{0.4\textwidth}
\centering
\begin{overpic}[width=0.52\textwidth]{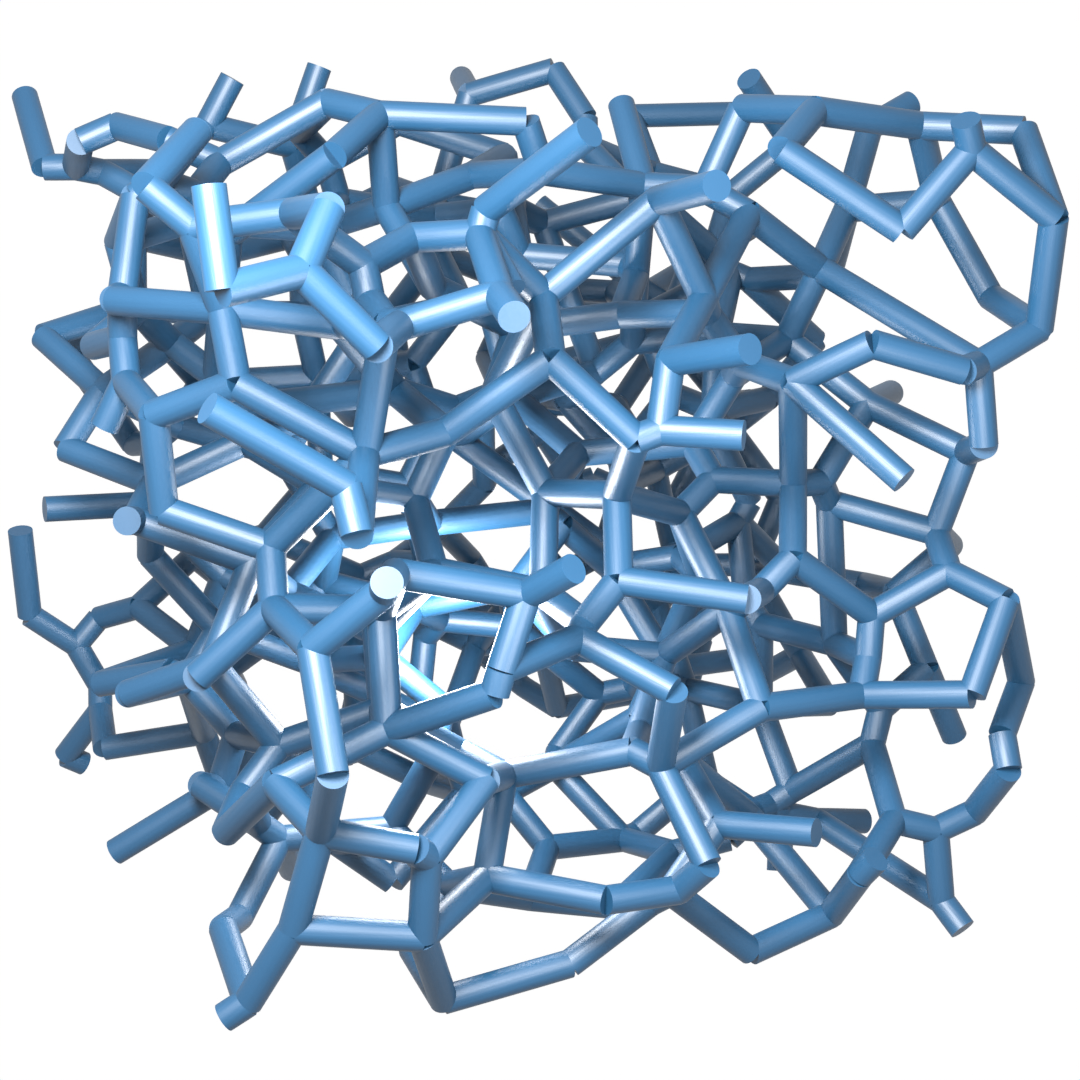}
\put(0,96){\textbf{a}}
\end{overpic}
\mbox{}\\[12mm] 
\begin{overpic}[width=0.49\textwidth]{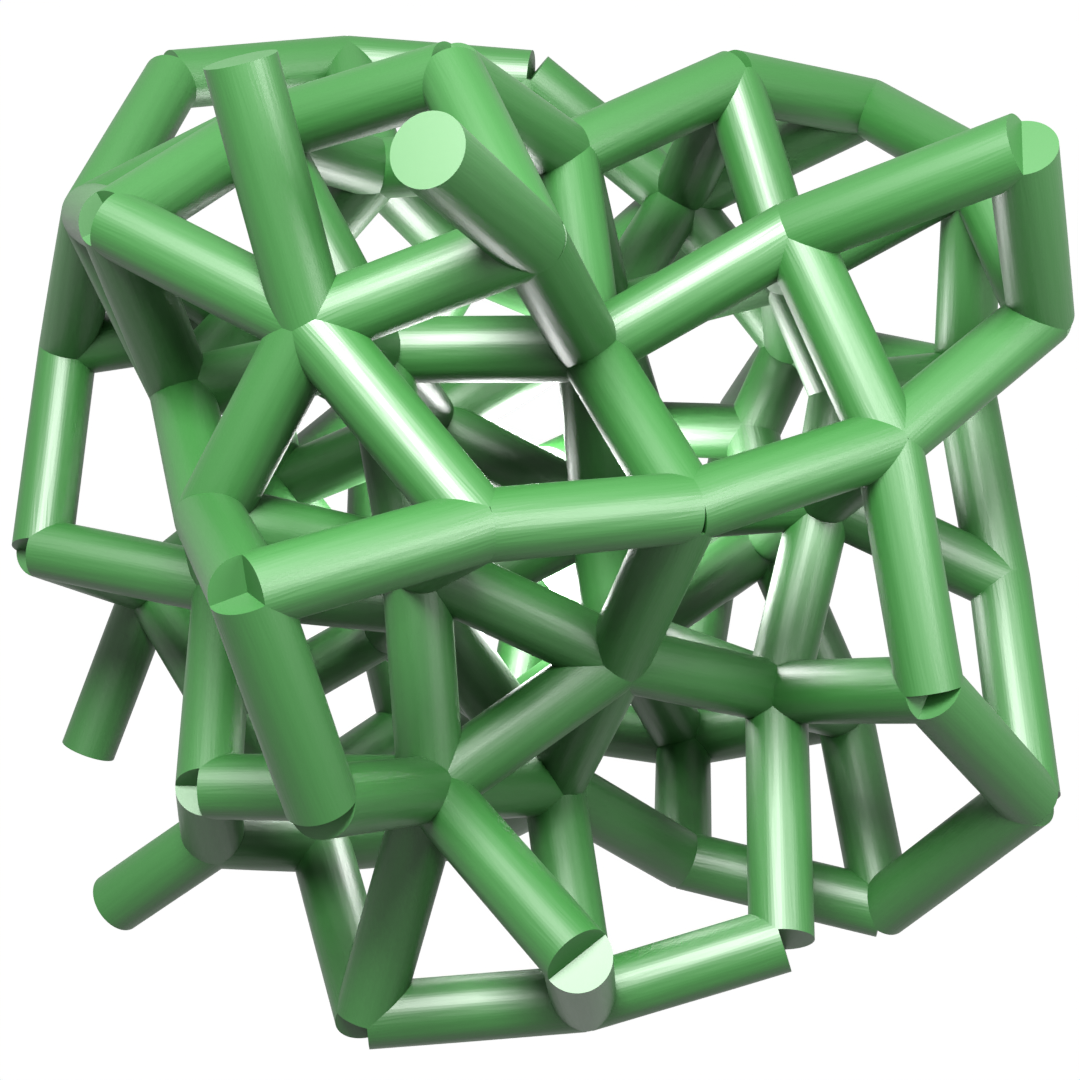}
\put(0,99){\textbf{d}}
\end{overpic}
\begin{overpic}[width=0.49\textwidth]{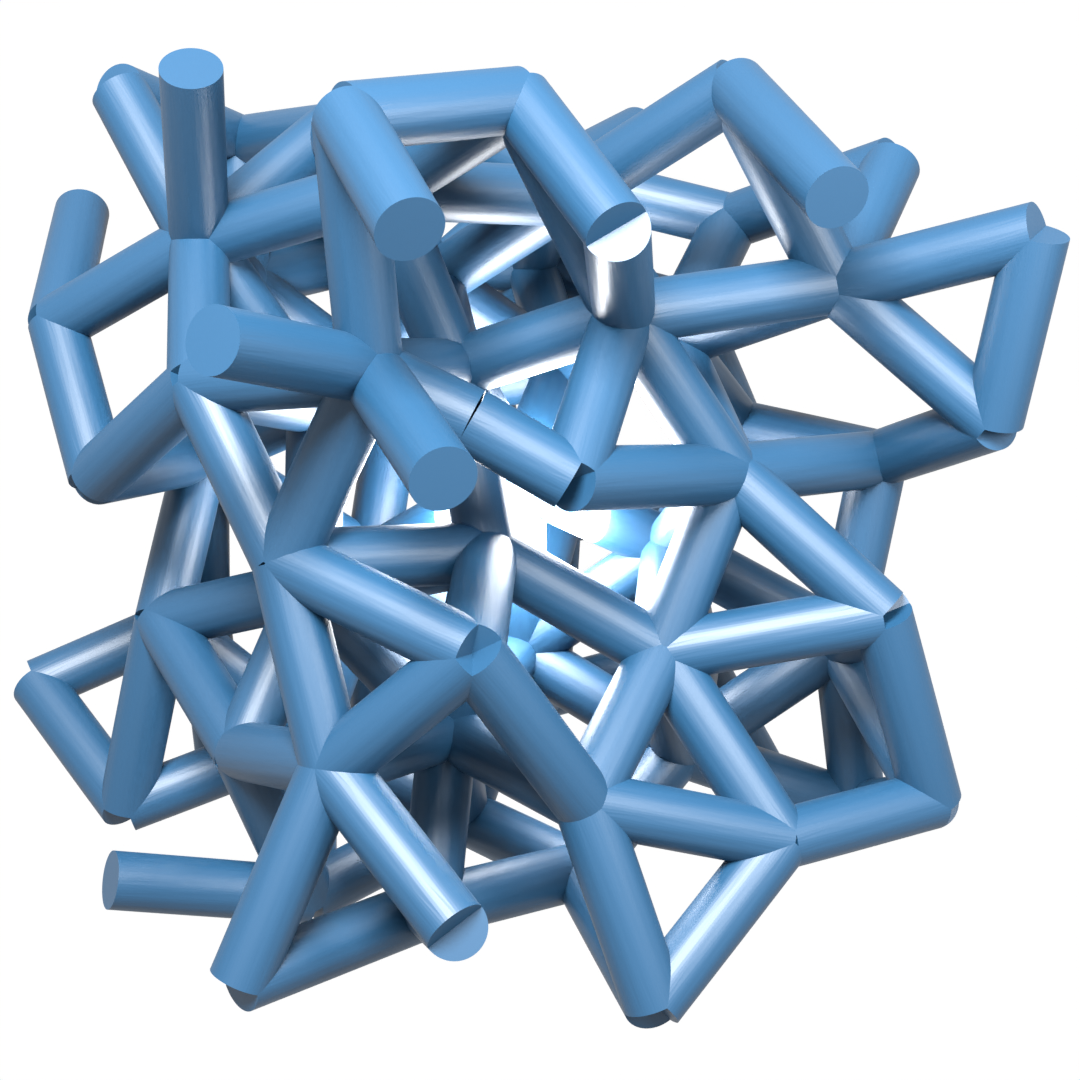}
\put(0,99){\textbf{e}}
\end{overpic}
\mbox{}\\[12mm] 
\begin{overpic}[width=0.49\textwidth]{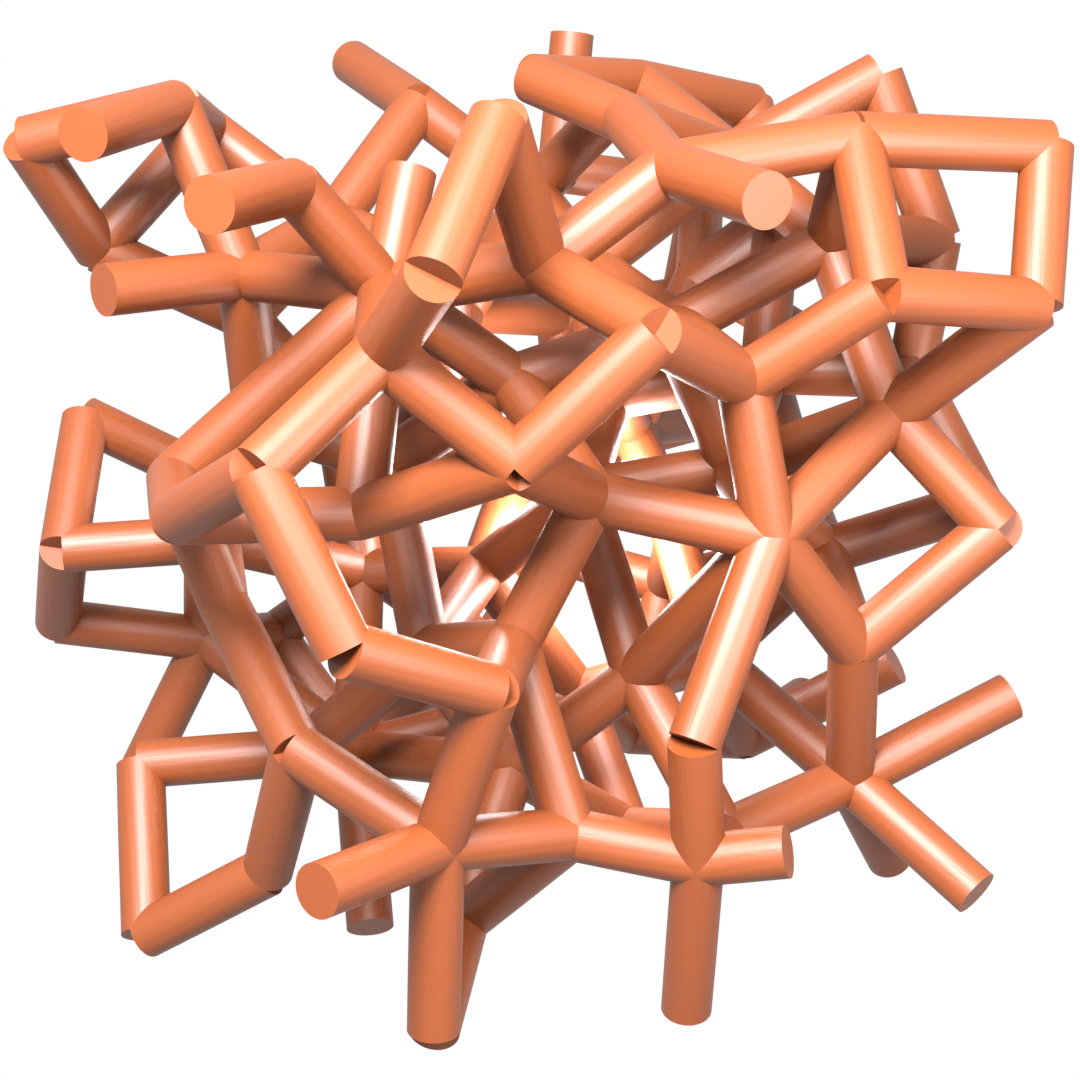}
\put(0,104){\textbf{h}}
\end{overpic}
\end{minipage}
\begin{minipage}{0.17\textwidth}
\centering
\begin{overpic}[width=\textwidth]{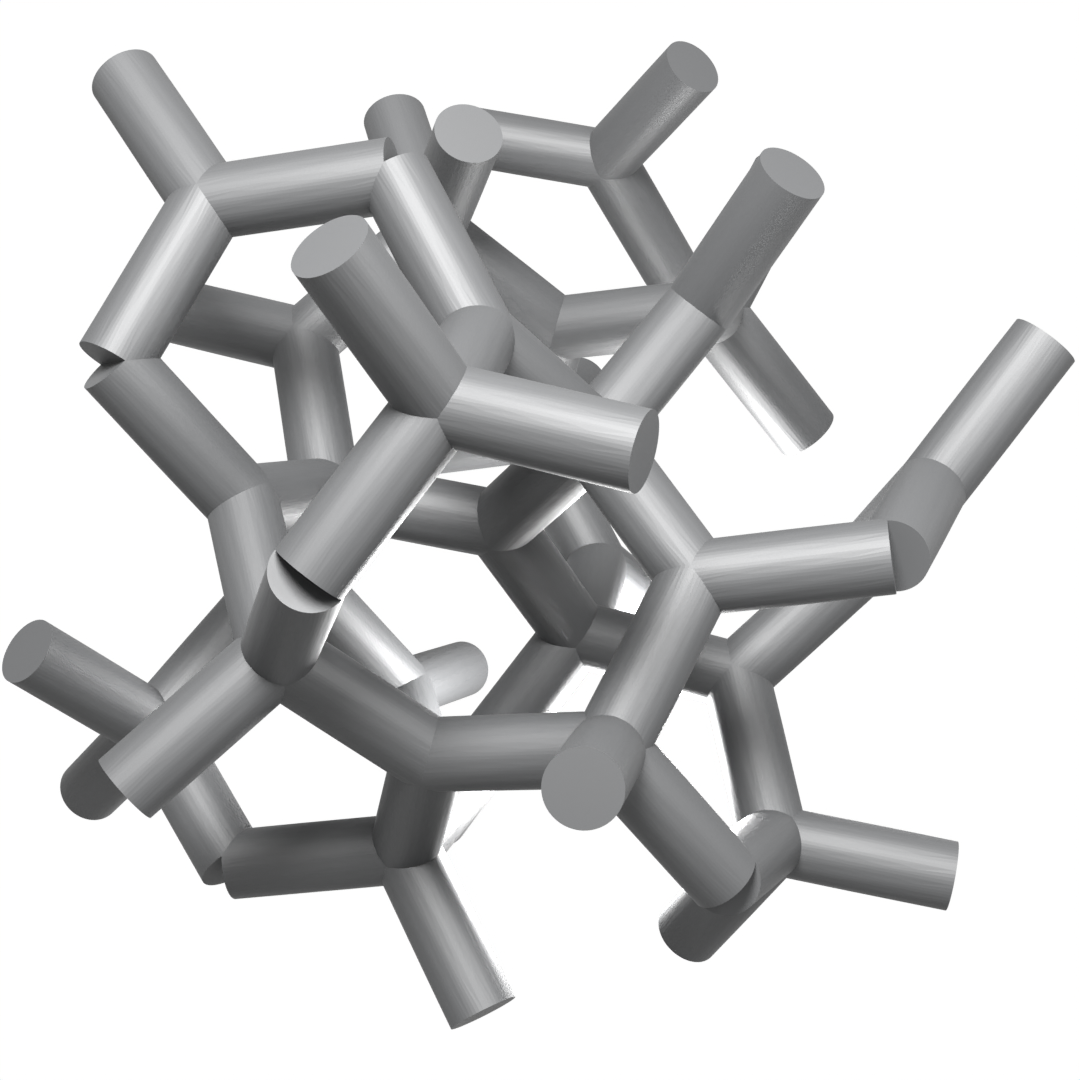}
\put(0,103){\textbf{b}}
\end{overpic}
\mbox{}\\[13mm] 
\begin{overpic}[width=\textwidth]{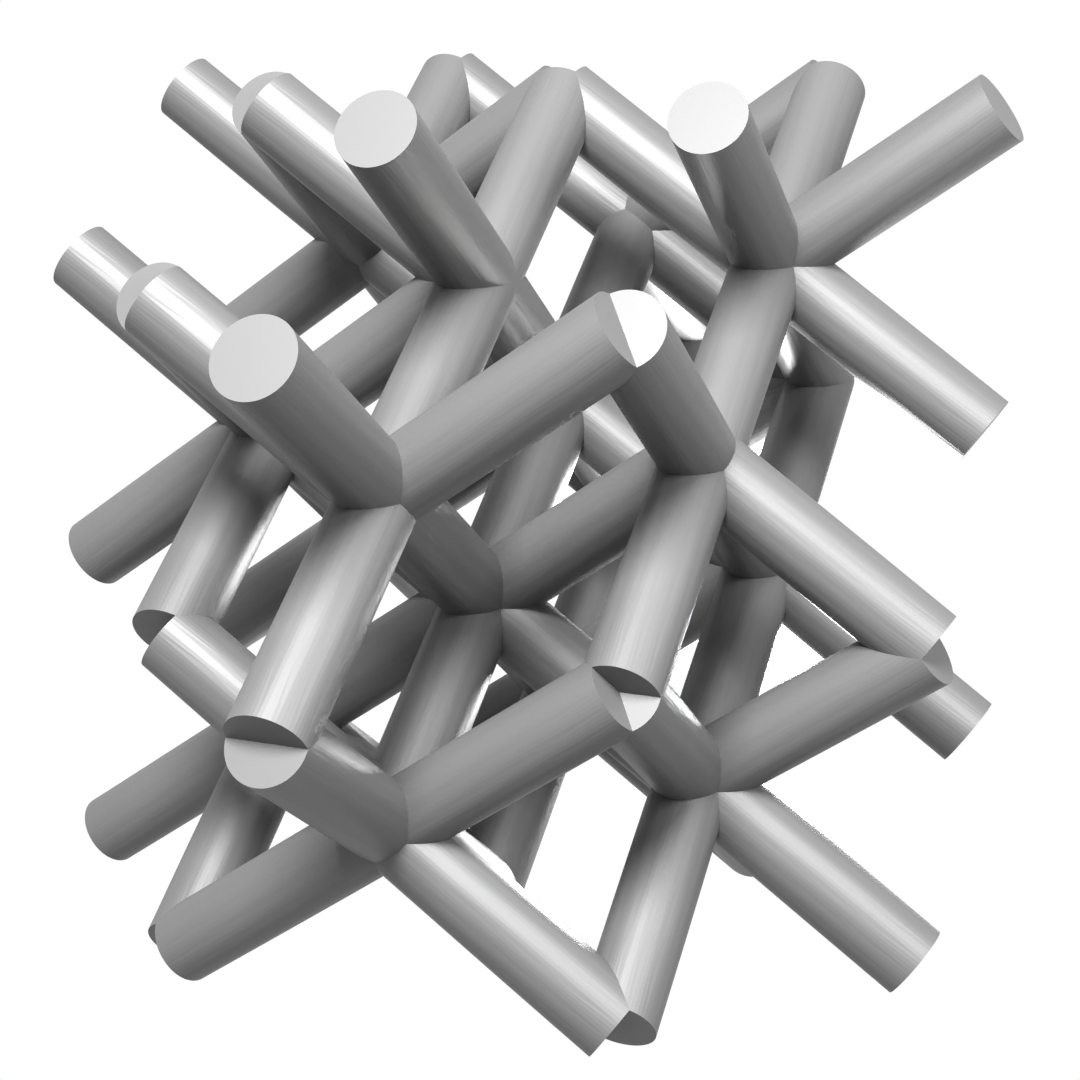}
\put(0,108){\textbf{f}}
\end{overpic}
\mbox{}\\[13mm] 
\begin{overpic}[width=\textwidth]{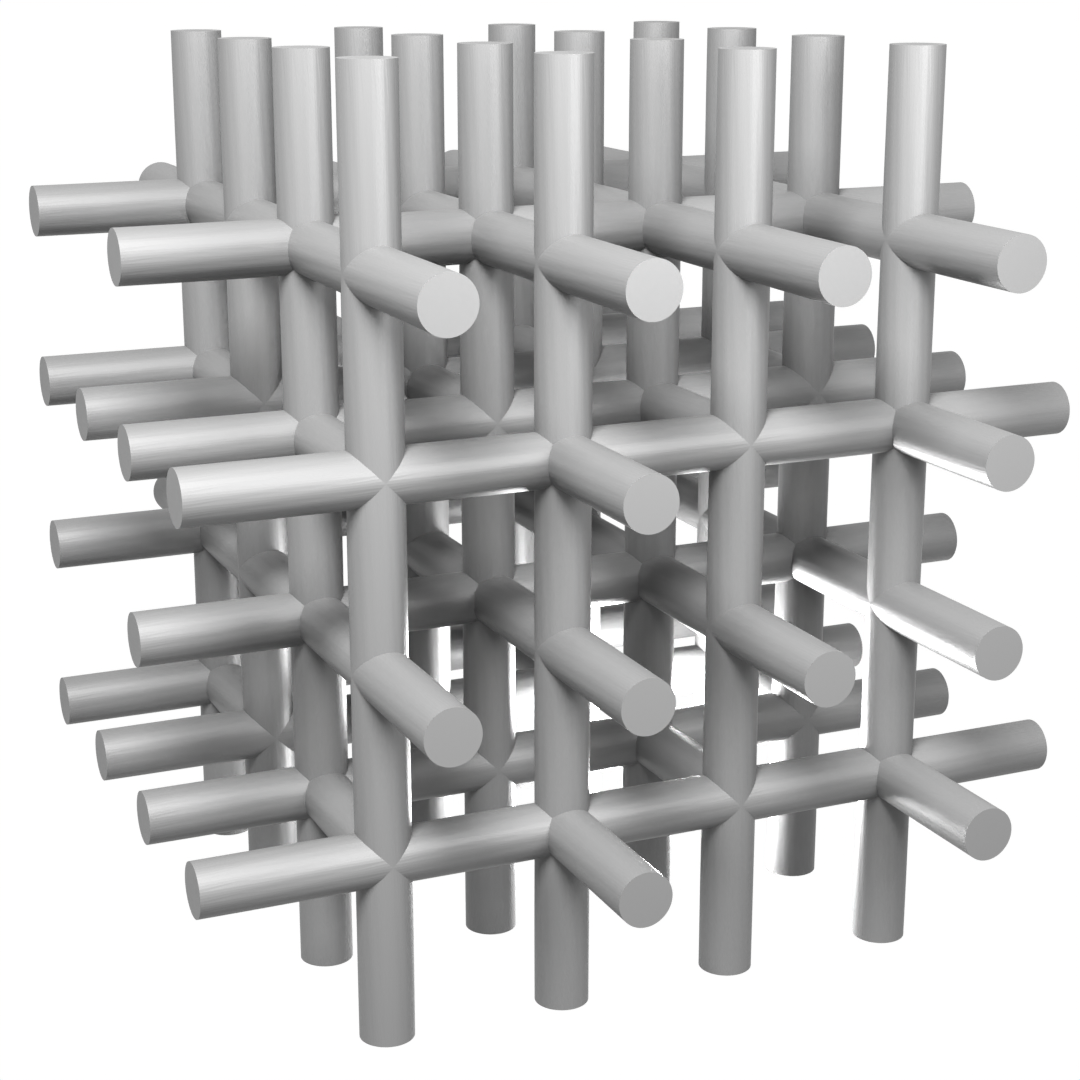}
\put(0,108){\textbf{k}}
\end{overpic}
\end{minipage}
\begin{minipage}{0.4\textwidth}
\centering
\scaledinset{l}{0.01}{b}{.87}{\textbf{c}}{\includegraphics[width=\textwidth]{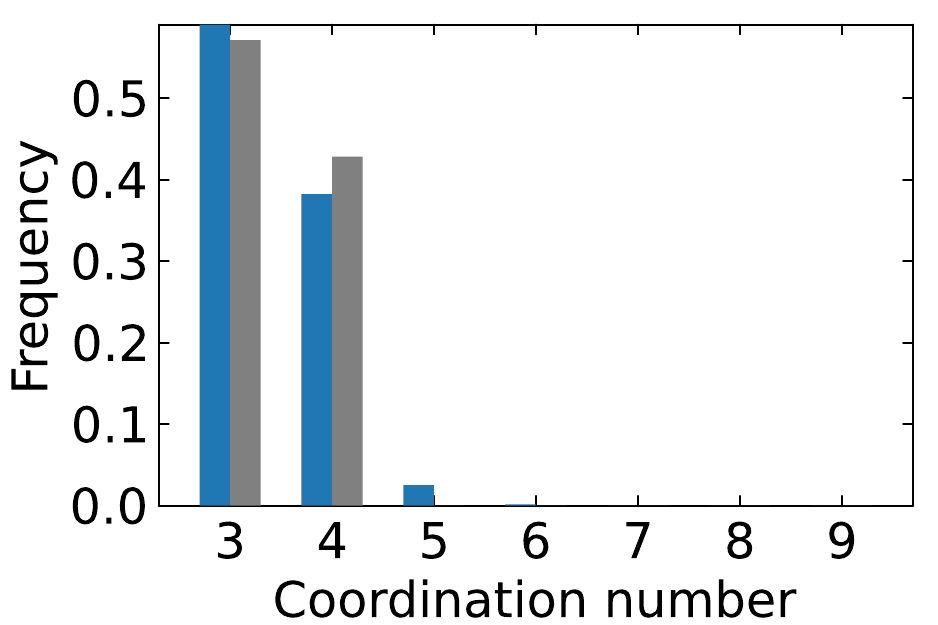}}
\scaledinset{l}{0.01}{b}{.87}{\textbf{g}}{\includegraphics[width=\textwidth]{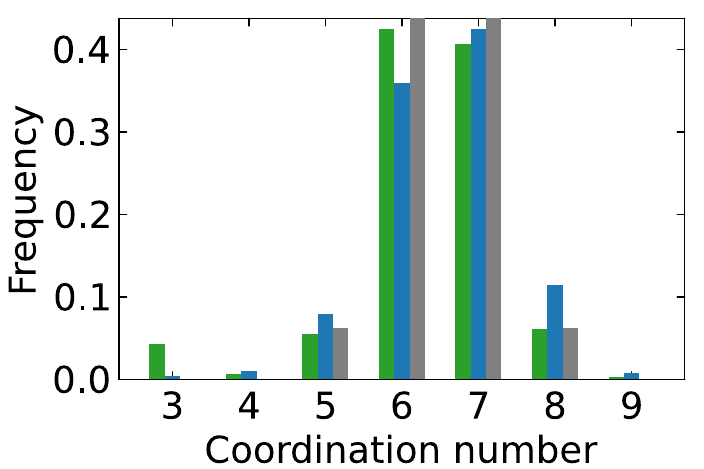}}
\scaledinset{l}{0.01}{b}{.87}{\textbf{l}}{\includegraphics[width=\textwidth]{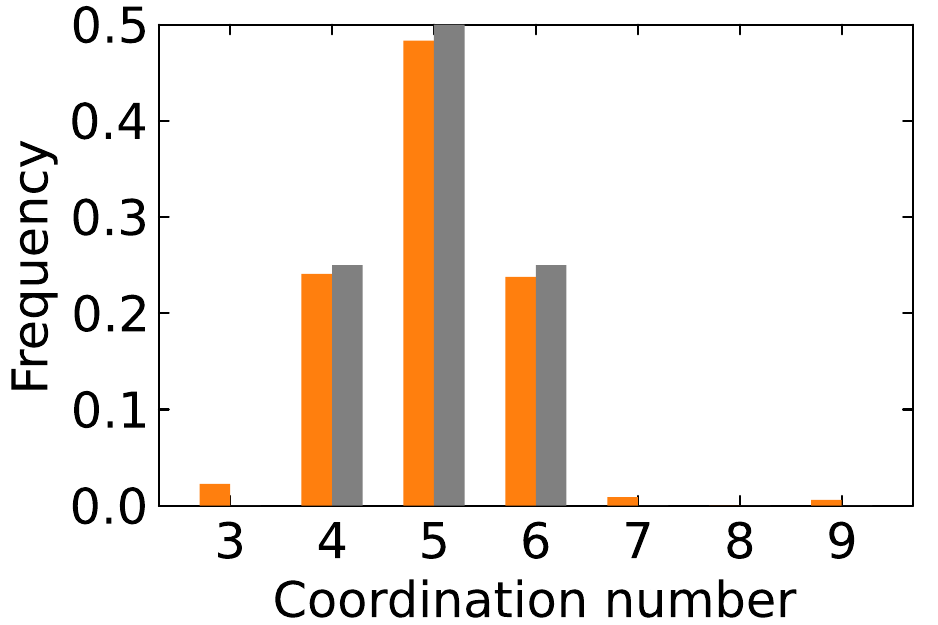}}
\end{minipage}
\caption{\label{fig:bio_periodic_coord_nr}
\textbf{a} A section of the skeletonized, disordered photonic network of the \textit{Pachyrhynchus congestus mirabilis} weevil gives rise to blue structural color (PCM blue). 
\textbf{b} The periodic \textbf{ctn} network has similar coordination number statistics as PCM blue, as shown in \textbf{c}.
\textbf{d} and \textbf{e} show sections of the StV green and  StV blue biological networks, respectively. \textbf{f} The periodic $\textbf{bcu}_\mathrm{mod}$ network has  coordination number statistics similar to those of  StV green and StV blue, as shown in \textbf{g}. The StA orange network in \textbf{h} and  the periodic $\textbf{pcu}_\mathrm{mod}$ network in \textbf{k} have similar coordination number statistics, as shown in \textbf{l}.
}
\end{figure*}

\subsection{Disordered biophotonic and initial periodic networks}
\label{sec:biophotonic}
Structural color arises from the interference of visible light in the presence of biophotonic nanostructures in many animals and plants \cite{vogler-neuling2023}. As dielectric contrast increases, these structures can form complete photonic band gaps (PBGs), which prevent light from entering at any angle \cite{joannopoulos2008}. This phenomenon is well-established in periodic systems, or photonic crystals. However, a thorough understanding of how a reduced photonic density of states emerges in disordered structures is still needed. Our goal is to statistically reproduce these disordered biophotonic networks, laying the groundwork for future work to explore how their structural characteristics affect the photonic density of states.  To achieve this, we will use the extended WWW algorithm introduced in Section~\ref{sec:WWW} and the order metrics from Section~\ref{sec:order_metrics}.

This study focuses on reproducing biophotonic networks based on existing 3D network data. This data was obtained using focused ion beam scanning electron microscopy (FIB-SEM).
FIB-SEM tomography was performed, followed by image processing and skeletonization \cite{djeghdi2022}. Djeghdi et al.\ analyzed the disordered photonic network in the wing scales of the \textit{Pachyrhynchus congestus mirabilis} weevil, which exhibit blue structural color (PCM blue), using this method. Figure~\ref{fig:bio_periodic_coord_nr}a illustrates a section of the skeletonized network. Bauernfeind et al.\ analyzed structurally colored wing scales in \textit{Sternotomis virescens} (StV green and StV blue, Figures \ref{fig:bio_periodic_coord_nr}d~and~\ref{fig:bio_periodic_coord_nr}e) \cite{bauernfeind2023} and \textit{Sternotomis amabilis} (StA orange, Figure~\ref{fig:bio_periodic_coord_nr}h) \cite{bauernfeind2024} longhorn beetles using the same tomography approach.

As discussed in Section~\ref{sec:established_www}, the coordination number statistic of the final disordered network is determined by the initial network in both the original WWW algorithm and our extension. Therefore, we selected periodic initial networks from the crystallographic database~\cite{rcsr2025} with coordination number distributions similar to those of the biological networks. To reproduce the PCM blue network, we started with the \textbf{ctn} network (Figure~\ref{fig:bio_periodic_coord_nr}b) because both networks have coordination numbers $Z\inl\approx 3.4\pm0.5$ (Figure~\ref{fig:bio_periodic_coord_nr}c). Filtering the database \cite{rcsr2025} for cubic space groups with $\inl\ge 195$ and valencies $Z\inl\in\{3,4\}$ yields the \textbf{ctn} network as one with the least complex unit cell, i.e., the smallest genus.

As discussed by Bauernfeind et al.\ \cite{bauernfeind2023} and analyzed in more detail in Section~\ref{sec:reproducing_biophotonic}, the StV green and StV blue networks (Figures \ref{fig:bio_periodic_coord_nr}d,e) have very similar order metrics. The color difference stems from the larger strut lengths of the green network, as evidenced by the difference in mean bond lengths of $d_\text{StV green}\eq\SI{240 \pm 18}{\nano\meter}$, compared to $d_\text{StV blue}\eq\SI{213 \pm 18}{\nano\meter}$. Regarding coordination number statistics $Z\inl\in\{5,6,7,8\}$, with $Z_\text{StV green}\eq6.3\inl\pm1.0$ and $Z_\text{StV blue}\eq6.6\inl\pm0.9$. None of the listed networks in ref.~\cite{rcsr2025} have valencies ranging from 5 to 8. Therefore, we constructed an initial network by removing some bonds from the body-centered cubic network \textbf{bcu} to obtain the modified network $\textbf{bcu}_\text{mod}$ (Figure~\ref{fig:bio_periodic_coord_nr}f). Figure~\ref{fig:bio_periodic_coord_nr}g compares the coordination number statistics of StV green, StV blue, and $\textbf{bcu}_\text{mod}$. Similarly, no networks are listed for the network StA orange (Figure~\ref{fig:bio_periodic_coord_nr}h) with $Z\inl\in\{3,4,5\}$ and $Z_\text{StA orange}\eq5.0 \inl\pm 0.9$.
We remove bonds from the primitive cubic network \textbf{pcu}, to obtain the modified network $\textbf{pcu}_\text{mod}$ (Figure~\ref{fig:bio_periodic_coord_nr}k). The coordination number statistics of $\textbf{pcu}_\text{mod}$ are comparable to those of the StA orange network ((Figure~\ref{fig:bio_periodic_coord_nr}l). 

\begin{figure*}
\centering
\begin{minipage}{0.25\textwidth}
\centering
\begin{overpic}[width=\textwidth]{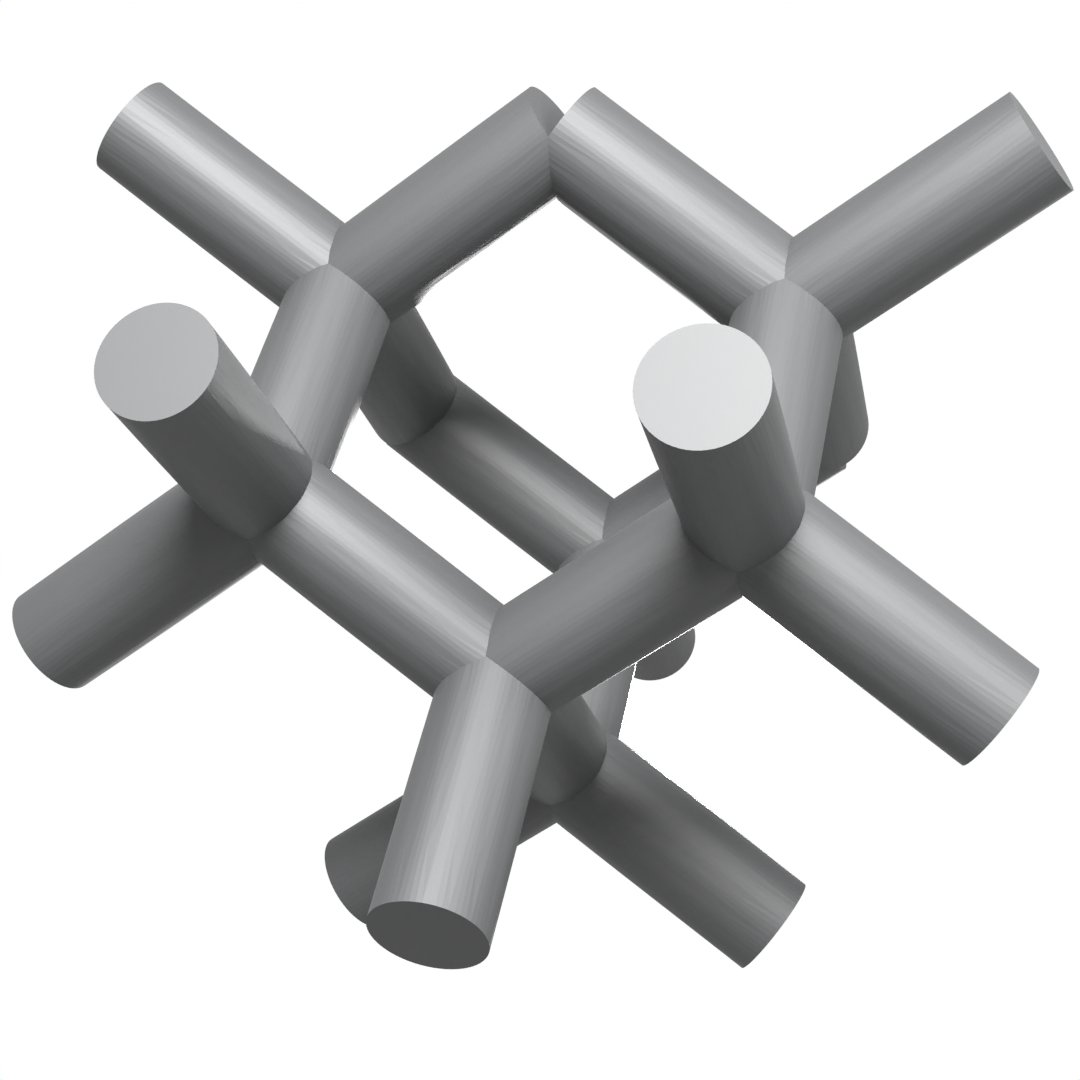}
\put(0,92){\textbf{a}}
\end{overpic}
\end{minipage}\hfill
%
\begin{minipage}{0.25\textwidth}
\centering
\begin{overpic}[width=\textwidth]{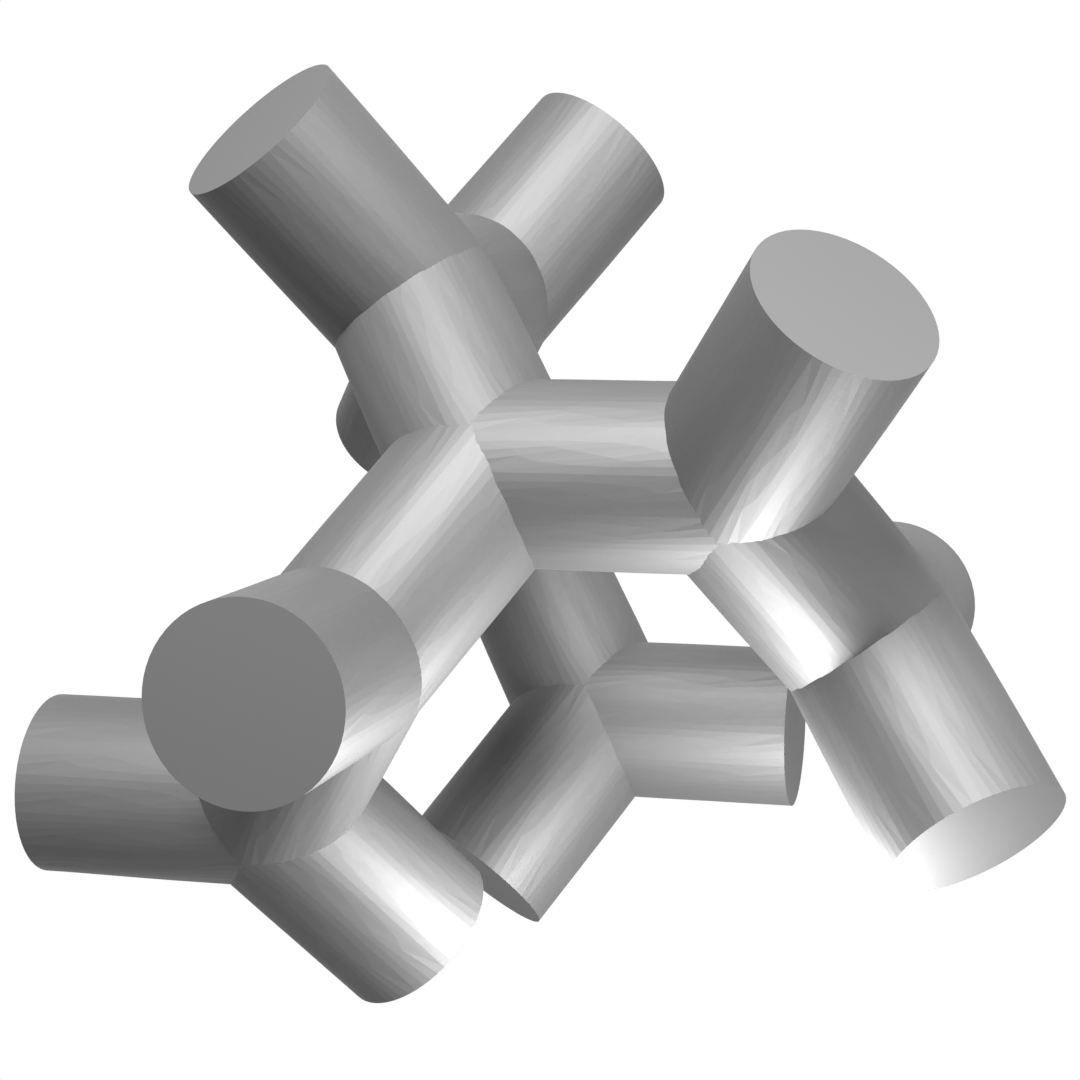}
\put(0,92){\textbf{b}}
\end{overpic}
\end{minipage}\hfill
%
\begin{minipage}{0.25\textwidth}
\centering
\begin{overpic}[width=\textwidth]{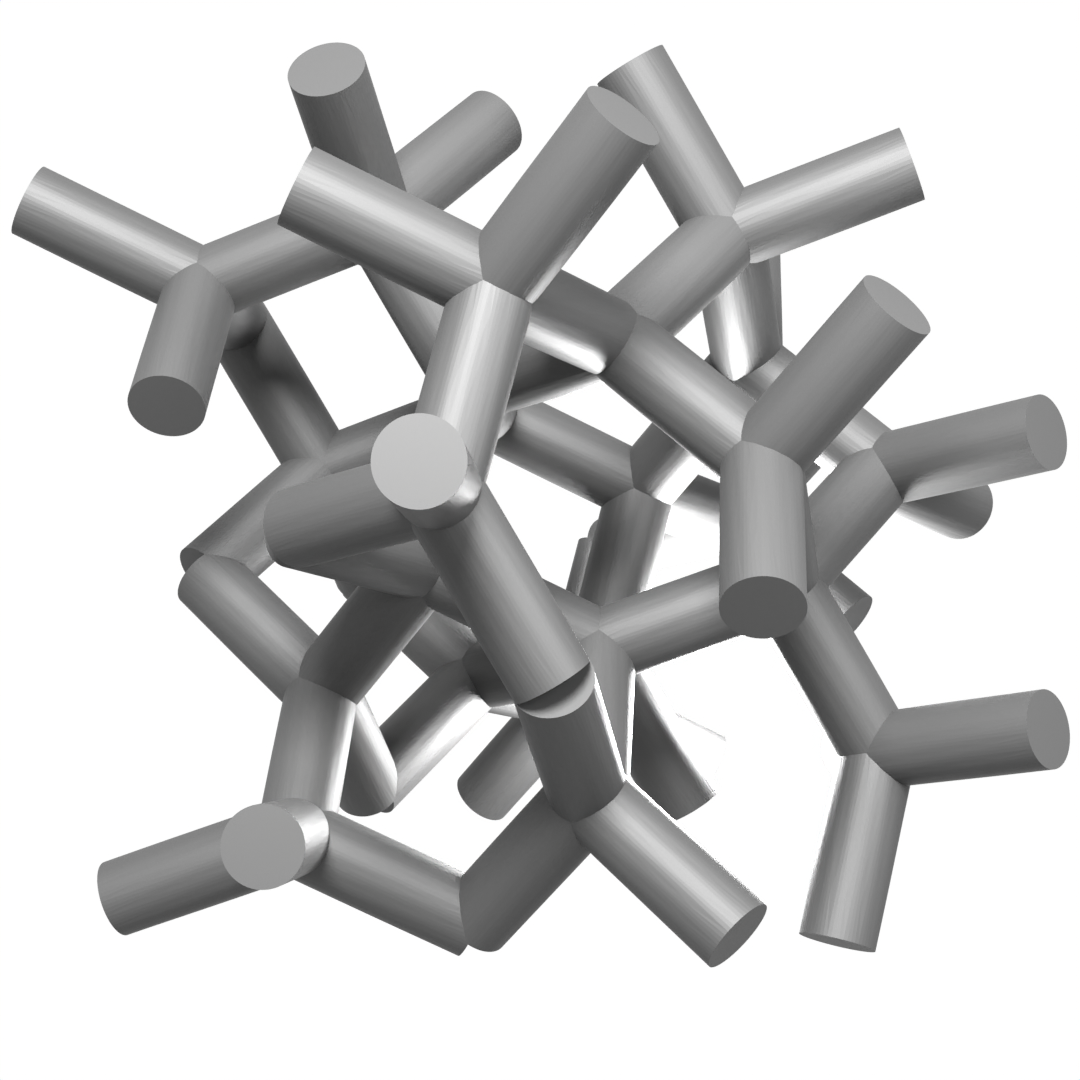}
\put(0,92){\textbf{c}}
\end{overpic}
\end{minipage}
\caption{\label{fig:dia_srs_lcs}
We applied the extended WWW algorithm to the periodic diamond \textbf{dia} (\textbf{a}), gyroid \textbf{srs} (\textbf{b}), and \textbf{lcs} (\textbf{c}) networks, which have wide photonic band gaps.
}
\end{figure*}

We also generated three types of regular disordered networks. The first two types are generated from the cubic diamond \textbf{dia} ($Z\eq4$, Figure~\ref{fig:dia_srs_lcs}a) and the gyroid \textbf{srs} ($Z\eq3$, Figure~\ref{fig:dia_srs_lcs}b). The WWW algorithm has previously been applied to these networks \cite{wooten1985,djordjevic1995, barkema2000, sellers2017}.
These networks are of high interest due to their ability to open significant photonic band gaps \cite{cersonsky2021} and their presence in biological structural color materials \cite{djeghdi2022, michielsen2007}.
The last type starts from the $4$-coordinated \textbf{lcs} network (Figure~\ref{fig:dia_srs_lcs}c) \add{which has never been investigated under WWW evolution}. This network was recently found to have a wide PBG similar to those of \textbf{dia} and \textbf{srs} \cite{cui2026}. Diamond and the \textbf{lcs} networks are interesting to compare because, despite having the same coordination number, their ring statistics and bond angles differ. In summary, we consider six initial networks ranging in valencies from three to eight. In the next section, we investigate how the input parameters of the extended WWW algorithm affect the resulting disordered networks.

\subsection{The effect of the parameters on the extended WWW algorithm}
\label{sec:effect_www_parameters}
Starting from the initial networks presented in Section~\ref{sec:biophotonic}, we analyze how the order metrics of the generated networks vary according to the inputs of the extended WWW algorithm. For each initial network, we generate between 450 (for $\textbf{bcu}_\mathrm{mod}$) and 3,750 (for \textbf{ctn}) disordered networks. The input parameter space of the modified WWW algorithm is spanned by $\beta$, $T_\mathrm{max}$, and $\Delta T$. We sample random bond-bending constants in the range of  $\beta \inl\in [0,10]$ to vary the ratio of disorder in bond angles and lengths. To sample the melting transition, i.e., the phase transition from crystalline to disordered, we vary the maximum heating temperature and the temperature gradient in units of the $\beta$-dependent melting temperature: $T_\mathrm{max}\inl\in[T_\mathrm{melt}/2, 2T_\mathrm{melt}]$, and $\Delta T \inl\in [T_\mathrm{melt}/4, 2T_\mathrm{melt}]$.

We will now discuss the results for the initial \textbf{ctn} network of 224 vertices in detail. Analogous statistics for the other five networks can be found in Appendix~\ref{sec:si_effect_www_inputs} (Figures \ref{fig:bcu_cn_5_6_7_8_nr_accepted_moves_isotropy}~to~\ref{fig:lcs_isotropy_topology}). The order metrics $\overline{q}_l$, $\sigma_{q_l}$ (Section~\ref{sec:similarity_primitives}), $A_\mathrm{v}$ (Equation \eqref{eqn:anisotropy_metric}), $\overline{s}$, and $\sigma_{s}$ (Section~\ref{sec:topology}) were excluded from the in-depth analysis due to their strong correlation with other metrics. The Steinhardt bond order parameters $\overline{q}_l$ and their standard deviations $\sigma_{q_l}$ are related to bond-angle disorder $\sigma_\theta$, but they are more difficult to interpret intuitively. The anisotropy metric $A_\mathrm{v}$ is strongly correlated with the analogous metric $A_\mathrm{b}$,  which contains more information about the mathematical network. Similarly, the ring sizes $\overline{s}$ and their standard deviations $\sigma_{s}$ are closely linked to the ring radii and their standard deviations.

\begin{figure*}
\begin{minipage}{0.48\textwidth}
\centering
\scaledinset{l}{0.01}{b}{.91}{\textbf{a}}{\includegraphics[width=\textwidth]{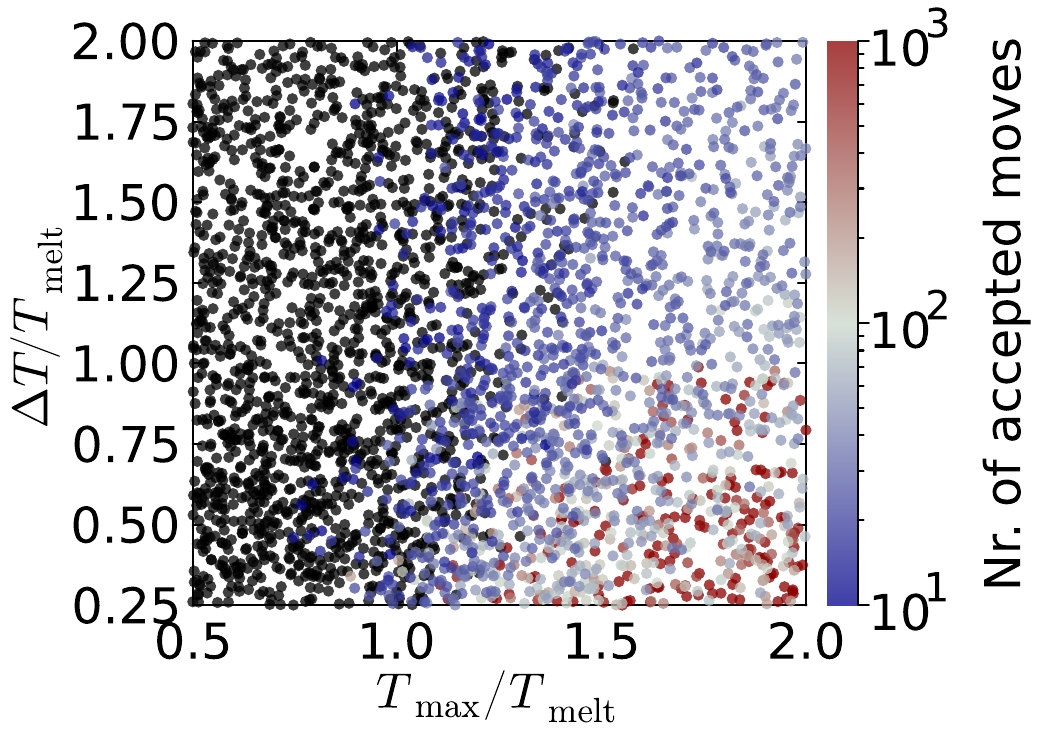}}
\end{minipage}
%
\begin{minipage}{0.48\textwidth}
\centering
\scaledinset{l}{0.01}{b}{.91}{\textbf{b}}{\includegraphics[width=\textwidth]{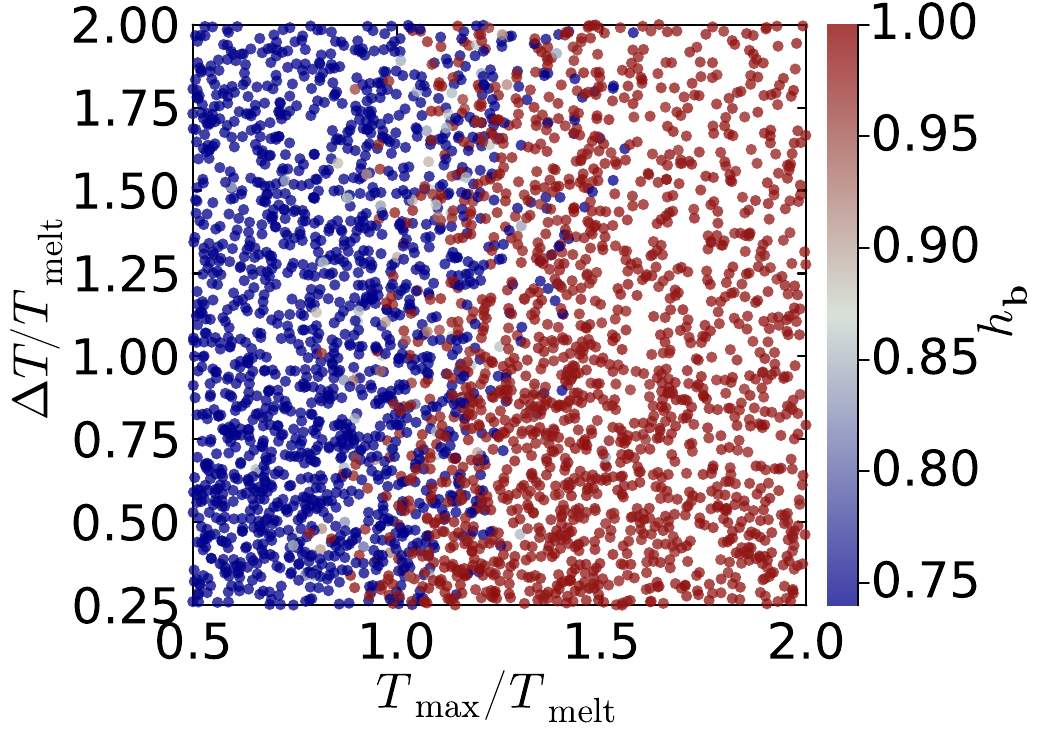}}
\end{minipage}
\caption{\label{fig:ctn_nr_accepted_moves_isotropy}
\add{The number of accepted Monte Carlo moves and the isotropy metric $h_\mathbf{b}$ for 3750 networks generated from the \textbf{ctn} network.}
\textbf{a} The number of accepted Monte Carlo moves \delete{in the \textbf{ctn} network} increases with increasing maximal heating temperature $T_\mathrm{max}$ and with decreasing heating gradient $\Delta T$. The black markers correspond to networks with ten or fewer accepted moves and show that the melting transition begins at $T_\mathrm{max} \inl\gtrsim T_\mathrm{melt}$. All values of $\beta \inl\in [0,10]$ are considered.
\textbf{b} The isotropy metric $h_\mathbf{b}$ effectively measures the melting transition of the initial \textbf{ctn} network.
}
\end{figure*}

Figure~\ref{fig:ctn_nr_accepted_moves_isotropy}a shows the melting transition of the initial \textbf{ctn} network. It plots the number of accepted Monte Carlo moves against the heating profile parameters $T_\mathrm{max}$, and $\Delta T$. In this and the following figures, we project the 3D WWW parameter space onto two resolved parameters while considering all values of the third parameter, here $\beta$. For low values of $T_\mathrm{max}$, no bond switches are accepted, and the network remains in its initial crystallographic state. The melting transition occurs close to the melting temperature of Equation~\eqref{eqn:t_melt}, $T_\mathrm{max} \inl\approx T_\mathrm{melt}$. This justifies our somewhat arbitrary definition of $T_\mathrm{melt}$ given our values of $\Delta T$. We also observe a slight shift of the melting transition to lower values of $T_\mathrm{max}$ as $\Delta T$ decreases. This behavior corresponds to an increase in accepted moves at low values of $\Delta T$. This reflects the fact that with smaller heating and cooling gradients, more Monte Carlo moves are performed at high temperatures, making melting more probable. The isotropy metrics introduced in Section~\ref{sec:isotropy}, such as bond orientation entropy $h_\mathbf{b}$, are also well-suited to measuring the melting transition as illustrated in Figure~\ref{fig:ctn_nr_accepted_moves_isotropy}b.
As expected, the crystallographic network with reduced isotropy quickly approaches a statistically isotropic configuration upon melting.

\begin{figure}
\centering
\includegraphics[width=0.48\linewidth]{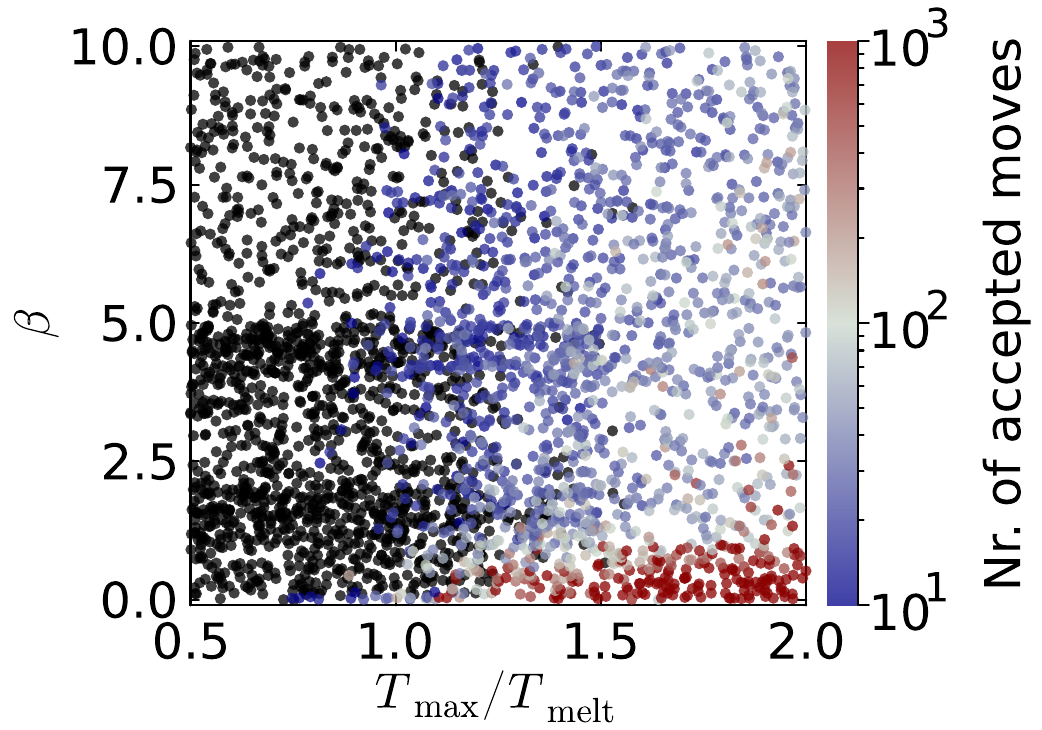}
\caption{\label{fig:ctn_nr_accepted_moves_beta} There is an inverse correlation between the bond-bending force constant $\beta$ and the number of accepted Monte Carlo moves when the network is heated above the melting transition $T_\mathrm{max} \gtrsim T_\mathrm{melt}$. All values of $\Delta T$ are considered. \add{The metrics of 3750 networks are displayed.}}
\end{figure}

Figure~\ref{fig:ctn_nr_accepted_moves_beta} shows the significant effect of the bond-bending constant $\beta$ on the number of accepted Monte Carlo moves when $T_\mathrm{max} \inl\gtrsim T_\mathrm{melt}$. At small values of $\beta \inl\lesssim 1$, the number of accepted moves exceeds $10^3$, whereas at higher $\beta$ values, only 10 to 100 moves are accepted. For small $\beta$ values, the generalized Keating strain energy (Equation~\eqref{eqn:keating_generalized}) is governed by the bond-stretching term $E_r$, which represents a two-body interaction. The three-body interaction bond-bending term $E_\theta$ can be neglected. Having a two-body interaction govern the dynamics appears to increase the number of favorable Monte Carlo moves. Even bond chains that are strongly flexed after the bond switch can be accepted if the bond distances relax close to the equilibrium length (Figure~\ref{fig:bond_switch}). In summary, the temperature gradient has a minor influence on the melting behavior, while the maximal temperature and bending constant have a more significant influence on the network statistics. Therefore, we focus on the projection onto these two parameters.

\begin{figure*}
\begin{minipage}{0.48\textwidth}
\centering
\scaledinset{l}{0.01}{b}{.91}{\textbf{a}}{\includegraphics[width=\textwidth]{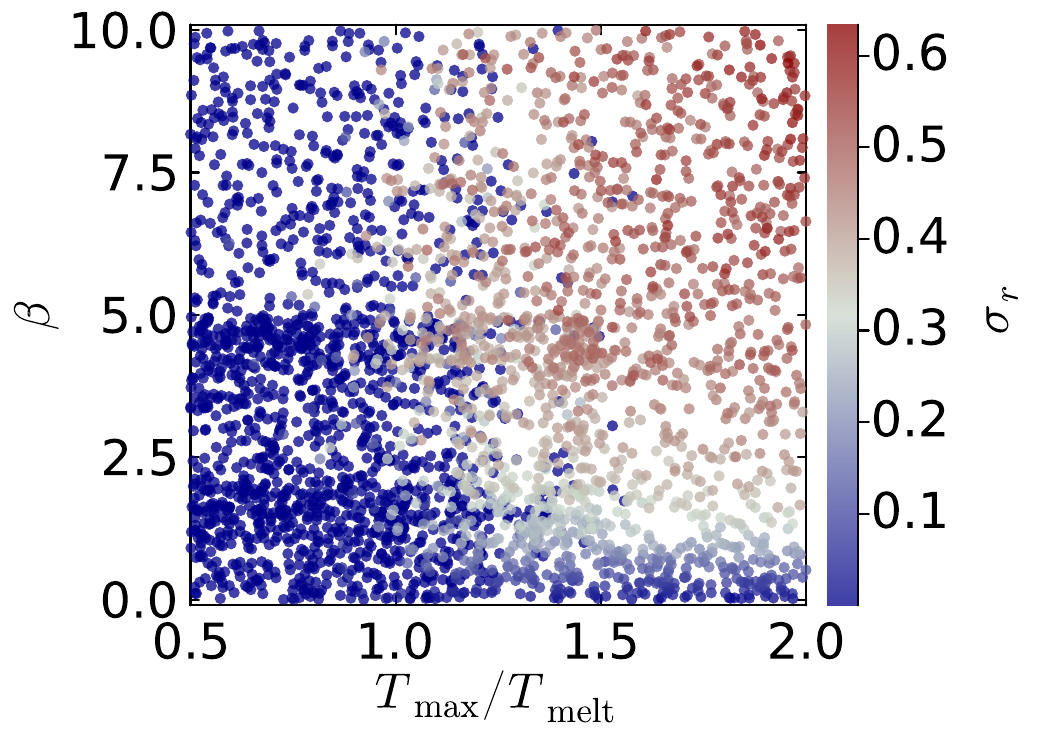}}
\end{minipage}
%
\begin{minipage}{0.48\textwidth}
\centering
\scaledinset{l}{0.01}{b}{.91}{\textbf{b}}{\includegraphics[width=\textwidth]{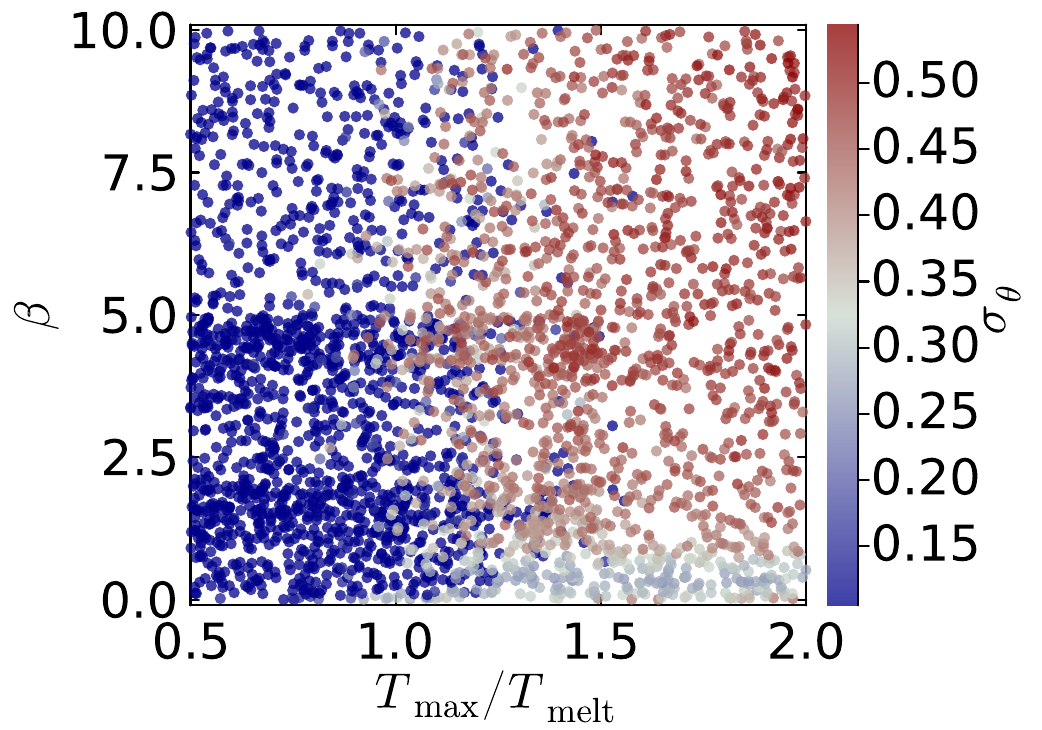}}
\end{minipage}
\caption{\label{fig:ctn_bond_length_angle}
Disorder in the primitives of networks generated from the initial \textbf{ctn} network. \add{The metrics of 3750 networks are displayed.}
\textbf{a} Disorder in the bond lengths, measured by $\sigma_r$, increases with a larger bond-bending force constant $\beta$ and a higher maximal heating temperature $T_\mathrm{max}$. 
\textbf{b} Similarly, disorder in bond angles $\sigma_\theta$ increases with $\beta$ and $T_\mathrm{max}$. For small $\beta$ values and large $T_\mathrm{max}$ values, $\sigma_r$ approaches the values of the crystalline network in its melted state. $\sigma_\theta$ is bounded from below by $\sigma_\theta \inl\gtrsim 0.25$. These plots consider all values of $\Delta T$.
}
\end{figure*}

Figure~\ref{fig:ctn_bond_length_angle} illustrates the impact of the algorithm's input parameters $T_\mathrm{max}$ and $\beta$ on the bond length and angle statistics of the generated networks. Similar to the increase in isotropy with increasing $T_\mathrm{max}$ (Figure~\ref{fig:ctn_nr_accepted_moves_isotropy}b), the degree of disorder in the bond lengths,  as measured by $\sigma_r$, generally increases with $T_\mathrm{max}$ (Figure~\ref{fig:ctn_bond_length_angle}a). Additionally, there is a strong positive correlation between $\beta$ and $\sigma_r$. As previously mentioned, the low values of $\sigma_r$ at small $\beta$ are due to the higher energy penalty for disorder in bond lengths compared to disorder in bond angles. Interestingly, for small values of $\beta \inl\lesssim 0.15$, $\sigma_r$ approaches 0, similar to the periodic \textbf{ctn} network. 
Counterintuitively, this low degree of disorder is achieved by performing a large number of Monte Carlo moves, as shown in Figure~\ref{fig:ctn_nr_accepted_moves_beta}. While the initially accepted moves appear to introduce defects into the crystalline system, resulting in higher $\sigma_r$ values, subsequent moves relax the strain of the defect states, reducing the degree of disorder.

Figure~\ref{fig:ctn_bond_length_angle}b shows a positive correlation between disorder in the bond angles $\sigma_\theta$ and $T_\mathrm{max}$, as well as with $\beta$. Unlike $\sigma_r$, bond angle standard deviation increases from $\sigma_\theta\eq0.106$ in the crystalline state to $\sigma_\theta\inl\gtrsim 0.25$ at the onset of melting. Lower values are not reached in the disordered state. Since high $\beta$ values impose a high energy penalty on disorder in bond angles, Figure~\ref{fig:ctn_nr_accepted_moves_beta} shows that only a few Monte Carlo moves are accepted at large $\beta$ values. As discussed above, this is due to three-body interactions in the bond-bending energy term $E_\theta$, which reduces the number of favorable bond switches. The small number of accepted Monte Carlo moves makes complete relaxation of the disorder introduced by the first moves unlikely. A more substantial decrease in $\sigma_\theta$ could be achieved through repeated heating and quenching of the network
(see Ref.\ \cite{barkema2000}). Starting from a disordered diamond-like network, Barkema et al.\ achieved disordered networks with $\sigma_\theta\inl\approx 0.175$. However, our disordered networks generated from the crystalline \textbf{dia} only achieved values of $\sigma_\theta\inl\approx 0.28$ (see Figure~\ref{fig:dia_bond_length_angle}b in Appendix~\ref{sec:si_effect_www_inputs}). Unlike \textbf{dia} with $\sigma_\theta\eq0$, small values of $\sigma_\theta$ are mode difficult to achieve from \textbf{ctn} due to its mixed coordination number $Z\inl\in \{3,4\}$.

\begin{figure*}
\begin{minipage}{0.48\textwidth}
\centering
\scaledinset{l}{0.01}{b}{.91}{\textbf{a}}{\includegraphics[width=\textwidth]{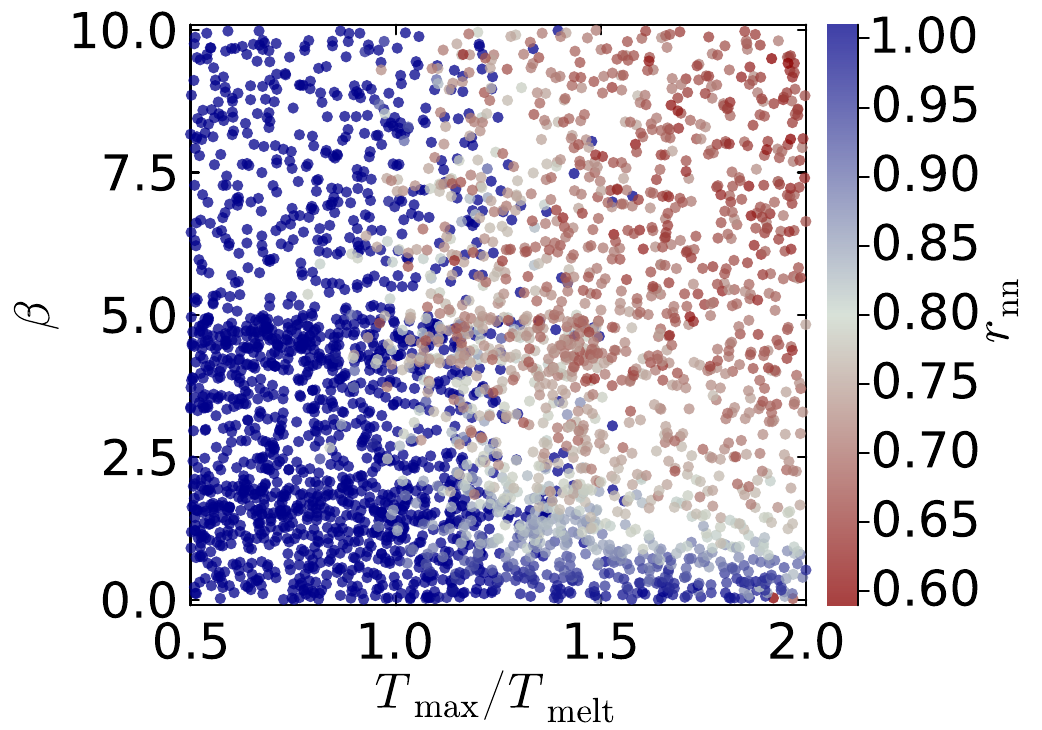}}
\end{minipage}
%
\begin{minipage}{0.48\textwidth}
\centering
\scaledinset{l}{0.01}{b}{.91}{\textbf{b}}{\includegraphics[width=\textwidth]{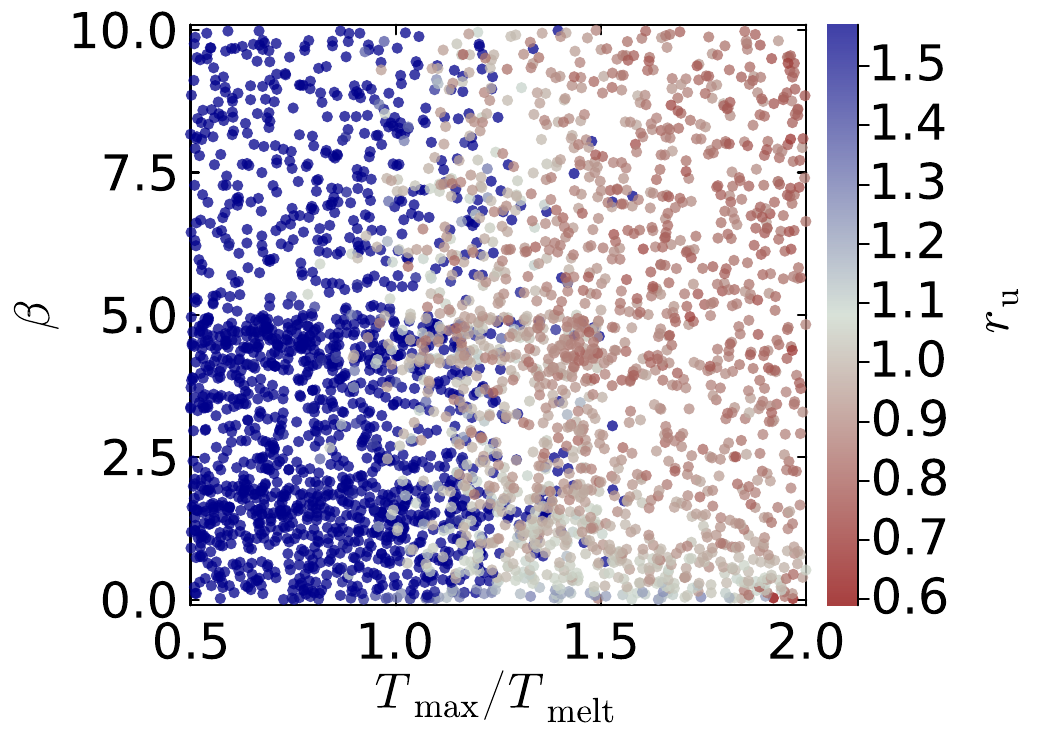}}
\end{minipage}
\begin{minipage}{0.48\textwidth}
\centering
\scaledinset{l}{0.01}{b}{.91}{\textbf{c}}{\includegraphics[width=\textwidth]{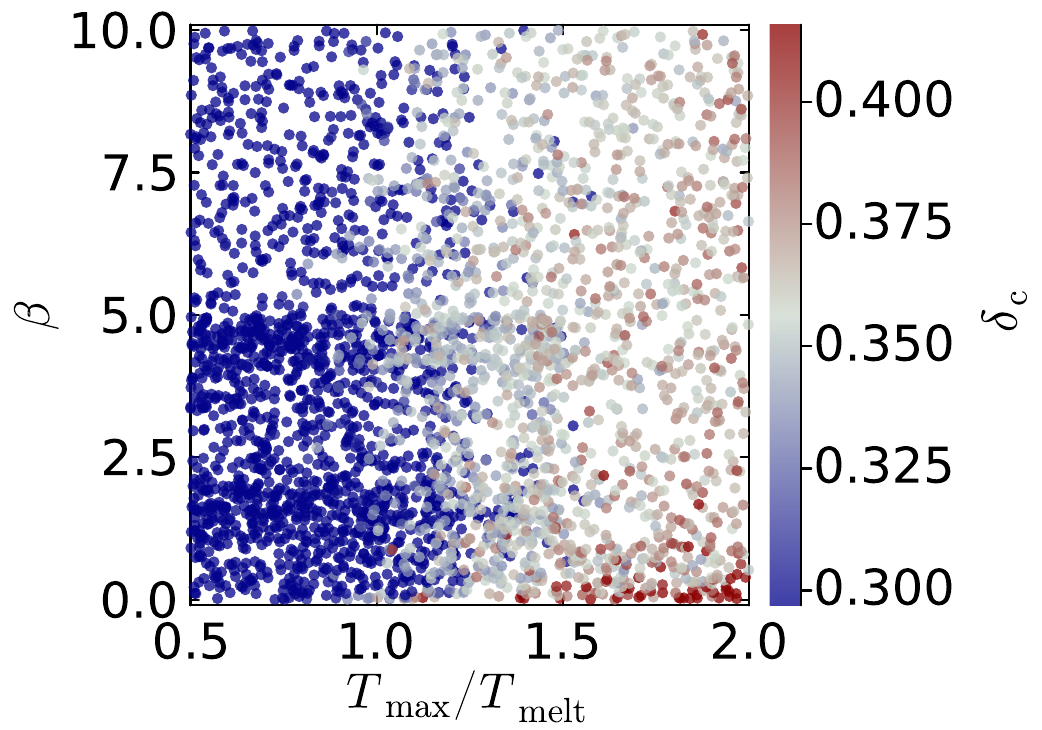}}
\end{minipage}
%
\begin{minipage}{0.48\textwidth}
\centering
\scaledinset{l}{0.01}{b}{.91}{\textbf{d}}{\includegraphics[width=\textwidth]{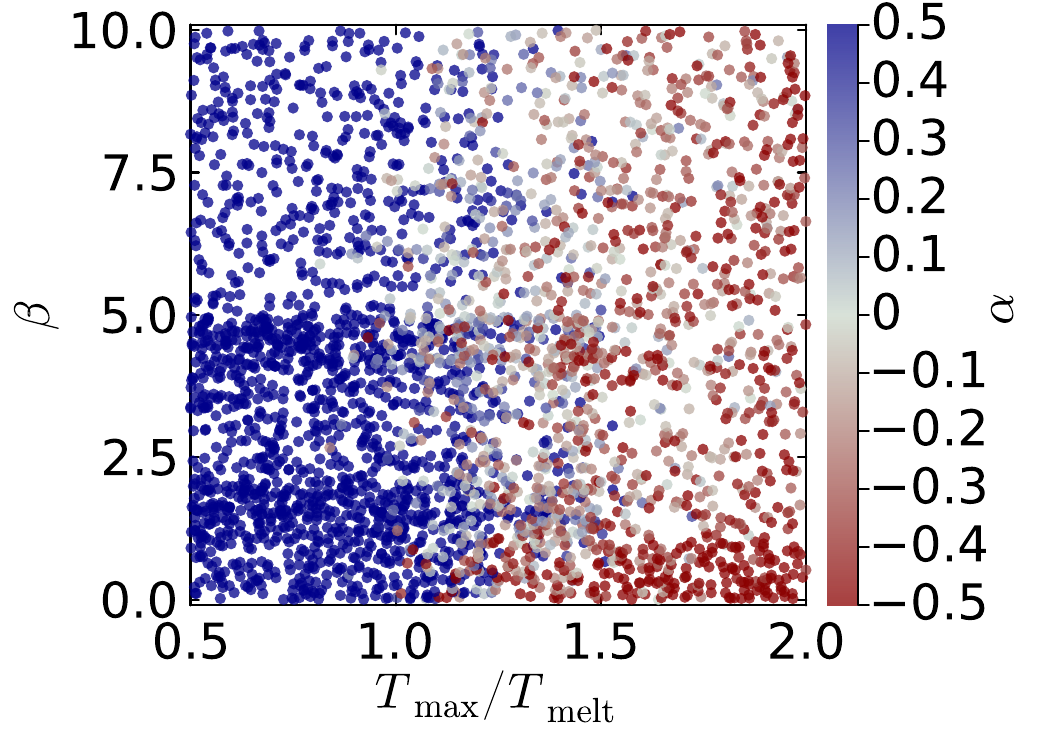}}
\end{minipage}
\caption{\label{fig:ctn_homogeneity_hyperuniformity}
The homogeneity metrics introduced in Section~\ref{sec:homogeneity} are plotted against the algorithm inputs $T_\mathrm{max}$ and $\beta$ for networks generated from the initial \textbf{ctn} network. Red colors indicate high disorder with respect to the corresponding metric. \add{The metrics of 3750 networks are displayed.}
\textbf{a} The \replace{coordinated neighbor distance $r_\mathrm{c}$}{nearest-neighbor distance $r_\mathrm{nn}$} captures the average distance to the nearest connected neighbor.
\textbf{b} The \replace{uncoordinated neighbor distance}{nearest-uncoordinated-neighbor distance}  $r_\mathrm{u}$ measures the average distance to the nearest neighbor that is not coordinated with the reference vertex.
\textbf{c} The critical pore radius $\delta_\mathrm{c}$ is the radius of the largest sphere that can percolate a network's void space.
\textbf{d} The hyperuniformity metric $\alpha$ is positive for hyperuniform networks.
}
\end{figure*}

Figure \ref{fig:ctn_bond_length_angle}
suggests that well-relaxed, disordered networks are obtained at low values of $\beta$ and when heated above the melting transition $T_\mathrm{max}\inl\gtrsim T_\mathrm{melt}$. The two-body bond-stretching interaction governing the system allows for a sufficiently high number of accepted Monte Carlo moves to gradually reduce the strain energy. This interpretation is supported by the homogeneity order metrics \replace{$r_\mathrm{c}$}{$r_\mathrm{nn}$} and $r_\mathrm{u}$ plotted in Figures~\ref{fig:ctn_homogeneity_hyperuniformity}a-b. These metrics capture the average distance to the closest vertex and the distance to the closest vertex with which no bond exists, thereby measuring small-scale clustering (Section~\ref{sec:homogeneity}). The most ordered networks appear at values of $\beta \inl\lesssim 0.5$. 

On the other hand, the critical pore radius $\delta_\mathrm{c}$ and the hyperuniformity metric $\alpha$ seem to contradict the high-order trend at low $\beta$ values (Figures~\ref{fig:ctn_homogeneity_hyperuniformity}c-d). A comparison of Figure~\ref{fig:ctn_homogeneity_hyperuniformity}c with Figure~\ref{fig:ctn_nr_accepted_moves_beta} shows that the critical pore radius $\delta_\mathrm{c}$ increases monotonically with the number of accepted Monte Carlo moves. Similarly, only networks around the melting transition are hyperuniform and exhibit $\alpha > 0$, in addition to crystalline and hyperuniform unmelted networks. These observations demonstrate that the Keating strain energy, which governs the Monte Carlo evolution, contains only local and direct space interactions. The critical pore radius measures pore sizes at all length scales, including those that extend beyond the immediate surroundings of individual vertices. Large pores hinder hyperuniformity, which describes large-scale homogeneity \cite{zhang_2017}. Algorithms in reciprocal space, such as collective coordinate control,  are necessary to optimize for hyperuniformity \cite{uche2006}. Nevertheless, our extended WWW algorithm generates hyperuniform disordered networks, particularly when $1.0 \inl\lesssim T_\mathrm{max}/T_\mathrm{melt} \inl\lesssim 1.3$. 

\begin{figure*}
\begin{minipage}{0.48\textwidth}
\centering
\scaledinset{l}{0.01}{b}{.91}{\textbf{a}}{\includegraphics[width=\textwidth]{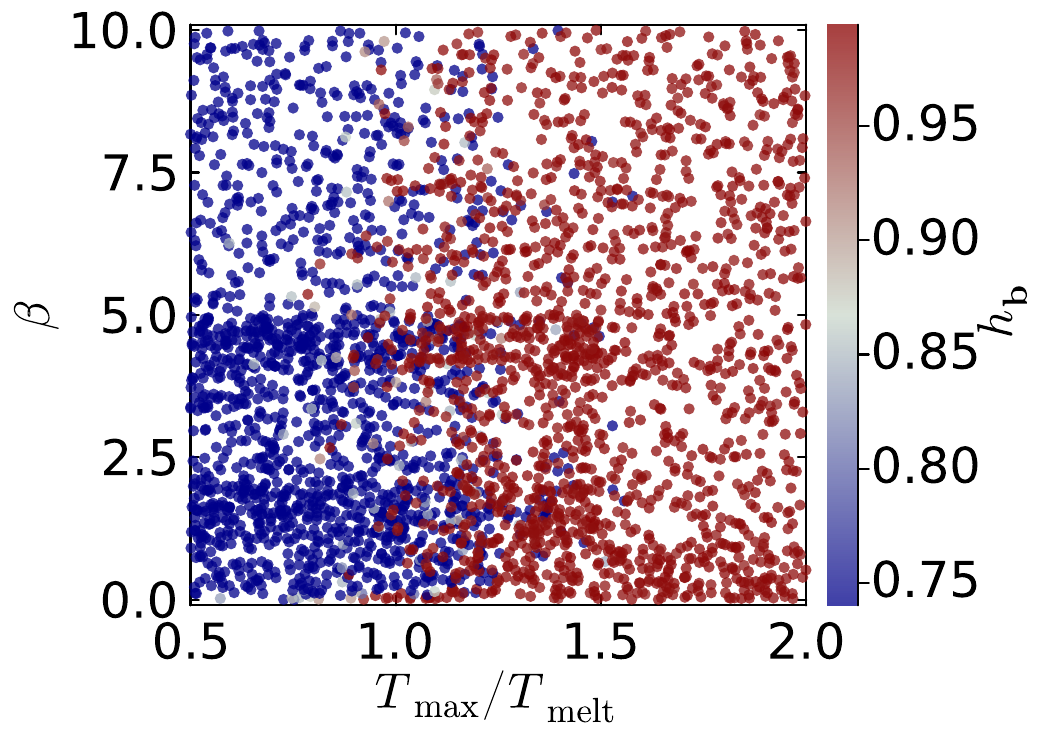}}
\end{minipage}
%
\begin{minipage}{0.48\textwidth}
\centering
\scaledinset{l}{0.01}{b}{.91}{\textbf{b}}{\includegraphics[width=\textwidth]{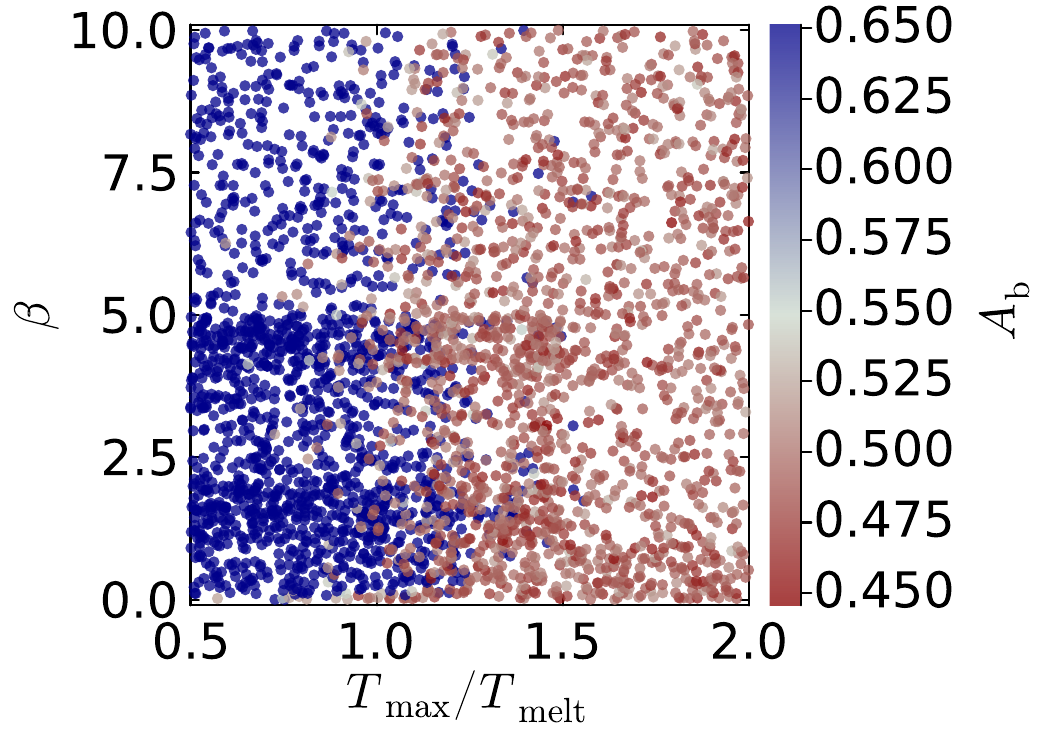}}
\end{minipage}
\begin{minipage}{0.48\textwidth}
\centering
\scaledinset{l}{0.01}{b}{.91}{\textbf{c}}{\includegraphics[width=\textwidth]{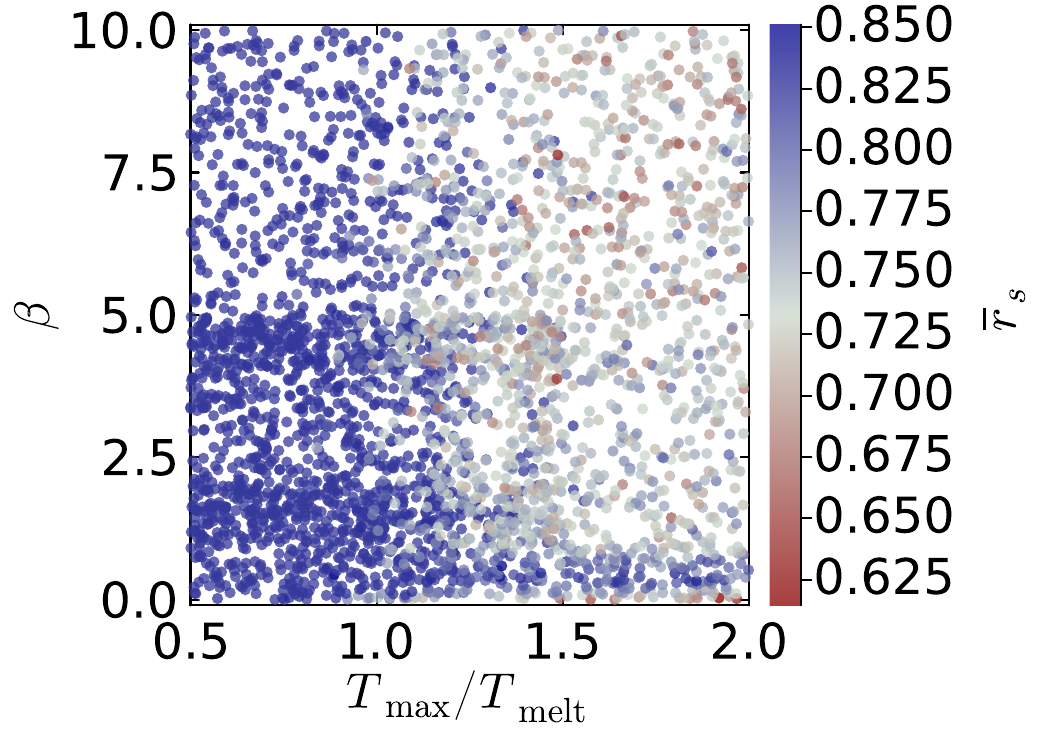}}
\end{minipage}
%
\begin{minipage}{0.48\textwidth}
\centering
\scaledinset{l}{0.01}{b}{.91}{\textbf{d}}{\includegraphics[width=\textwidth]{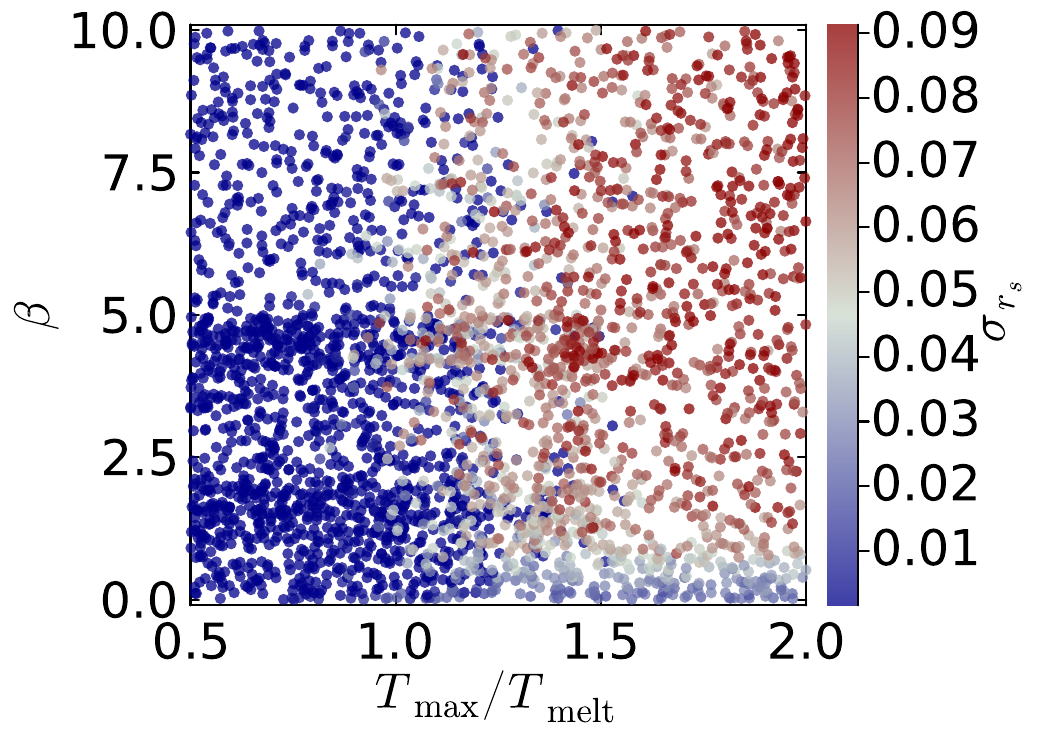}}
\end{minipage}
\caption{\label{fig:ctn_isotropy_topology}
The isotropy and topology metrics of Sections \ref{sec:isotropy}~\ref{sec:topology} as a function of the inputs to the network generation algorithm, $T_\mathrm{max}$ and $\beta$, for networks generated from the initial \textbf{ctn} network. Red colors indicate high disorder with respect to the corresponding metric. \add{The metrics of 3750 networks are displayed.}
\textbf{a} The isotropy metric bond orientation entropy $h_\mathbf{b}$ approaches 1 for isotropic networks.
\textbf{b} The \add{bond} structure factor anisotropy metric $A_\mathrm{b}$ is below $0.6$ for isotropic networks.
\textbf{c} The mean ring radius $\overline{r}_s$  is generally smaller for disordered networks because their vertices tend to cluster, resulting in fewer bonds per ring.
\textbf{d} The ring radius standard deviation $\sigma_{r_s}$ quantifies the variation in the diameters of very strong rings. Small values correspond to little disorder in ring sizes.
}
\end{figure*}

To conclude, we use the isotropy and topology metrics presented in Sections \ref{sec:isotropy}~and~\ref{sec:topology}, respectively, to analyze how the WWW parameters affect the resulting networks. Figures \ref{fig:ctn_isotropy_topology}a-b show how the isotropy metrics $h_\mathbf{b}$ and $A_\mathrm{b}$ fluctuate based on the WWW inputs $\beta$ and $T_\mathrm{max}$. Both metrics indicate the melting transition, switching from anisotropic to isotropic as soon as the first Monte Carlo moves are accepted. The behavior of the topological metrics $\overline{r}_s$ and $\sigma_{r_s}$ aligns with that of the local order metrics  $\sigma_r$, $\sigma_\theta$, \replace{$r_\mathrm{c}$}{$r_\mathrm{nn}$}, and $r_\mathrm{u}$. These metrics exhibit maximal order in the disordered state when there are high numbers of bond switches at low values of $\beta$. In this regime, the rings vary minimally in radius. The large average ring radii observed in this regime indicate homogeneously distributed vertices and bonds. Small ring radii, on the other hand, are caused by vertex clustering.
Based on these observations, we infer that the ring metrics capture small-scale order. Large-scale characteristics are only measured by the critical pore radius and the hyperuniformity metric $\alpha$.

\subsection{Order metric prediction using a neural network}
\label{sec:neural_network}
When using the extended WWW algorithm to target a specific set of order metrics, it is helpful to have a computationally efficient method of predicting the outcome based on the inputs. This eliminates the need for trial-and-error network generation, which takes several minutes per network on a single core of an 11\textsuperscript{th} Gen Intel Core i5 processor running at $\SI{2.4}{\giga\hertz}$. We can write the mapping from algorithm inputs $\textbf{x}_\mathrm{in}$ to order metrics $\textbf{x}_\mathrm{metric}$ as
\begin{align*}
    f{:}\; \mathbb{R}^{N_\mathrm{in}} \to \mathbb{R}^{N_\mathrm{metric}}, \quad \mathbf{x}_\mathrm{in} \mapsto \mathbf{x}_\mathrm{metric},
\end{align*}
where $f$ is the stochastic function describing the action of the extended WWW algorithm. In our case, after fixing the initial network, $N_\mathrm{in}\eq3$ ($\beta$, $T_\mathrm{max}$, $\Delta T$) and $N_\mathrm{metric}\eq42$.\footnote{The list of 42 order metrics is given in Table~\ref{tab:list_of_metrics}, Appendix~\ref{sec:list_of_symbols}.}

\begin{figure}
\centering
\includegraphics[width=0.3 \linewidth]{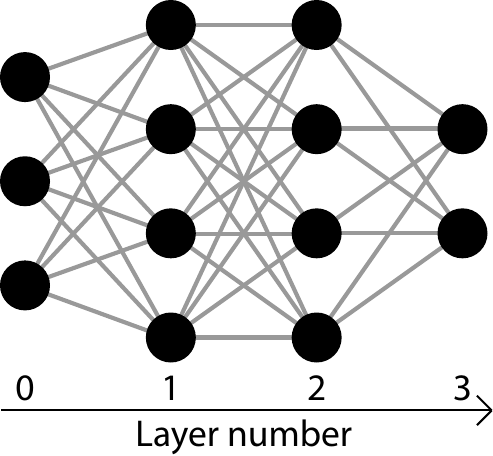}
\caption{\label{fig:neural_network} Example of a three-layer ($N_\mathrm{layers}\eq 3$) feedforward neural network. The input layer is counted as 0. The black circles represent neurons and the gray lines depict the data flowing to the right. Our neural network has $N_\mathrm{in}\eq3$ inputs: $\beta$, $T_\mathrm{max}$, and $\Delta T$. The number of outputs $N_\text{out}$ equals the number of order metrics or PCs that the network is trained on. The depicted network has $N_\text{out}\eq2$ outputs. Between input and output layers, we illustrate two hidden layers, each with $N_\mathrm{neurons}\eq4$ neurons. $N_\mathrm{layers}$, $N_\mathrm{neurons}$, and $N_\text{out}$ are hyperparameters that we optimize.}
\end{figure}

We solved the multivariate regression problem of finding an estimator
\begin{align*}
    f^*{:}\; \mathbf{x}_\mathrm{in} \mapsto \mathbf{x}_\mathrm{metric}^\text{predict}
\end{align*}
of the true function $f$ by training a fully connected feedforward neural network on a dataset of generated networks. We divided the dataset of 3,750 networks, generated from the initial periodic \textbf{ctn}, into three sets: training (70\%), validation (15\%), and testing (15\%). We implemented and trained the network using the PyTorch library \cite{paszke2019}. Figure~\ref{fig:neural_network} illustrates our network architecture. The input and output dimensionalities determine the number of neurons in the 0\textsuperscript{th} input layer and the final output layer,  which has an index of $N_\mathrm{layers}$ \footnote{Following the convention of counting layers after the input layer, we indexed the input layer as zero and the final layer as $N_\mathrm{layers}$.}. Between these two layers, we placed $N_\mathrm{layers}\inl-1$ layers, each with $N_\mathrm{neurons}$ neurons. Each neuron in layers $0$ to $N_\mathrm{layers}\inl-1$ has a rectified linear unit (ReLU) activation function.

\begin{table}
\centering
\begin{tabular}{lll}
\hline
 & \textbf{Order metric} $x_{\mathrm{metric}, i}$ \quad & \textbf{Weight} $w_i$ \\
\hline\rule{-2pt}{1em}
\multirow{5}{*}{\textbf{Network primitives}} \quad
    & $\sigma_r$ & 0.15 \\
    & $\sigma_\theta$ & 0.15 \\
    & $\overline{q}_l$ for all $l\in \{0,1,\dots,12\}$ & $0.05/13$ \\
    & $\sigma_{q_l}$ for all $l\in \{0,1,\dots,12\}$ & $0.05/13$ \\
    & $h_\phi$ & 0.05 \\
\hline\rule{-2pt}{1em}
\multirow{4}{*}{\textbf{Homogeneity}} 
    & $r_\mathrm{nn}$ & 0.05 \\
    & $r_\mathrm{u}$ & 0.05 \\
    & $\delta_\mathrm{c}$ & 0.10 \\
    & $\alpha$ & 0.02 \\
\hline\rule{-2pt}{1em}
\multirow{3}{*}{\textbf{Isotropy}} 
    & $h_\mathbf{b}$ & 0.10 \\
    & $A_\mathrm{v}$ & 0.05 \\
    & $A_\mathrm{b}$ & 0.05 \\
\hline\rule{-2pt}{1em}
\multirow{6}{*}{\textbf{Topology}} 
    & $\overline{Z}$ & 0.01 \\
    & $\sigma_Z$ & 0.01 \\
    & $\overline{s}$ & 0.005 \\
    & $\sigma_s$ & 0.005 \\
    & $\overline{r}_s$ & 0.05 \\
    & $\sigma_{r_s}$ & 0.05 \\
\hline
\end{tabular}
\caption{
\label{tab:weighted_mse}
We use a weighted mean-squared error loss function with given weights for all order metrics to evaluate the neural network's prediction accuracy. These weights are selected to balance the varying ranges of the order metrics and treat different order metric groups equally (see Section~\ref{sec:order_metrics}). }
\end{table}

We use a weighted mean-squared error loss function to evaluate the performance of our neural network and adjust its hyperparameters. The loss function weights $w_i$ are selected to balance the order metric groups and to compensate for their varying value ranges (Table~\ref{tab:weighted_mse}). 
Network accuracy is measured by the average loss on the test dataset. The best-performing network is achieved through preprocessing of the order metric data. First, we scale each order metric data component of $\mathbf{x}_\mathrm{metric}$ to a mean value of 0 and a standard deviation of 1. Then, we reduce its dimensionality using principal component analysis (PCA) \cite{jolliffe2002, schittenkopf1997, tipping2002}. The optimal number of principal components (PCs) $N_\mathrm{PC}$ depends on the level of correlation and noise in the data. The more correlated the dimensions of the data are, the fewer PCs are needed to incorporate a sufficient amount of information. Furthermore, in noisy data, PCs above a certain threshold represent more noise than useful information. Therefore, dimensionality reduction prevents overfitting. There are several methods to estimate the optimal number of PCs. One method is to set a cumulative percentage threshold of total variation, and another is to use cross-validation \cite{jolliffe2002}. We use an unweighted mean-squared error loss function to train on preprocessed order metric data. Then, we select the number of PCs that yields the best performance on the test data, as measured by a weighted mean-squared error loss function.

We perform hyperparameter tuning through random sampling. The hyperparameters $N_\mathrm{layers}$ and $N_\mathrm{neurons}$ determine the network's architecture. Other hyperparameters include the learning rate $\eta$ and the batch size $N_\mathrm{batch}$, which influence the training process \footnote{For more information about neural network training parameters, please refer to the PyTorch documentation \cite{pytorch2025}.}. Furthermore, we train on data with and without preprocessing (see above). For preprocessed data, we vary the number of PCs $N_\mathrm{PC}$. The lowest test loss is achieved with 
\begin{align*}
    \eta = 0.005 && , &&
    N_\mathrm{batch} = 1 && , &&
    N_\mathrm{layers} = 4 && , &&
    N_\mathrm{neurons} = 72 && \text{and} &&
    N_\mathrm{PC} = 10 \quad .
\end{align*}

\begin{figure*}
\begin{minipage}{0.48\textwidth}
\centering
\scaledinset{l}{0.01}{b}{1.00}{\textbf{a}}{\includegraphics[width=\textwidth]{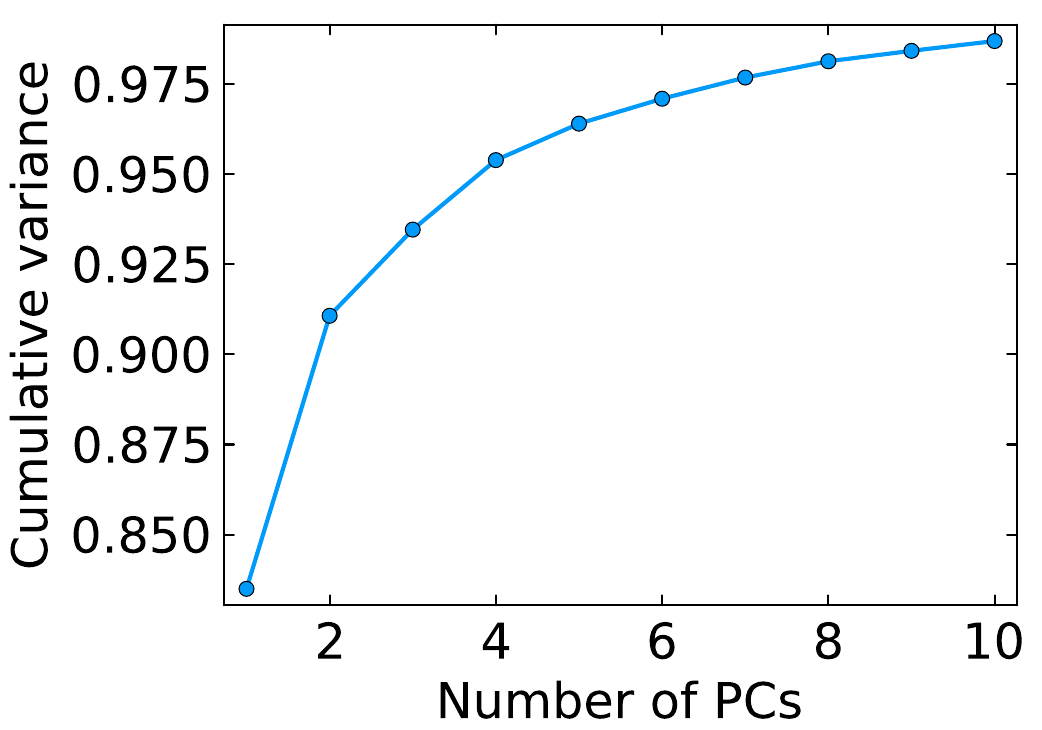}}
\end{minipage}
%
\begin{minipage}{0.48\textwidth}
\centering
\scaledinset{l}{0.01}{b}{.83}{\textbf{b}}{\includegraphics[width=\textwidth]{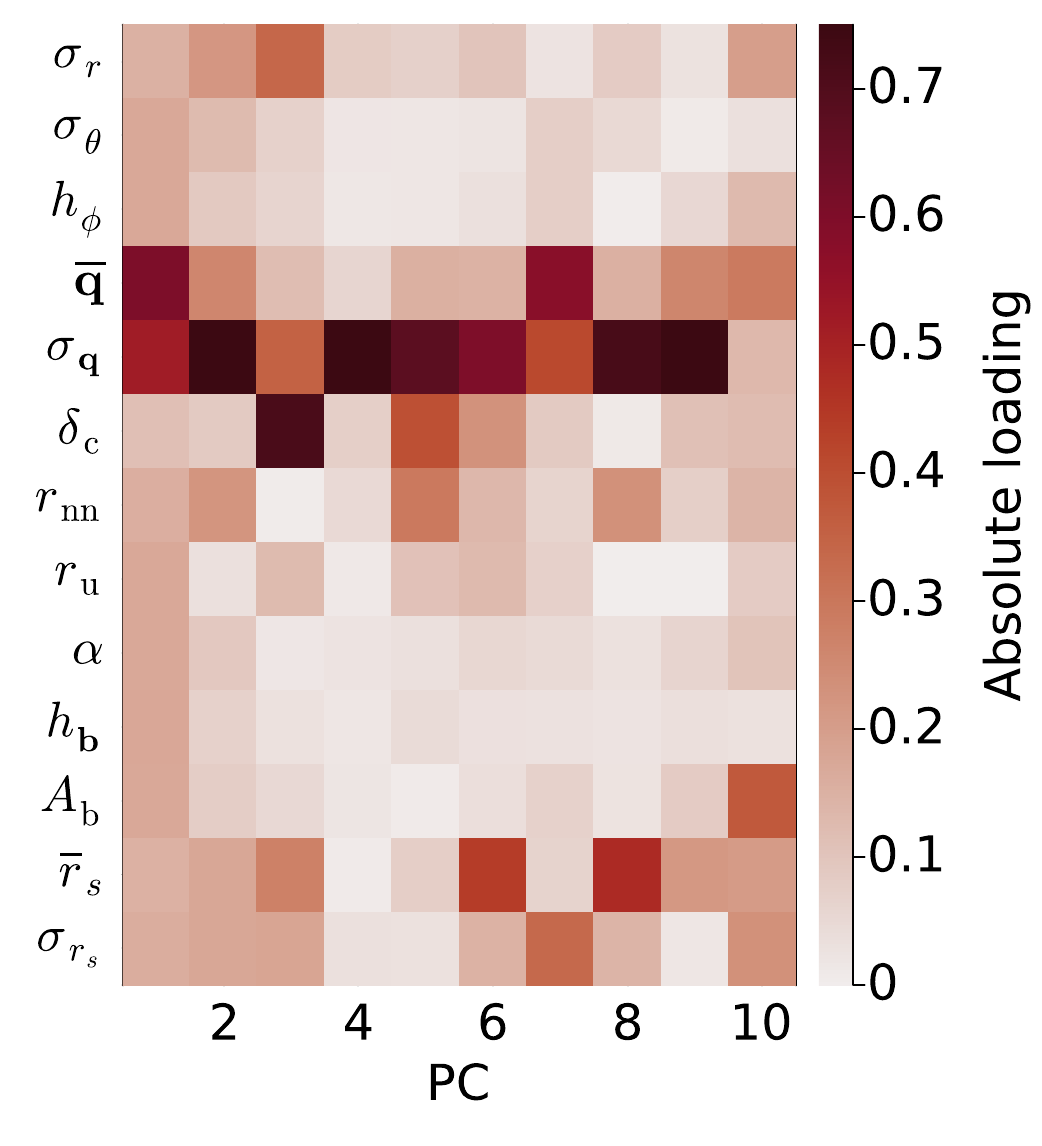}}
\end{minipage}
\caption{\label{fig:pca_info}
Training the neural network using ten PCs representing 42 order metrics yields higher prediction accuracy than using the full dataset.
\textbf{a} The first PC explains 83\% of the variance in the order metrics. The first ten PCs explain 99\% of the variance and produce the best network performance, suggesting that the remaining 1\% consist primarily of noise.
\textbf{b} Absolute values of the PCA loadings when considering all order metrics. For the Steinhardt parameters, we depict the norm of the vectors $\mathbf{\overline{q}}$ and $\mathbf{\sigma_{q}}$, whose components are the loadings of $\overline{q}_l$ and $\sigma_{q_l}$, $l\in \{0, 1, ..., 12\}$. Combining 13 metrics into one norm results in their prominent magnitudes in the heatmap. Besides $\mathbf{\overline{q}}$ and $\mathbf{\sigma_{q}}$, the first PC is loaded with approximately equal contributions from all order metrics. Higher PCs reveal correlations between the order metrics.  For example, the second PC shows correlations between the small-length-scale metrics. Since $Z$ and $\sigma_z$ remain constant throughout the dataset, their vanishing PC loadings are not shown.
}
\end{figure*}

Figure~\ref{fig:pca_info} shows the cumulative explained variance as a function of the number of PCs, along with the absolute values of their order metric loadings. The first PC explains 83\% of the total variance in the order metrics (Figure~\ref{fig:pca_info}a) and has nearly equal loadings for all metrics (Figure~\ref{fig:pca_info}b), indicating strong correlations. The only deviations are the loadings of the Steinhardt parameters and their standard deviations. For clarity, we combine them into single values, denoted by $\mathbf{\overline{q}}$ and $\mathbf{\sigma_{q}}$. These are the norms of the loading vectors in the two subspaces spanned by the $q_l$ and $\sigma_{q_l}$, $l\in \{0, 1, ..., 12\}$. We find that $\sigma_{q_4}$, $\sigma_{q_7}$, $\sigma_{q_{10}}$, and $\sigma_{q_{11}}$ predominantly contribute to the high loadings of $\mathbf{\sigma_{q}}$ (Figure~\ref{fig:pca_info_steinhardt} in Appendix~\ref{sec:si_pca_order_metrics}). For the \textbf{ctn} network, these metrics indicate the transition from the crystalline to the disordered phase. Higher-order PCs can be assigned to groups of metrics representing different network features. The second PC contains information about small-length-scale features, and the third PC explains intermediate-scale characteristics. The fourth PC contains metrics related to Steinhardt's local bond order parameters (Figure~\ref{fig:pca_info_steinhardt} in Appendix~\ref{sec:si_pca_order_metrics}), while higher PCs focus on rings and isotropy.

\begin{figure*}
\begin{minipage}{0.48\textwidth}
\centering
\scaledinset{l}{0.01}{b}{.91}{\textbf{a}}{\includegraphics[width=\textwidth]{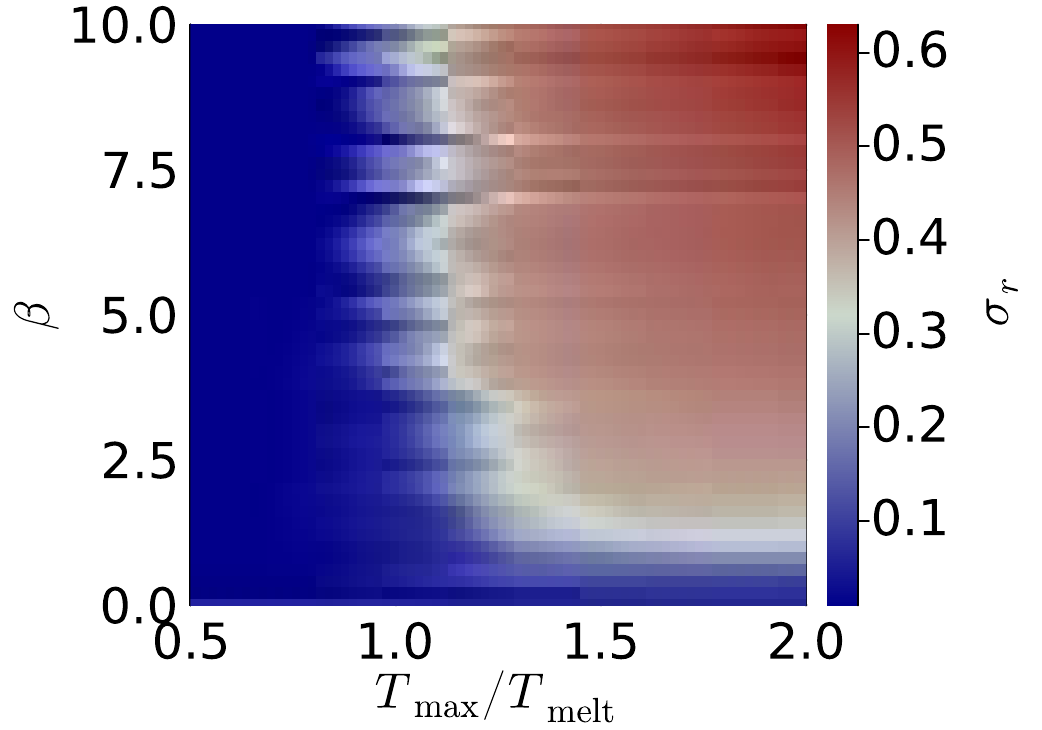}}
\end{minipage}
%
\begin{minipage}{0.48\textwidth}
\centering
\scaledinset{l}{0.01}{b}{.91}{\textbf{b}}{\includegraphics[width=\textwidth]{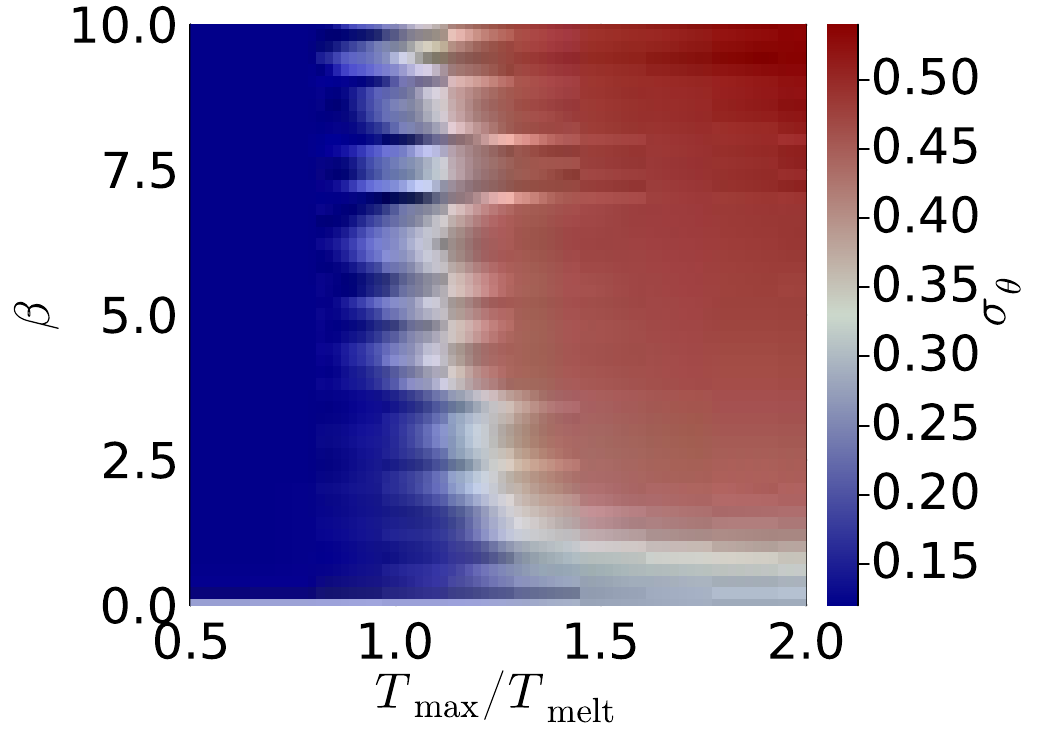}}
\end{minipage}
\caption{\label{fig:ctn_predict_primitives}
Neural network predictions of the order metrics describing primitives for networks generated from the initial \textbf{ctn} network. The noise in the transition from crystalline state to melted networks originates from energy fluctuations in the definition of the melting temperature (Equation~\eqref{eqn:t_melt}). An intermediate temperature gradient $\Delta T \eq T_\mathrm{max}/2$ was set because it has a negligible effect on the order metrics (see Section~\ref{sec:effect_www_parameters}).
\textbf{a} The bond length standard deviation prediction can be compared to the actual data in Figure~\ref{fig:ctn_bond_length_angle}a. 
\textbf{b} The bond angle standard deviation prediction can be compared to the actual data in Figure~\ref{fig:ctn_bond_length_angle}b.
}
\end{figure*}

\begin{figure*}
\begin{minipage}{0.48\textwidth}
\centering
\scaledinset{l}{0.01}{b}{.91}{\textbf{a}}{\includegraphics[width=\textwidth]{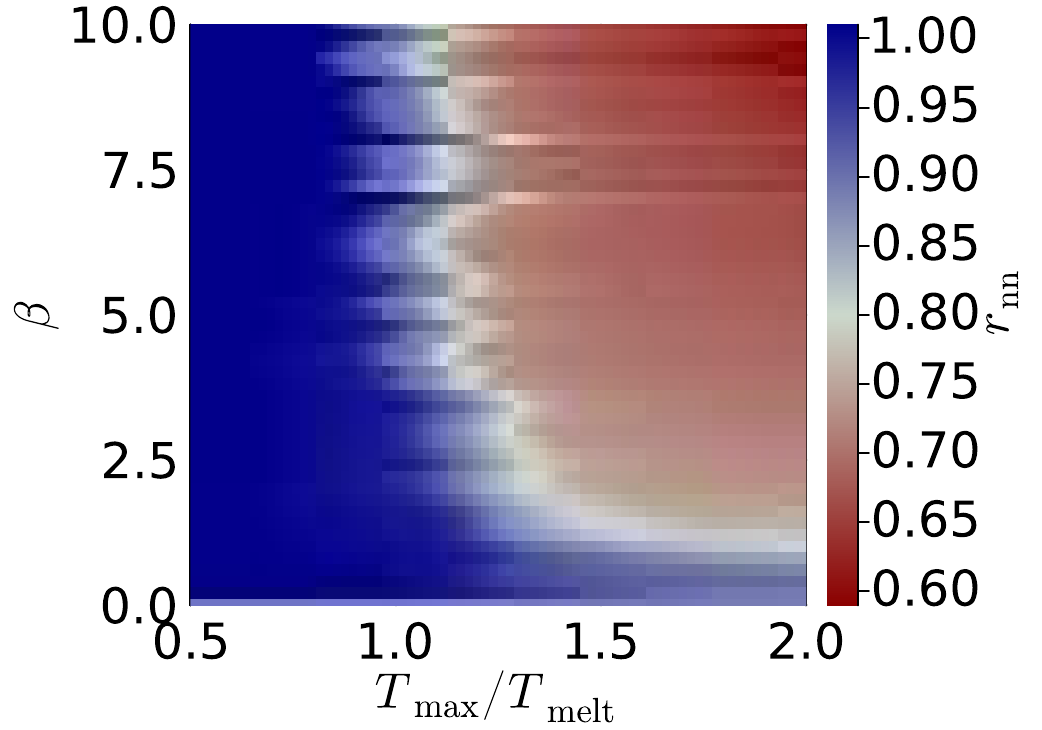}}
\end{minipage}
%
\begin{minipage}{0.48\textwidth}
\centering
\scaledinset{l}{0.01}{b}{.91}{\textbf{b}}{\includegraphics[width=\textwidth]{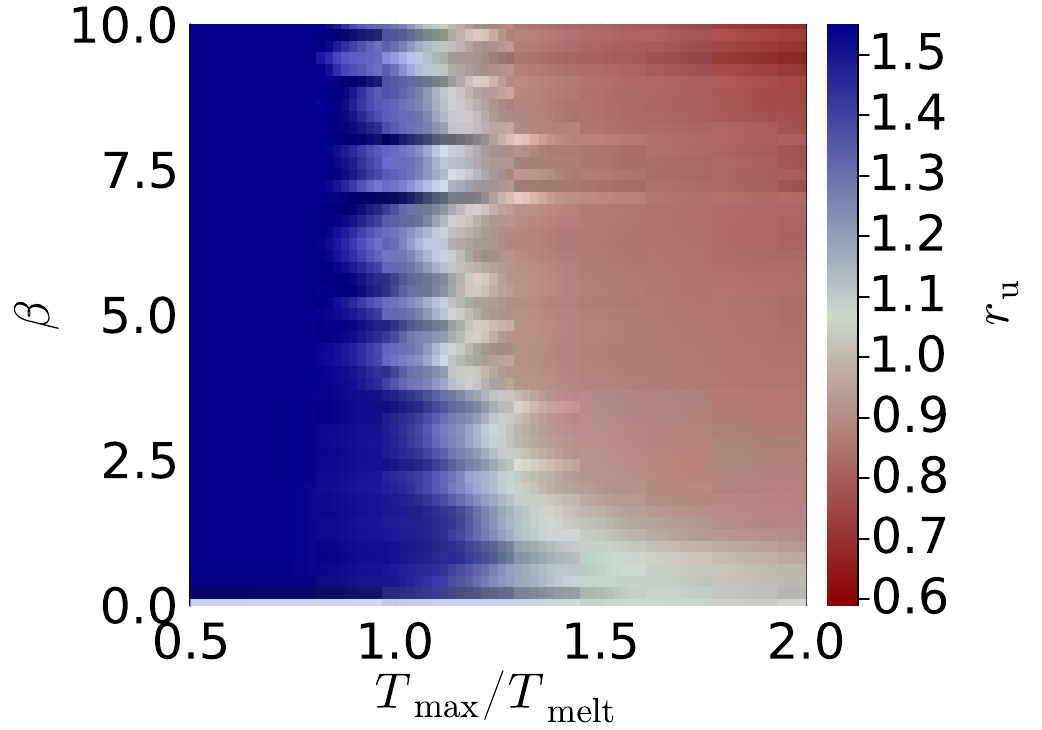}}
\end{minipage}
\begin{minipage}{0.48\textwidth}
\centering
\scaledinset{l}{0.01}{b}{.91}{\textbf{c}}{\includegraphics[width=\textwidth]{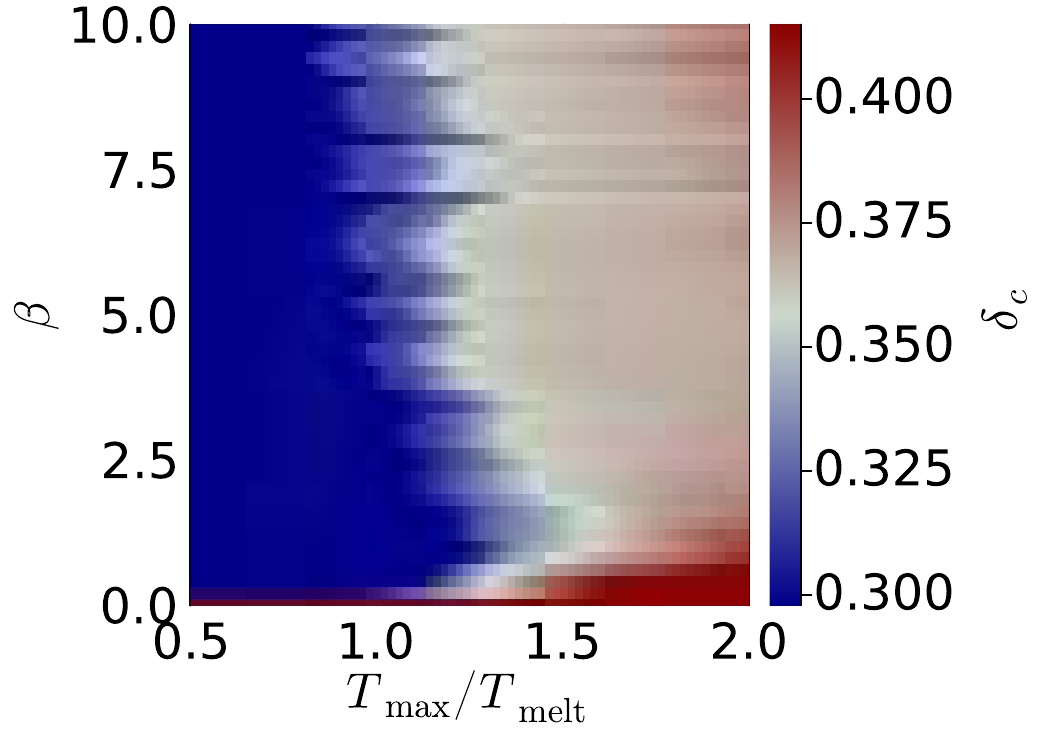}}
\end{minipage}
%
\begin{minipage}{0.48\textwidth}
\centering
\scaledinset{l}{0.01}{b}{.91}{\textbf{d}}{\includegraphics[width=\textwidth]{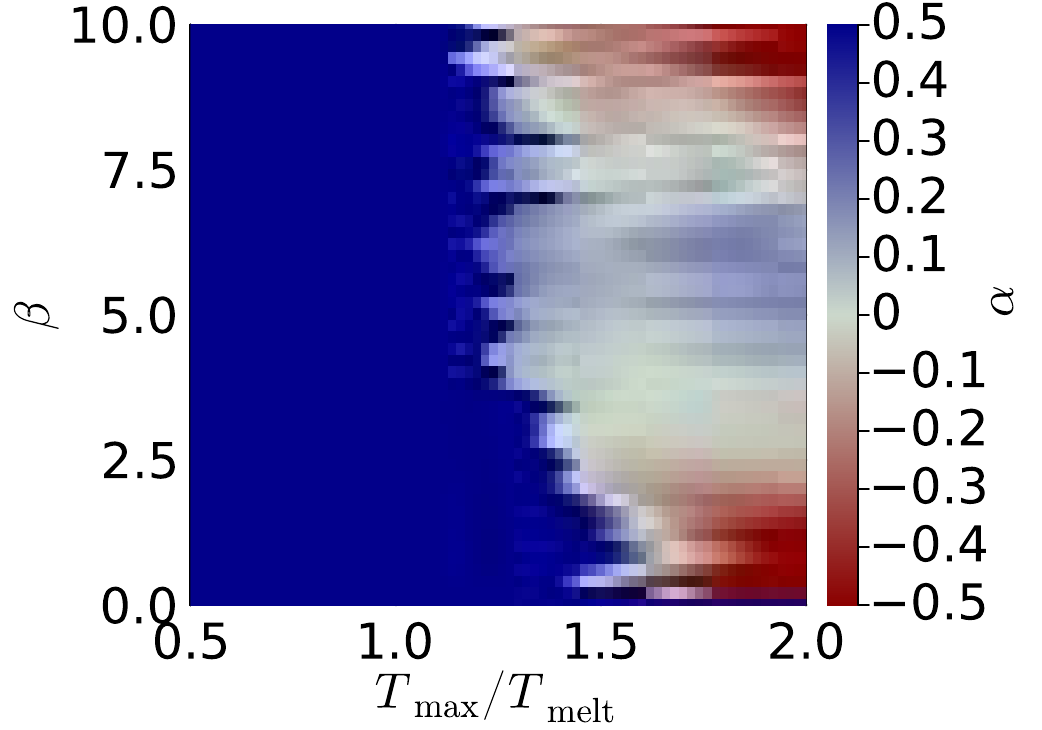}}
\end{minipage}
\caption{\label{fig:ctn_predict_homogeneity}
Neural network predictions of the order metrics describing homogeneity for networks generated from the initial \textbf{ctn} network. The noise in the transition from crystalline state to melted state originates from energy fluctuations in the definition of the melting temperature (Equation~\eqref{eqn:t_melt}). An intermediate temperature gradient $\Delta T \eq T_\mathrm{max}/2$ was set because it has a negligible effect on the order metrics (see Section~\ref{sec:effect_www_parameters}).
\textbf{a} The \replace{vertex homogeneity}{nearest-neighbor distance} prediction can be compared to the actual data in Figure~\ref{fig:ctn_homogeneity_hyperuniformity}a.
\textbf{b} The \replace{uncoordinated neighbor distance}{nearest-uncoordinated-neighbor distance}  prediction can be compared to the actual data in Figure~\ref{fig:ctn_homogeneity_hyperuniformity}b.
\textbf{c} The critical pore radius prediction can be compared to the actual data in Figure~\ref{fig:ctn_homogeneity_hyperuniformity}c.
\textbf{d} The hyperuniformity metric $\alpha$ prediction can be compared to the actual data in Figure~\ref{fig:ctn_homogeneity_hyperuniformity}d.
}
\end{figure*}

\begin{figure*}
\begin{minipage}{0.48\textwidth}
\centering
\scaledinset{l}{0.01}{b}{.91}{\textbf{a}}{\includegraphics[width=\textwidth]{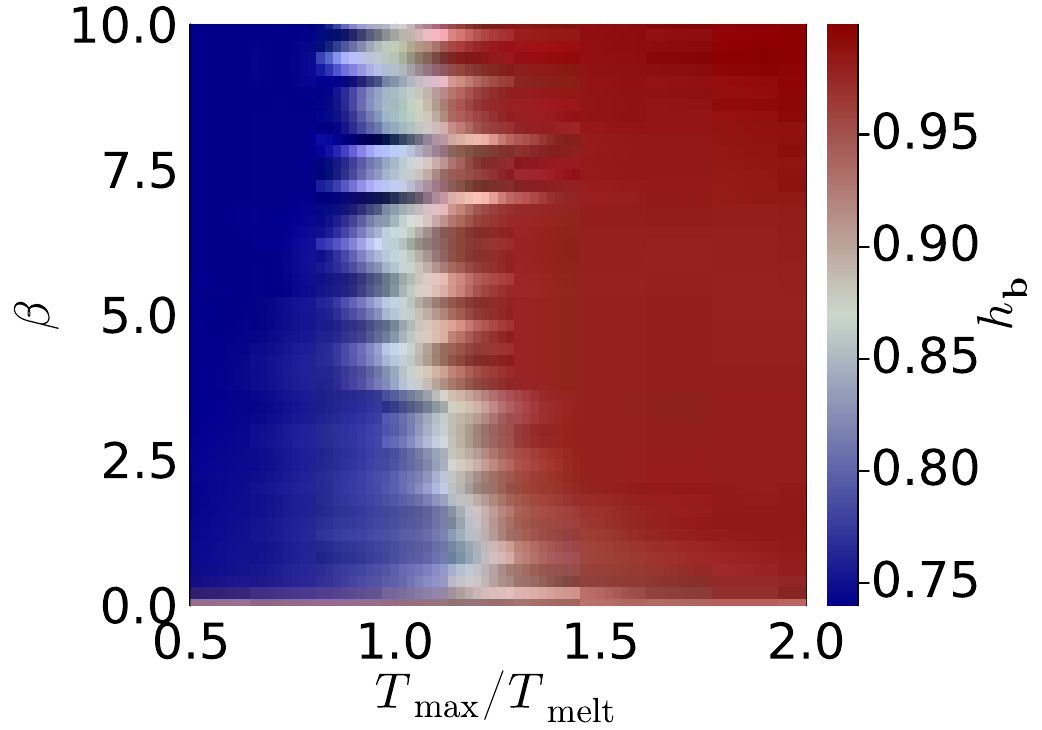}}
\end{minipage}
%
\begin{minipage}{0.48\textwidth}
\centering
\scaledinset{l}{0.01}{b}{.91}{\textbf{b}}{\includegraphics[width=\textwidth]{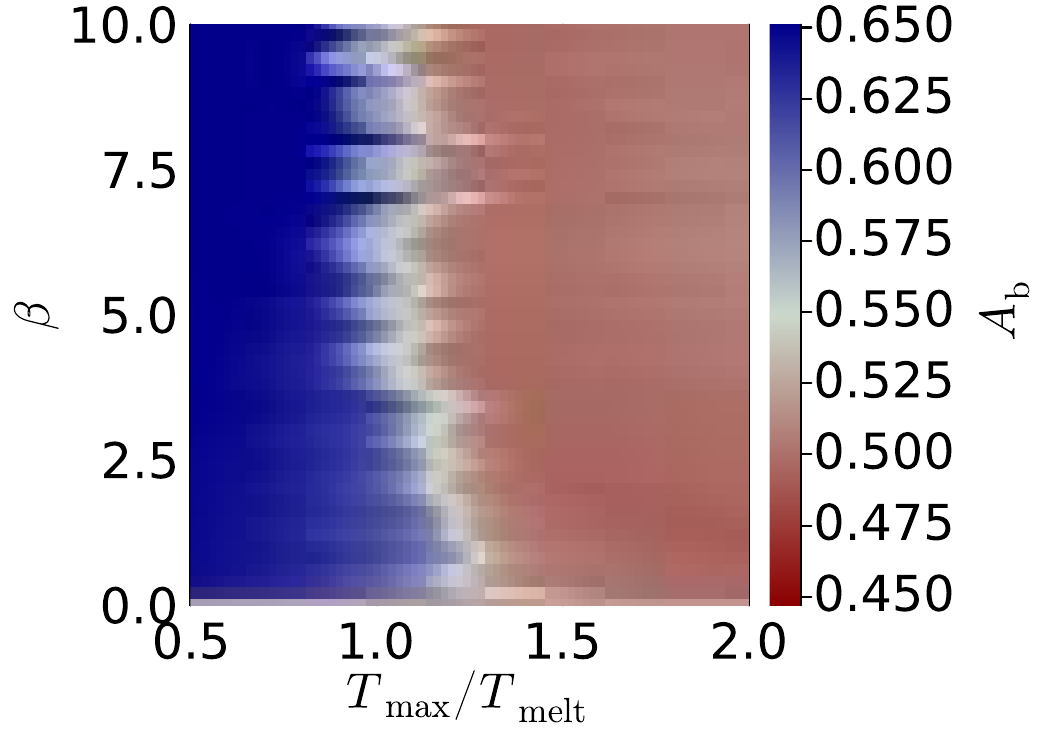}}
\end{minipage}

\begin{minipage}{0.48\textwidth}
\centering
\scaledinset{l}{0.01}{b}{.91}{\textbf{c}}{\includegraphics[width=\textwidth]{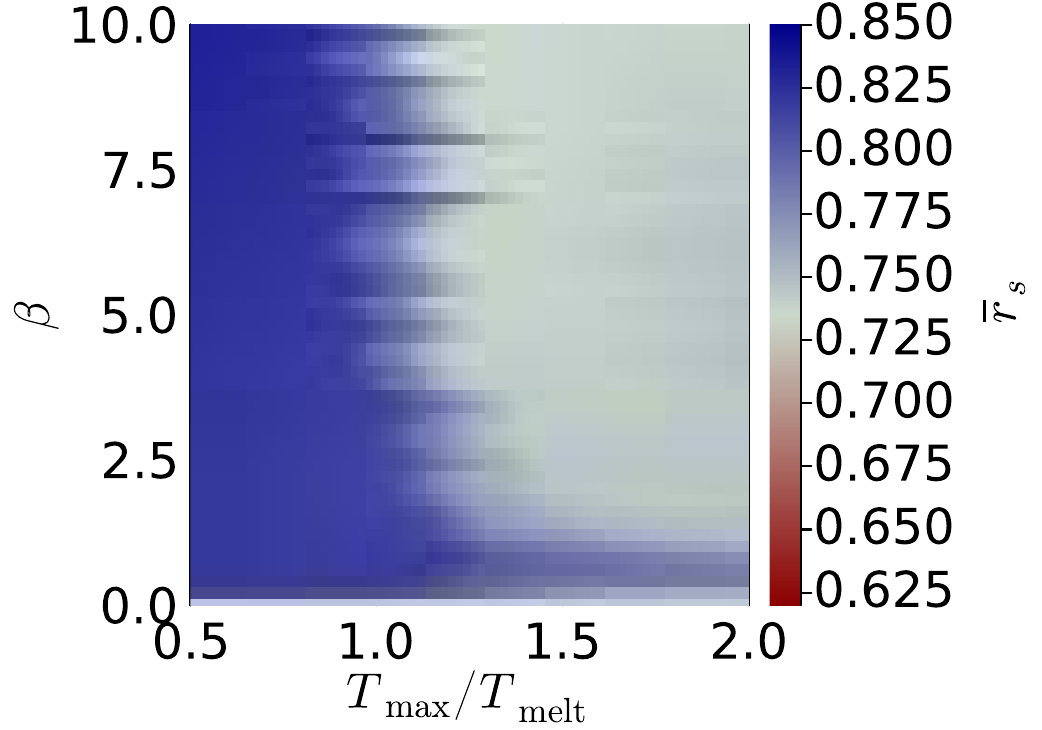}}
\end{minipage}
%
\begin{minipage}{0.48\textwidth}
\centering
\scaledinset{l}{0.01}{b}{.91}{\textbf{d}}{\includegraphics[width=\textwidth]{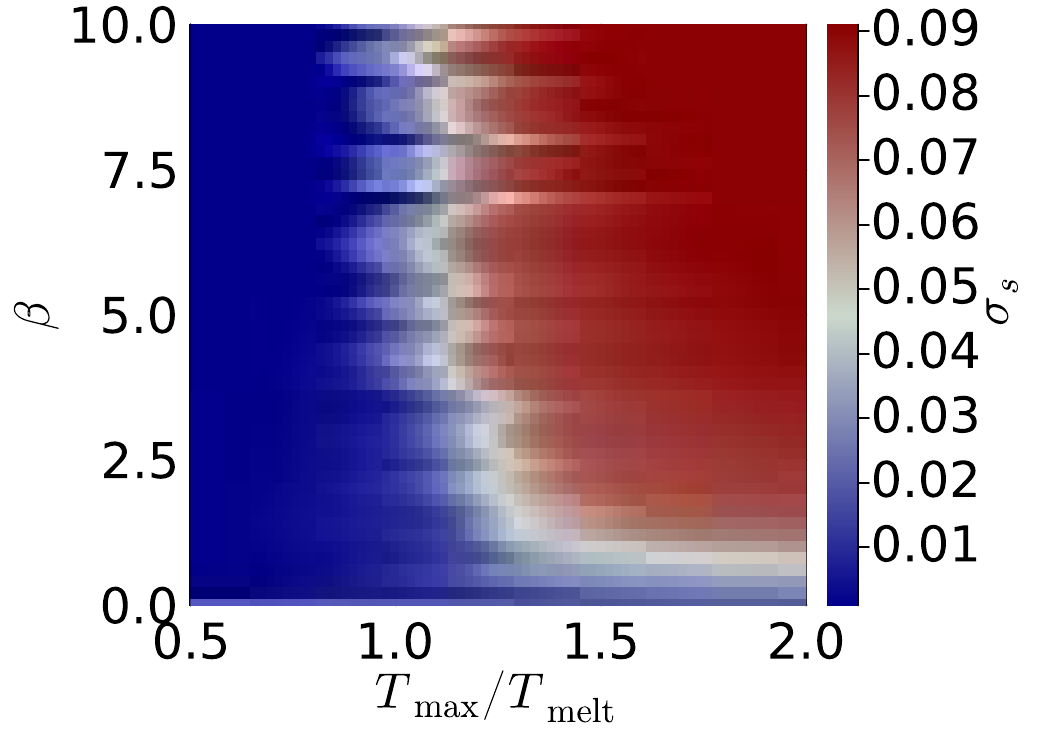}}
\end{minipage}
\caption{\label{fig:ctn_predict_isotropy_topology}
Neural network predictions of the order metrics describing isotropy and topology for networks generated from the initial \textbf{ctn} network. The noise in the transition from crystalline state to melted state originates from energy fluctuations in the definition of the melting temperature Equation~\eqref{eqn:t_melt}). An intermediate temperature gradient $\Delta T \eq T_\mathrm{max}/2$ was set because it has a negligible effect on the order metrics (see Section~\ref{sec:effect_www_parameters}).
\textbf{a} The bond orientation entropy prediction can be compared to the actual data in Figure~\ref{fig:ctn_isotropy_topology}a. 
\textbf{b} The \add{bond} structure factor anisotropy metric prediction can be compared to the actual data in Figure~\ref{fig:ctn_isotropy_topology}b.
\textbf{c} The mean ring radius prediction can be compared to the actual data in Figure~\ref{fig:ctn_isotropy_topology}c.
\textbf{d} The ring radius standard deviation prediction can be compared to the actual data in Figure~\ref{fig:ctn_isotropy_topology}d. }
\end{figure*}

Figures \ref{fig:ctn_predict_primitives}, \ref{fig:ctn_predict_homogeneity}, and \ref{fig:ctn_predict_isotropy_topology} show the order metric predictions of the tuned and trained neural network. Comparing these figures with the true order metrics in Figures \ref{fig:ctn_bond_length_angle}, \ref{fig:ctn_homogeneity_hyperuniformity}, and \ref{fig:ctn_isotropy_topology} shows that the order and disorder regimes are accurately reproduced for all metrics. In all figures, the melting transition at $T_\mathrm{max} \inl\approx T_\mathrm{melt}$ is clearly visible. As with the true data, the metrics can be divided according to their behavior in the $T_\mathrm{max} \inl\gtrsim T_\mathrm{melt}$ and $\beta \inl\lesssim 0.5$ input parameter regime. Order metrics that measure small length scales indicate high order, while large-scale homogeneity metrics $\delta_\mathrm{c}$ and $\alpha$ indicate high disorder. An interesting intermediate regime exists for $\beta \inl\lesssim 1.5$ and $ T_\mathrm{max}\inl\approx 1.5 T_\mathrm{melt}$. In this regime, networks are expected to become disordered and isotropic yet remain hyperuniform.
\add{Furthermore, for $2.5\inl\lesssim \beta \inl\lesssim 7.5$ and $ T_\mathrm{max}\inl\gtrsim 1.3 T_\mathrm{melt}$, the neural network predicts networks that are slightly hyperuniform with $0\inl\lesssim \alpha \inl\lesssim 0.2$. A comparison with Figure~\ref{fig:ctn_homogeneity_hyperuniformity}d suggests that this prediction is the effect of strong scattering of $\alpha$ values in this parameter regime.}

The coefficient of determination is a measure of a neural network's prediction accuracy. Since the output of the extended WWW algorithm exhibits strong statistical fluctuations that hinder accurate predictions, the coefficient quantifies the variance in the order metrics $\mathbf{x}_\mathrm{metric}$ for the same input values $\mathbf{x}_\mathrm{in}$.
The coefficient of determination \cite{hughes1971}
\begin{align*}
    R^2 = 1 - \frac{\sum_i (x_{\mathrm{metric}, i}-x_{\mathrm{metric}, i}^\text{predict})^2}{\sum_i (x_{\mathrm{metric}, i}-\overline{x}_{\mathrm{metric}})^2}
\end{align*}
represents the proportion of the variance in the data explained by the prediction.
In our case, it is calculated for each order metric individually. The index $i$ runs over all networks generated in the test set. $x_{\mathrm{metric}, i}^\text{predict}$ is the predicted order metric for network $i$. The corresponding true metric is $x_{\mathrm{metric}, i}$, and $\overline{x}_{\mathrm{metric}}$ is the mean order metric value in the test dataset. $R^2\eq 1$ for perfect predictions and $R^2 \inl\leq 0$ for random predictions.

\begin{figure}
\centering
\includegraphics[width=0.48 \linewidth]{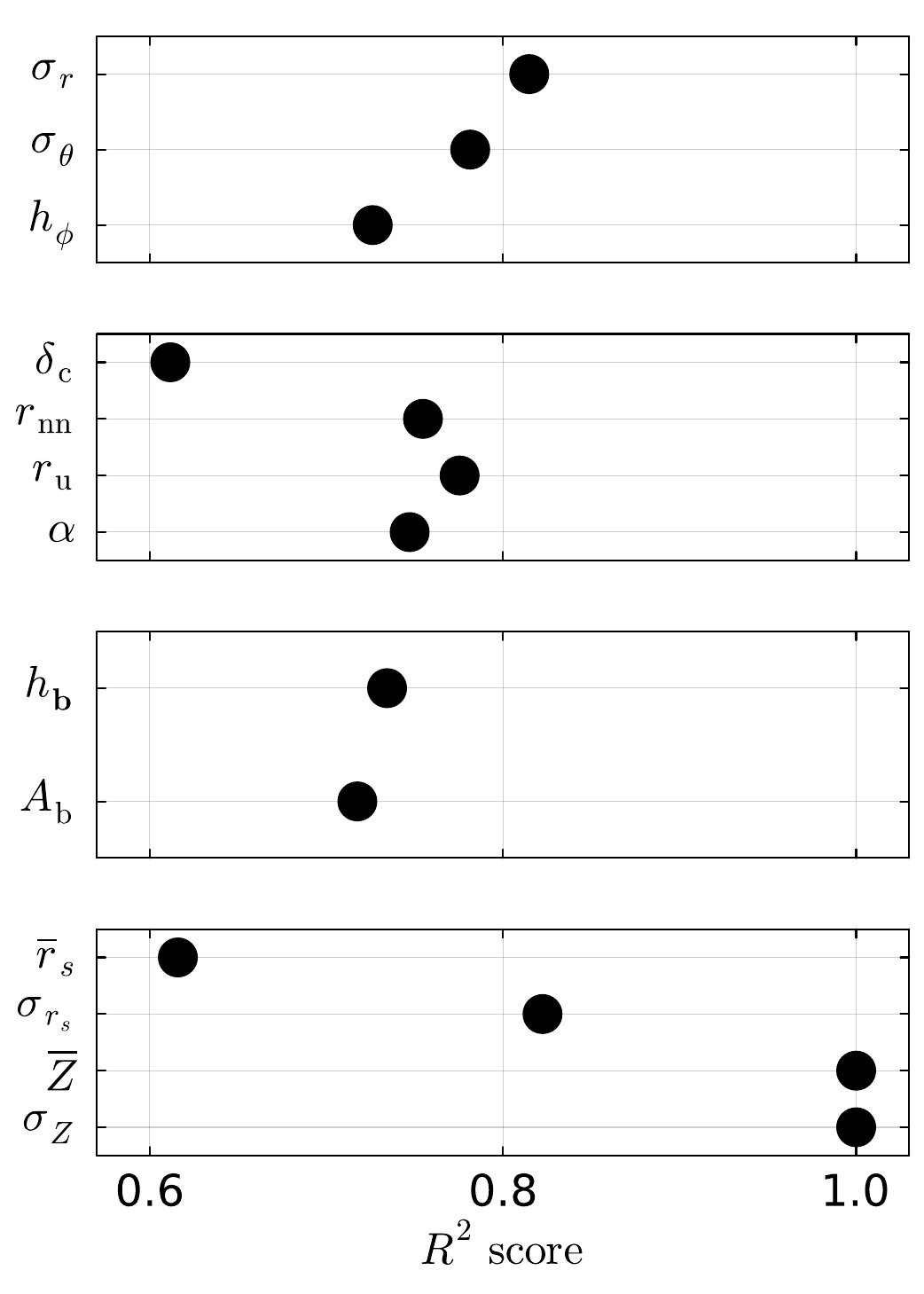}
\caption{\label{fig:ctn_predict_r2} The coefficient of determination $R^2$ measures the accuracy of a neural network's predictions. It indirectly quantifies the variance in the order metrics of the generated networks when using the same generation algorithm inputs. \add{Here, $R^2$ is calculated based on the true and predicted metrics of 562 networks contained in the test dataset.}}
\end{figure}

Figure~\ref{fig:ctn_predict_r2} plots $R^2$ for all considered order metrics. Since the coordination number is constant for all networks, $R^2\eq 1$ for the corresponding metrics $\overline{Z}$ and $\sigma_Z$. In addition to the coordination number metrics, the neural network achieves maximum determination with the small-scale order metrics $\sigma_r$, $\sigma_\theta$, and $\sigma_{r_s}$.
The high values for \replace{$R^2(\sigma_r)\eq 0.82$}{$R^2(\sigma_r)\eq 0.81$} and \replace{$R^2(\sigma_\theta)\eq 0.79$}{$R^2(\sigma_\theta)\eq 0.78$} are explained by the fact that bond lengths and angles are directly controlled by the generalized Keating energy (Equation~\eqref{eqn:keating_generalized}), which governs the network evolution. However, the high degree of control over disorder in ring radii, quantified by \replace{$R^2(\sigma_{r_s})\eq 0.83$}{$R^2(\sigma_{r_s})\eq 0.82$}, is less evident, especially since mean ring radii exhibit low determination with $R^2(\overline{r}_s)\eq 0.62$. This suggests little correlation between $\overline{r}_s$ and $\sigma_{r_s}$. This is supported by the relatively independent PC loadings of the two variables (Figure~\ref{fig:pca_info}b).

Besides the mean ring radius, the critical pore radius has the lowest coefficient of determination \replace{$R^2(\delta_\mathrm{c})\eq 0.6$}{$R^2(\delta_\mathrm{c})\eq 0.61$}. This metric captures homogeneity at all length scales and is sensitive to pore opening. Since the Keating energy does not control this process, there is significant variance in $\delta_\mathrm{c}$ despite identical input parameters. This leads to inaccurate neural network predictions.
However, we have now developed the necessary tools to reproduce the biological networks introduced in Section~\ref{sec:biophotonic}.

\subsection{Reproducing disordered biophotonic networks}
\label{sec:reproducing_biophotonic}
In this section, we identify generated networks that resemble the biological geometries introduced in Section~\ref{sec:biophotonic} by comparing their respective order metrics, as discussed in Section~\ref{sec:order_metrics}. 
Section~\ref{sec:neural_network} assessed the neural network's order metric prediction accuracy using a weighted mean square error loss function with weights $w_i$, provided in Table~\ref{tab:weighted_mse}. We define an \textit{order distance metric} between two networks with the same functional form to quantify their structural  similarity,
\begin{align}
d_\mathrm{order}=\left[ 
\sum_{i=1}^{N_\mathrm{metric}}
w_i\left(x_{\mathrm{metric}, i}^\text{net 2} -x_{\mathrm{metric}, i}^\text{net 1} \right)^2 
\right]^{\frac{1}{2}} 
\quad .
\label{eqn:order_distance_metric}
\end{align}

\begin{figure*}
\begin{minipage}{0.48\textwidth}
\centering
\scaledinset{l}{0.01}{b}{1.05}{\textbf{a}}{\includegraphics[width=0.6\textwidth]{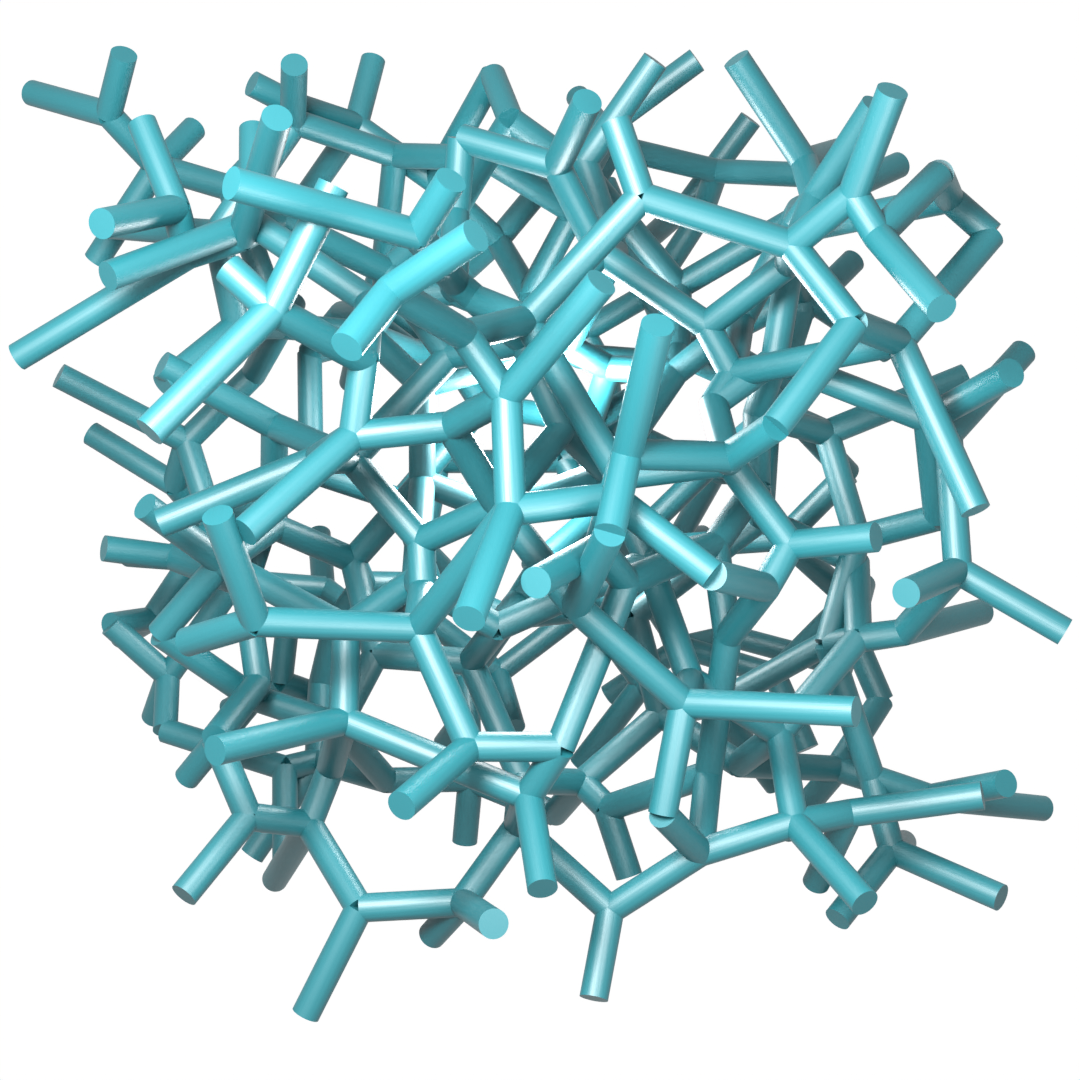}}
\end{minipage}
%
\begin{minipage}{0.48\textwidth}
\centering
\scaledinset{l}{0.01}{b}{.72}{\textbf{b}}{\includegraphics[width=\textwidth]{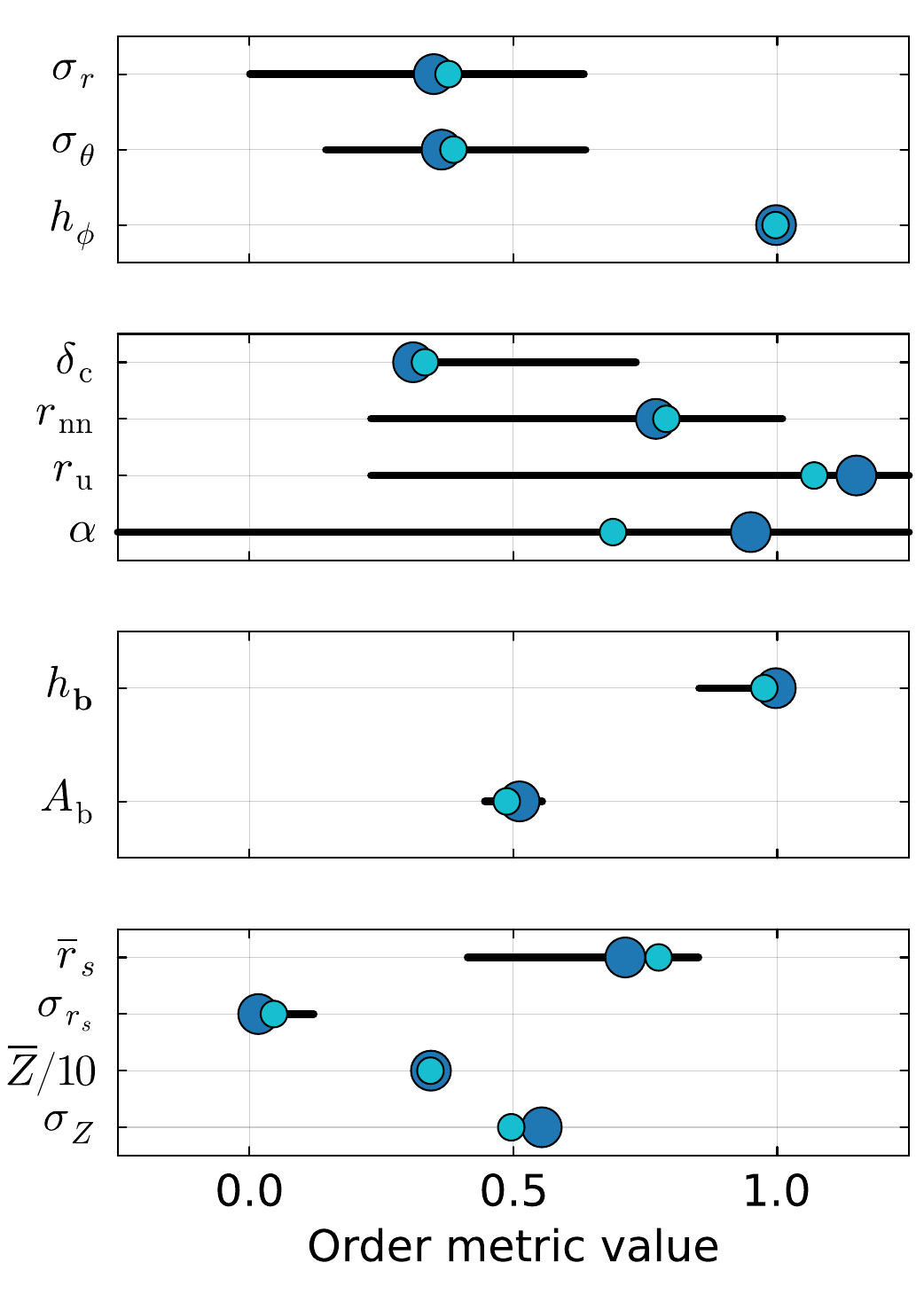}}
\end{minipage}
\caption{\label{fig:pachy_ctn_generated}
\textbf{a} The network with the shortest order metric distance to the PCM blue biophotonic network (Figure~\ref{fig:bio_periodic_coord_nr}a) was generated from an initial \textbf{ctn} network.
\textbf{b} The order metric values of the network generated in \textbf{a} (small light blue markers) closely match those of the PCM blue biophotonic network (large dark blue markers). The black bars represent the order metrics ranges achieved by modifying the initial \textbf{ctn} network with at least ten accepted Monte Carlo moves.
}
\end{figure*}

We now identify the generated networks with the smallest order distance metric compared to biological networks. Figure~\ref{fig:pachy_ctn_generated}a shows the network $\textbf{ctn}_\mathrm{PCM\ blue}$, which is the generated network closest to PCM blue. It is obtained by evolving the \textbf{ctn} network with 224 vertices with WWW parameters
\begin{align*}
\beta=6.3359 && , && T_\mathrm{max}=8.607=1.334 \, T_\mathrm{melt} && \text{and} && \Delta T = 11.717 =1.543 \, T_\mathrm{melt}
\quad .
\end{align*}
The original \textbf{ctn} network was modified with 87 accepted Monte Carlo moves. Since each bond switch involves a chain of four vertices, each vertex was included in approximately 1.5 bond switches.

Figure~\ref{fig:pachy_ctn_generated}b shows the high level of agreement between the biological and the generated network. This is evident through the order metric comparison between $\textbf{ctn}_\mathrm{PCM\ blue}$ and PCM blue. A comparison of Figure~\ref{fig:pachy_ctn_generated}b with Figures \ref{fig:ctn_bond_length_angle}~and~\ref{fig:ctn_homogeneity_hyperuniformity} shows that $\textbf{ctn}_\mathrm{PCM\ blue}$ is an intermediate disorder network with respect to $\sigma_r$ and $\sigma_\theta$, which describe network primitives. However, with respect to the homogeneity metrics $\delta_\mathrm{c}$, $r_\mathrm{u}$, and especially $\alpha$, the order metric values of PCM blue approach the limits of what can be achieved with the extended WWW algorithm. This is because the Keating strain energy does not penalize disorder in these homogeneity metrics. Nevertheless, $\textbf{ctn}_\mathrm{PCM\ blue}$ achieves homogeneity values similar to those of PCM blue.

\begin{figure*}
\begin{minipage}{0.48\textwidth}
\centering
\scaledinset{l}{0.01}{b}{1.05}{\textbf{a}}{\includegraphics[width=0.6\textwidth]{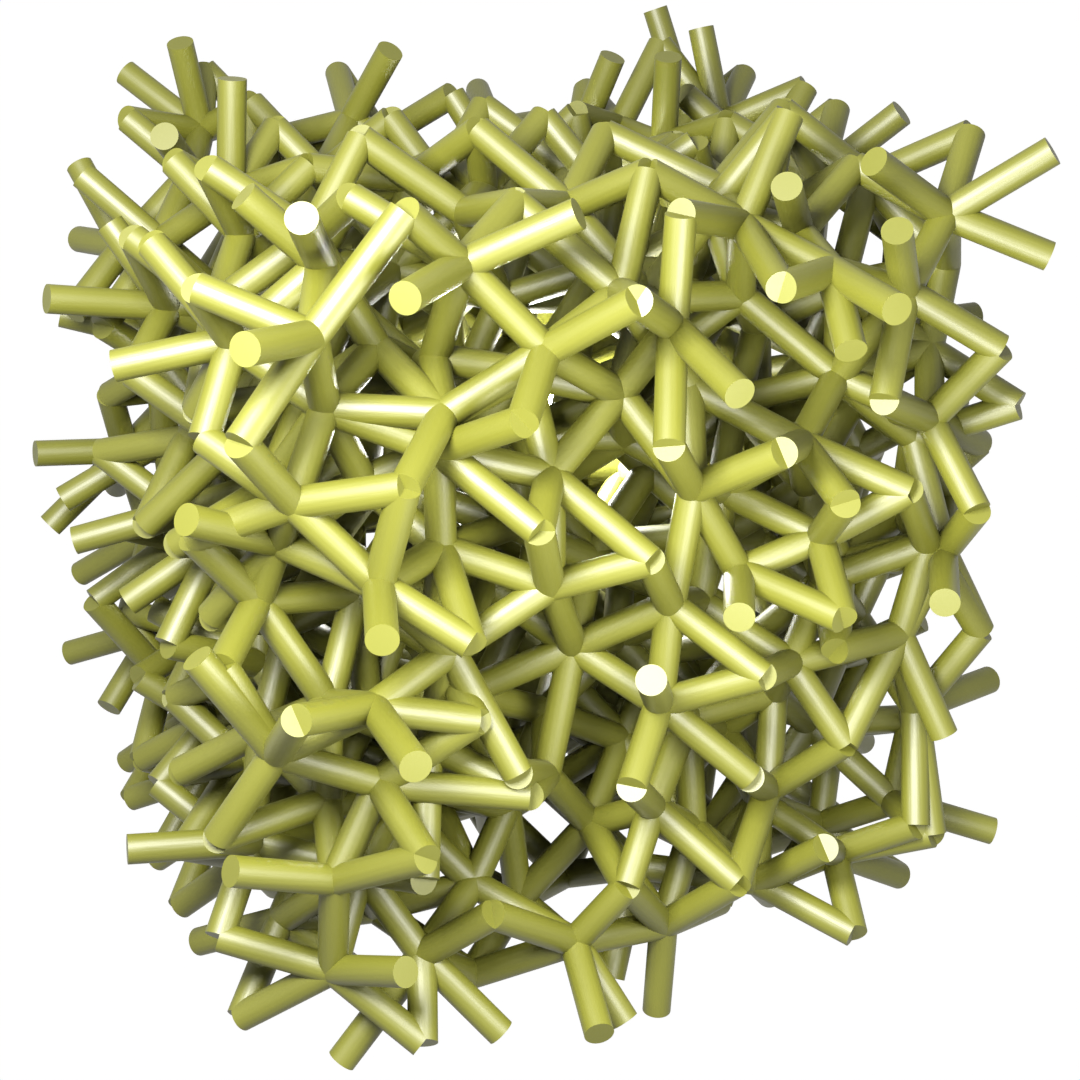}}
\scaledinset{l}{0.01}{b}{1.05}{\textbf{b}}{\includegraphics[width=0.6\textwidth]{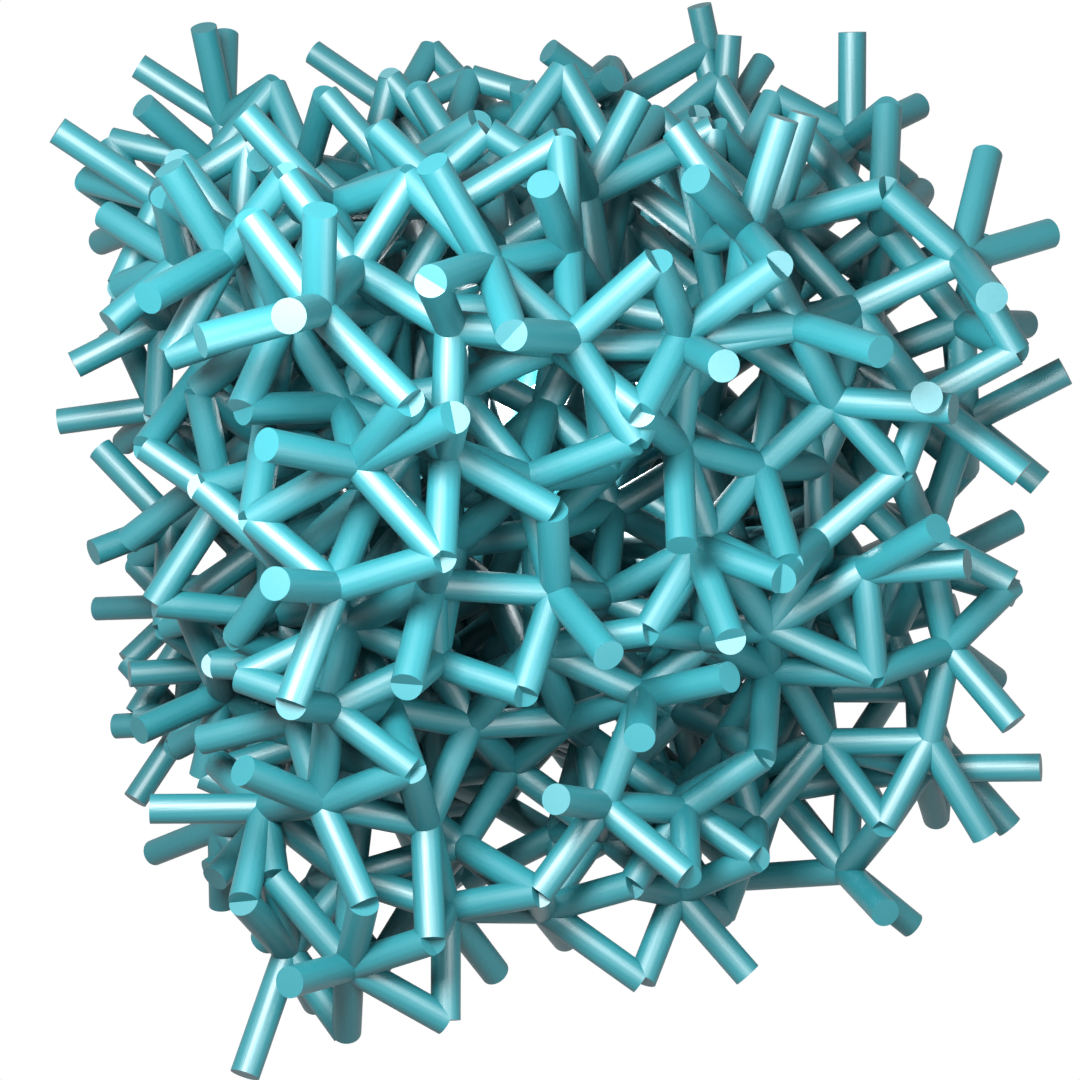}}
\end{minipage}
%
\begin{minipage}{0.48\textwidth}
\centering
\scaledinset{l}{0.01}{c}{.47}{\textbf{c}}{\includegraphics[width=\textwidth]{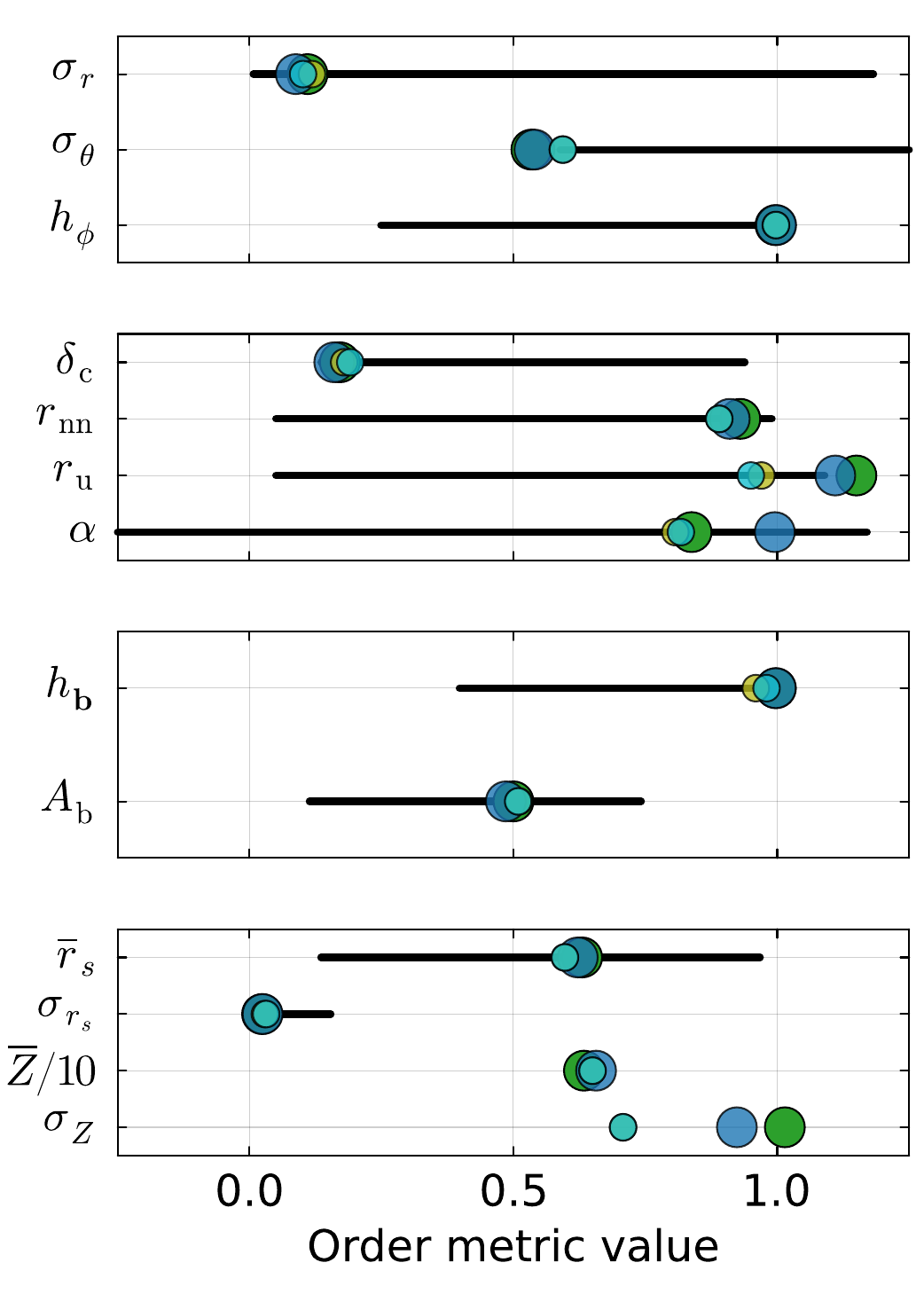}}
\end{minipage}
\caption{\label{fig:stern_vir_bcu_generated}
\textbf{a} The generated network $\textbf{bcu}_\mathrm{StV\ green}$ has the smallest order metric distance to the StV green biological network (Figure~\ref{fig:bio_periodic_coord_nr}d). It was generated from an initial $\textbf{bcu}_\mathrm{mod}$ network.
\textbf{b} The generated network $\textbf{bcu}_\mathrm{StV\ blue}$ is the most similar to StV green (Figure~\ref{fig:bio_periodic_coord_nr}e).
\textbf{c} The order metric values of networks $\textbf{bcu}_\mathrm{StV\ green}$  (\textbf{a}, small olive green markers) and $\textbf{bcu}_\mathrm{StV\ blue}$  (\textbf{b}, small light blue markers) closely match the biophotonic network metrics of StV green (large green markers) and StV blue (large dark blue markers). The black bars represent the ranges of order metrics achieved by modifying the initial $\textbf{bcu}_\mathrm{mod}$ network with at least ten accepted Monte Carlo moves.
}
\end{figure*}

Furthermore, we reproduce the biophotonic structures of the beetle Sternotomini virescens (StV green and StV blue in Figure~\ref{fig:bio_periodic_coord_nr}d-e). The StV green structure is best reproduced by the $\textbf{bcu}_\mathrm{StV\ green}$ network (Figure~\ref{fig:stern_vir_bcu_generated}a), generated from the periodic $\textbf{bcu}_\mathrm{mod}$ network with 432 vertices, with WWW parameters
\add{
\begin{align*}
\beta=0.0007 && , && T_\mathrm{max}=0.2790=0.973 \, T_\mathrm{melt} && \text{and} && \Delta T = 0.5608 =1.955 \, T_\mathrm{melt}
\quad .
\end{align*}}
These parameters resulted in \replace{340}{247} accepted Monte Carlo moves, with every vertex involved in approximately \replace{3.1}{2.3} accepted moves.
Similarly, network $\textbf{bcu}_\mathrm{StV\ blue}$ (Figure~\ref{fig:stern_vir_bcu_generated}b) -- the network closest to StV blue -- was generated with
\begin{align*}
\beta=0.0012 && , && T_\mathrm{max}=0.2314=0.678 \, T_\mathrm{melt} && \text{and} && \Delta T = 0.4296 =1.260 \, T_\mathrm{melt}
\quad .
\end{align*}
This resulted in 329 accepted Monte Carlo moves, three per vertex. 

Figure~\ref{fig:stern_vir_bcu_generated}c shows good agreement between the order metrics of the generated networks $\textbf{bcu}_\mathrm{StV\ green}$ and $\textbf{bcu}_\mathrm{StV\ blue}$, and the corresponding biological networks. The most significant discrepancy is in the coordination number standard deviation $\sigma_Z$, which is determined by the initial network. Figure~\ref{fig:bio_periodic_coord_nr}k shows that our choice of the initial network $\textbf{bcu}_\mathrm{mod}$ has a simplified coordination number histogram that neglects the high- and low-coordination number tails of the biological networks. 

Besides $\sigma_Z$, the homogeneity metrics $r_\mathrm{u}$ and $\alpha$ differ significantly between generated and biological samples. The \replace{coordinated neighbor distance $r_\mathrm{c}$}{nearest-neighbor distance $r_\mathrm{nn}$} is strongly influenced by the bond-stretching term $E_r$ in Equation~\eqref{eqn:keating_dimensionless}. However, the bond-stretching term has no effect on $r_\mathrm{u}$, the average distance to the closest vertex with which no bond exists. Therefore, the network's Monte Carlo evolution can produce local clusters of unconnected vertices, reducing $r_\mathrm{u}$. Figures \ref{fig:ctn_homogeneity_hyperuniformity}b~and~\ref{fig:ctn_predict_homogeneity}b show that networks generated with small $\beta$ values have the greatest disorder and homogeneity with respect to $r_\mathrm{u}$. Examples of these networks are $\textbf{bcu}_\mathrm{StV\ green}$ and $\textbf{bcu}_\mathrm{StV\ blue}$. As discussed in Section~\ref{sec:effect_www_parameters}, similarly to $r_\mathrm{u}$, the hyperuniformity metric is not directly affected by the Keating energy. Therefore, $\alpha$ generally decreases with every accepted Monte Carlo move.

\begin{figure*}
\begin{minipage}{0.48\textwidth}
\centering
\scaledinset{l}{0.01}{b}{1.05}{\textbf{a}}{\includegraphics[width=0.6\textwidth]{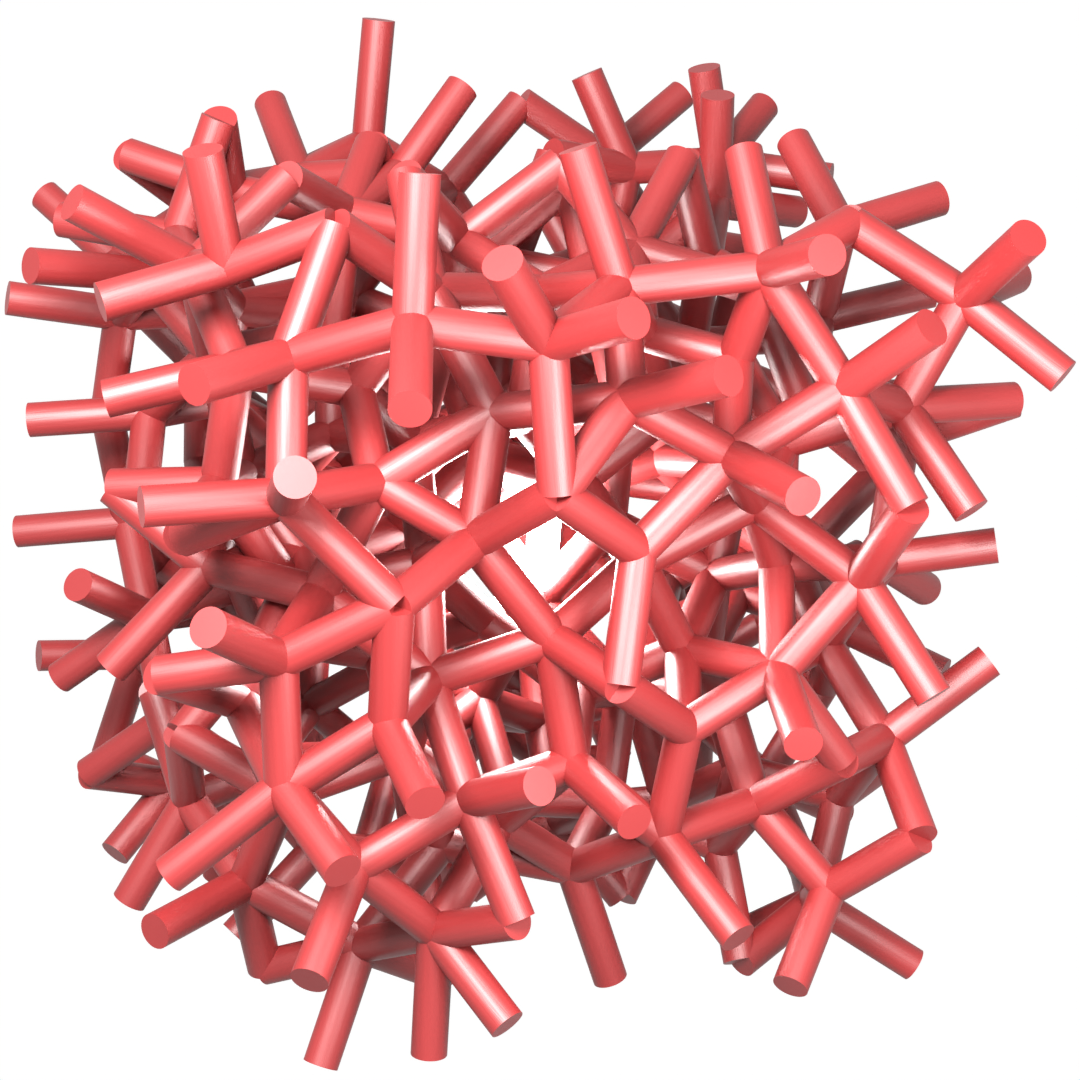}}
\end{minipage}
%
\begin{minipage}{0.48\textwidth}
\centering
\scaledinset{l}{0.01}{b}{.72}{\textbf{b}}{\includegraphics[width=\textwidth]{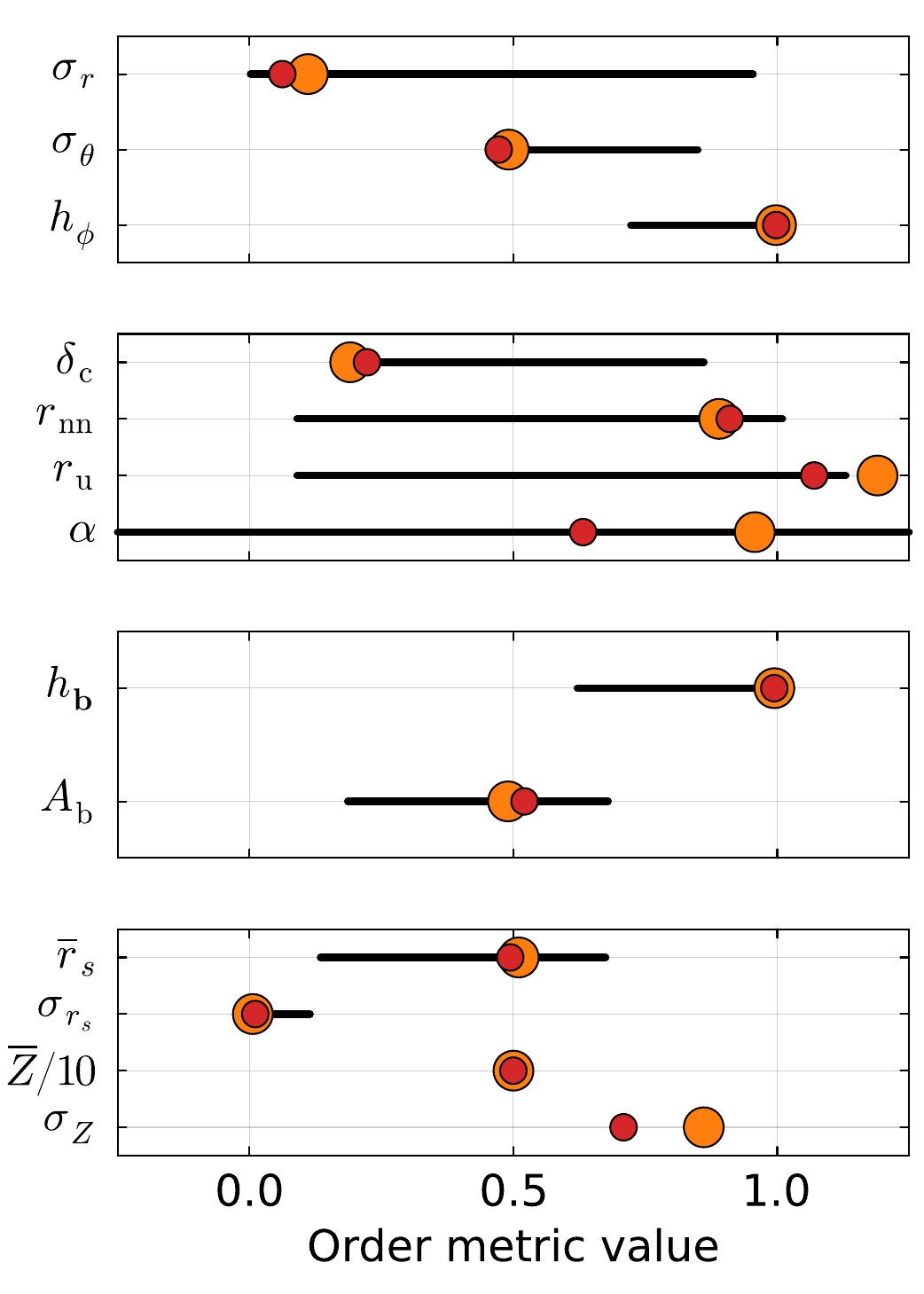}}
\end{minipage}
\caption{\label{fig:stern_ama_bcu_generated}
\textbf{a} The network with the smallest order metric distance to the StA orange biological network (Figure~\ref{fig:bio_periodic_coord_nr}h) was generated from an initial $\textbf{pcu}_\text{mod}$ network.
\textbf{b} The order metric values of the network in \textbf{a} (small red markers) closely match those of the StA orange  biophotonic network (large orange markers). The black bars represent the range of order metrics achieved by modifying the initial $\textbf{pcu}_\text{mod}$ network with at least ten accepted Monte Carlo moves.
}
\end{figure*}

The fourth biological network StA orange, introduced in Section~\ref{sec:biophotonic}, was reproduced from the initial $\textbf{pcu}_\text{mod}$ network with the algorithm inputs
\begin{align*}
\beta=0.0978 && , && T_\mathrm{max}=0.1788=1.034 \, T_\mathrm{melt} && \text{and} && \Delta T = 0.1858 =1.074 \, T_\mathrm{melt}
\quad .
\end{align*}
These inputs resulted in 689 accepted Monte Carlo moves, meaning each vertex was involved in an average of 13 bond switches. Figure~\ref{fig:stern_ama_bcu_generated}a shows the resulting disordered network, and Figure~\ref{fig:stern_ama_bcu_generated}b compares its order metric values to those of the biological network. Similar trends to those observed in the reproduction of the StV networks in Figure~\ref{fig:stern_vir_bcu_generated}c are seen here. While the small-scale and isotropy order metrics are well reproduced, the intermediate and large-scale homogeneity metrics \add{including hyperuniformity} differ significantly between biological and generated networks. As previously discussed, this discrepancy is due to the Keating energy, which is limited to short length scales in its current form.

\add{Interestingly, we found that all analyzed biological networks are hyperuniform. With $\alpha$ values between 0.838 for StV green and 0.995 for StV blue, all networks fall into the weakest hyperuniformity class III, close to class II \cite{torquato2018}. While hyperuniformity supports the formation of photonic band gaps \cite{florescu2009, froufe2016, rothammer2021, siedentop2024}, these require refractive index contrasts of at least 1.9 in 3D \cite{joannopoulos2008}. The investigated biophotonic networks have refractive index contrasts of about 1.55 \cite{djeghdi2022, bauernfeind2023, bauernfeind2024} for which the relation between hyperuniformity and structural color remains unclear. This represents an important direction for future investigation.}

Despite minor deviations in the homogeneity metrics, we conclude that our extended WWW algorithm successfully reproduces all disordered biophotonic networks. We achieved this by matching the coordination number statistics of the target networks with those of an initially crystalline network. Additionally, we created network datasets by randomly sampling the algorithm's input parameters. The resulting disordered networks span a wide range of order metric values, as apparent from all figures in Sections \ref{sec:effect_www_parameters} and \ref{sec:reproducing_biophotonic}. Although the generalized Keating energy provides significant control over short-length-scale metrics, higher variance is observed in longer-ranged metrics related to pore sizes and hyperuniformity.

\section{Conclusion}
Computer-generated disordered spatial networks are essential to many fields in the natural and social sciences. We modified the Keating strain energy to adapt the existing WWW algorithm to networks with arbitrary coordination number statistics\add{, in particular to networks with valencies higher than four}. \replace{We also adjusted}{For the first time, we tuned }the degree \add{and type} of disorder in the generated networks by \replace{modifying}{simultaneously adjusting} the bond-bending force constant \add{in the Keating energy} and the temperature profile governing the Monte Carlo evolution \delete{in the Keating energy}. To analyze the effect of these adjustments on the resulting networks, we introduced an extensive set of order metrics covering network primitives, homogeneity, isotopy, and topology. \add{This set provides a novel statistical classification of structural disorder in networks.}

As a case study, we statistically reproduced four disordered biophotonic networks responsible for structural color in various beetles. Starting with six crystalline networks, we randomly sampled the algorithm inputs to generate datasets containing thousands of disordered networks. Next, we trained a feedforward neural network to predict order metrics derived from the algorithm's inputs. \add{The neural network predictions enable the efficient generation of disordered networks with targeted average structural properties.} Our results show that the Keating energy enables precise control over short-range order metrics by quantifying disorder in bond lengths and angles. However, we observed higher statistical variance in long-range metrics that measure pore sizes and hyperuniformity. \add{We found that all biophotonic networks are hyperuniform.}

This work paves the way for studying the relationship between the structure of disordered spatial networks and their properties. Optical simulations on the generated network datasets are an essential next step in understanding structural color in disordered photonic materials. Observing the effects of individual order metrics on the reflectance spectra obtained with finite-difference time-domain simulations will reveal the relevant correlations between structural and optical properties. \add{While hyperuniformity has been related to the opening of photonic band gaps (PBGs) at high refractive index contrasts \cite{florescu2009, froufe2016, rothammer2021, siedentop2024}, the connection to structural color at lower refractive index contrasts is not yet understood.}

Further understanding can be gained from the photonic densities of states (PDOS). The PDOS can be obtained by voxelizing the photonic network and applying a plane-wave method \cite{froufe2016}, or by decorating the material with dipoles and using a Green's matrix method \cite{dal_negro2016}. We expect to gain insight into the emergence of PBGs, defect states within them, and localized edge modes \cite{yu2020}. Studying emission in disordered networks could lead to random lasing and improve our understanding of generating and controlling complex lasing spectra \cite{sapienza2022}. Additionally, training a neural network on the resulting datasets will enable us to predict optical properties from structural information. This offers a highly efficient alternative to optical simulations.

Beyond photonic materials, the extended WWW algorithm and the order metrics presented in this work can be applied to disordered spatial networks in any field of research.
\add{As an example that is closer to the initial application of the WWW algorithm, our method could support the investigation of amorphous oxide semiconductors such as indium oxide or ternary oxides like In-Ga-O \cite{medvedeva_2017}. These materials exhibit favorable properties, such as mechanical flexibility and high carrier mobility. As for disordered photonic systems, structural data of amorphous oxide semiconductors is sparse, and many of their structure-property relations are not well understood. Their coordination numbers can reach six, and valency distributions can be obtained from density functional theory. From this starting point, statistical populations of networks can be generated with our extended WWW algorithm and used to simulate the connection between structural, mechanical and electrical properties.}

\replace{Generated}{This example illustrates that} datasets \add{generated with our extended WWW algorithm} can be tailored to a specific application by selecting an appropriate initial network and adjusting the temperature profile. Additional terms can be added to the generalized strain energy to emphasize the desired order metrics. Including reciprocal space metrics in the strain energy could provide direct control over wave scattering phenomena, but would increase computational complexity.

\medskip
\textbf{Acknowledgments} \par 
The authors thank Niklas R. Schwarz for the data preprocessing of the \textit{Sternotomis amabilis} beetle photonic network. The authors thank Viola V. Vogler-Neuling for helpful discussions about biophotonic materials and Luis S. Froufe-Pérez for valuable input about neural networks. 
This study was supported by a European Research Council (ERC) Advanced grant (PrISMoID, 833895), the Adolphe Merkle Foundation, and the Swiss National Science Foundation (SNSF) through the National Center of Competence in Research Bio-Inspired Materials (grant no. 51NF40-182881).

\medskip
\textbf{Data Availability Statement}
The data that support the findings of this study are openly available on Zenodo \cite{hemmann_zenodo_2026}. The source code supporting this work is available on GitHub \cite{hemmann_github_2026}.

\medskip

%
\bibliographystyle{MSP}

\bibliography{refs}

\appendix
\noindent
{\Huge \textbf{Appendices}}

\section{List of order metrics}
\label{sec:list_of_symbols}

\begin{table}[h]
\centering
\begin{tabular}{lll}
\hline
 & \textbf{Order metric} & \textbf{Meaning} \\
\hline\rule{-2pt}{1em}
\multirow{5}{*}{\textbf{Network primitives}} \quad
    & $\sigma_r$ & Bond length standard deviation \\
    & $\sigma_\theta$ & Bond angle standard deviation \\
    & $\overline{q}_l$ for all $l\in \{0,1,\dots,12\}$ & Mean Steinhardt local bond order parameter \\
    & $\sigma_{q_l}$ for all $l\in \{0,1,\dots,12\}$ & \makecell[l]{Steinhardt local bond order parameter \\ standard deviation} \\
    & $h_\phi$ & Dihedral angle entropy \\
\hline\rule{-2pt}{1em}
\multirow{4}{*}{\textbf{Homogeneity}} 
    & \replace{$r_\mathrm{c}$}{$r_\mathrm{nn}$} & \replace{Coordinated neighbor distance}{Nearest-neighbor distance} \\
    & $r_\mathrm{u}$ & \replace{Uncoordinated neighbor distance}{Nearest-uncoordinated-neighbor distance} \\
    & $\delta_\mathrm{c}$ & Critical pore radius \\
    & $\alpha$ & Hyperuniformity metric \\
\hline\rule{-2pt}{1em}
\multirow{3}{*}{\textbf{Isotropy}} 
    & $h_\mathbf{b}$ & Bond orientation entropy \\
    & $A_\mathrm{v}$ & Vertex anisotropy metric \\
    & $A_\mathrm{b}$ & Bond anisotropy metric \\
\hline\rule{-2pt}{1em}
\multirow{6}{*}{\textbf{Topology}} 
    & $\overline{Z}$ & Mean coordination number \\
    & $\sigma_Z$ & Coordination number standard deviation \\
    & $\overline{s}$ & Mean ring size \\
    & $\sigma_s$ & Ring size standard deviation \\
    & $\overline{r}_s$ & Mean ring radius \\
    & $\sigma_{r_s}$ & Ring radius standard deviation \\
\hline
\end{tabular}
\caption{
\label{tab:list_of_metrics}
A complete list of all order metrics for spatial networks defined in Section~\ref{sec:order_metrics}. 
}
\end{table}

\section{Effect of input parameters on the extended WWW algorithm}
\label{sec:si_effect_www_inputs}
In the main article, we analyze how the input parameters affect networks derived from the initial crystalline \textbf{ctn} network using the extended WWW algorithm . Here, we provide a similar analysis for networks generated from the other five other initial networks introduced in Section~\ref{sec:biophotonic} of the main text: $\textbf{bcu}_\text{mod}$, $\textbf{pcu}_\text{mod}$, \textbf{dia}, \textbf{srs}, and \textbf{lcs}.

\newpage
\subsection{$\textbf{bcu}_\text{mod}$}

\begin{figure*}[h]
\begin{minipage}{0.48\textwidth}
\centering
\scaledinset{l}{0.01}{b}{.91}{\textbf{a}}{\includegraphics[width=\textwidth]{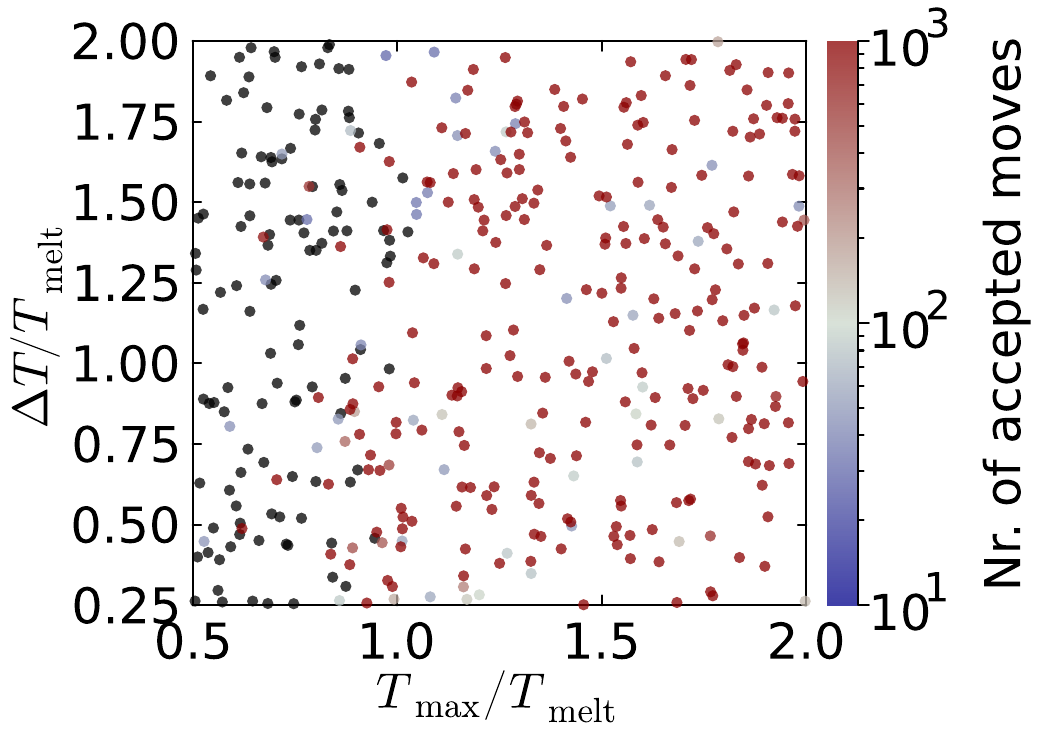}}
\end{minipage}
%
\begin{minipage}{0.48\textwidth}
\centering
\scaledinset{l}{0.01}{b}{.91}{\textbf{b}}{\includegraphics[width=\textwidth]{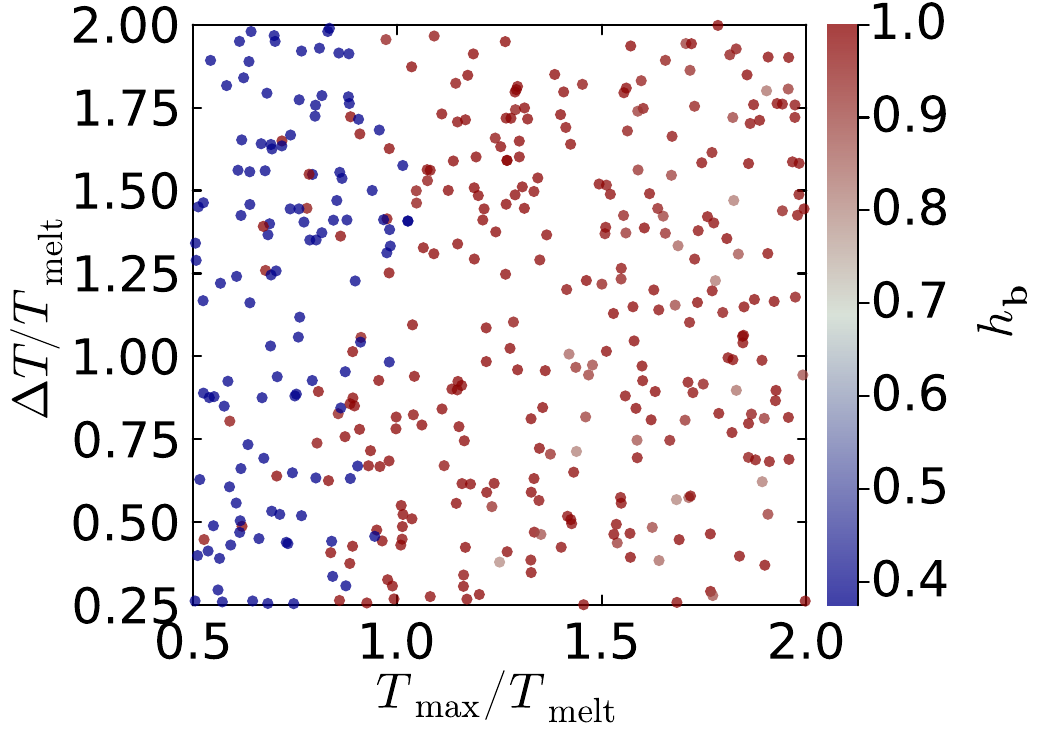}}
\end{minipage}
\caption{\label{fig:bcu_cn_5_6_7_8_nr_accepted_moves_isotropy}
Melting transition for networks generated from the initial $\textbf{bcu}_\mathrm{mod}$ network. \add{The metrics of 454 networks are displayed.}
\textbf{a} Number of accepted Monte Carlo moves plotted against $T_\mathrm{max}$ and $\Delta T$. Black markers correspond to networks with 10 or fewer accepted moves.
\textbf{b} Bond orientation entropy $h_\mathbf{b}$.
}
\end{figure*}

\begin{figure}[h]
\includegraphics[width=0.48\linewidth]{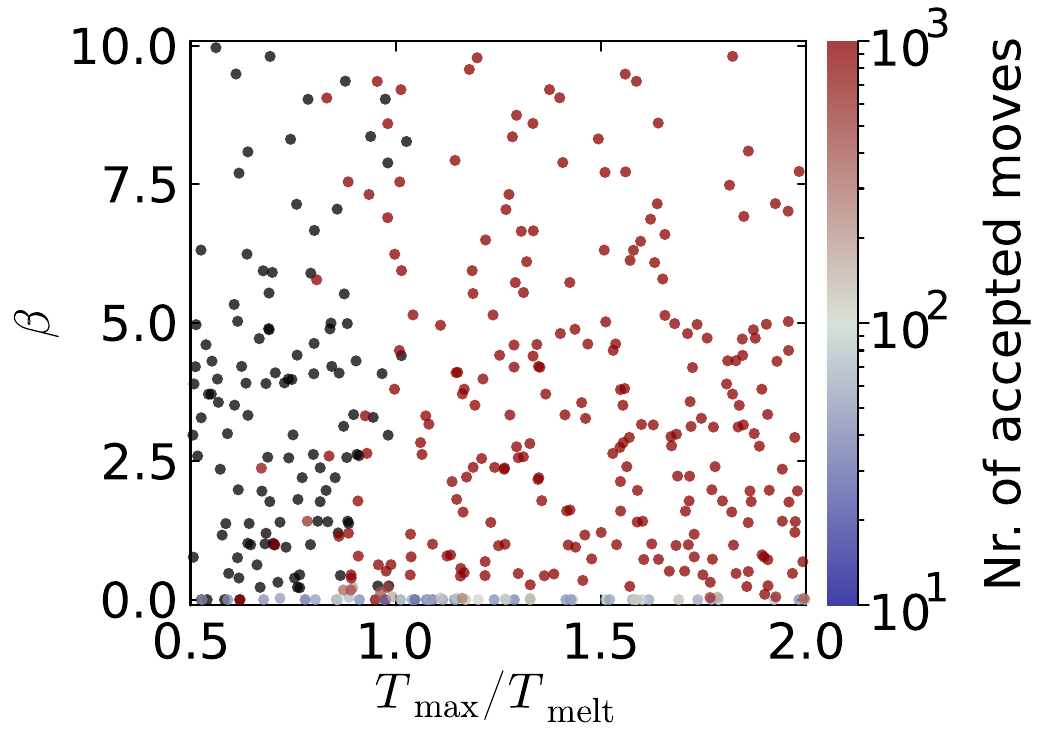}
\caption{\label{fig:bcu_cn_5_6_7_8_nr_accepted_moves_beta} Number of accepted Monte Carlo moves for the initial $\textbf{bcu}_\mathrm{mod}$ plotted against $T_\mathrm{max}$ and $\Delta T$. The black markers correspond to networks with ten or fewer accepted moves. \add{The metrics of 454 networks are displayed.}}
\end{figure}

\begin{figure*}[h]
\begin{minipage}{0.48\textwidth}
\centering
\scaledinset{l}{0.01}{b}{.91}{\textbf{a}}{\includegraphics[width=\textwidth]{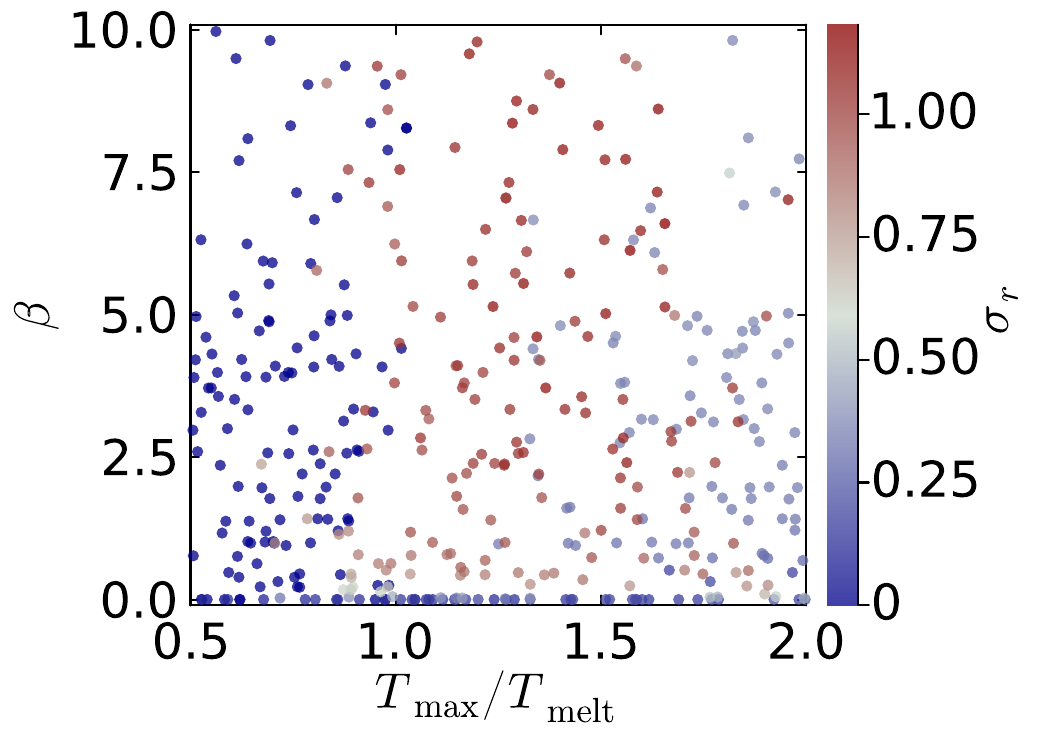}}
\end{minipage}
%
\begin{minipage}{0.48\textwidth}
\centering
\scaledinset{l}{0.01}{b}{.91}{\textbf{b}}{\includegraphics[width=\textwidth]{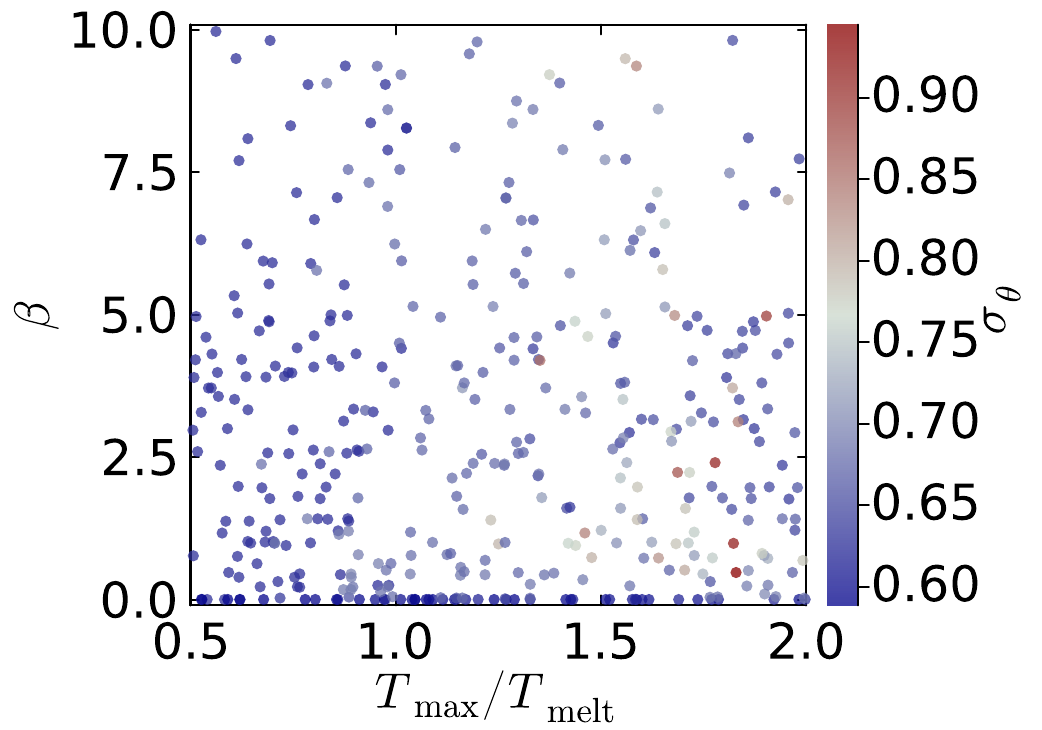}}
\end{minipage}
\caption{\label{fig:bcu_cn_5_6_7_8_bond_length_angle}
Network primitive metrics for networks generated from the initial $\textbf{bcu}_\mathrm{mod}$ plotted against the algorithm inputs $T_\mathrm{max}$ and $\beta$. Red colors indicate high disorder with respect to the corresponding metric. \add{The metrics of 454 networks are displayed.}
\textbf{a} Bond length standard deviation $\sigma_r$. 
\textbf{b} Bond angle standard deviation $\sigma_\theta$.
}
\end{figure*}

\begin{figure*}[h]
\begin{minipage}{0.48\textwidth}
\centering
\scaledinset{l}{0.01}{b}{.91}{\textbf{a}}{\includegraphics[width=\textwidth]{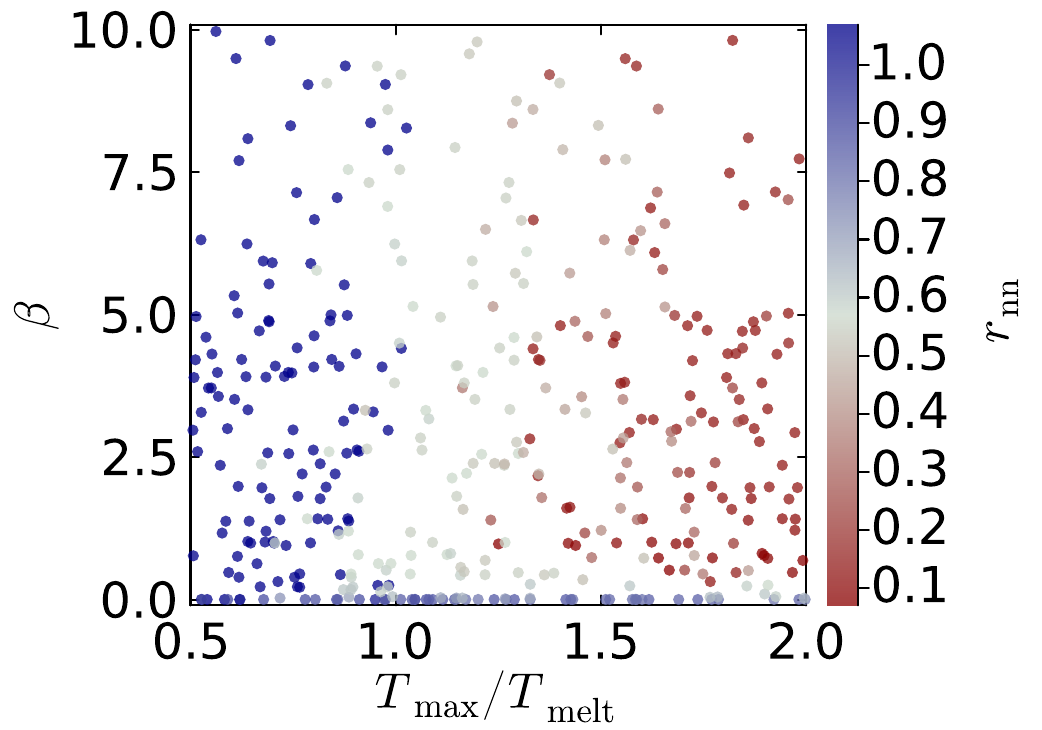}}
\end{minipage}
%
\begin{minipage}{0.48\textwidth}
\centering
\scaledinset{l}{0.01}{b}{.91}{\textbf{b}}{\includegraphics[width=\textwidth]{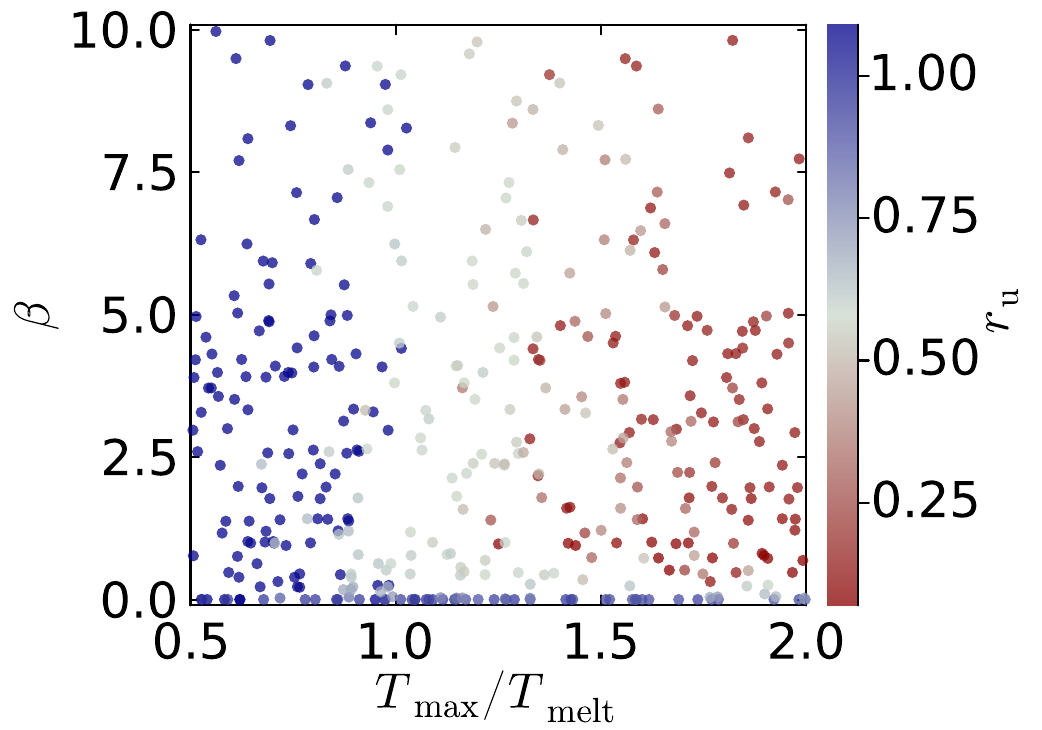}}
\end{minipage}
\begin{minipage}{0.48\textwidth}
\centering
\scaledinset{l}{0.01}{b}{.91}{\textbf{c}}{\includegraphics[width=\textwidth]{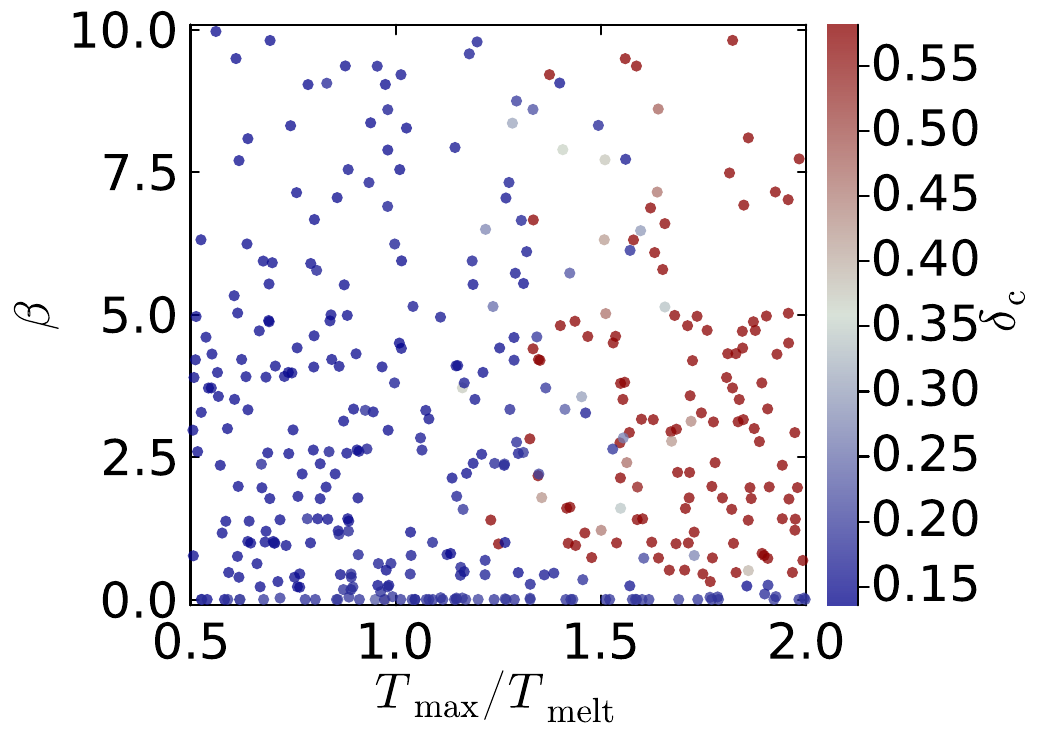}}
\end{minipage}
%
\begin{minipage}{0.48\textwidth}
\centering
\scaledinset{l}{0.01}{b}{.91}{\textbf{d}}{\includegraphics[width=\textwidth]{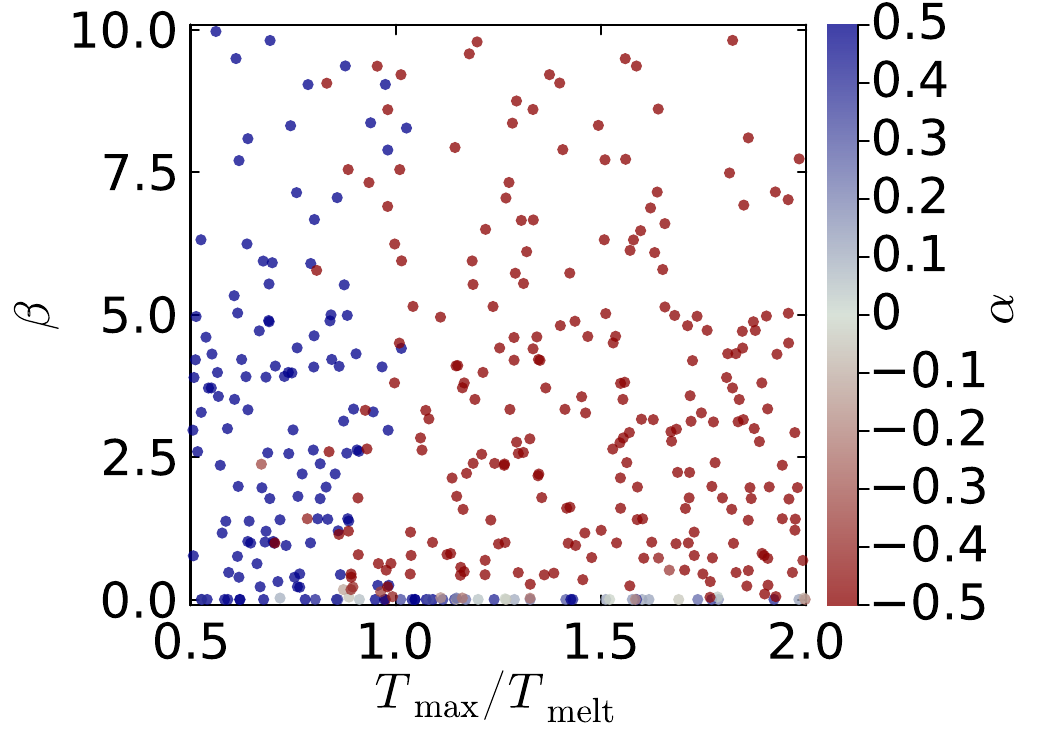}}
\end{minipage}
\caption{\label{fig:bcu_cn_5_6_7_8_homogeneity_hyperuniformity}
Homogeneity metrics for networks generated from the initial $\textbf{bcu}_\mathrm{mod}$ plotted against the algorithm inputs $T_\mathrm{max}$ and $\beta$. Red colors indicate high disorder with respect to the corresponding metric. \add{The metrics of 454 networks are displayed.}
\textbf{a} \replace{Coordinated neighbor distance $r_\mathrm{c}$}{Nearest-neighbor distance $r_\mathrm{nn}$}.
\textbf{b} \replace{Uncoordinated neighbor distance}{Nearest-uncoordinated-neighbor distance} $r_\mathrm{u}$.
\textbf{c} Critical pore radius $\delta_\mathrm{c}$.
\textbf{d} Hyperuniformity metric $\alpha$.
}
\end{figure*}

\begin{figure*}[h]
\begin{minipage}{0.48\textwidth}
\centering
\scaledinset{l}{0.01}{b}{.91}{\textbf{a}}{\includegraphics[width=\textwidth]{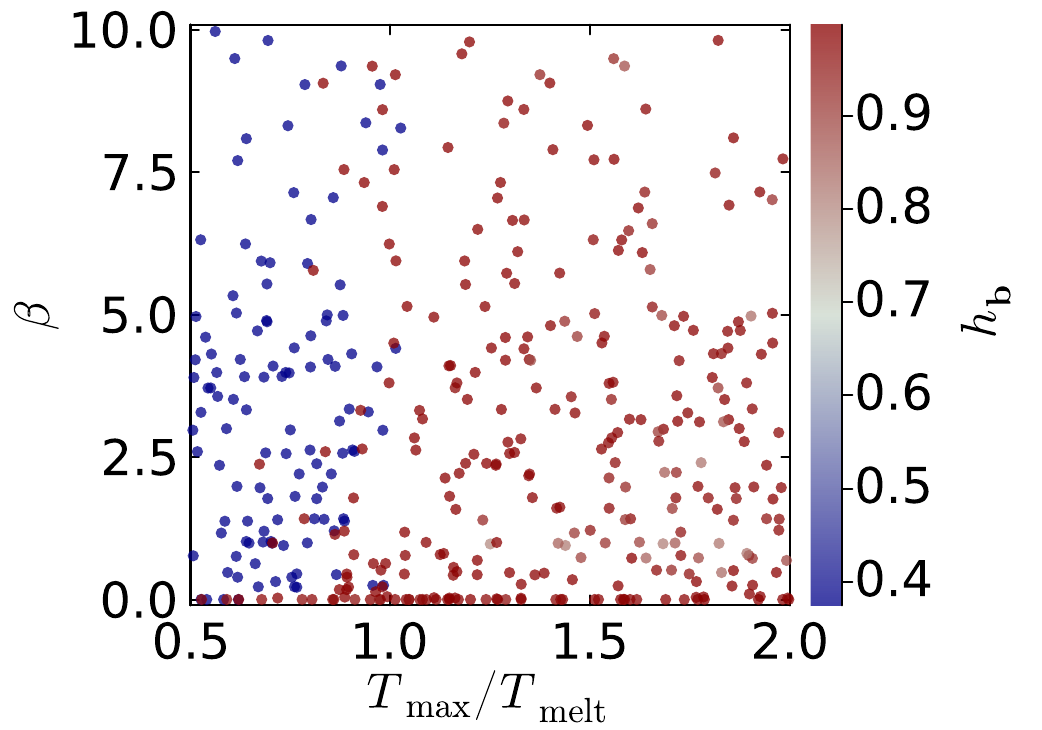}}
\end{minipage}
%
\begin{minipage}{0.48\textwidth}
\centering
\scaledinset{l}{0.01}{b}{.91}{\textbf{b}}{\includegraphics[width=\textwidth]{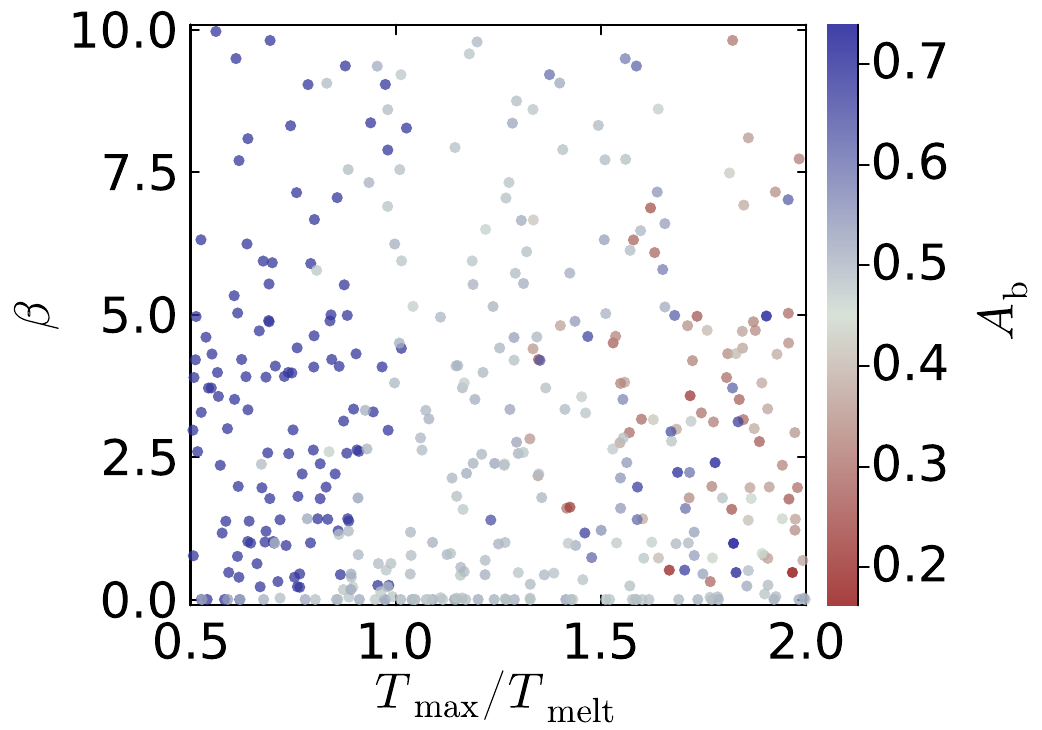}}
\end{minipage}
\begin{minipage}{0.48\textwidth}
\centering
\scaledinset{l}{0.01}{b}{.91}{\textbf{c}}{\includegraphics[width=\textwidth]{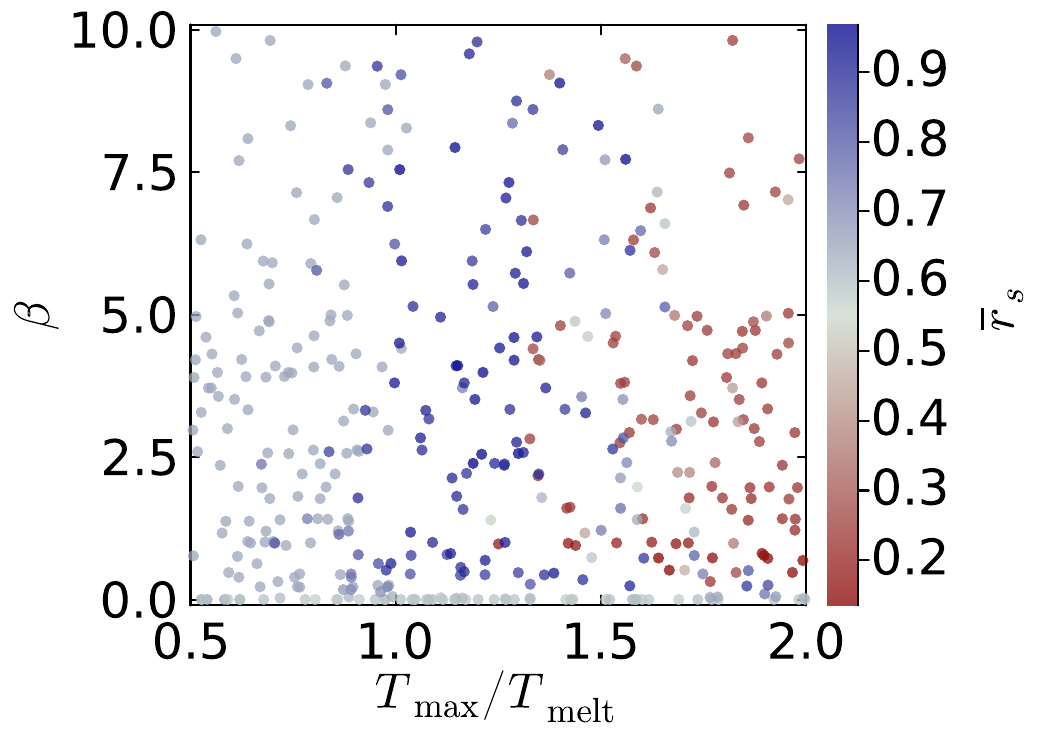}}
\end{minipage}
%
\begin{minipage}{0.48\textwidth}
\centering
\scaledinset{l}{0.01}{b}{.91}{\textbf{d}}{\includegraphics[width=\textwidth]{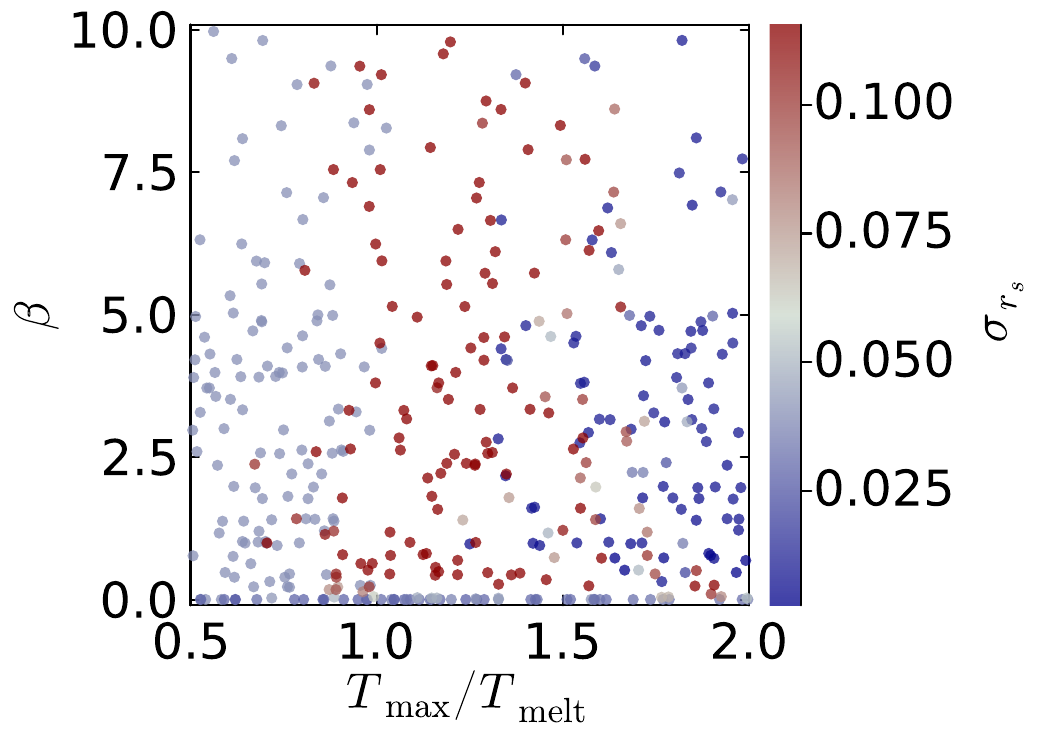}}
\end{minipage}
\caption{\label{fig:bcu_cn_5_6_7_8_isotropy_topology}
Isotropy and topology metrics for networks generated from the initial $\textbf{bcu}_\mathrm{mod}$ as a function of the network generation algorithm inputs $T_\mathrm{max}$ and $\beta$. Red colors indicate high disorder with respect to the corresponding metric. \add{The metrics of 454 networks are displayed.}
\textbf{a} Bond orientation entropy $h_\mathbf{b}$.
\textbf{b} \add{Bond} structure factor anisotropy metric $A_\mathrm{b}$.
\textbf{c} Mean ring radius $\overline{r}_s$.
\textbf{d} Ring radius standard deviation $\sigma_{r_s}$.
}
\end{figure*}

\clearpage

\subsection{$\textbf{pcu}_\text{mod}$}

\begin{figure*}[h]
\begin{minipage}{0.48\textwidth}
\centering
\scaledinset{l}{0.01}{b}{.91}{\textbf{a}}{\includegraphics[width=\textwidth]{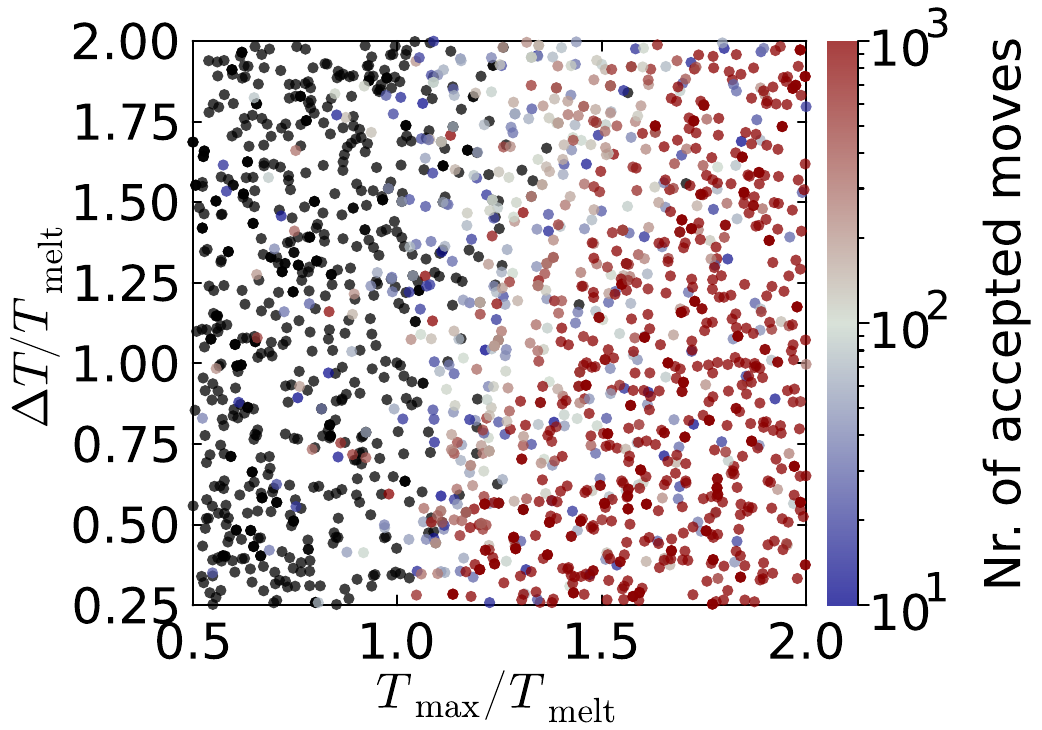}}
\end{minipage}
%
\begin{minipage}{0.48\textwidth}
\centering
\scaledinset{l}{0.01}{b}{.91}{\textbf{b}}{\includegraphics[width=\textwidth]{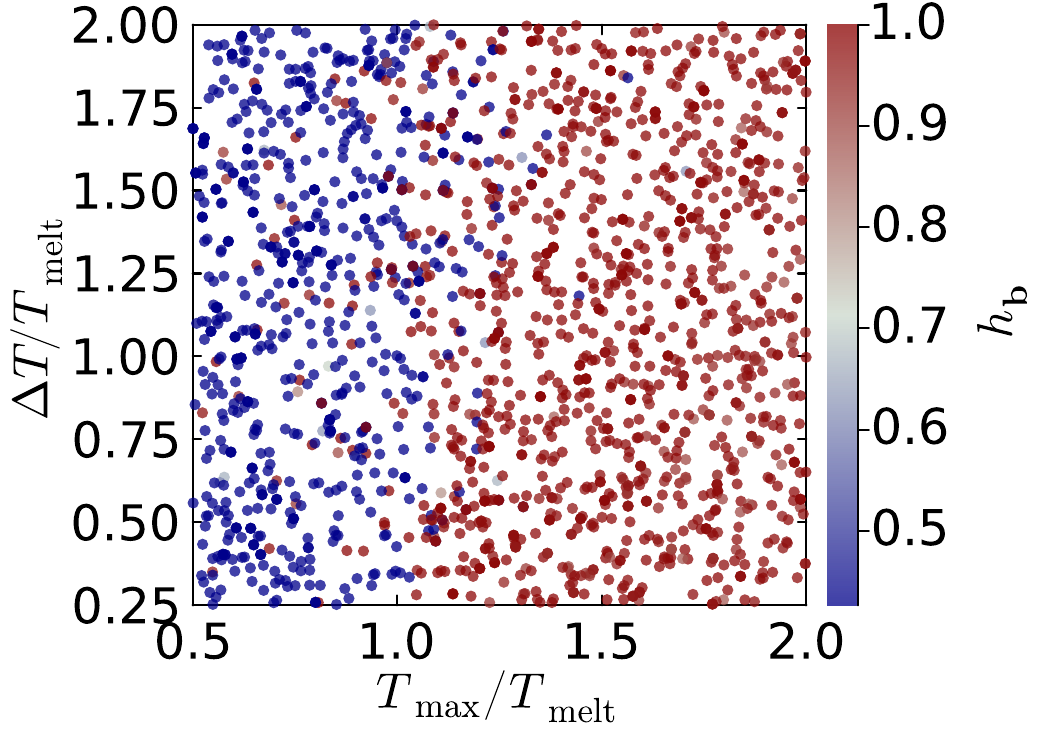}}
\end{minipage}
\caption{\label{fig:pcu_cn_4_5_6_nr_accepted_moves_isotropy}
Illustration of the melting transition for networks generated from the initial $\textbf{pcu}_\mathrm{mod}$ network. \add{The metrics of 1815 networks are displayed.}
\textbf{a} Number of accepted Monte Carlo moves plotted against $T_\mathrm{max}$ and $\Delta T$. The black markers correspond to networks with ten or fewer accepted moves.
\textbf{b} Bond orientation entropy $h_\mathbf{b}$.
}
\end{figure*}

\begin{figure}[h]
\includegraphics[width=0.48\linewidth]{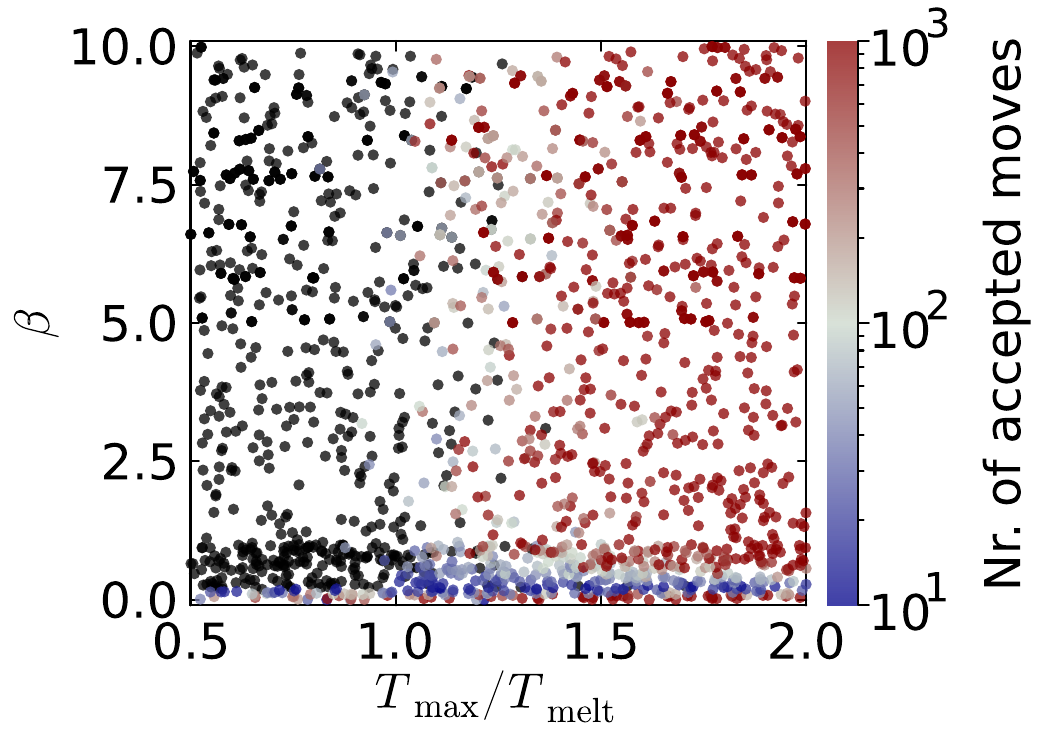}
\caption{\label{fig:pcu_cn_4_5_6_nr_accepted_moves_beta} Number of accepted Monte Carlo moves for the initial $\textbf{pcu}_\mathrm{mod}$ plotted against $T_\mathrm{max}$ and $\Delta T$. The black markers correspond to networks with ten or fewer accepted moves. \add{The metrics of 1815 networks are displayed.}}
\end{figure}

\begin{figure*}[h]
\begin{minipage}{0.48\textwidth}
\centering
\scaledinset{l}{0.01}{b}{.91}{\textbf{a}}{\includegraphics[width=\textwidth]{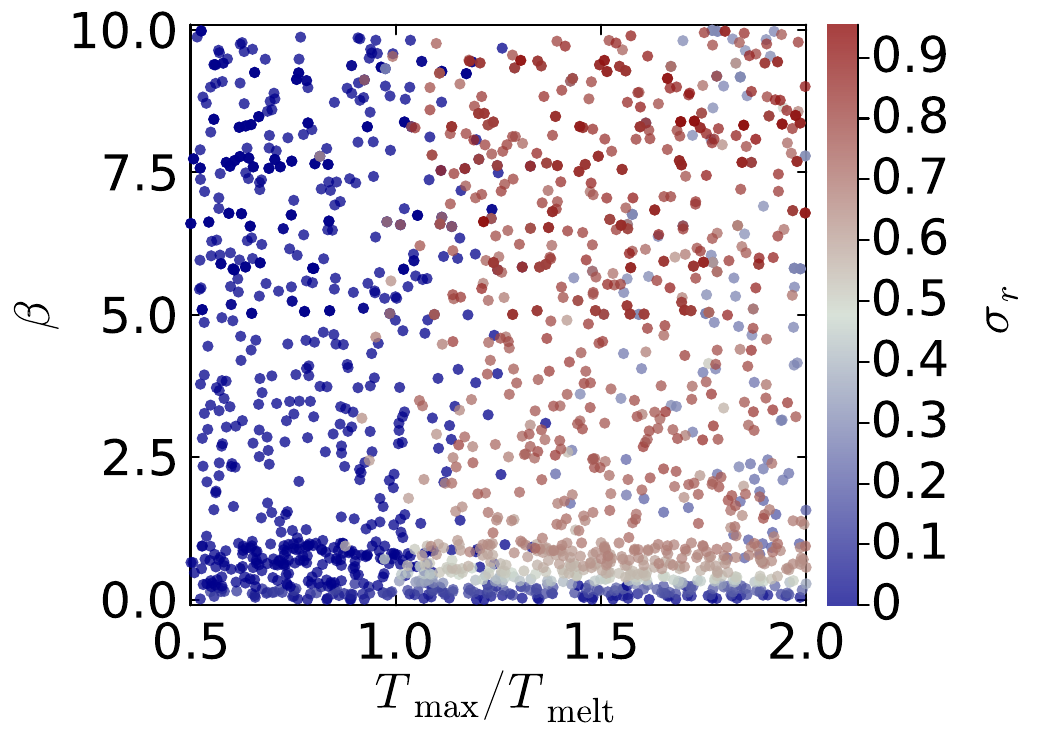}}
\end{minipage}
%
\begin{minipage}{0.48\textwidth}
\centering
\scaledinset{l}{0.01}{b}{.91}{\textbf{b}}{\includegraphics[width=\textwidth]{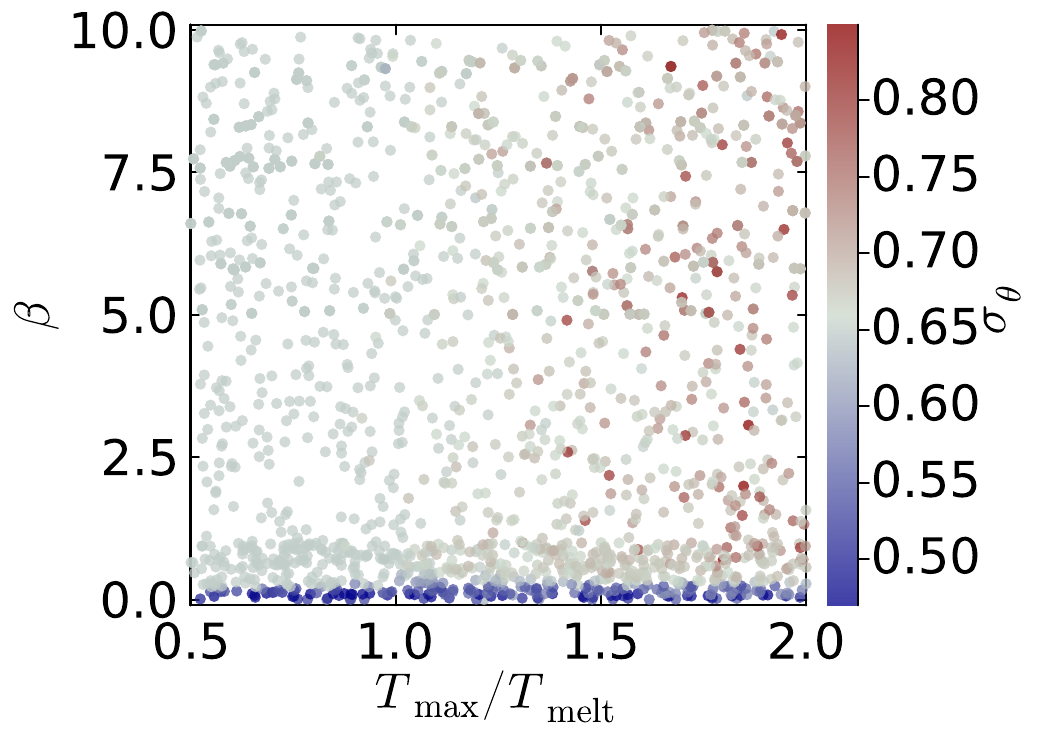}}
\end{minipage}
\caption{\label{fig:pcu_cn_4_5_6_bond_length_angle}
Network primitive metrics for networks generated from the initial $\textbf{pcu}_\mathrm{mod}$ plotted against the algorithm inputs $T_\mathrm{max}$ and $\beta$. Red colors indicate high disorder with respect to the corresponding metric. \add{The metrics of 1815 networks are displayed.}
\textbf{a} Bond length standard deviation $\sigma_r$. 
\textbf{b} Bond angle standard deviation $\sigma_\theta$.
}
\end{figure*}

\begin{figure*}[h]
\begin{minipage}{0.48\textwidth}
\centering
\scaledinset{l}{0.01}{b}{.91}{\textbf{a}}{\includegraphics[width=\textwidth]{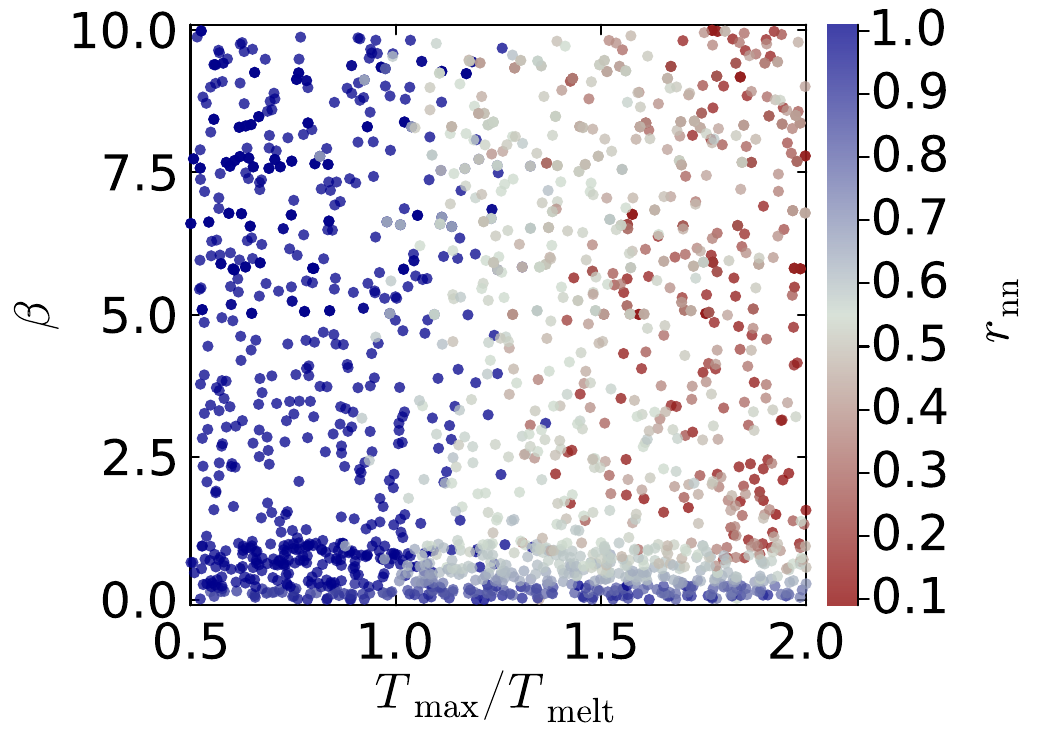}}
\end{minipage}
%
\begin{minipage}{0.48\textwidth}
\centering
\scaledinset{l}{0.01}{b}{.91}{\textbf{b}}{\includegraphics[width=\textwidth]{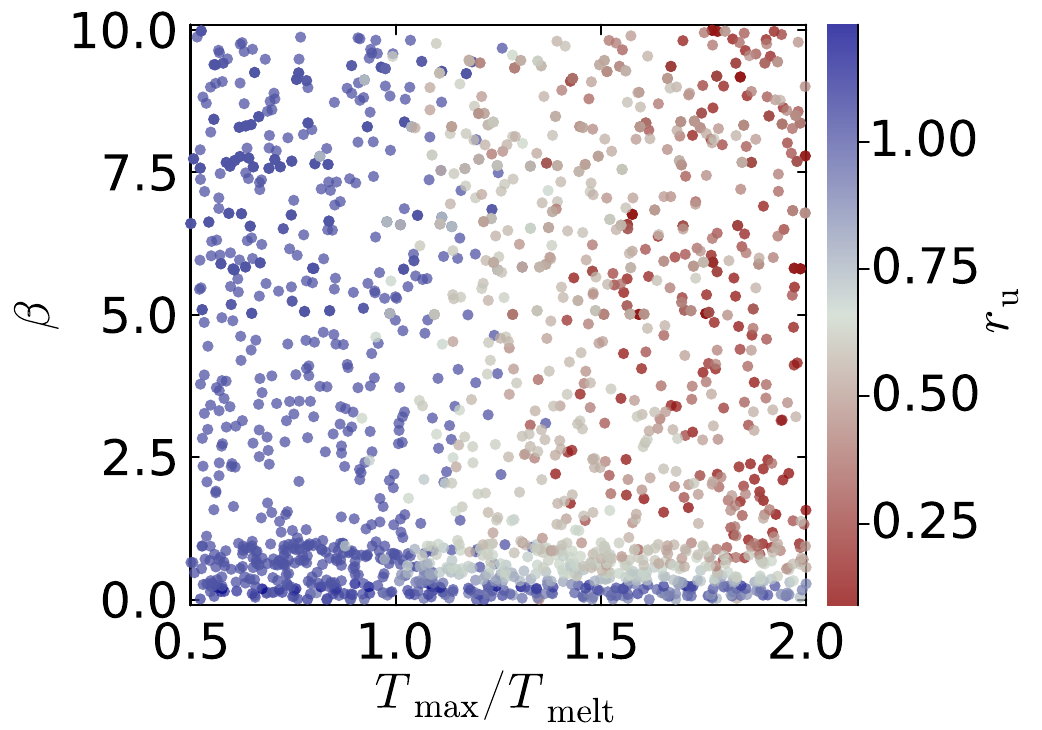}}
\end{minipage}
\begin{minipage}{0.48\textwidth}
\centering
\scaledinset{l}{0.01}{b}{.91}{\textbf{c}}{\includegraphics[width=\textwidth]{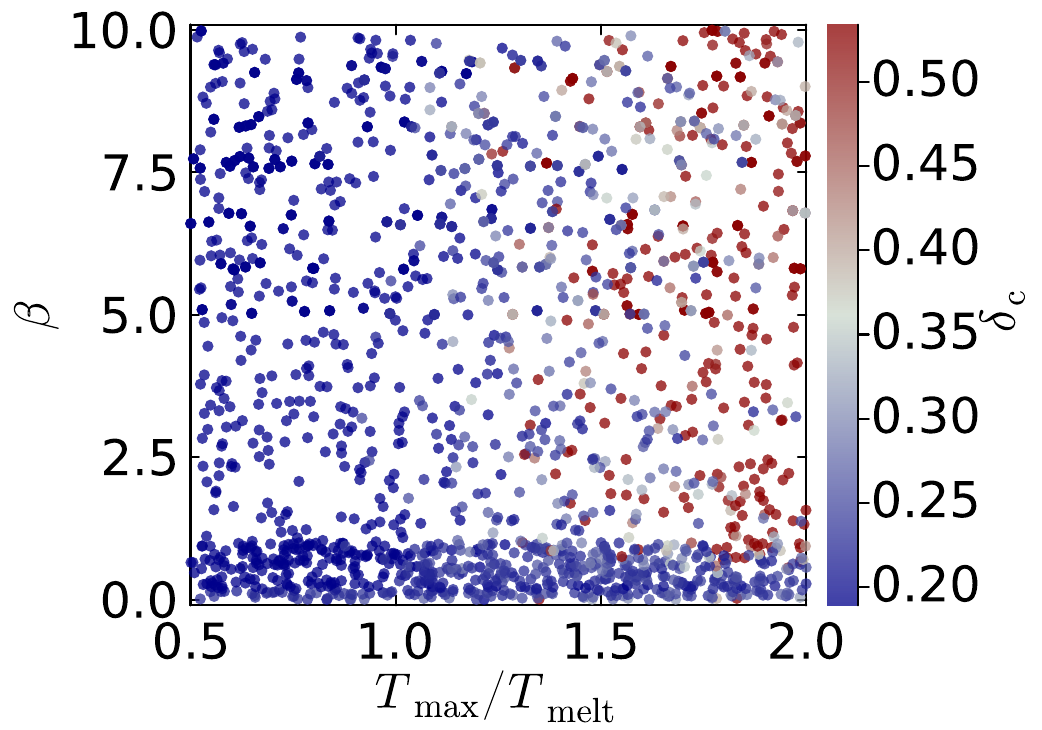}}
\end{minipage}
%
\begin{minipage}{0.48\textwidth}
\centering
\scaledinset{l}{0.01}{b}{.91}{\textbf{d}}{\includegraphics[width=\textwidth]{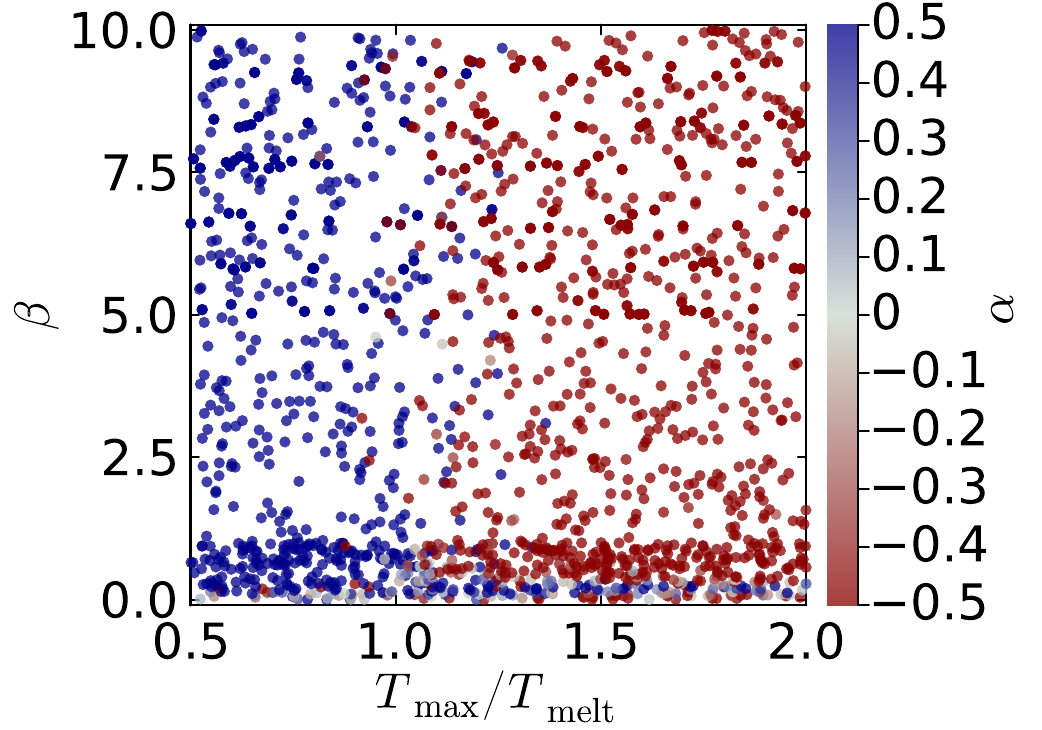}}
\end{minipage}
\caption{\label{fig:pcu_cn_4_5_6_homogeneity_hyperuniformity}
Homogeneity metrics for networks generated from the initial $\textbf{pcu}_\mathrm{mod}$ plotted against the algorithm inputs $T_\mathrm{max}$ and $\beta$. Red colors indicate high disorder with respect to the corresponding metric. \add{The metrics of 1815 networks are displayed.}
\textbf{a} \replace{Coordinated neighbor distance $r_\mathrm{c}$}{Nearest-neighbor distance $r_\mathrm{nn}$}.
\textbf{b} \replace{Uncoordinated neighbor distance}{Nearest-uncoordinated-neighbor distance} $r_\mathrm{u}$.
\textbf{c} Critical pore radius $\delta_\mathrm{c}$.
\textbf{d} Hyperuniformity metric $\alpha$.
}
\end{figure*}

\begin{figure*}[h]
\begin{minipage}{0.48\textwidth}
\centering
\scaledinset{l}{0.01}{b}{.91}{\textbf{a}}{\includegraphics[width=\textwidth]{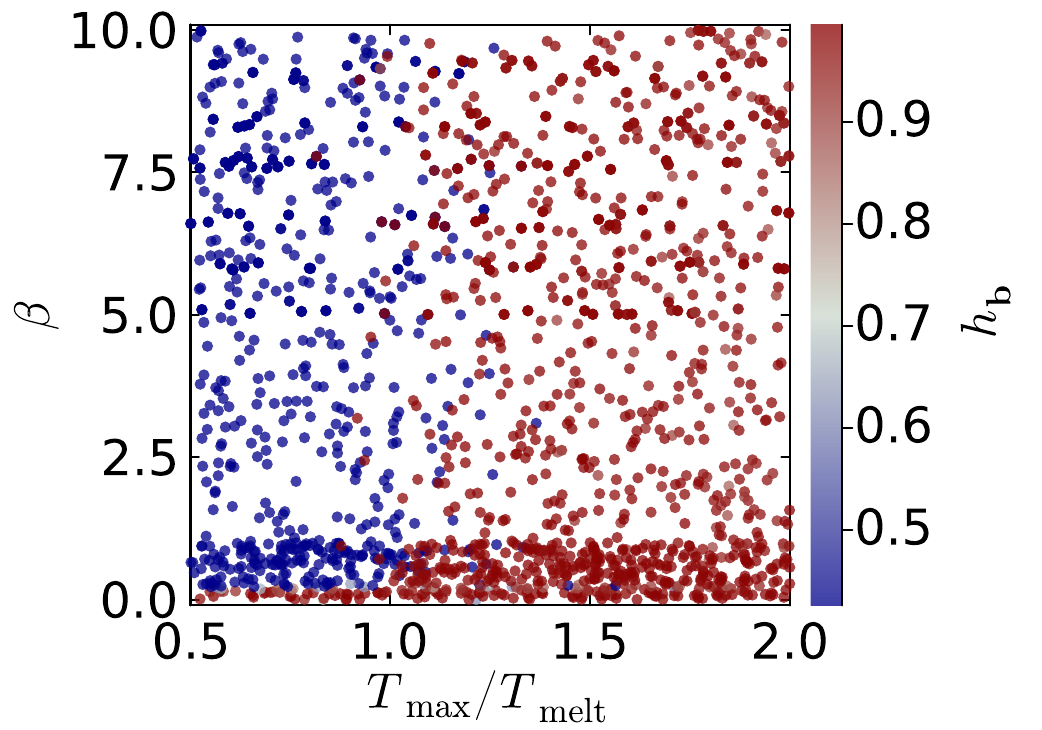}}
\end{minipage}
%
\begin{minipage}{0.48\textwidth}
\centering
\scaledinset{l}{0.01}{b}{.91}{\textbf{b}}{\includegraphics[width=\textwidth]{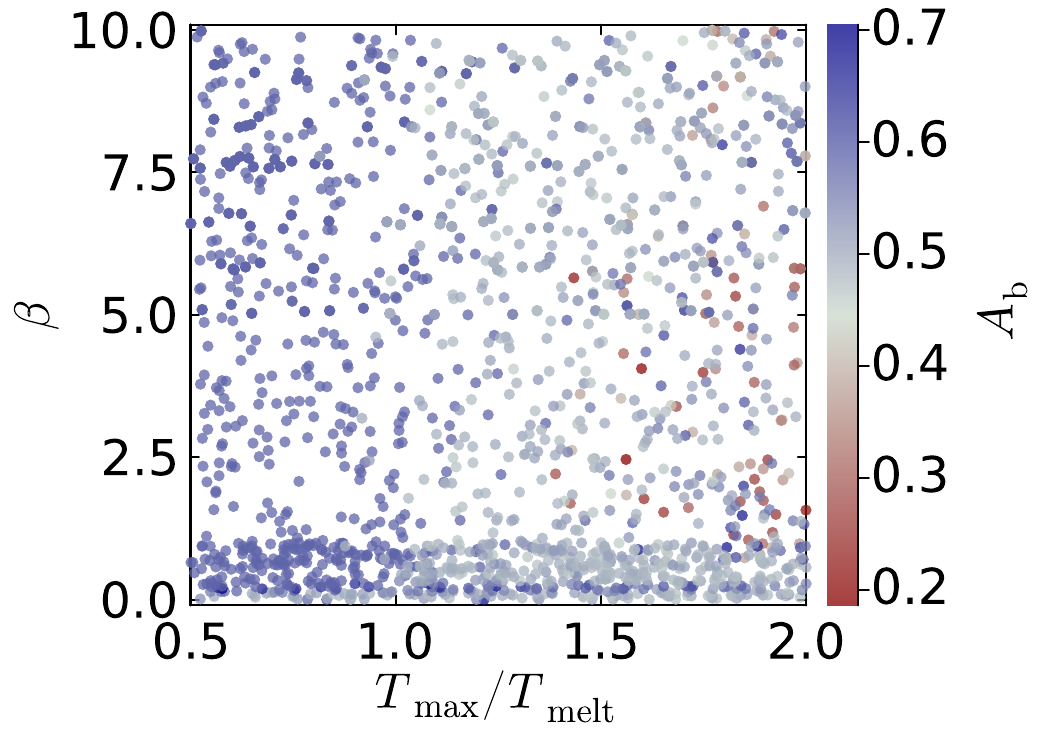}}
\end{minipage}
\begin{minipage}{0.48\textwidth}
\centering
\scaledinset{l}{0.01}{b}{.91}{\textbf{c}}{\includegraphics[width=\textwidth]{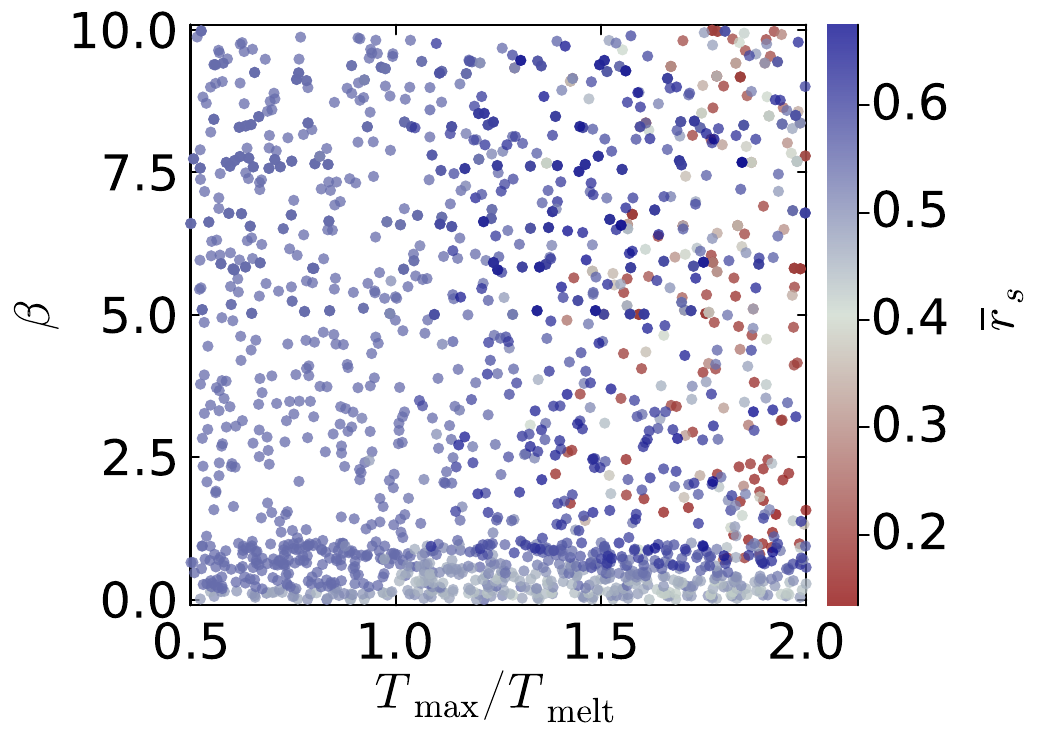}}
\end{minipage}
%
\begin{minipage}{0.48\textwidth}
\centering
\scaledinset{l}{0.01}{b}{.91}{\textbf{d}}{\includegraphics[width=\textwidth]{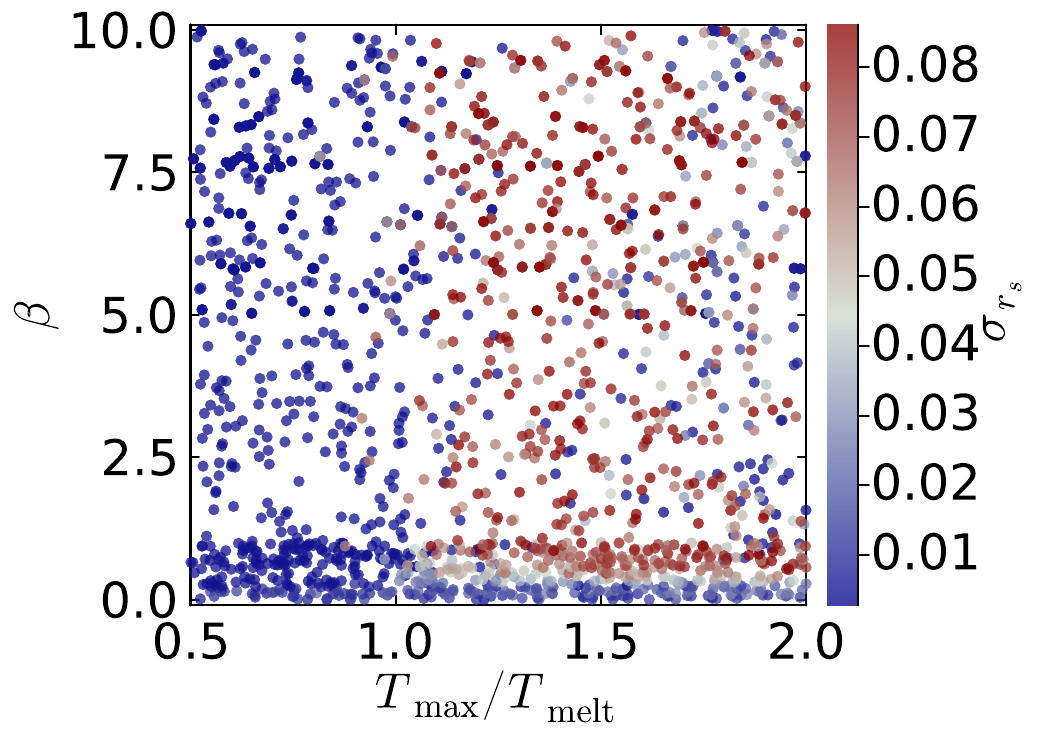}}
\end{minipage}
\caption{\label{fig:pcu_cn_4_5_6_isotropy_topology}
Isotropy and topology metrics for networks generated from the initial $\textbf{pcu}_\mathrm{mod}$ as a function of the network generation algorithm inputs $T_\mathrm{max}$ and $\beta$. Red colors indicate high disorder with respect to the corresponding metric. \add{The metrics of 1815 networks are displayed.}
\textbf{a} Bond orientation entropy $h_\mathbf{b}$.
\textbf{b} \add{Bond} structure factor anisotropy metric $A_\mathrm{b}$.
\textbf{c} Mean ring radius $\overline{r}_s$.
\textbf{d} Ring radius standard deviation $\sigma_{r_s}$.
}
\end{figure*}

\clearpage

\subsection{\textbf{dia}}

\begin{figure*}[h]
\begin{minipage}{0.48\textwidth}
\centering
\scaledinset{l}{0.01}{b}{.91}{\textbf{a}}{\includegraphics[width=\textwidth]{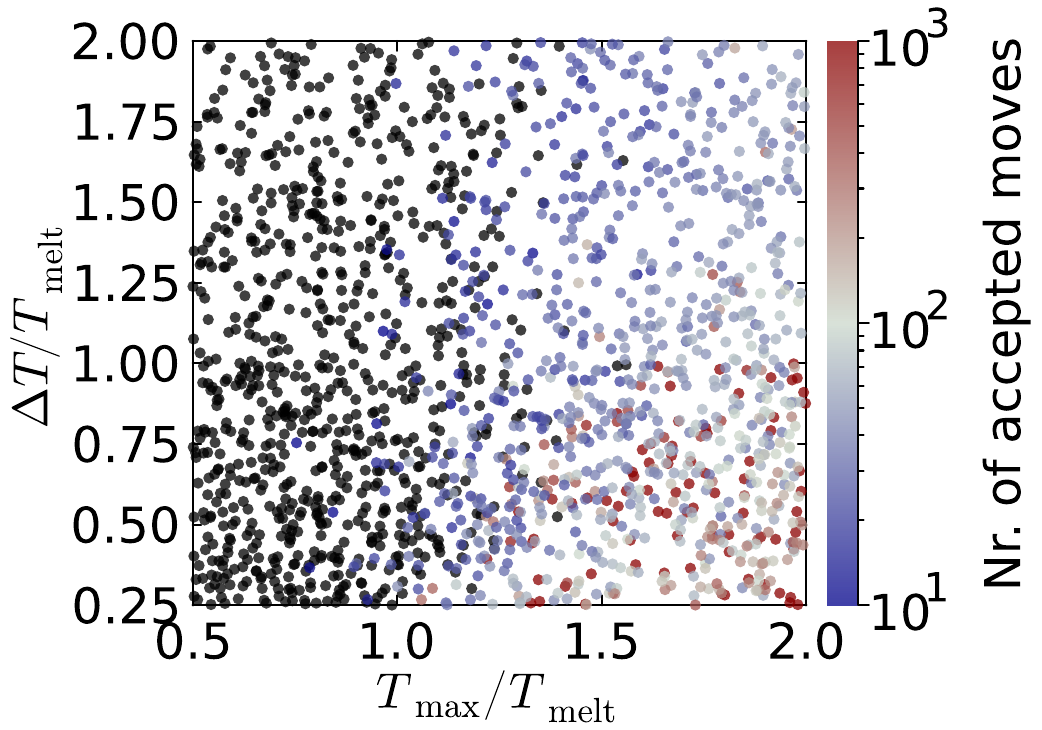}}
\end{minipage}
%
\begin{minipage}{0.48\textwidth}
\centering
\scaledinset{l}{0.01}{b}{.91}{\textbf{b}}{\includegraphics[width=\textwidth]{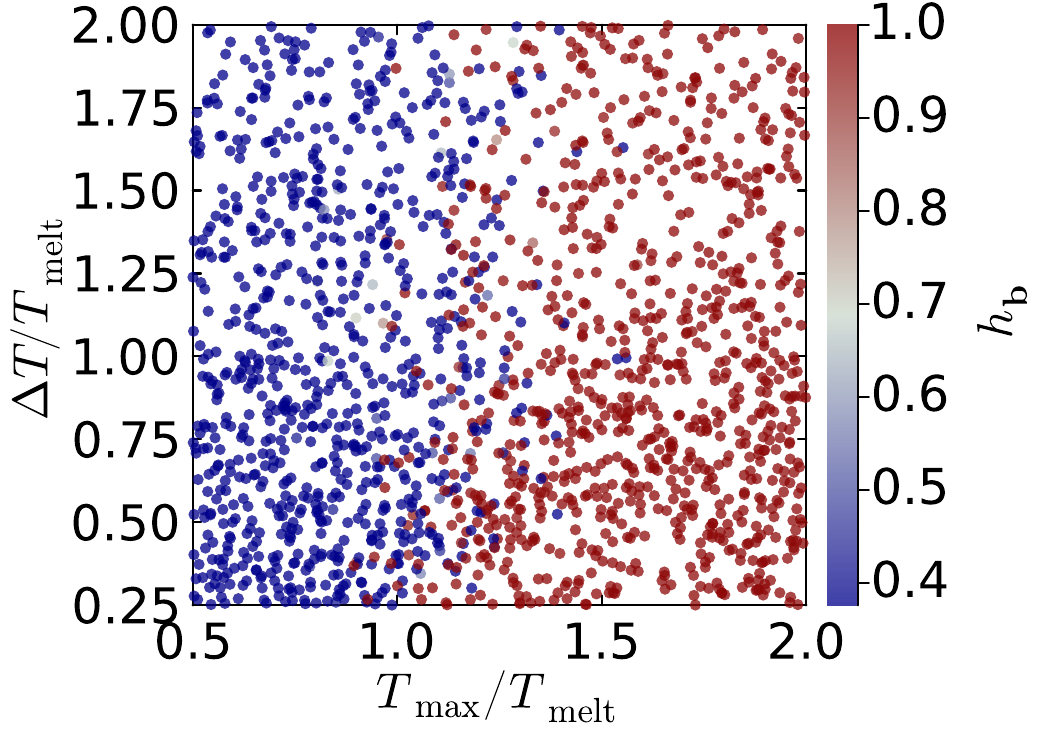}}
\end{minipage}
\caption{\label{fig:dia_nr_accepted_moves_isotropy}
Melting transition for networks generated from the initial \textbf{dia} network. \add{The metrics of 2250 networks are displayed.}
\textbf{a} Number of accepted Monte Carlo moves plotted against $T_\mathrm{max}$ and $\Delta T$. The black markers correspond to networks with ten or fewer accepted moves.
\textbf{b} Bond orientation entropy $h_\mathbf{b}$.
}
\end{figure*}

\begin{figure}[h]
\includegraphics[width=0.48\linewidth]{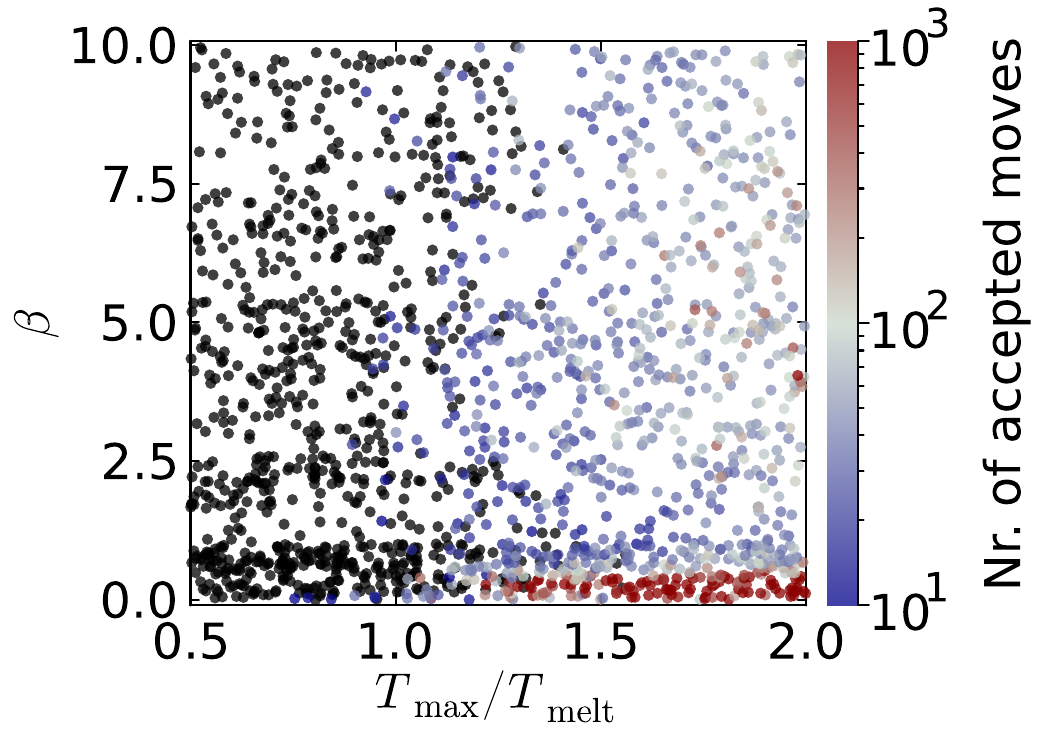}
\caption{\label{fig:dia_nr_accepted_moves_beta} Number of accepted Monte Carlo moves for the initial \textbf{dia} plotted against $T_\mathrm{max}$ and $\Delta T$. The black markers correspond to networks with ten or fewer accepted moves. \add{The metrics of 2250 networks are displayed.}}
\end{figure}

\begin{figure*}[h]
\begin{minipage}{0.48\textwidth}
\centering
\scaledinset{l}{0.01}{b}{.91}{\textbf{a}}{\includegraphics[width=\textwidth]{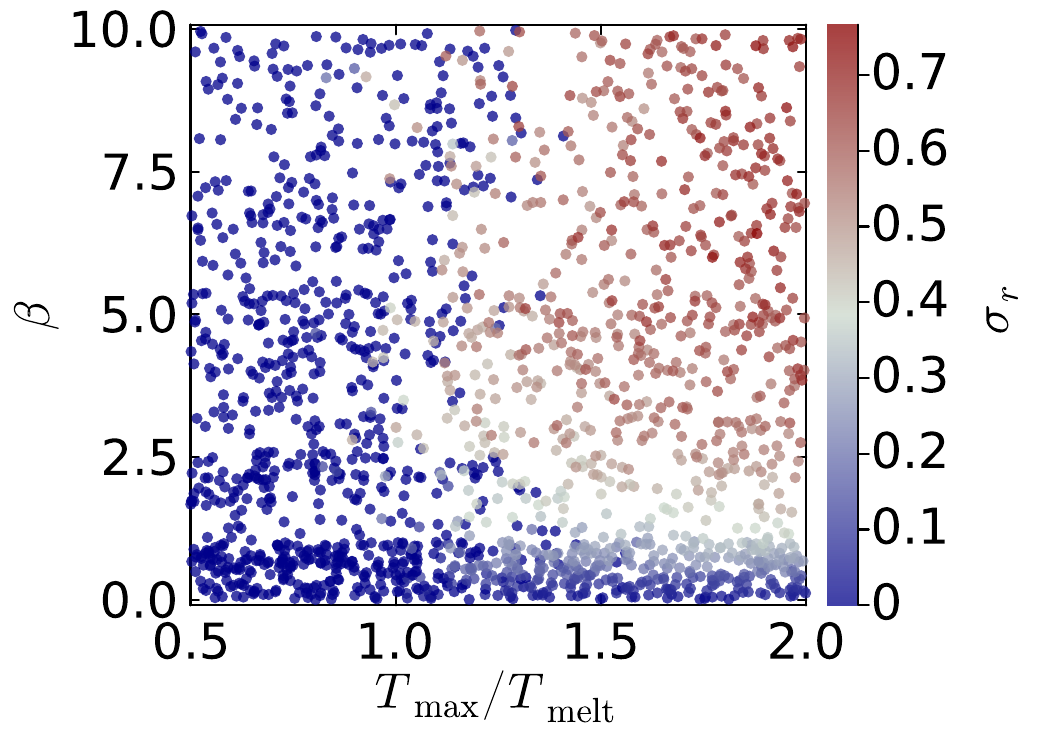}}
\end{minipage}
%
\begin{minipage}{0.48\textwidth}
\centering
\scaledinset{l}{0.01}{b}{.91}{\textbf{b}}{\includegraphics[width=\textwidth]{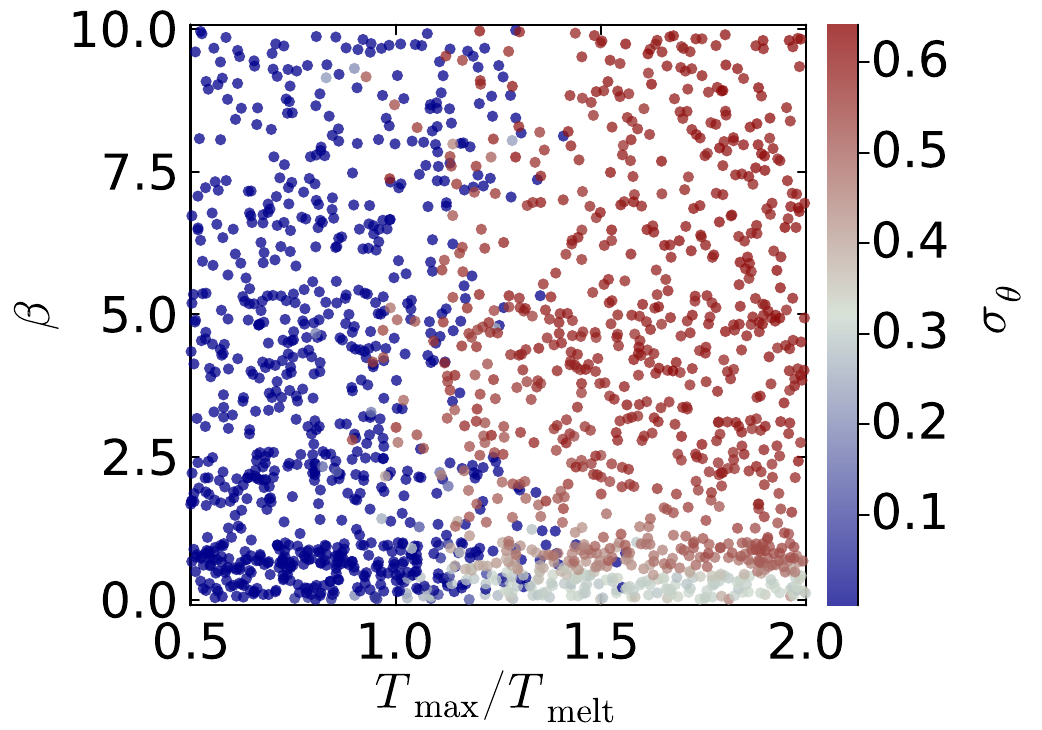}}
\end{minipage}
\caption{\label{fig:dia_bond_length_angle}
Network primitive metrics for networks generated from the initial \textbf{dia} plotted against the algorithm inputs $T_\mathrm{max}$ and $\beta$. Red colors indicate high disorder with respect to the corresponding metric. \add{The metrics of 2250 networks are displayed.}
\textbf{a} Bond length standard deviation $\sigma_r$. 
\textbf{b} Bond angle standard deviation $\sigma_\theta$.
}
\end{figure*}

\begin{figure*}[h]
\begin{minipage}{0.48\textwidth}
\centering
\scaledinset{l}{0.01}{b}{.91}{\textbf{a}}{\includegraphics[width=\textwidth]{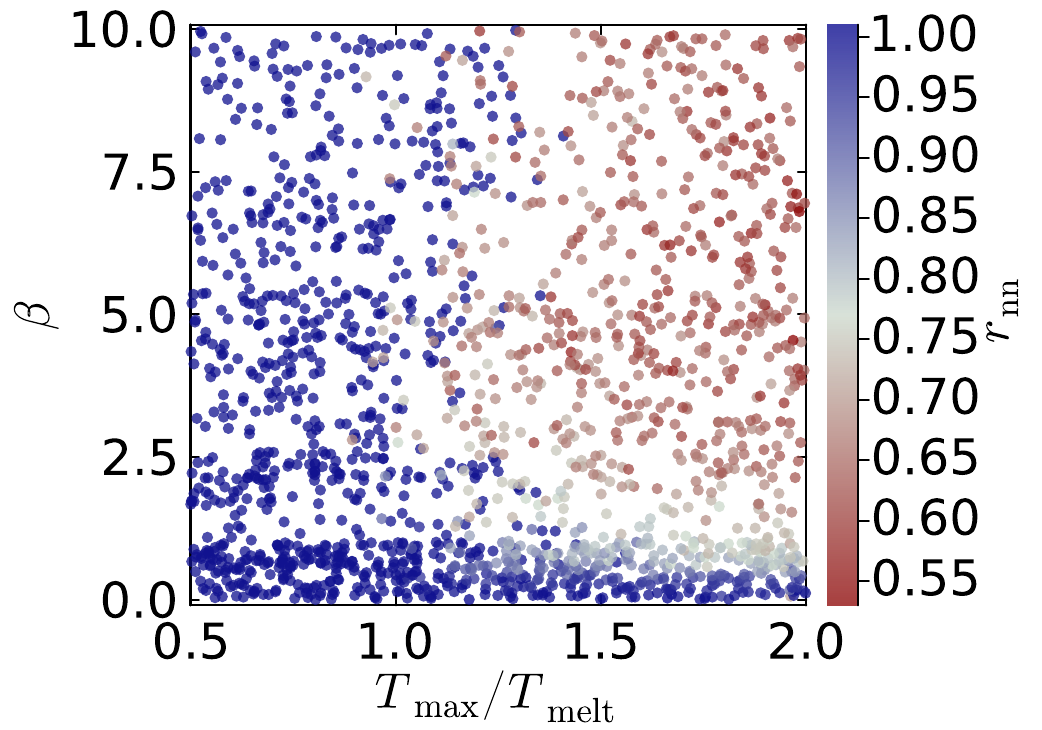}}
\end{minipage}
%
\begin{minipage}{0.48\textwidth}
\centering
\scaledinset{l}{0.01}{b}{.91}{\textbf{b}}{\includegraphics[width=\textwidth]{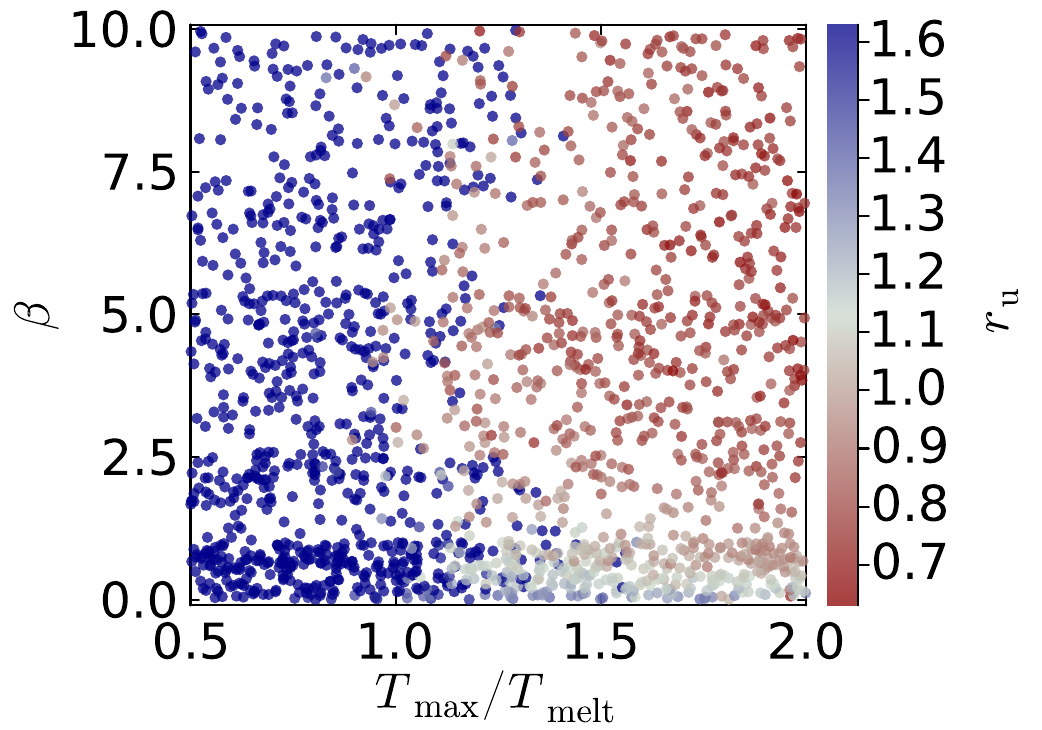}}
\end{minipage}
\begin{minipage}{0.48\textwidth}
\centering
\scaledinset{l}{0.01}{b}{.91}{\textbf{c}}{\includegraphics[width=\textwidth]{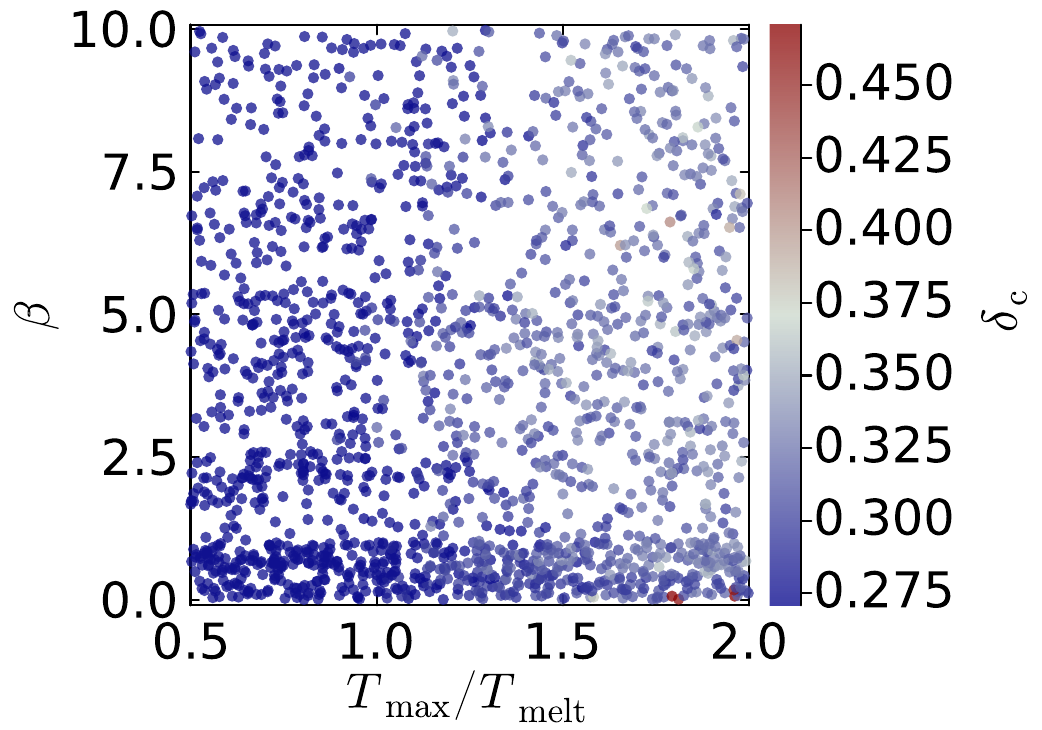}}
\end{minipage}
%
\begin{minipage}{0.48\textwidth}
\centering
\scaledinset{l}{0.01}{b}{.91}{\textbf{d}}{\includegraphics[width=\textwidth]{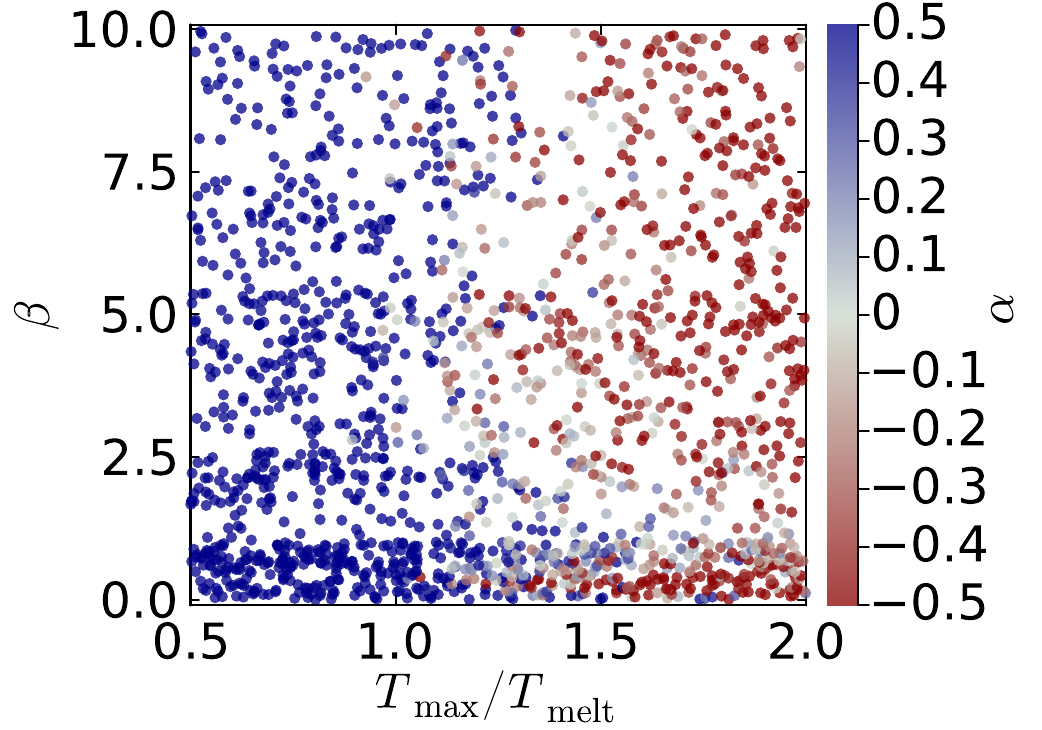}}
\end{minipage}
\caption{\label{fig:dia_homogeneity_hyperuniformity}
Homogeneity metrics for networks generated from the initial \textbf{dia} plotted against the algorithm inputs $T_\mathrm{max}$ and $\beta$. Red colors indicate high disorder with respect to the corresponding metric. \add{The metrics of 2250 networks are displayed.}
\textbf{a} \replace{Coordinated neighbor distance $r_\mathrm{c}$}{Nearest-neighbor distance $r_\mathrm{nn}$}.
\textbf{b} \replace{Uncoordinated neighbor distance}{Nearest-uncoordinated-neighbor distance} $r_\mathrm{u}$.
\textbf{c} Critical pore radius $\delta_\mathrm{c}$.
\textbf{d} Hyperuniformity metric $\alpha$.
}
\end{figure*}

\begin{figure*}[h]
\begin{minipage}{0.48\textwidth}
\centering
\scaledinset{l}{0.01}{b}{.91}{\textbf{a}}{\includegraphics[width=\textwidth]{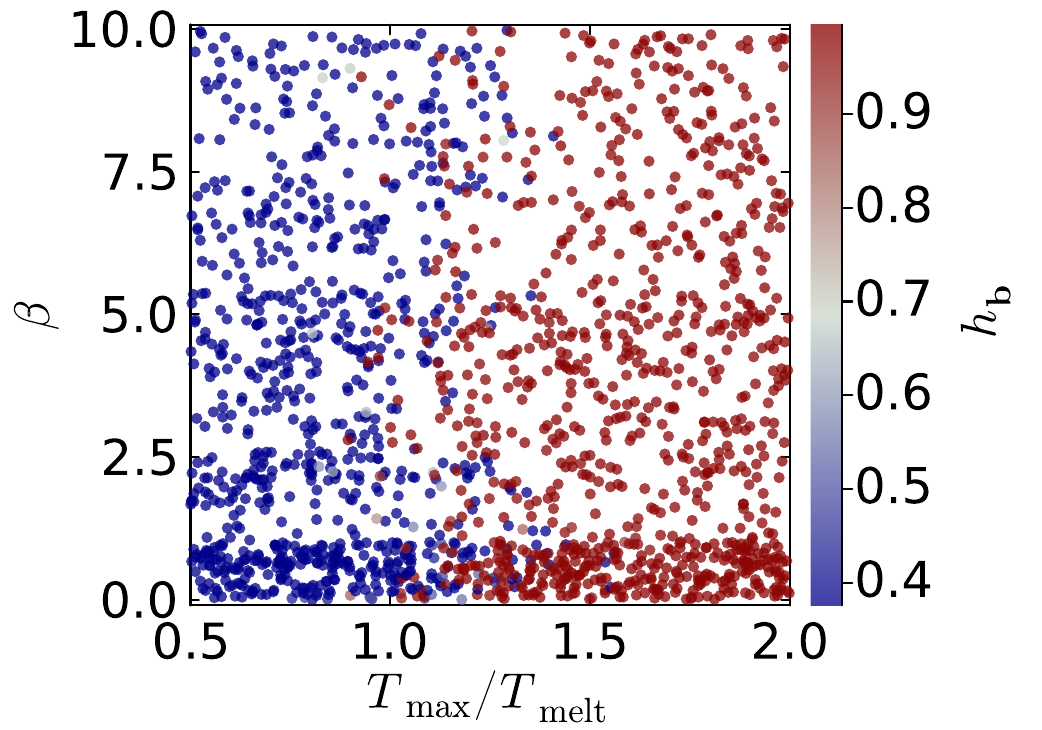}}
\end{minipage}
%
\begin{minipage}{0.48\textwidth}
\centering
\scaledinset{l}{0.01}{b}{.91}{\textbf{b}}{\includegraphics[width=\textwidth]{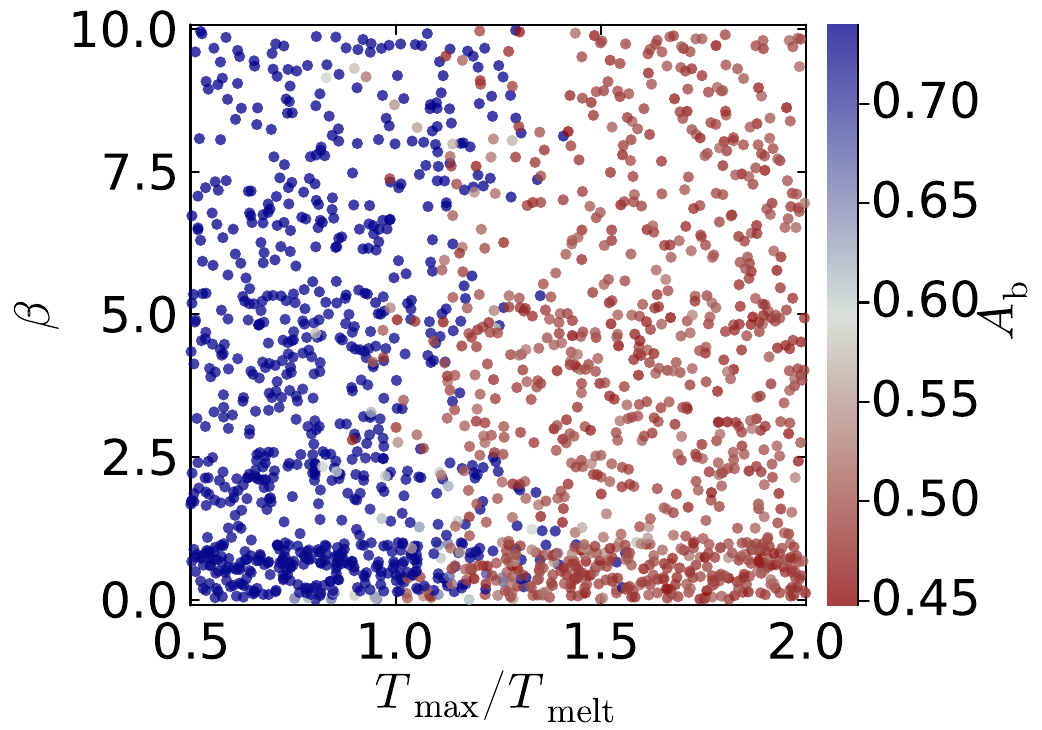}}
\end{minipage}
\begin{minipage}{0.48\textwidth}
\centering
\scaledinset{l}{0.01}{b}{.91}{\textbf{c}}{\includegraphics[width=\textwidth]{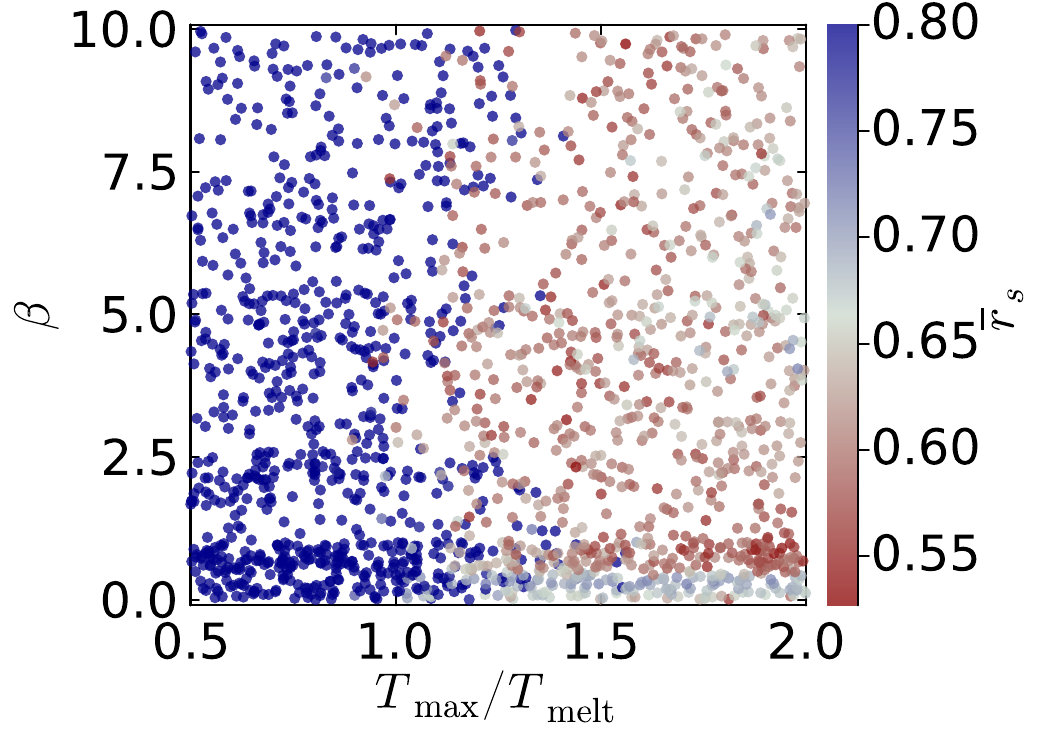}}
\end{minipage}
%
\begin{minipage}{0.48\textwidth}
\centering
\scaledinset{l}{0.01}{b}{.91}{\textbf{d}}{\includegraphics[width=\textwidth]{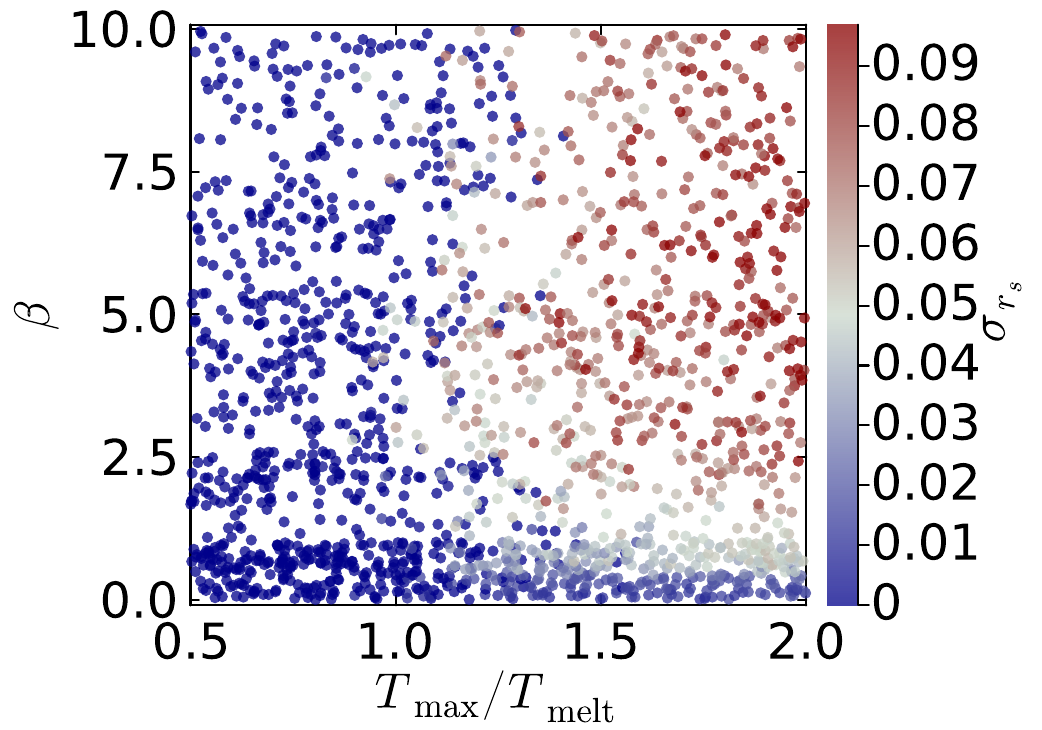}}
\end{minipage}
\caption{\label{fig:dia_isotropy_topology}
Isotropy and topology metrics for networks generated from the initial \textbf{dia} as a function of the network generation algorithm inputs $T_\mathrm{max}$ and $\beta$. Red colors indicate high disorder with respect to the corresponding metric. \add{The metrics of 2250 networks are displayed.}
\textbf{a} Bond orientation entropy $h_\mathbf{b}$.
\textbf{b} \add{Bond} structure factor anisotropy metric $A_\mathrm{b}$.
\textbf{c} Mean ring radius $\overline{r}_s$.
\textbf{d} Ring radius standard deviation $\sigma_{r_s}$.
}
\end{figure*}

\clearpage

\subsection{\textbf{srs}}

\begin{figure*}[h]
\begin{minipage}{0.48\textwidth}
\centering
\scaledinset{l}{0.01}{b}{.91}{\textbf{a}}{\includegraphics[width=\textwidth]{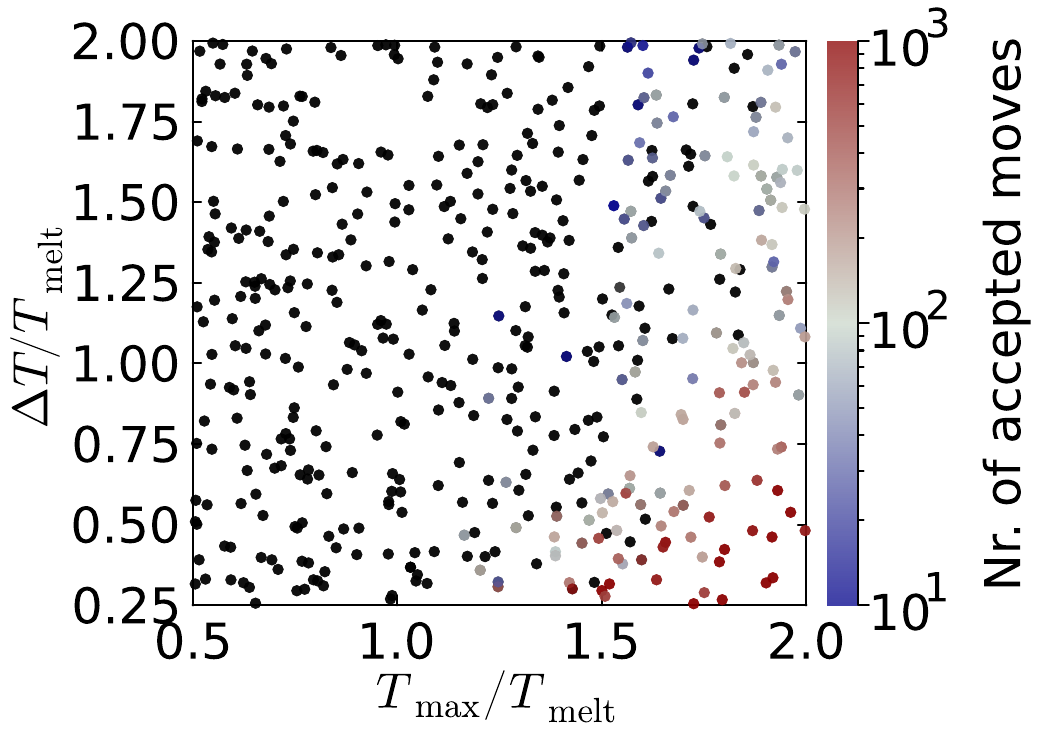}}
\end{minipage}
%
\begin{minipage}{0.48\textwidth}
\centering
\scaledinset{l}{0.01}{b}{.91}{\textbf{b}}{\includegraphics[width=\textwidth]{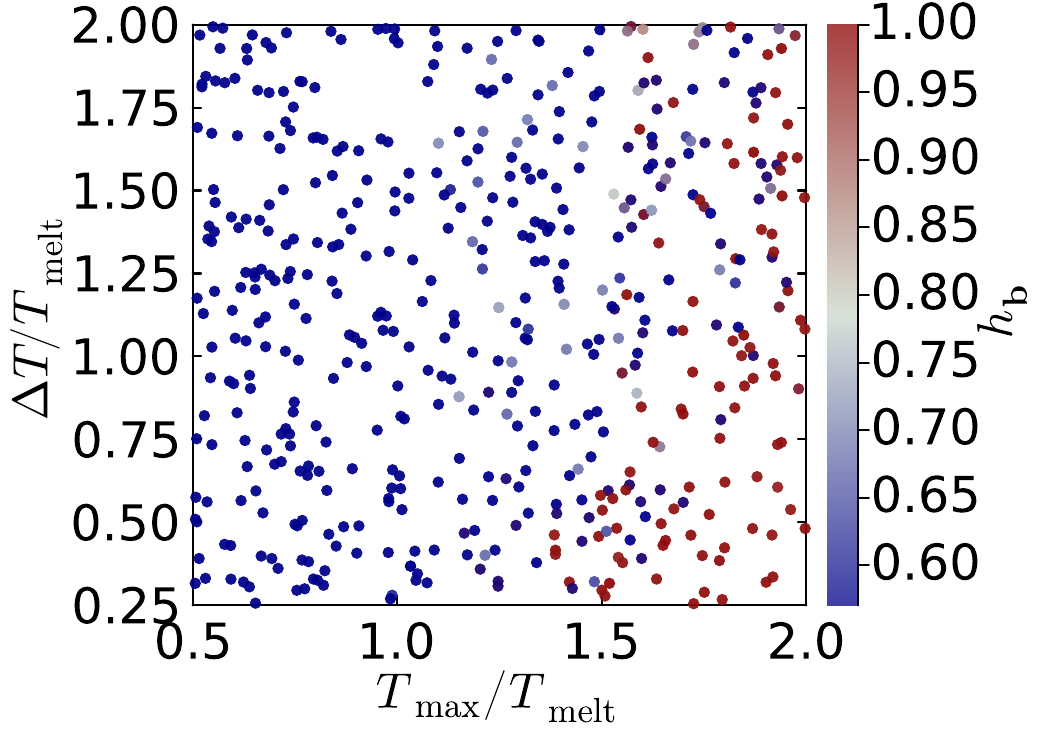}}
\end{minipage}
\caption{\label{fig:srs_nr_accepted_moves_isotropy}
Melting transition for networks generated from the initial \textbf{srs}. \add{The metrics of 1000 networks are displayed.}
\textbf{a} Number of accepted Monte Carlo moves plotted against $T_\mathrm{max}$ and $\Delta T$. The black markers correspond to networks with 10 or fewer accepted moves.
\textbf{b} Bond orientation entropy $h_\mathbf{b}$.
}
\end{figure*}

\begin{figure}[h]
\includegraphics[width=0.48\linewidth]{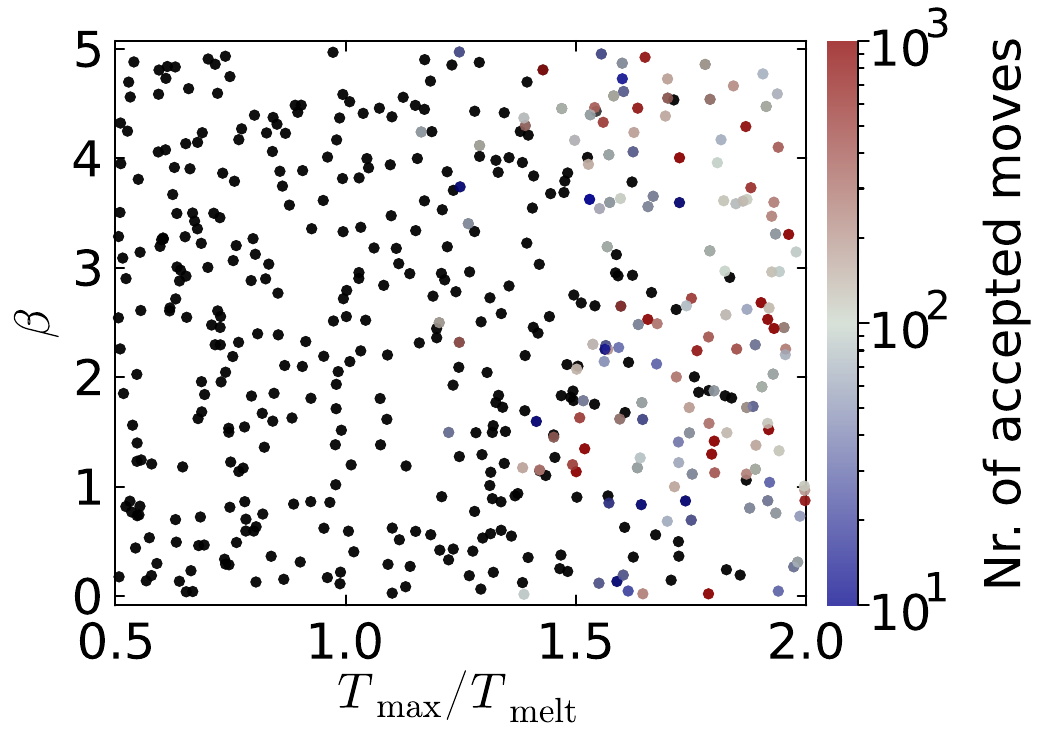}
\caption{\label{fig:srs_nr_accepted_moves_beta} Number of accepted Monte Carlo moves for the initial \textbf{srs} plotted against $T_\mathrm{max}$ and $\Delta T$. The black markers correspond to networks with ten or fewer accepted moves. \add{The metrics of 1000 networks are displayed.}}
\end{figure}

\begin{figure*}[h]
\begin{minipage}{0.48\textwidth}
\centering
\scaledinset{l}{0.01}{b}{.91}{\textbf{a}}{\includegraphics[width=\textwidth]{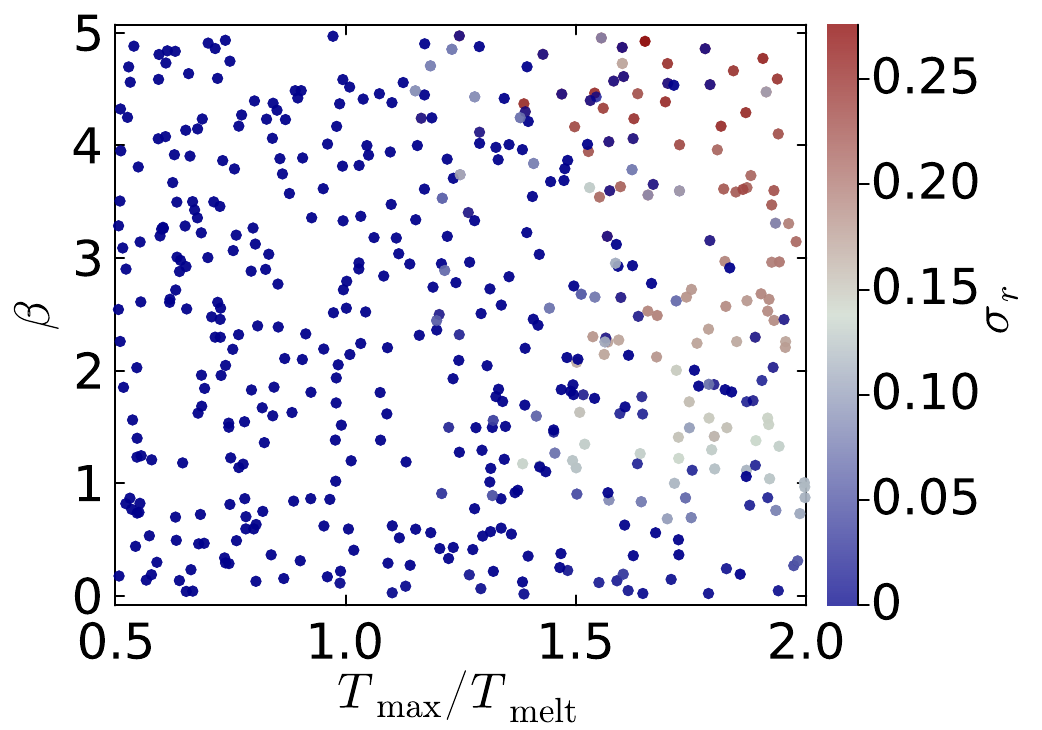}}
\end{minipage}
%
\begin{minipage}{0.48\textwidth}
\centering
\scaledinset{l}{0.01}{b}{.91}{\textbf{b}}{\includegraphics[width=\textwidth]{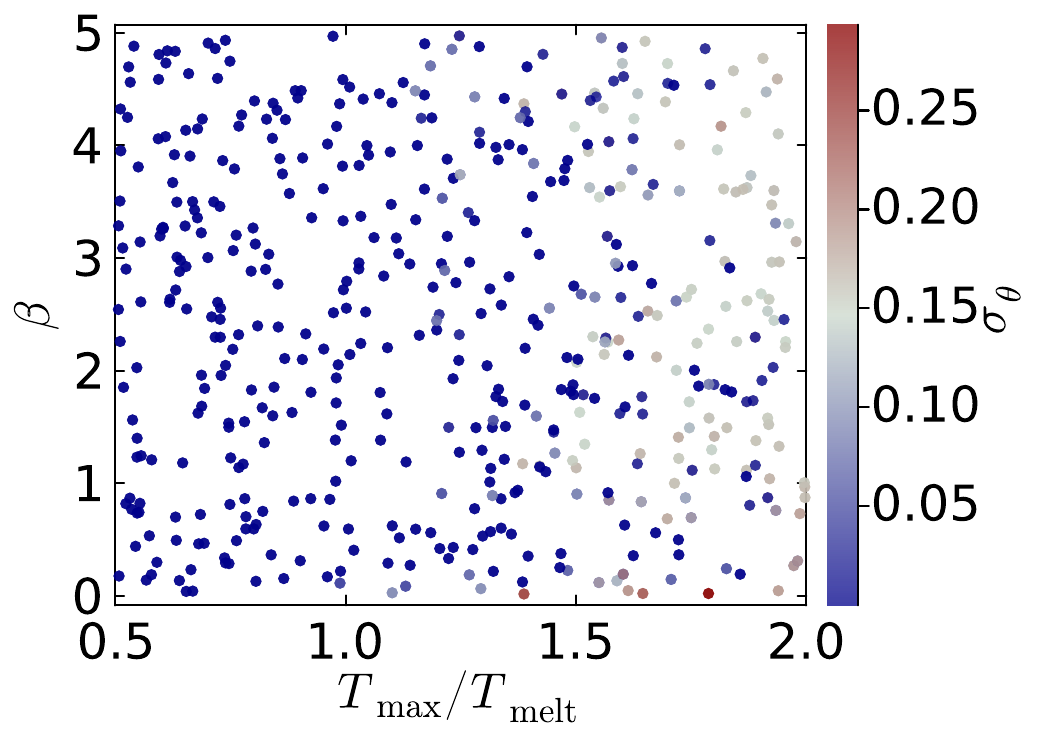}}
\end{minipage}
\caption{\label{fig:srs_bond_length_angle}
Network primitive metrics for networks generated from the initial \textbf{srs} plotted against the algorithm inputs $T_\mathrm{max}$ and $\beta$. Red colors indicate high disorder with respect to the corresponding metric. \add{The metrics of 1000 networks are displayed.}
\textbf{a} Bond length standard deviation $\sigma_r$. 
\textbf{b} Bond angle standard deviation $\sigma_\theta$.
}
\end{figure*}

\begin{figure*}[h]
\begin{minipage}{0.48\textwidth}
\centering
\scaledinset{l}{0.01}{b}{.91}{\textbf{a}}{\includegraphics[width=\textwidth]{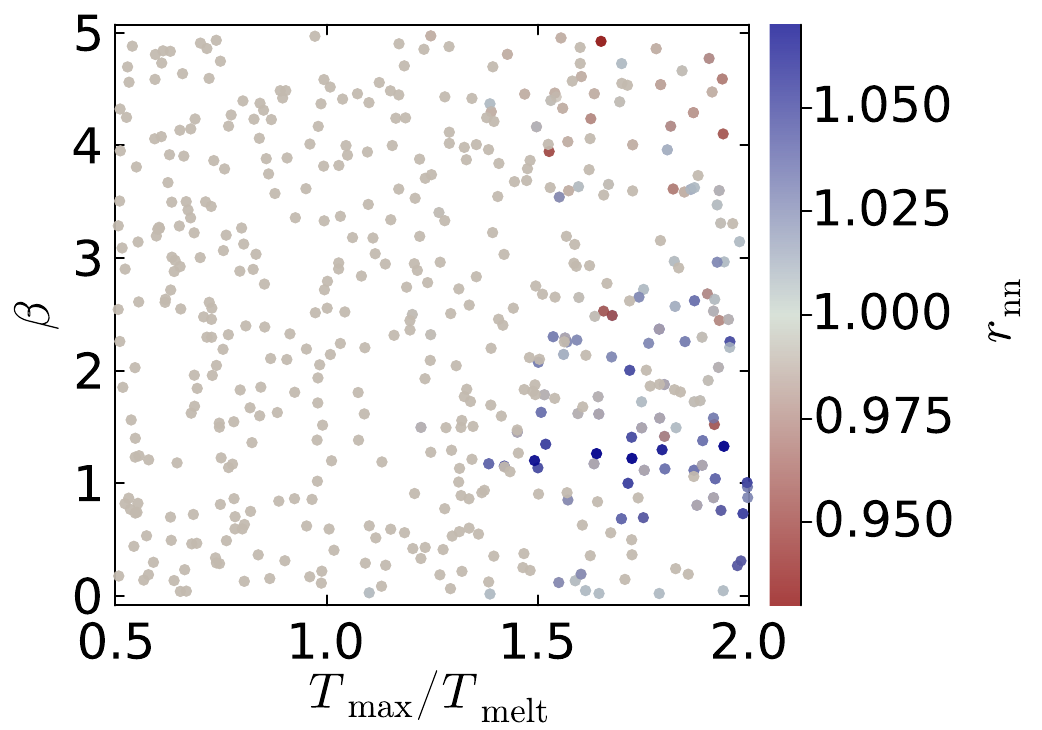}}
\end{minipage}
%
\begin{minipage}{0.48\textwidth}
\centering
\scaledinset{l}{0.01}{b}{.91}{\textbf{b}}{\includegraphics[width=\textwidth]{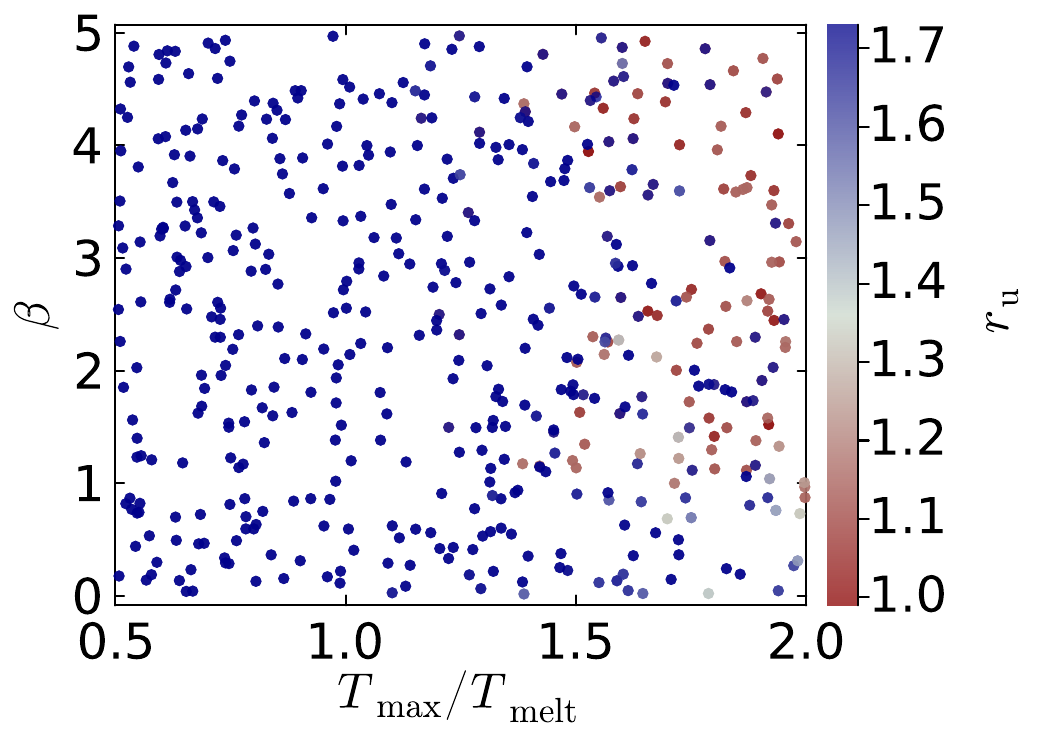}}
\end{minipage}
\begin{minipage}{0.48\textwidth}
\centering
\scaledinset{l}{0.01}{b}{.91}{\textbf{c}}{\includegraphics[width=\textwidth]{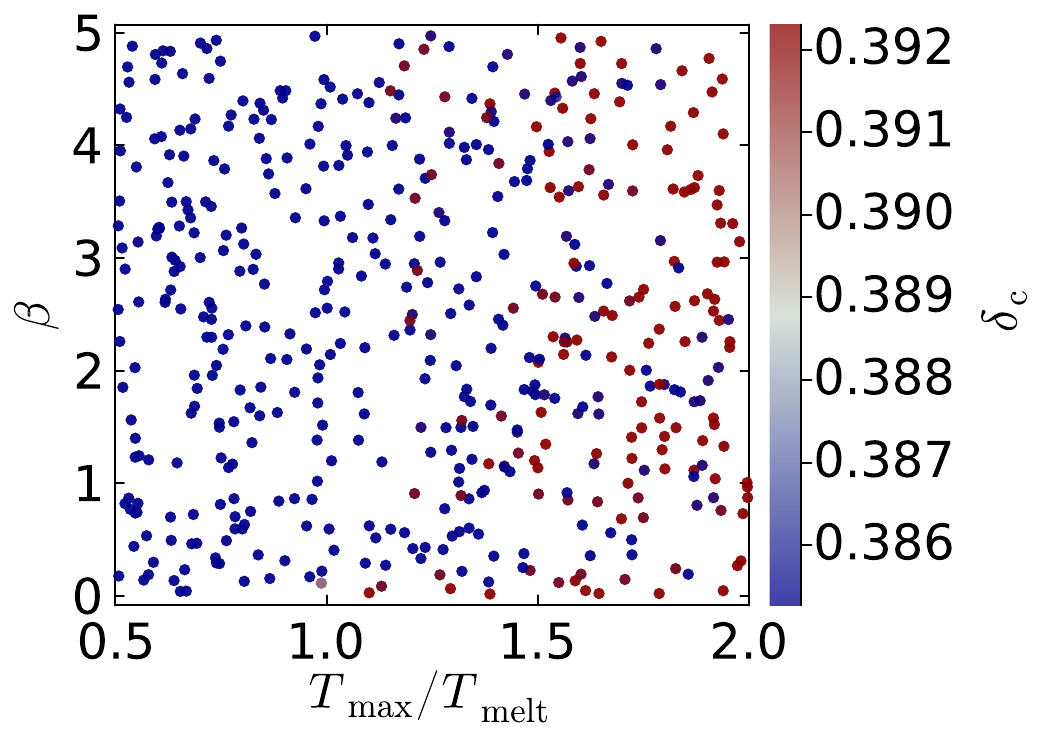}}
\end{minipage}
%
\begin{minipage}{0.48\textwidth}
\centering
\scaledinset{l}{0.01}{b}{.91}{\textbf{d}}{\includegraphics[width=\textwidth]{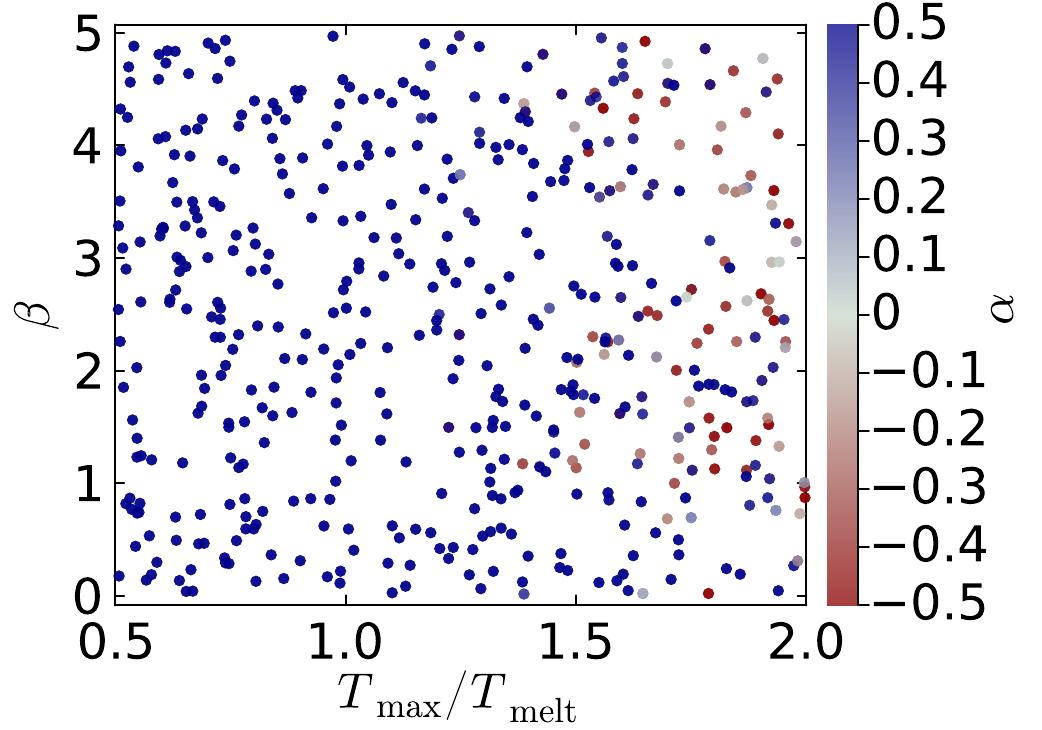}}
\end{minipage}
\caption{\label{fig:srs_homogeneity_hyperuniformity}
Homogeneity metrics for networks generated from the initial \textbf{srs} plotted against the algorithm inputs $T_\mathrm{max}$ and $\beta$. Red colors indicate high disorder with respect to the corresponding metric. \add{The metrics of 1000 networks are displayed.}
\textbf{a} \replace{Coordinated neighbor distance $r_\mathrm{c}$}{Nearest-neighbor distance $r_\mathrm{nn}$}.
\textbf{b} \replace{Uncoordinated neighbor distance}{Nearest-uncoordinated-neighbor distance} $r_\mathrm{u}$.
\textbf{c} Critical pore radius $\delta_\mathrm{c}$.
\textbf{d} Hyperuniformity metric $\alpha$.
}
\end{figure*}

\begin{figure*}[h]
\begin{minipage}{0.48\textwidth}
\centering
\scaledinset{l}{0.01}{b}{.91}{\textbf{a}}{\includegraphics[width=\textwidth]{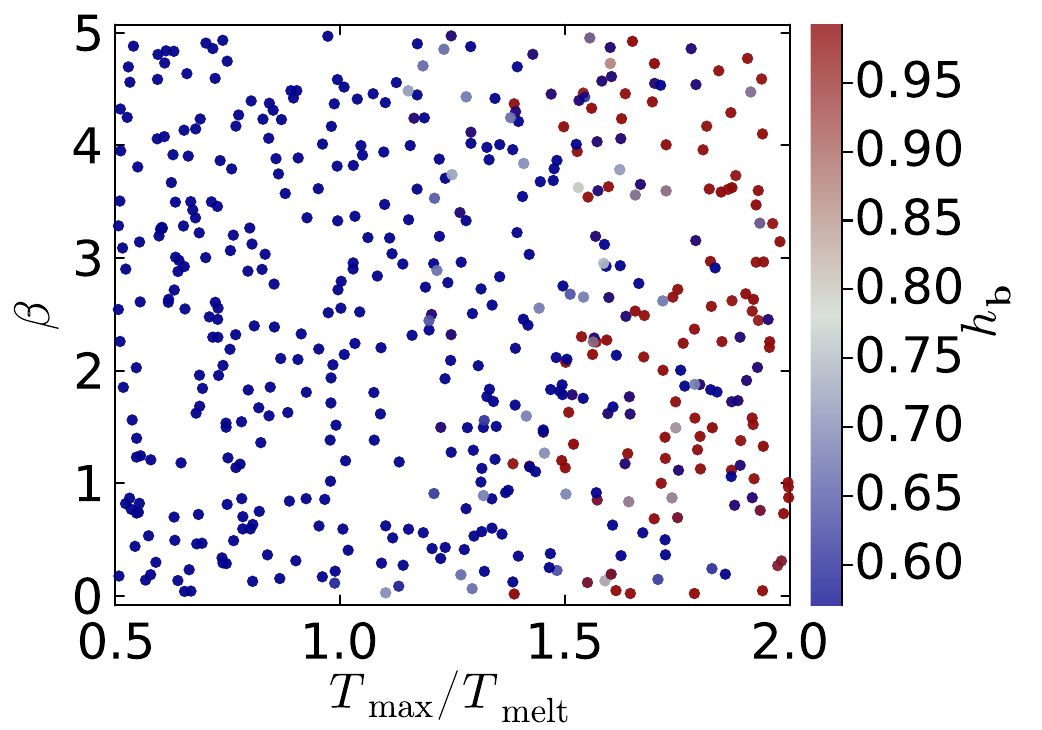}}
\end{minipage}
%
\begin{minipage}{0.48\textwidth}
\centering
\scaledinset{l}{0.01}{b}{.91}{\textbf{b}}{\includegraphics[width=\textwidth]{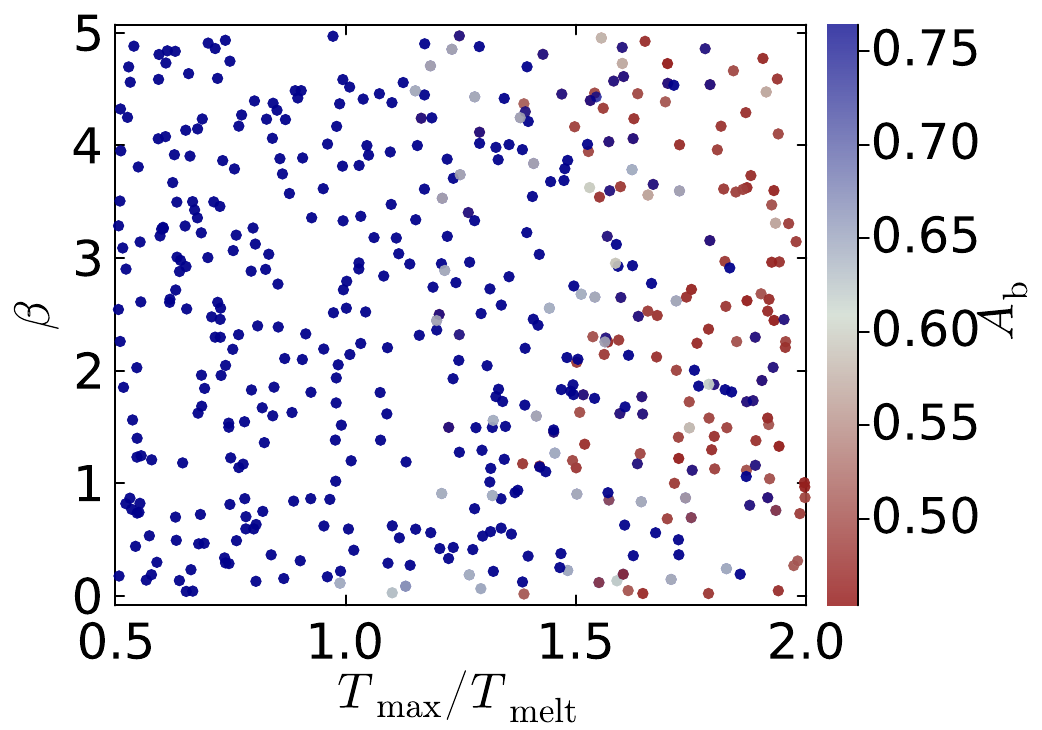}}
\end{minipage}
\begin{minipage}{0.48\textwidth}
\centering
\scaledinset{l}{0.01}{b}{.91}{\textbf{c}}{\includegraphics[width=\textwidth]{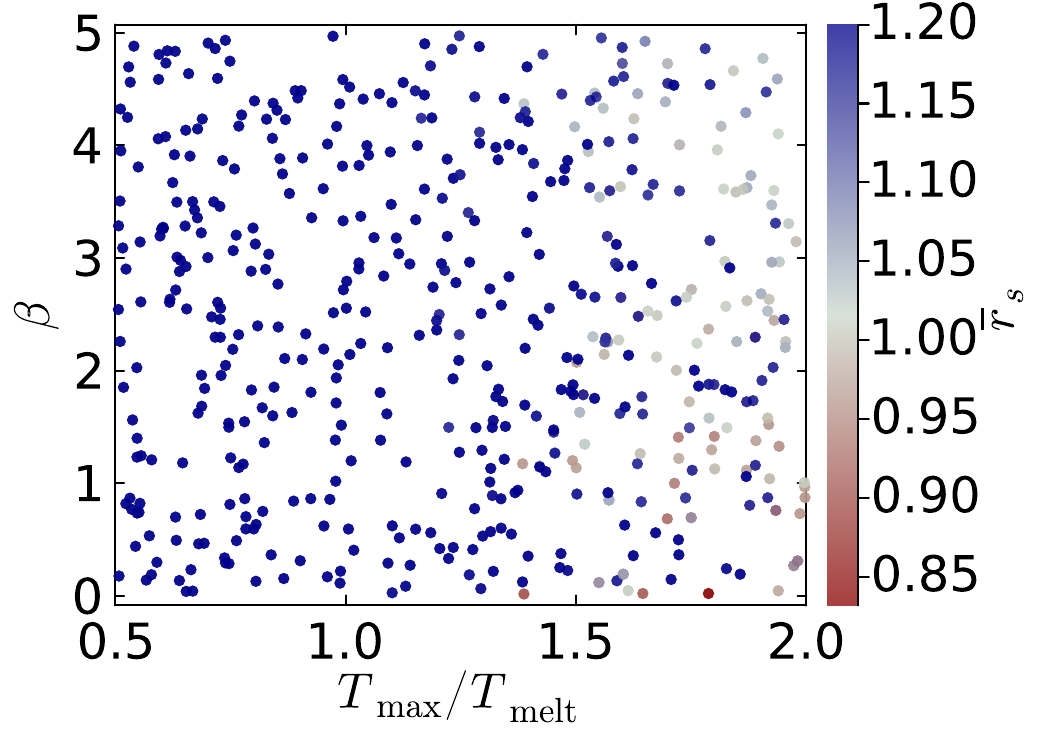}}
\end{minipage}
%
\begin{minipage}{0.48\textwidth}
\centering
\scaledinset{l}{0.01}{b}{.91}{\textbf{d}}{\includegraphics[width=\textwidth]{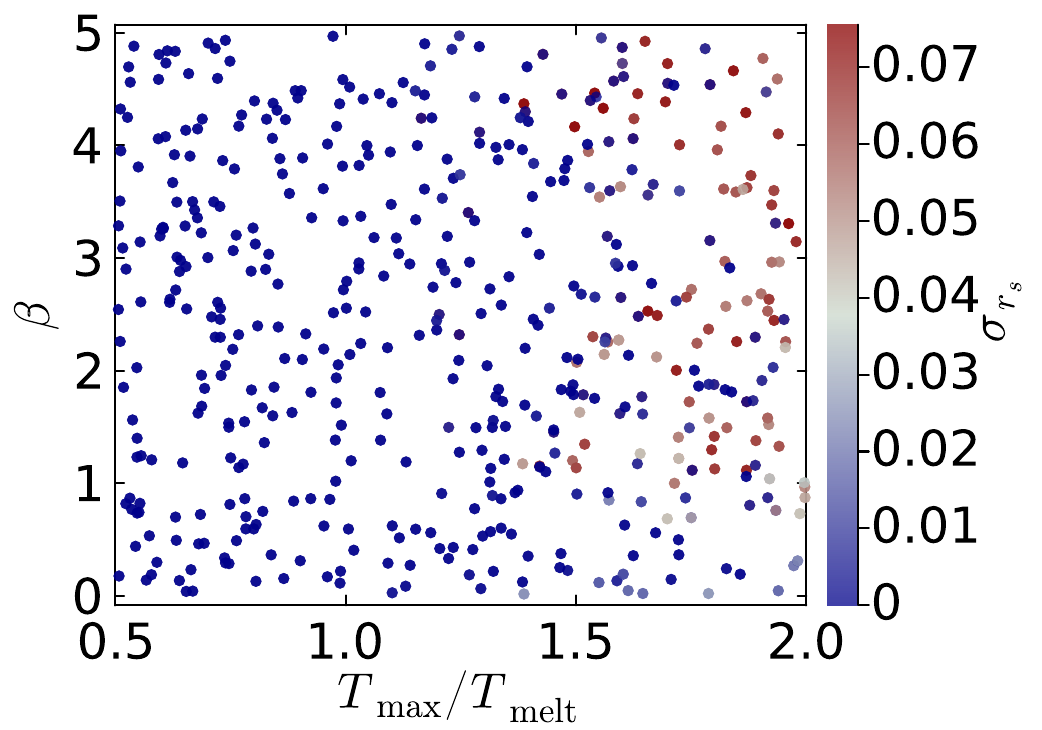}}
\end{minipage}
\caption{\label{fig:srs_isotropy_topology}
Isotropy and topology metrics for networks generated from the initial \textbf{srs} as a function of the network generation algorithm inputs $T_\mathrm{max}$ and $\beta$. Red colors indicate high disorder with respect to the corresponding metric. \add{The metrics of 1000 networks are displayed.}
\textbf{a} Bond orientation entropy $h_\mathbf{b}$.
\textbf{b} \add{Bond} structure factor anisotropy metric $A_\mathrm{b}$.
\textbf{c} Mean ring radius $\overline{r}_s$.
\textbf{d} Ring radius standard deviation $\sigma_{r_s}$.
}
\end{figure*}

\clearpage

\subsection{\textbf{lcs}}

\begin{figure*}[h]
\begin{minipage}{0.48\textwidth}
\centering
\scaledinset{l}{0.01}{b}{.91}{\textbf{a}}{\includegraphics[width=\textwidth]{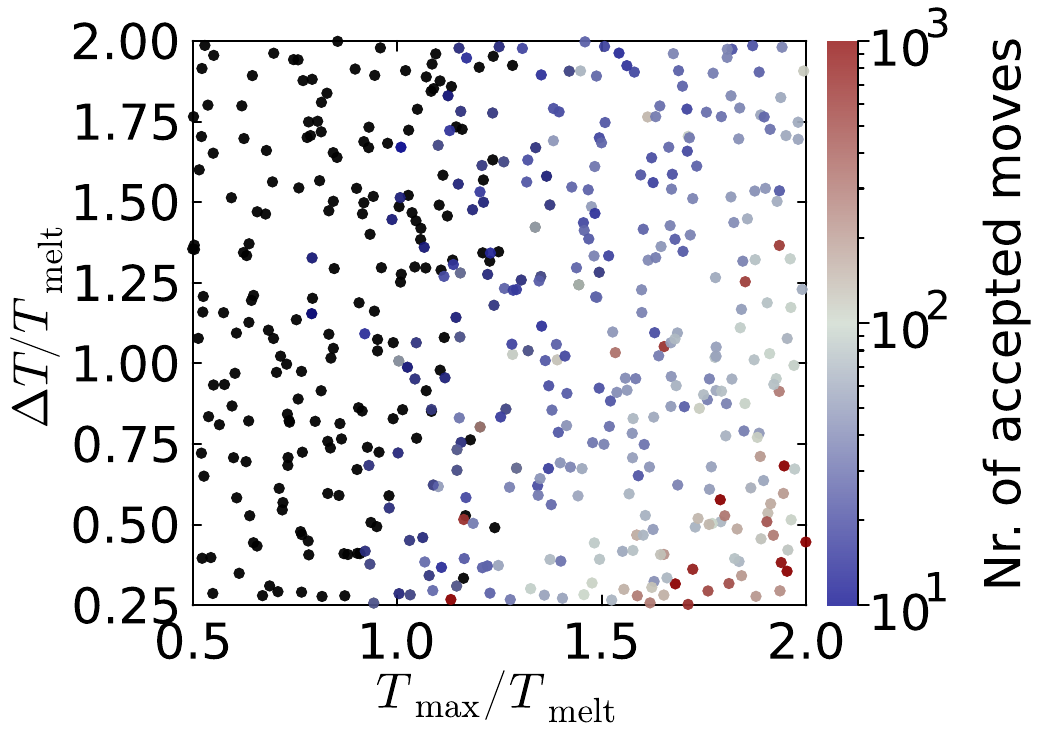}}
\end{minipage}
%
\begin{minipage}{0.48\textwidth}
\centering
\scaledinset{l}{0.01}{b}{.91}{\textbf{b}}{\includegraphics[width=\textwidth]{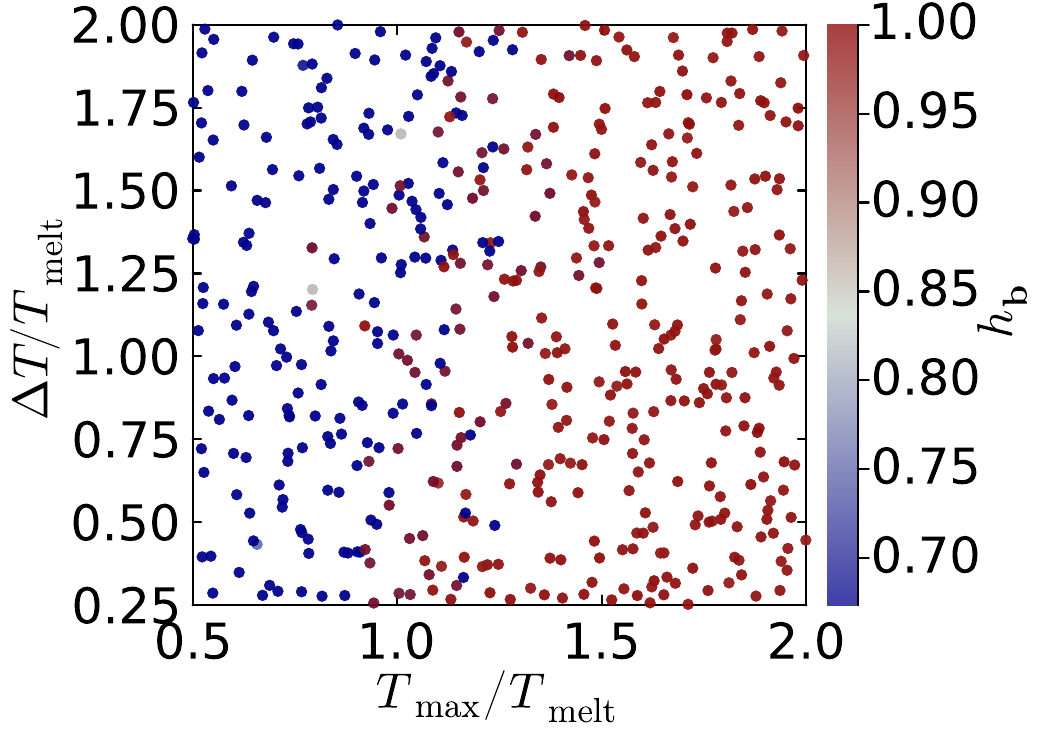}}
\end{minipage}
\caption{\label{fig:lcs_nr_accepted_moves_isotropy}
Melting transition for networks generated from the initial \textbf{lcs}. \add{The metrics of 1000 networks are displayed.}
\textbf{a} Number of accepted Monte Carlo moves plotted against $T_\mathrm{max}$ and $\Delta T$. The black markers correspond to networks with ten or fewer accepted moves.
\textbf{b} Bond orientation entropy $h_\mathbf{b}$.
}
\end{figure*}

\begin{figure}[h]
\includegraphics[width=0.48\linewidth]{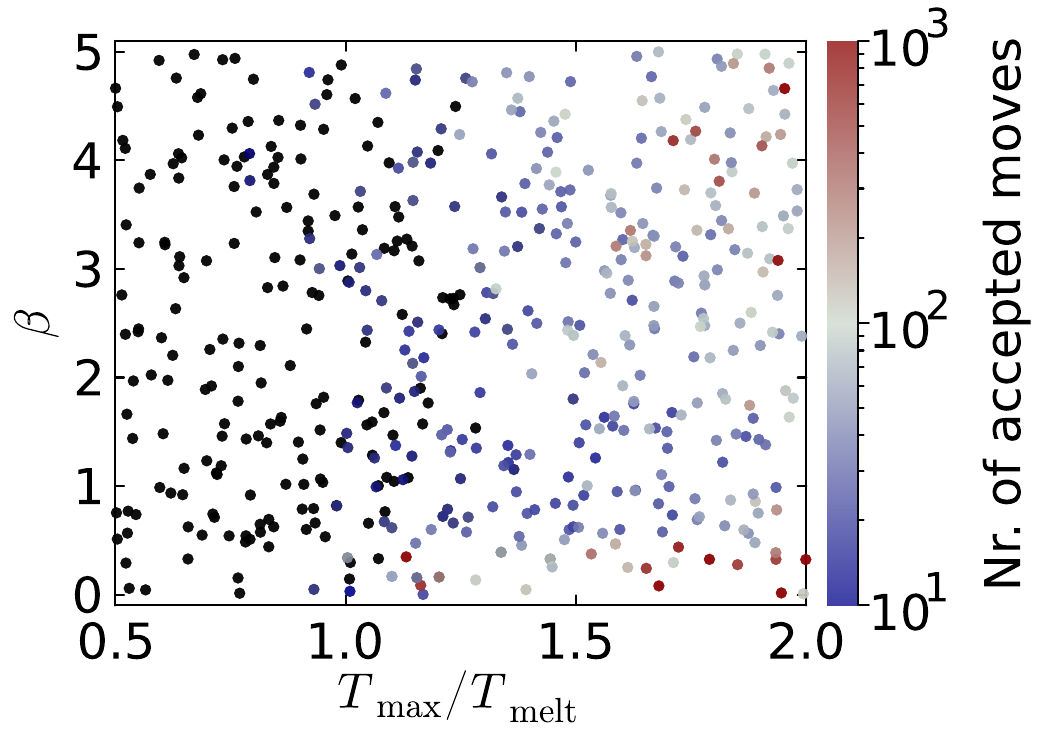}
\caption{\label{fig:lcs_nr_accepted_moves_beta} Number of accepted Monte Carlo moves for the initial \textbf{lcs} plotted against $T_\mathrm{max}$ and $\Delta T$. The black markers correspond to networks with ten or fewer accepted moves. \add{The metrics of 1000 networks are displayed.}}
\end{figure}

\begin{figure*}[h]
\begin{minipage}{0.48\textwidth}
\centering
\scaledinset{l}{0.01}{b}{.91}{\textbf{a}}{\includegraphics[width=\textwidth]{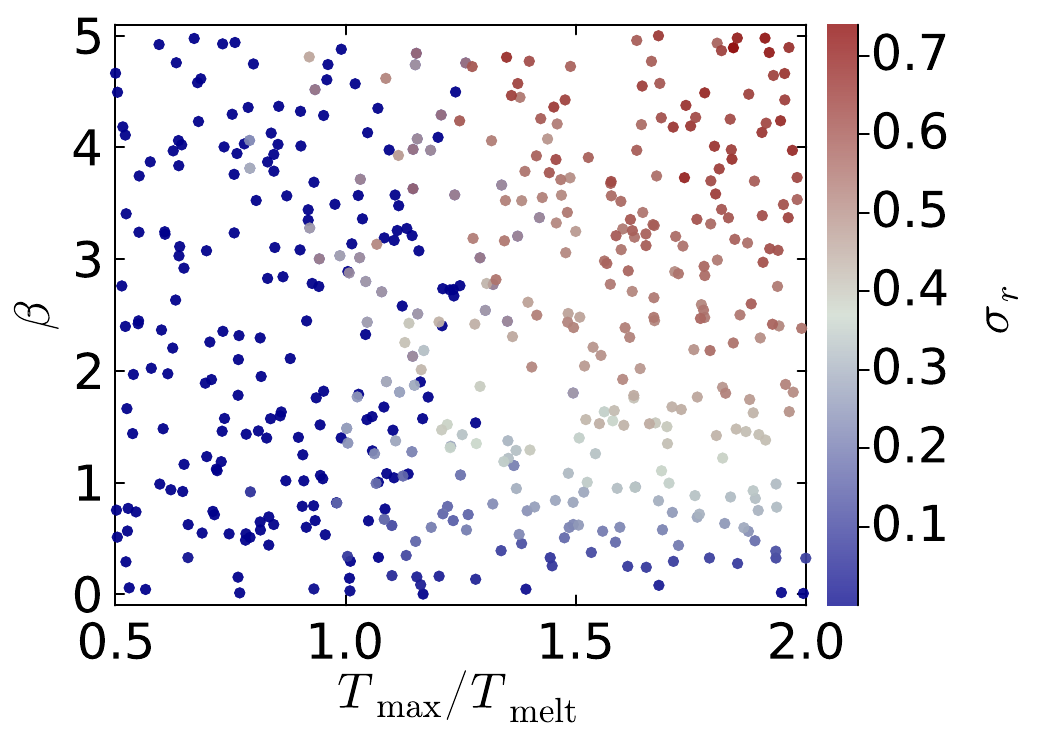}}
\end{minipage}
%
\begin{minipage}{0.48\textwidth}
\centering
\scaledinset{l}{0.01}{b}{.91}{\textbf{b}}{\includegraphics[width=\textwidth]{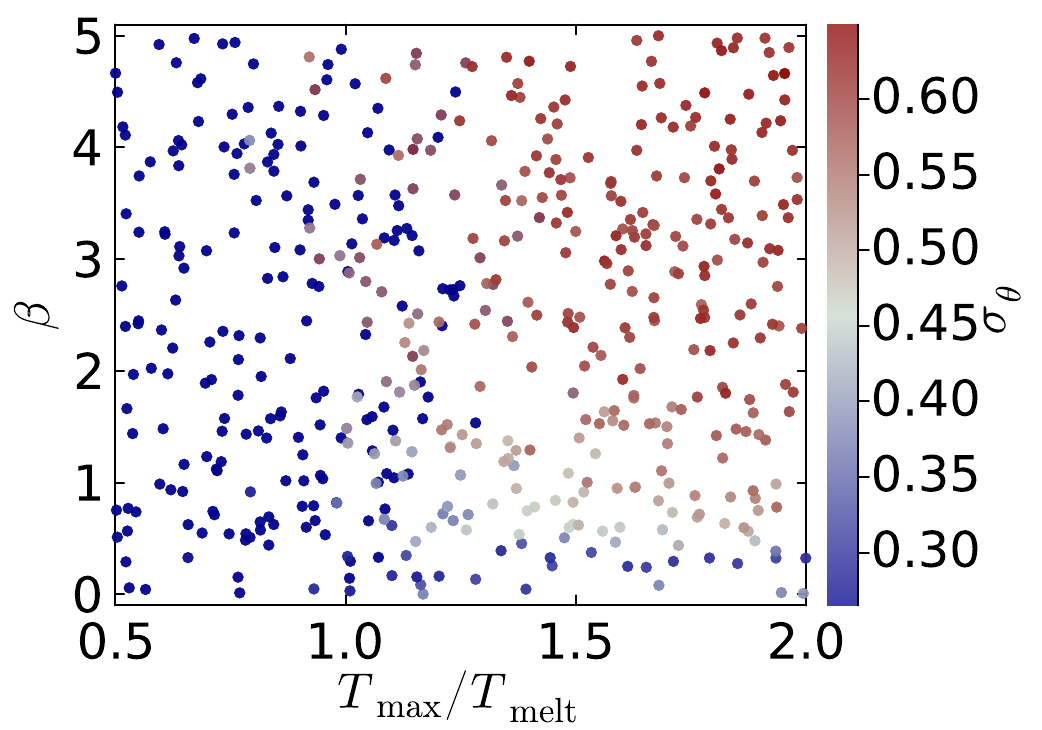}}
\end{minipage}
\caption{\label{fig:lcs_bond_length_angle}
Network primitive metrics for networks generated from the initial \textbf{lcs} plotted against the algorithm inputs $T_\mathrm{max}$ and $\beta$. Red colors indicate high disorder with respect to the corresponding metric. \add{The metrics of 1000 networks are displayed.}
\textbf{a} Bond length standard deviation $\sigma_r$. 
\textbf{b} Bond angle standard deviation $\sigma_\theta$.
}
\end{figure*}

\begin{figure*}[h]
\begin{minipage}{0.48\textwidth}
\centering
\scaledinset{l}{0.01}{b}{.91}{\textbf{a}}{\includegraphics[width=\textwidth]{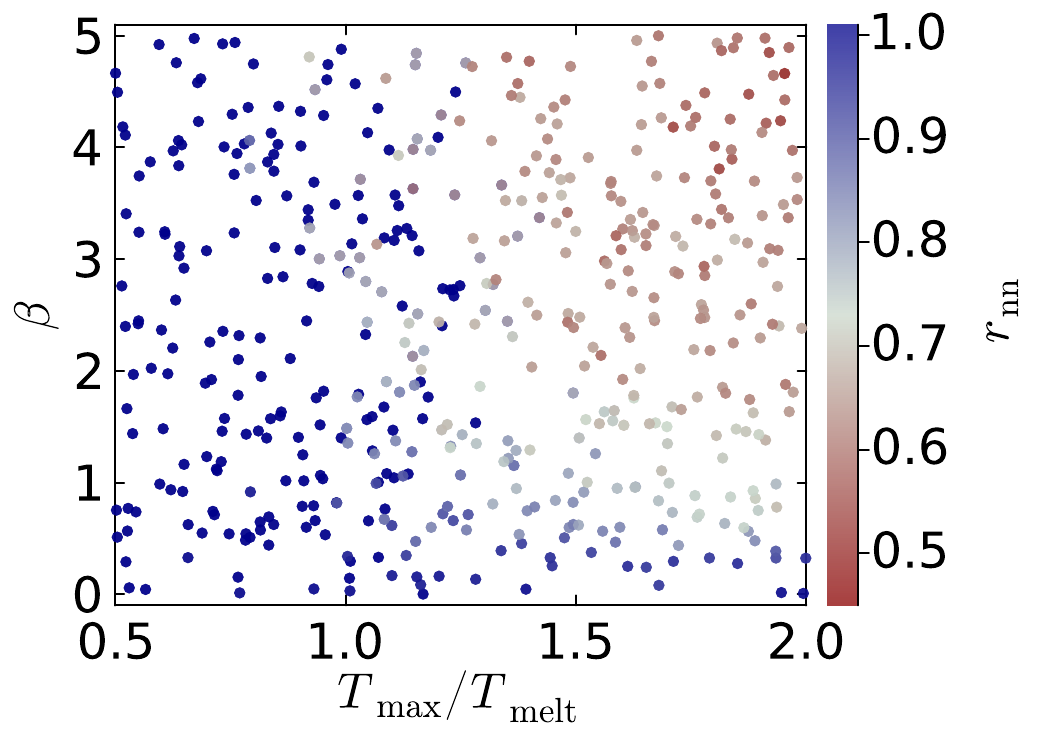}}
\end{minipage}
%
\begin{minipage}{0.48\textwidth}
\centering
\scaledinset{l}{0.01}{b}{.91}{\textbf{b}}{\includegraphics[width=\textwidth]{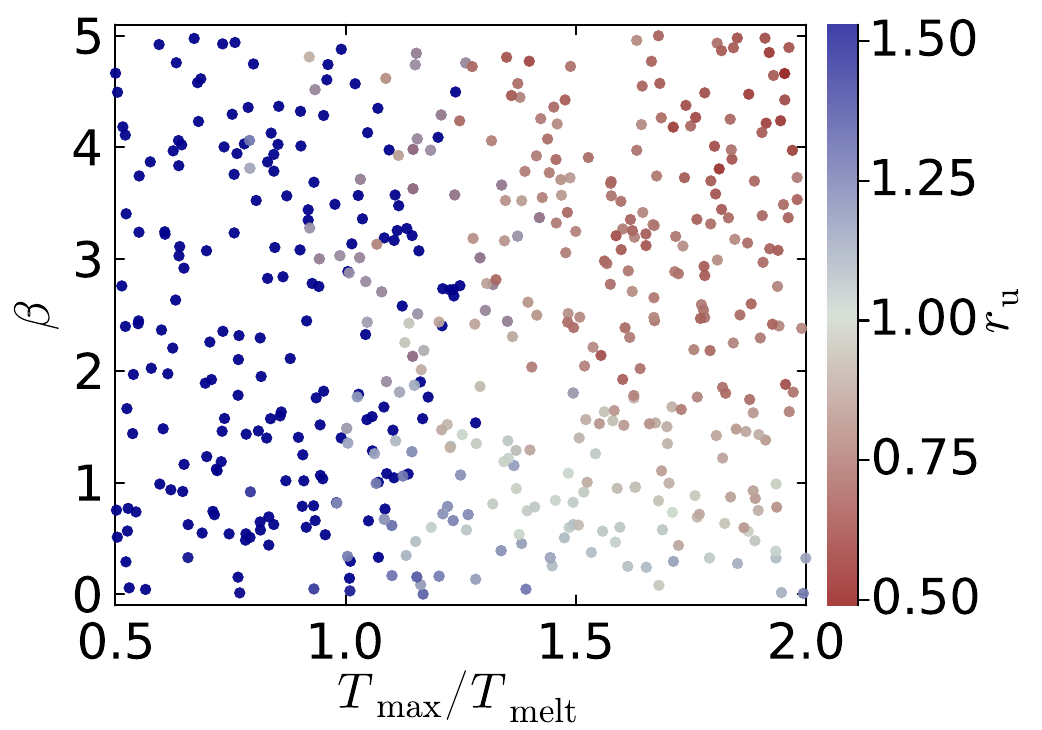}}
\end{minipage}
\begin{minipage}{0.48\textwidth}
\centering
\scaledinset{l}{0.01}{b}{.91}{\textbf{c}}{\includegraphics[width=\textwidth]{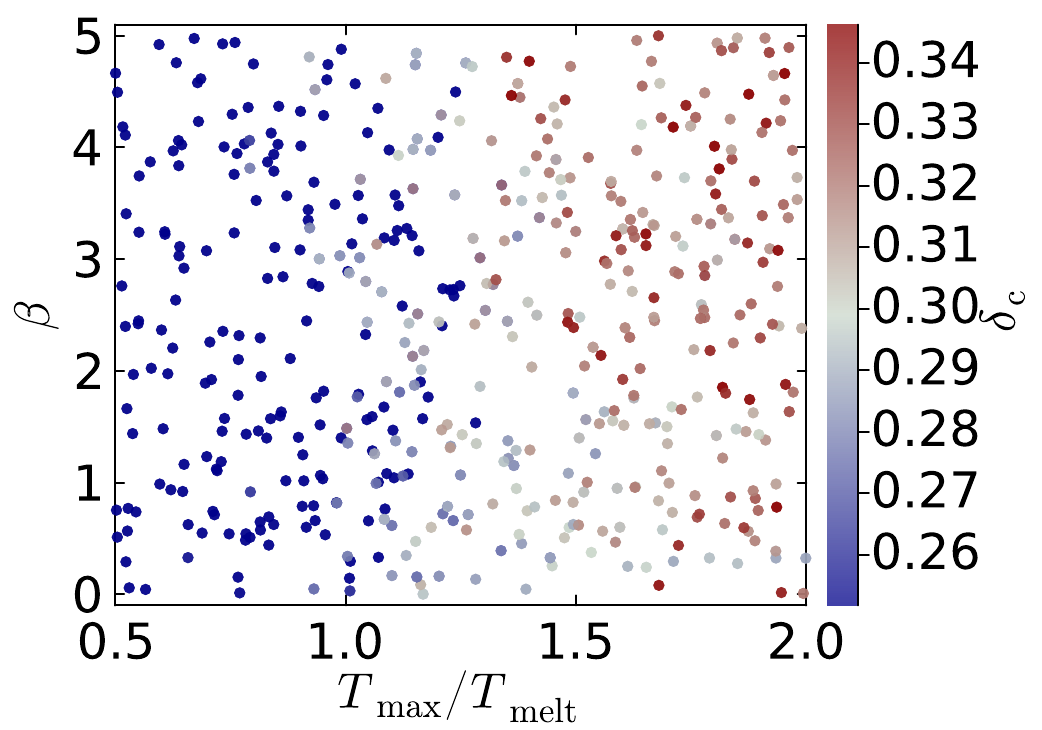}}
\end{minipage}
%
\begin{minipage}{0.48\textwidth}
\centering
\scaledinset{l}{0.01}{b}{.91}{\textbf{d}}{\includegraphics[width=\textwidth]{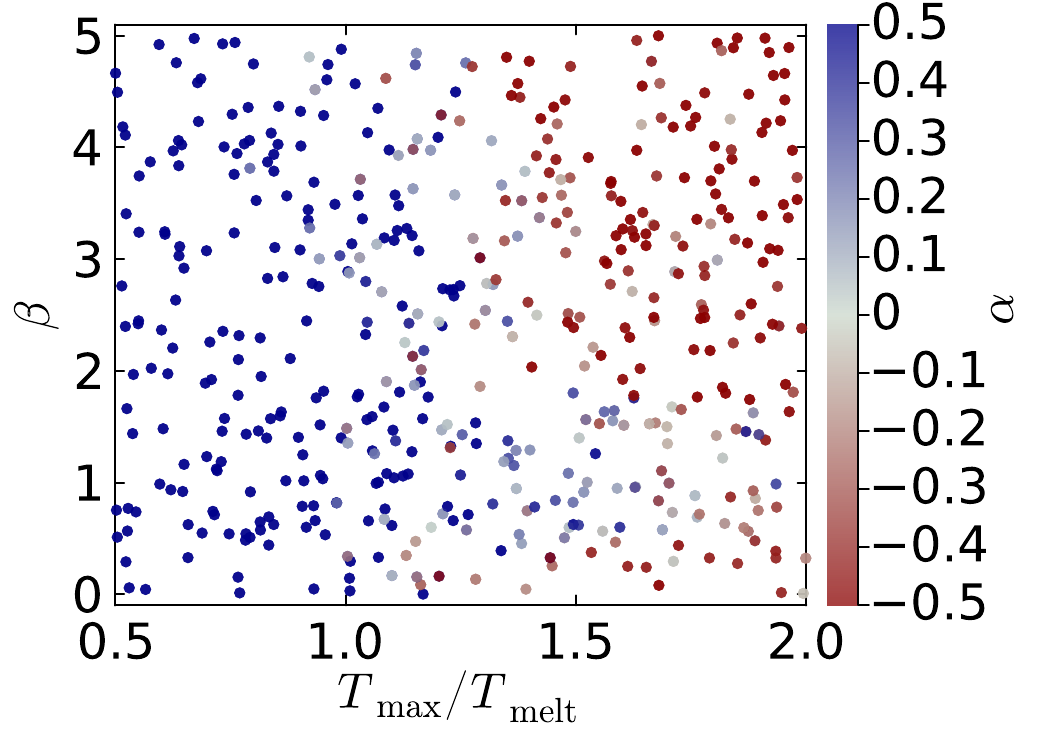}}
\end{minipage}
\caption{\label{fig:lcs_homogeneity_hyperuniformity}
Homogeneity metrics for networks generated from the initial \textbf{lcs} plotted against the algorithm inputs $T_\mathrm{max}$ and $\beta$. Red colors indicate high disorder with respect to the corresponding metric. \add{The metrics of 1000 networks are displayed.}
\textbf{a} \replace{Coordinated neighbor distance $r_\mathrm{c}$}{Nearest-neighbor distance $r_\mathrm{nn}$}.
\textbf{b} \replace{Uncoordinated neighbor distance}{Nearest-uncoordinated-neighbor distance} $r_\mathrm{u}$.
\textbf{c} Critical pore radius $\delta_\mathrm{c}$.
\textbf{d} Hyperuniformity metric $\alpha$.
}
\end{figure*}

\begin{figure*}[h]
\begin{minipage}{0.48\textwidth}
\centering
\scaledinset{l}{0.01}{b}{.91}{\textbf{a}}{\includegraphics[width=\textwidth]{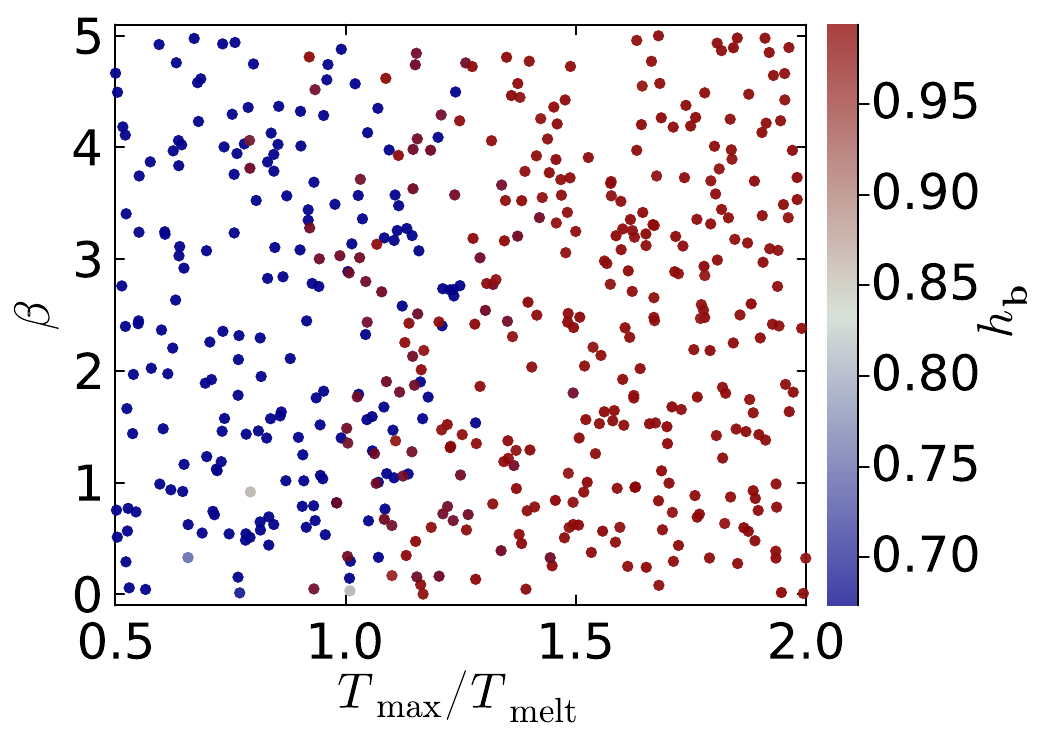}}
\end{minipage}
%
\begin{minipage}{0.48\textwidth}
\centering
\scaledinset{l}{0.01}{b}{.91}{\textbf{b}}{\includegraphics[width=\textwidth]{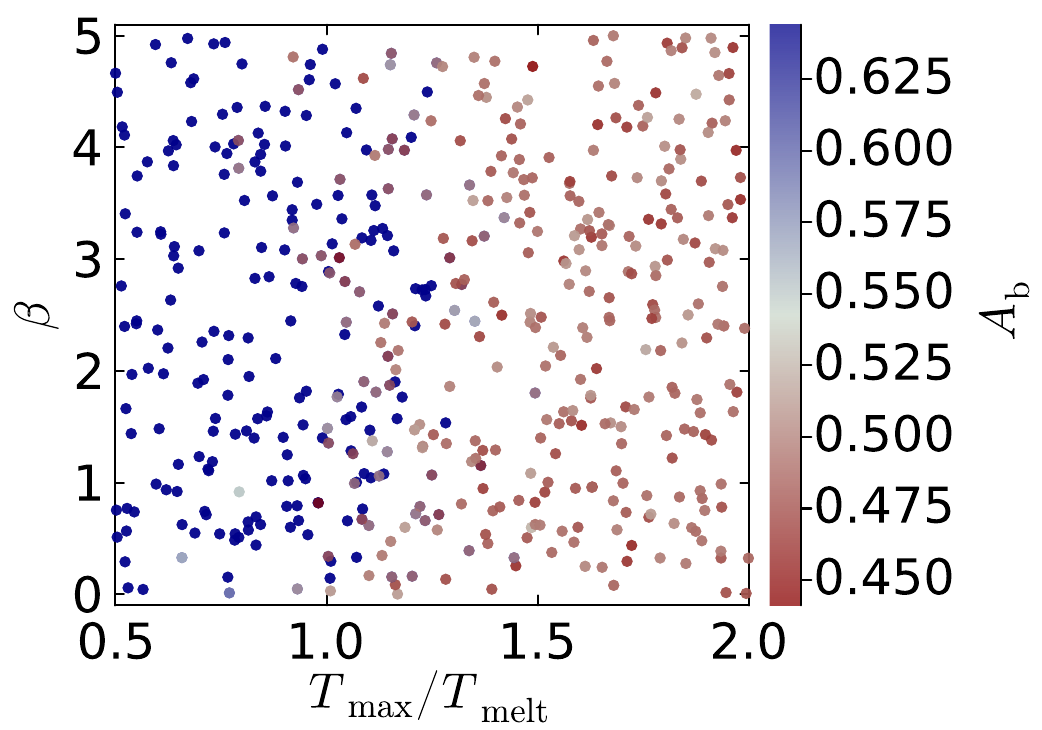}}
\end{minipage}
\begin{minipage}{0.48\textwidth}
\centering
\scaledinset{l}{0.01}{b}{.91}{\textbf{c}}{\includegraphics[width=\textwidth]{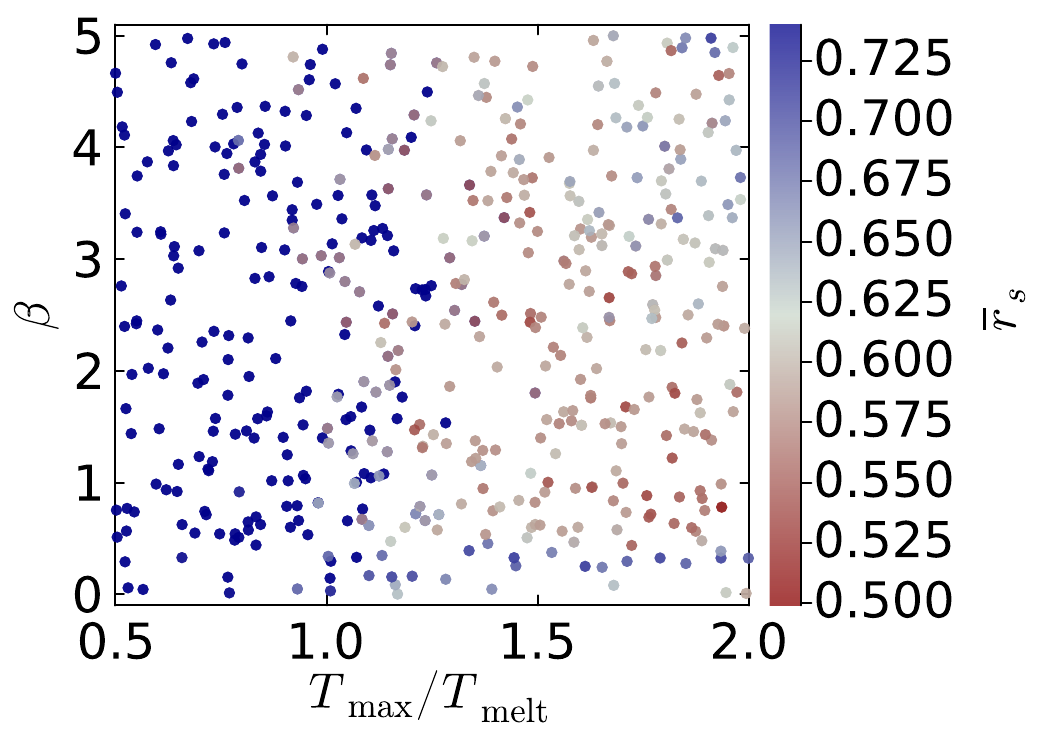}}
\end{minipage}
%
\begin{minipage}{0.48\textwidth}
\centering
\scaledinset{l}{0.01}{b}{.91}{\textbf{d}}{\includegraphics[width=\textwidth]{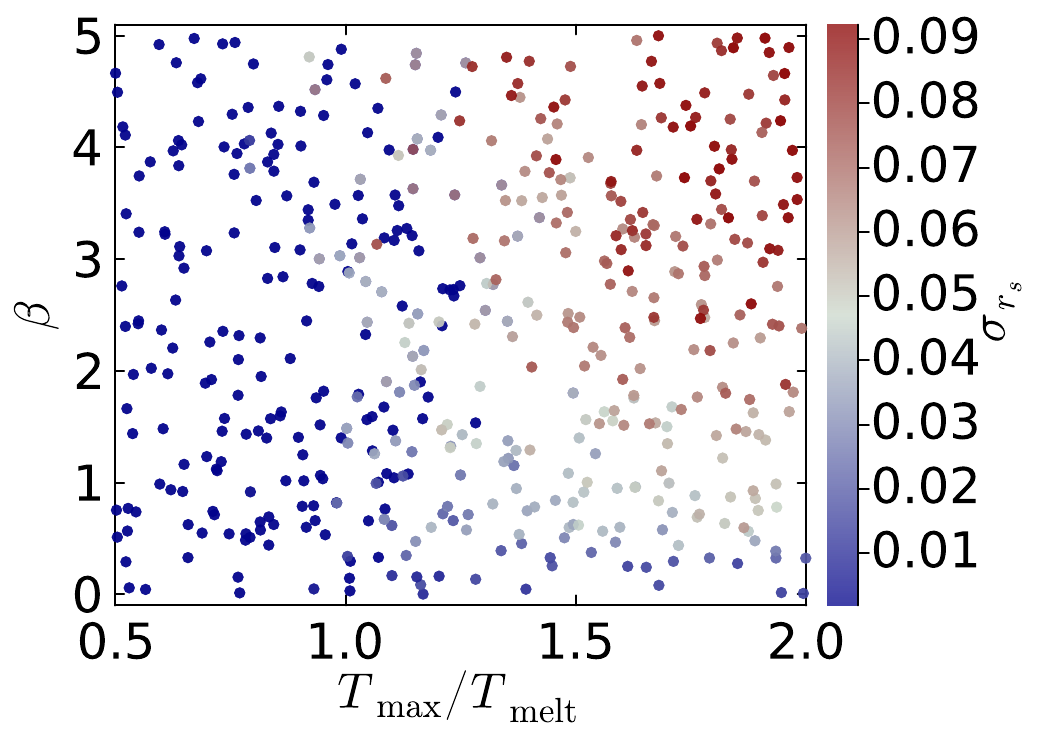}}
\end{minipage}
\caption{\label{fig:lcs_isotropy_topology}
Isotropy and topology metrics for networks generated from the initial \textbf{lcs} as a function of the network generation algorithm inputs $T_\mathrm{max}$ and $\beta$. Red colors indicate high disorder with respect to the corresponding metric. \add{The metrics of 1000 networks are displayed.}
\textbf{a} Bond orientation entropy $h_\mathbf{b}$.
\textbf{b} \add{Bond} structure factor anisotropy metric $A_\mathrm{b}$.
\textbf{c} Mean ring radius $\overline{r}_s$.
\textbf{d} Ring radius standard deviation $\sigma_{r_s}$.
}
\end{figure*}

\clearpage

\section{Principal component analysis of generated order metrics}
\label{sec:si_pca_order_metrics}
Here, we provide more information on applying principal component analysis to order metric data from networks generated from the \textbf{ctn} network (Section~\ref{sec:neural_network} in the main text).

\begin{figure}[h]
\centering
\includegraphics[width=0.48 \linewidth]{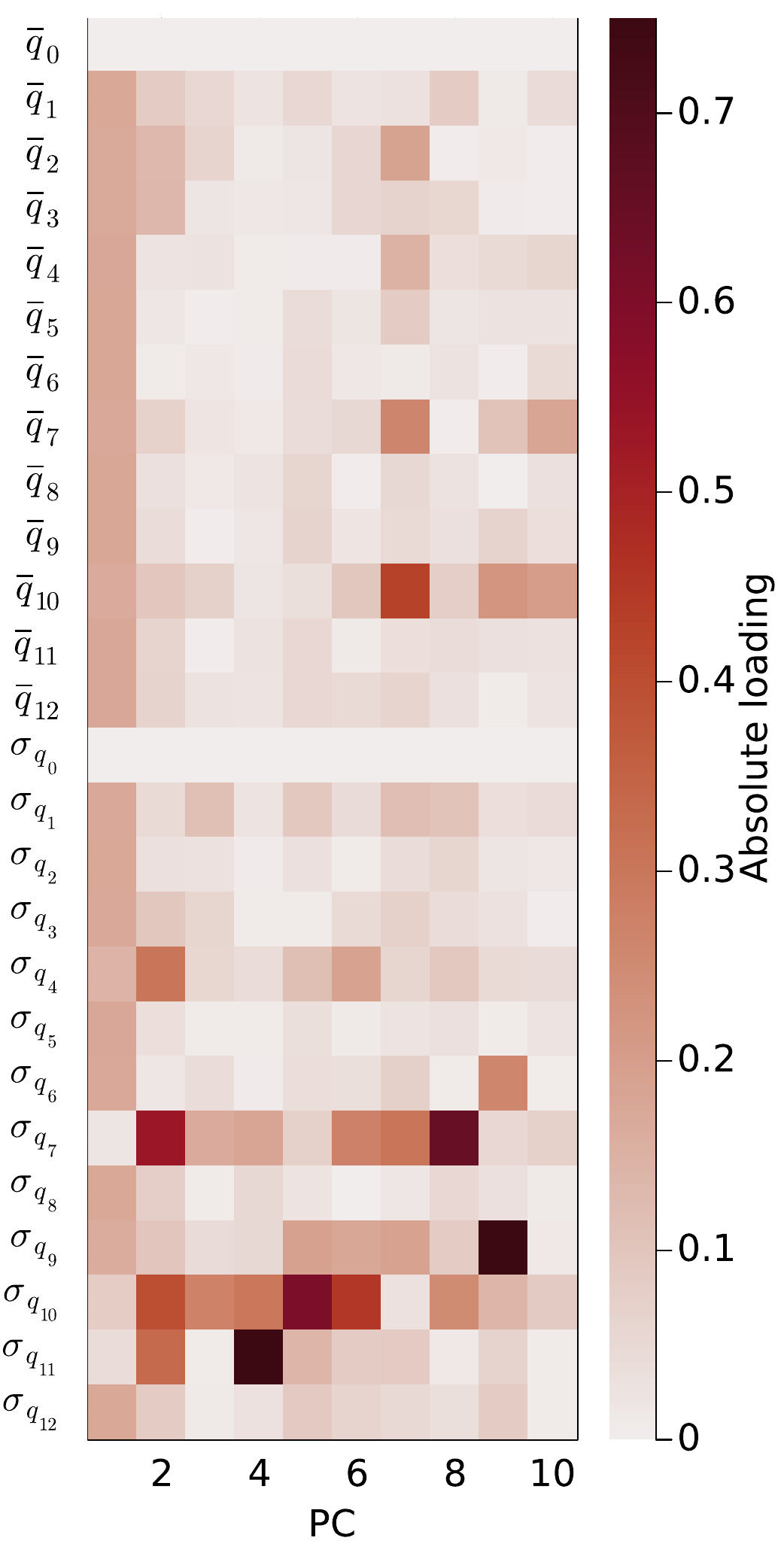}
\caption{\label{fig:pca_info_steinhardt} Absolute values of the PCA loadings with the mean Steinhardt local bond order parameters $\overline{q}_l$ and their standard deviations $\sigma_{q_l}$. This illustration shows the contributions of the combined loadings $\mathbf{\overline{q}}$ and $\mathbf{\sigma_{q}}$ in Figure~\ref{fig:pca_info}b.}
\end{figure}

\end{document}